\documentstyle[natbib,psfig]{article}

\psfigurepath{}
\psfull
\pssilent

\newcommand{\msun}{M_{\odot}}
\newcommand{\yr}{{\rm\,yr}}

\newcommand{\eg}{{\it e.g.\ }}
\newcommand{\ie}{{\it  i.e.\ }}

\newcommand{\etc}{{\it etc.\ }}

\newcommand{\au}{{\rm\,AU}}
\newcommand{\aui}{\hbox{\au$^{-1}$}}
\newcommand{\mten}{\! \! \! \! \! \! \! \! \! \!}
\newcommand{\sss}{\scriptscriptstyle}

\newcommand{\degree}{^{\circ}}

\newcommand{\qe}{q_{\sss E}}

\newcommand{\og}{\tilde{\omega}}
\newcommand{\Og}{\tilde{\Omega}}
\newcommand{\ig}{\tilde{\imath}}
\newcommand{\xg}{\tilde{x}}
\newcommand{\yg}{\tilde{y}}
\newcommand{\zg}{\tilde{z}}
\def\aud2{\au~day$^{-2}$}
\def\h2o{\mbox{$\rm H_{2}O$}}
\def\co2{\mbox{$\rm CO_{2}$}}
\def\1a{\mbox{$1/a$}}
\def\las{\mathrel{\hbox{\rlap{\hbox{\lower3pt\hbox{$\sim$}}}\hbox{\raise2pt\hbox{$<$}}}}}
\def\gas{\mathrel{\hbox{\rlap{\hbox{\lower3pt\hbox{$\sim$}}}\hbox{\raise2pt\hbox{$>$}}}}}
\def\lta{\las}
\def\gta{\gas}
\newcommand\half{{\textstyle{1\over2}}}

\begin{document}
\thispagestyle{empty}
\begin{center}

{\bf The Evolution of Long-Period Comets}\\ \bigskip
Paul Wiegert\footnote{Current address: Department of Physics and Astronomy, York University, North York, Ontario, M3J 1P3 Canada. Email: wiegert@aries.phys.yorku.ca}\\
Department of Astronomy\\
University of Toronto\\
Toronto, Ontario\\
M5S 3H8 Canada\\
\bigskip and\\ \bigskip
Scott Tremaine\\
Canadian Institute for Theoretical Astrophysics\\
and Canadian Institute for Advanced Research\\
Program in Cosmology and Gravity\\
University of Toronto\\
Toronto, Ontario\\
M5S 3H8 Canada\\
Email: tremaine@cita.utoronto.ca

\vfill

\noindent submitted to {\sl Icarus}, May 1997.
\vspace{2in}

\end{center}

\newpage

\centerline{\bf Abstract} \bigskip

We study the evolution of long-period comets by numerical integration
of their orbits, a more realistic approach than the Monte Carlo and
analytic methods previously used to study this problem. We follow the
comets from their origin in the Oort cloud until their final escape or
destruction, in a model solar system consisting of the Sun, the four
giant planets and the Galactic tide. We also examine the effects of
non-gravitational forces and the gravitational forces from a
hypothetical solar companion or
circumsolar disk.  We confirm the conclusion of Oort and other
investigators that the observed distribution of long-period comet
orbits does not match the expected steady-state distribution unless
there is fading or some similar process that depletes the population
of older comets. We investigate several simple fading laws. We can
match the observed orbit distribution if the fraction of comets
remaining observable after $m$ apparitions is $\propto m^{-0.6 \pm
0.1}$ (close to the fading law originally proposed by Whipple 1962);
or if approximately 95\% of comets live for only a few ($\sim 6$)
returns and the remainder last indefinitely.  Our results also yield
statistics such as the expected perihelion distribution, distribution
of aphelion directions, frequency of encounters with the giant
planets and the rate of production of Halley-type comets.

\section{Introduction}

Comets can be classified on the basis of their orbital period $P$ into
long-period (LP) comets with $P>200\yr$, and short-period (SP) comets
with $P<200\yr$; short-period comets are further subdivided into
Halley-type comets with $20\yr<P<200\yr$ and Jupiter-family comets
with $P<20\yr$ \cite[]{carval92}. The boundary between SP and LP
comets corresponds to a semimajor axis $a=(200)^{2/3}\au=34.2\au$,
which is useful because: (i) it
distinguishes between comets whose aphelia lie within or close to the
planetary system, and those that venture beyond; (ii) an orbital
period of $200\yr$ corresponds roughly to the length of time over
which routine telescopic observations have been taken---the sample of
comets with longer periods is much less complete; (iii) the planetary
perturbations suffered by comets with periods longer than $200\yr$ are
uncorrelated on successive perihelion passages (see footnote 2 below).

LP comets are believed to come from the Oort cloud \cite[]{oor50}, a
roughly spherical distribution of some $10^{12}$ comets with semimajor
axes between $10^{3.5}$ and $10^{4.5}\au$. The Oort cloud is probably
formed from planetesimals ejected from the outer Solar System by
planetary perturbations. LP comets---and perhaps some or all
Halley-family comets---are Oort-cloud comets that have evolved into
more tightly bound orbits under the influence of planetary and other
perturbations. Jupiter-family comets probably come from a quite
different source, the Kuiper belt outside Neptune, and will not be
discussed in this paper.

The observed distribution of the $\sim 700$ known LP comets is
determined mainly by celestial mechanics, although physical
evolution of the comets (e.g. fading or disruption during perihelion
passage near the Sun) and observational selection effects (comets with
large perihelion distance are undetectable) also play major roles. The
aim of this paper is to construct models of the orbital evolution of
LP comets and to compare these models to the observed distribution of
orbital elements.

This problem was first examined by \cite{oor50}, who focused on the
distribution of energy or inverse semimajor axis. He found that he could match
the observed energy distribution satisfactorily, with two caveats: (i) he had
to assume an {\it ad hoc} disruption probability $k=0.014$ per perihelion
passage; (ii) five times too many comets were present in a spike (the
``Oort spike'') near zero energy. Since most of the comets in the
Oort spike are on their first passage close to the Sun, he argued that they
may contain volatile ices (\eg CO, CO$_2$) that create a large bright coma for
the new comet, but are substantially or completely depleted in the
process. When the comet subsequently returns (assuming it has avoided ejection
and other loss mechanisms), it will be much fainter and may escape
detection. Most of the decrease in brightness would occur during the first
perihelion passage, and the brightness would level off as the most volatile
components of the comet's inventory are lost.  This ``fading hypothesis'' has
played a central role in all subsequent attempts to compare the observed and
predicted energy distributions of LP comets.

In \S~\ref{pa:observ} we examine the observed distribution of LP comet
orbits. The basic theoretical model of LP comet evolution is reviewed in
\S~\ref{pa:dynamics}. The simulation algorithm is described
in \S~\ref{pa:algorithm}, and the results are presented in
\S~\ref{pa:results}. The simulations and results 
are described in more detail by \cite{wie96}.

\section{Observations} \label{pa:observ}

The 1993 edition of the {\it Catalogue of Cometary Orbits} \cite[]{marwil93}
lists 1392 apparitions of 855 individual comets, of which 681 are LP
comets. The catalog includes, where possible, the comet's osculating orbital
elements at or near perihelion. When studying  LP comets it is often 
simpler to work with the elements of the orbit on which the comet
approached the planetary system (the ``original'' elements). These can be
calculated from the orbit determined near perihelion by integrating the
comet's trajectory backwards until it is outside the planetary
system. Original elements are normally quoted in the frame of the Solar System barycenter. Marsden and Williams list 289 LP comets that
have been observed well enough (quality classes 1 and 2) that reliable
original elements can be computed.

The difference between the original elements and the elements near perihelion
is generally small for the inclination $i$\footnote{Angular 
elements without subscripts are measured relative to the ecliptic. We shall
also use elements measured relative to the Galactic plane, which we denote by
a tilde \ie $\ig$, $\Og$ and $\og$.}, perihelion distance $q$, argument of
perihelion $\omega$ and longitude of the ascending node $\Omega$; when
examining the distribution of these elements we will therefore work with the
entire sample ($N=681$) of LP comets. The original semimajor axis
and eccentricity are generally quite different from the values of these
elements near perihelion, so when we examine these elements we shall use only
the smaller sample ($N=289$) for which original elements are available.

\subsection{Semimajor axis}

The energy per unit mass of a small body orbiting a point mass $M_\odot$ is
$-\half G\msun/a$, where $a$ is the semimajor axis. This is not precisely the
energy per unit mass of a comet orbiting the Sun---the expression neglects
contributions from the planets and the Galaxy---but provides a useful measure
of a comet's binding energy. For simplicity, we often use the inverse
semimajor axis $x\equiv 1/a$ as a measure of orbital energy. The boundary
between SP and LP comets is at $x = (200$~yr$)^{-2/3}= 0.029$~{\aui}.

\begin{figure}[p]
\centerline{\hbox{\psfig{figure=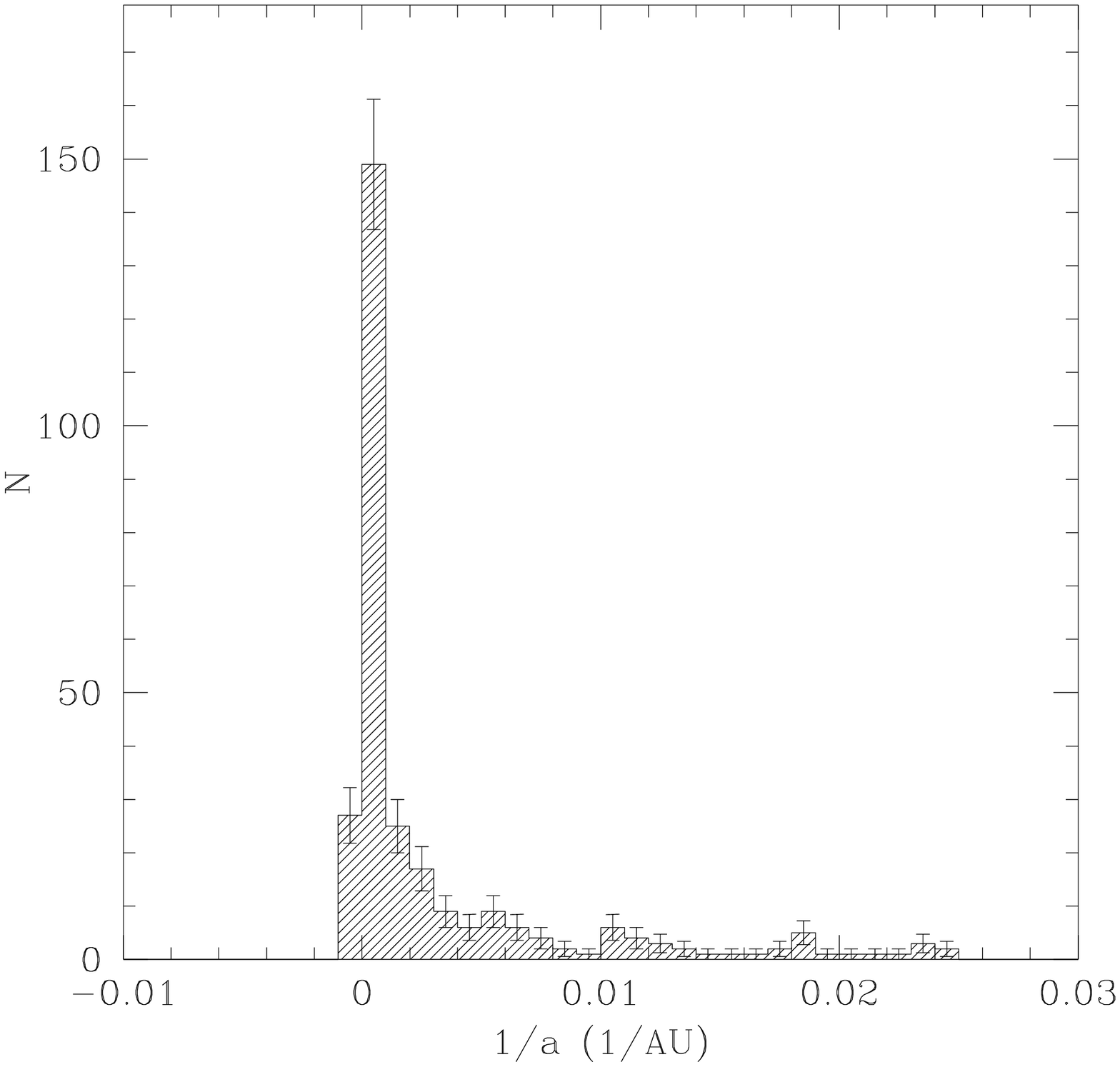,height=3in}
                  \psfig{figure=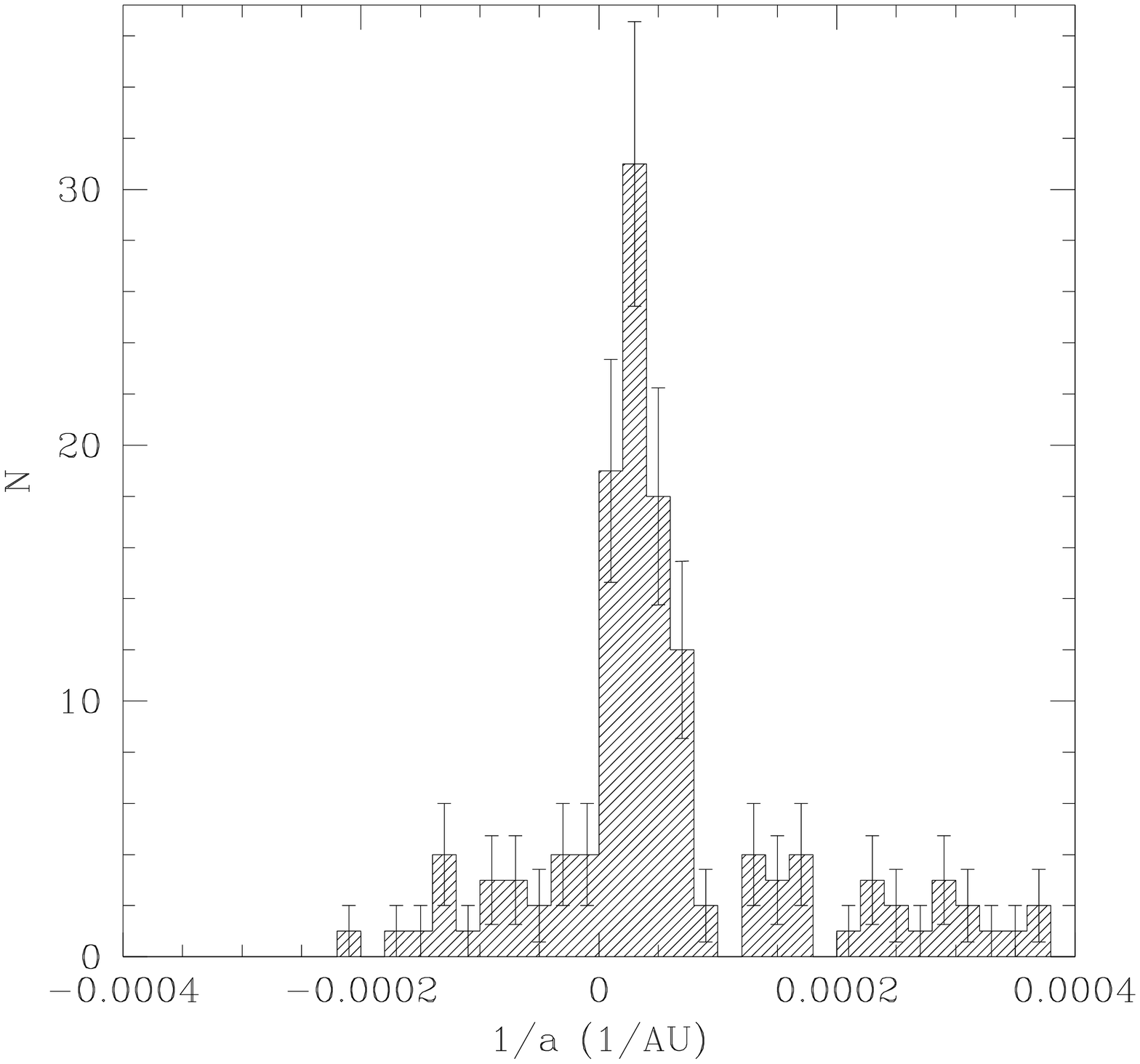,height=3in}}}
\centerline{\hspace*{1.45in}$(a)$\hfill$(b)$\hspace{1.3in}}
\centerline{\hbox{\psfig{figure=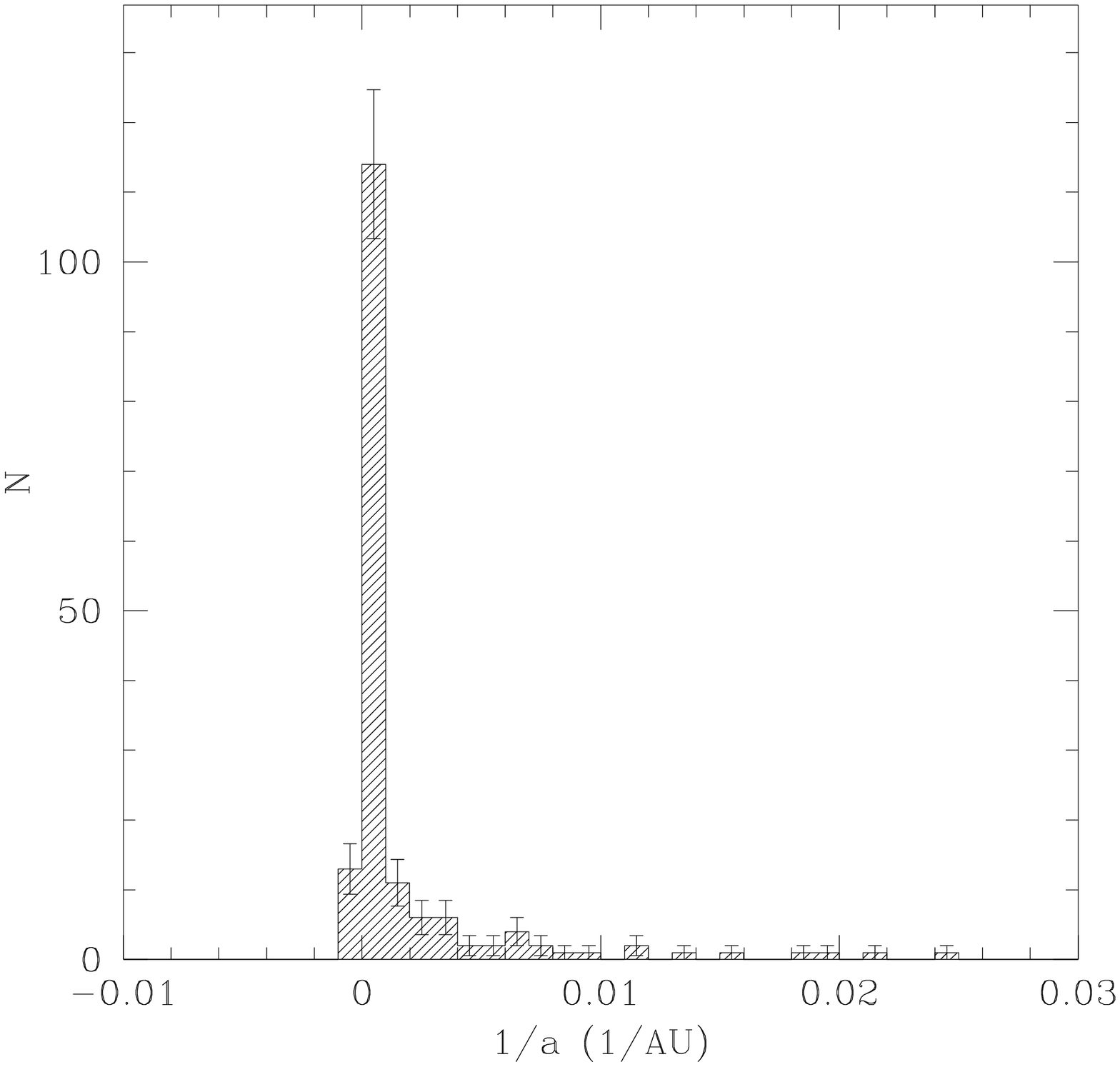,height=3in}
                  \psfig{figure=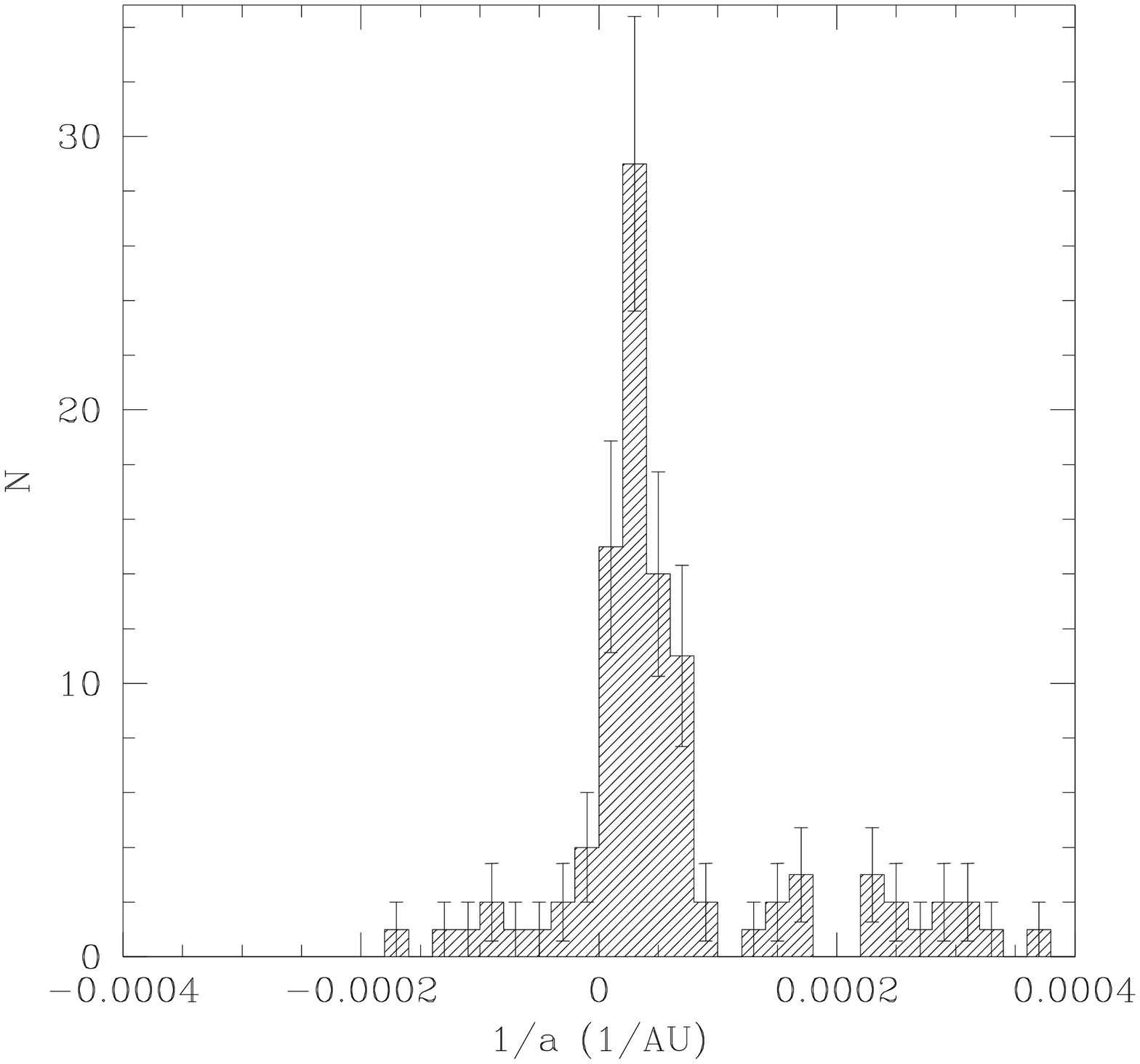,height=3in}}}
\centerline{\hspace*{1.45in}$(c)$\hfill$(d)$\hspace{1.3in}}
\caption{Distribution of original inverse semimajor axes of 289 LP
comets at two different magnifications (panels a,b) and for the
170 LP comets with the most accurate (Class 1)
orbits (panels c,d). Data taken from \protect\cite{marwil93}. There is
no obvious difference between the top and bottom panels, suggesting
that observational errors in the inverse semimajor axes are
unimportant.}
\label{fi:energy}
\end{figure}

Figure~\ref{fi:energy} displays histograms of $x=1/a$ for the 289 LP comets
with known original orbits, at two different magnifications.  The error bars
on this and all other histograms are $\pm 1$ standard deviation ($\sigma$)
assuming Poisson statistics ($\sigma = N^{1/2}$), unless stated otherwise.

The sharp spike in the distribution for $x\lta10^{-4}\au^{-1}$ (the ``Oort
spike'') was interpreted by \cite{oor50} as evidence for a population of
comets orbiting the Sun at large ($a \gta 10~000$~\au) distances, a population
which has come to be known as the Oort cloud. Comets in the spike are mostly
``new'' comets, on their first passage into the inner planetary system from
the Oort cloud.

\begin{figure}[p]
\centerline{\hbox{\psfig{figure=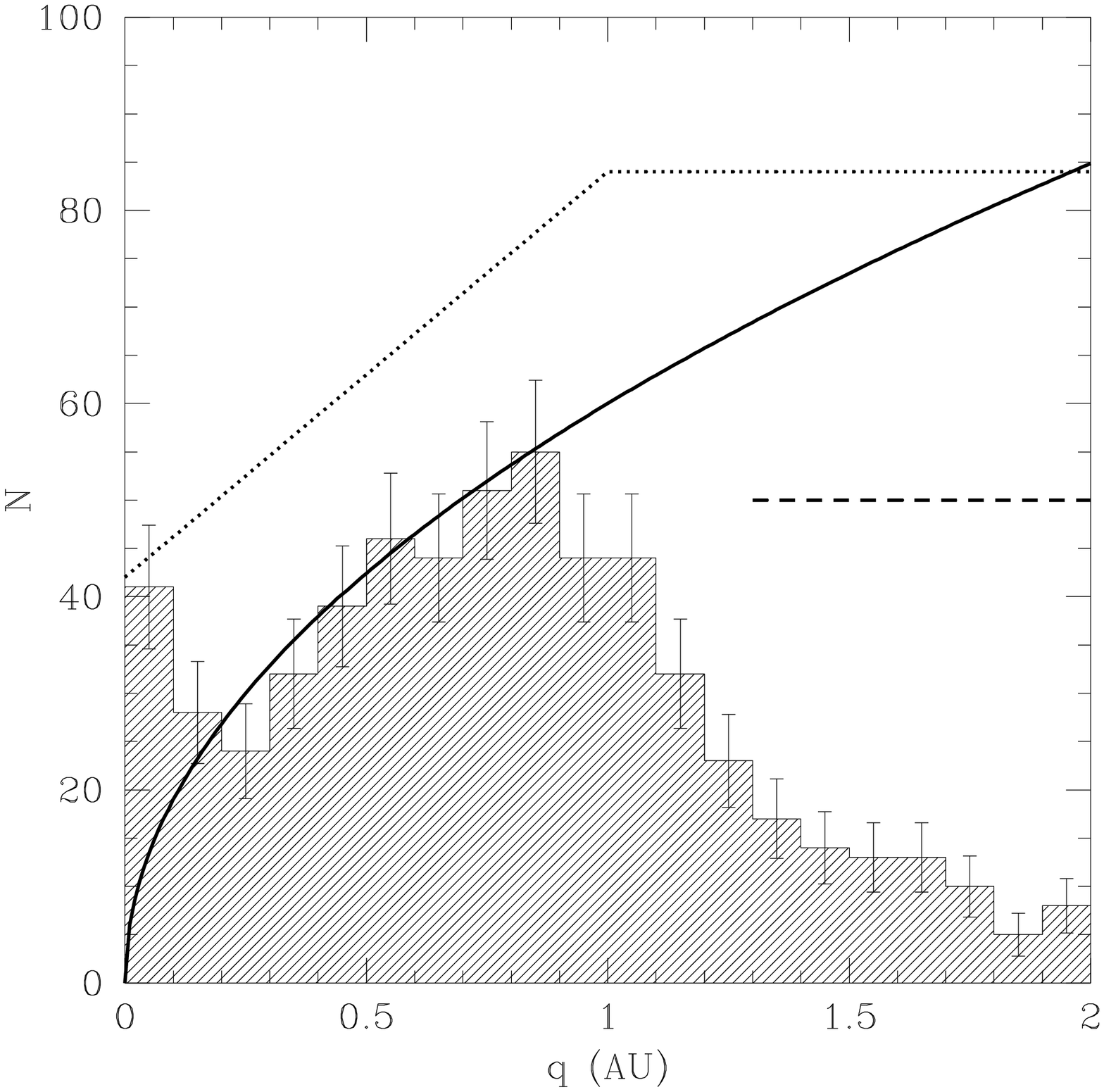,height=3in}
                  \psfig{figure=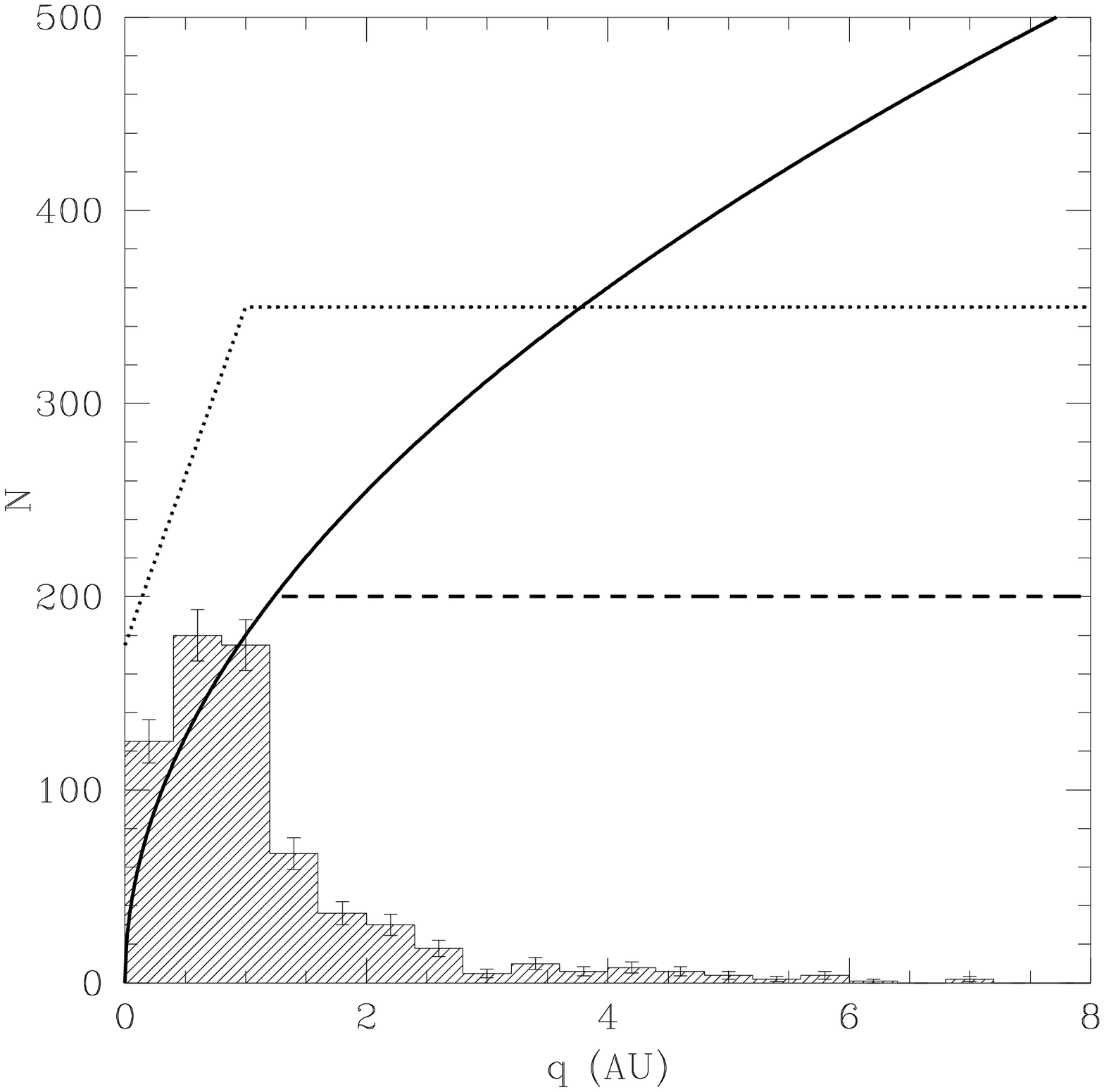,height=3in}}}
\centerline{\hspace*{1.45in}$(a)$\hfill$(b)$\hspace{1.3in}}
\caption{ Number $N$ versus perihelion distance $q$ for 681 LP comets,
on two different scales. Data taken from \protect\cite{marwil93}. The
solid line is the estimated intrinsic distribution from
\protect\cite{krepit78}, the dotted line is from
\protect\cite{eve67a}, and the dashed line is from
\protect\cite{showol82}. The
appropriate normalizations are difficult to determine for these
curves, and are chosen somewhat arbitrarily.}

\label{fi:lp_q}
\end{figure}

\subsection{Perihelion distance} \label{pa:perih}

Figure~\ref{fi:lp_q} shows the number of known LP comets versus
perihelion distance $q$.  The peak near 1~{\au} is due to observational
bias: comets appear brighter when nearer both the Sun and the
Earth. \label{pa:dndq} The intrinsic distribution $N(q)$ (defined so that
$N(q)dq$ is the number of detected {\it and} undetected LP comets with
perihelion in the interval $[q,q+dq]$) is difficult to
determine. \cite{eve67a} concluded that $N(q) \propto 0.4 + 0.6q$ for
$q<1\au$, and that for $q>1\au$, $N(q)$ is poorly constrained, probably lying
between a flat profile and one increasing linearly with $q$. \cite{krepit78}
also found the intrinsic distribution of $q$ to be largely indeterminate at $q
> 1$~\au, but preferred a model in which $N(q)\propto q^{1/2}$ over the range
$0 < q < 4$~{\au}. \cite{showol82} estimated $\int_0^q N(q)dq\propto 500q-175$
for $q>1.3\au$. 

These analyses also yield the completeness of the observed sample as a
function of $q$. Everhart estimates that only 20\% of observable
comets with $q<4${\au} are detected; the corresponding fraction in
Shoemaker and Wolfe is 28\%.  Kres\'{a}k and Pittich estimate that
60\% of comets with $q \le 1$~\au\ are detected, dropping to only 2\%
at $q = 4$~{\au}. Clearly the sample of LP comets is seriously
incomplete beyond $q=1$~{\au}, and the incompleteness is strongly
dependent on $q$. In comparing the data to our simulations we must
therefore impose a $q$-dependent selection function on our simulated
LP comets. We shall generally do this in the crudest possible way, by
declaring that our simulated comets are ``visible'' if and only if
$q<q_{\rm v}$, where $q_{\rm v}$ is taken to be 3~{\au}. This choice is
unrealistically large---probably $q_{\rm v}=1.5$~{\au} would be better---but
we find no evidence that other orbital elements are correlated with
perihelion distance in the simulations, and the larger cutoff improves
our statistics. We shall use the term ``apparition'' to denote a
perihelion passage with $q<q_{\rm v}$.

We have also explored a more elaborate model for selection effects
based on work by Everhart (1967a,b; \nocite{eve67a,eve67b} see
Wiegert 1996 
\nocite{wie96} for details). In this model the probability $p_{\rm v}$ that
an apparition is visible is given by
\begin{equation}
p_{\rm v}(q) = \left\{
\begin{array}{ll}
0 & \mbox{if $q > 2.5$~\au},\\
2.5 -(q/1~\au) & \mbox{if $1.5 \le q \le 2.5$~\au} \\
1  & \mbox{if $q <$ 1.5~\au} \\
\end{array}
\right. \label{eq:discoverprob2}
\end{equation}
The use of this visibility probability in our simulations makes very little
change in the distribution of orbital elements (except, of course, for the
distribution of perihelion distance). For the sake of brevity we shall mostly
discuss simulations using the simpler visibility criterion $q<q_{\rm v}=3\au$. 

\subsection{Inclination}

\begin{figure}[p]
\centerline{\hbox{\psfig{figure=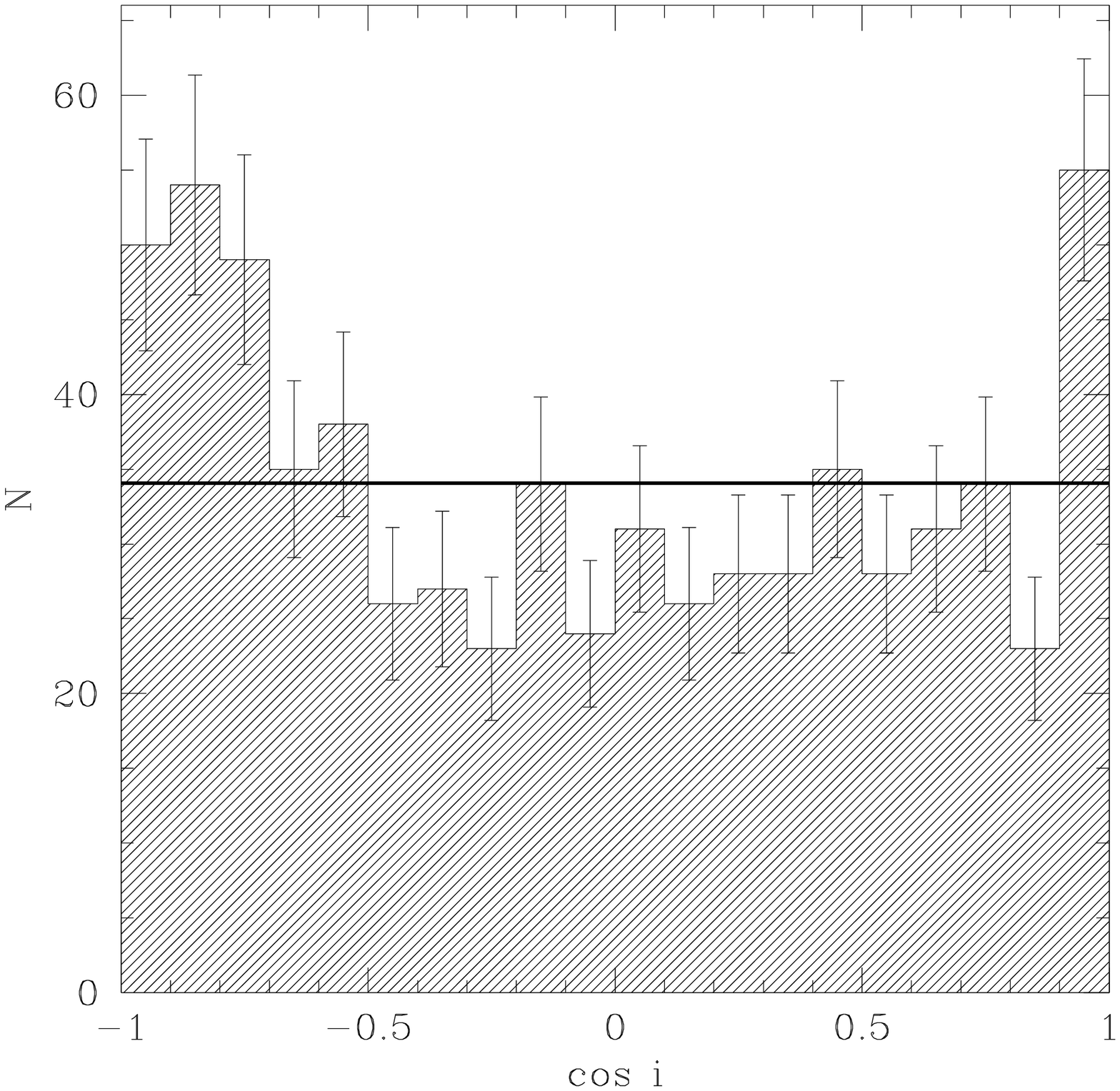,height=3in}
                  \psfig{figure=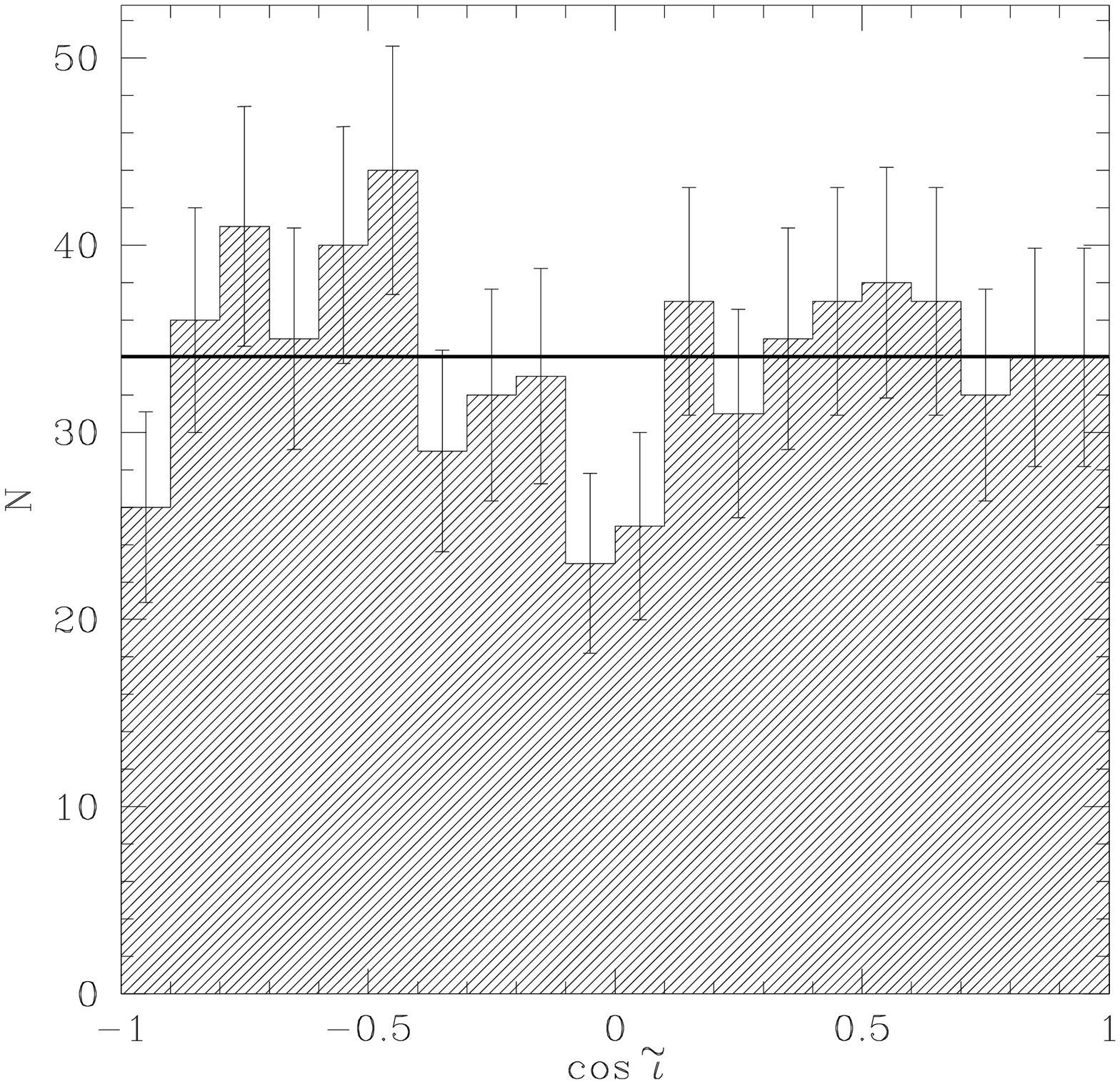,height=3in}}}
\centerline{\hspace*{1.45in}$(a)$\hfill$(b)$\hspace{1.3in}}
\caption{The distribution of the cosine of the inclination for the 681
 LP comets in (a) ecliptic coordinates, and (b) Galactic
 coordinates. A spherically symmetric distribution is indicated by the
 flat line. Data taken from \protect\cite{marwil93}.}
\label{fi:cosi}
\end{figure}

Figure~\ref{fi:cosi} shows the distribution of the cosine of the
inclination for the LP comets.  A spherically symmetric distribution
would be flat in this figure, as indicated by the heavy
line. \cite{eve67a} argued that inclination-dependent selection
effects are minor.

The inclination distribution in ecliptic coordinates is inconsistent
with spherical symmetry: the $\chi^2$ and Kolmogorov-Smirnov
statistics indicate probabilities of only $10^{-4}$ and $10^{-6}$
respectively that the distribution in Fig.~\ref{fi:cosi}(a)
distribution is flat.  Some of the discrepancy may arise from the
Kreutz sun-grazer group ($\sim 20$ members), which contributes at
$\cos i \sim -0.8$ and $\cos \ig \sim -0.4$.  The remaining excess at
$| \cos i| \sim 1$ may result from bias towards the ecliptic plane in
comet searches, or from an intrinsically anisotropic distribution of
LP comets resulting from individual stellar passages through the Oort
cloud.

The inclination distribution in Galactic coordinates may have a gap
near zero inclination, possibly reflecting the influence of the
Galactic tide (\S~\ref{pa:tide}), or confusion from the dense stellar
background in the Galactic plane.

\subsection{Longitude of ascending node}

\begin{figure}[p]
\centerline{\hbox{\psfig{figure=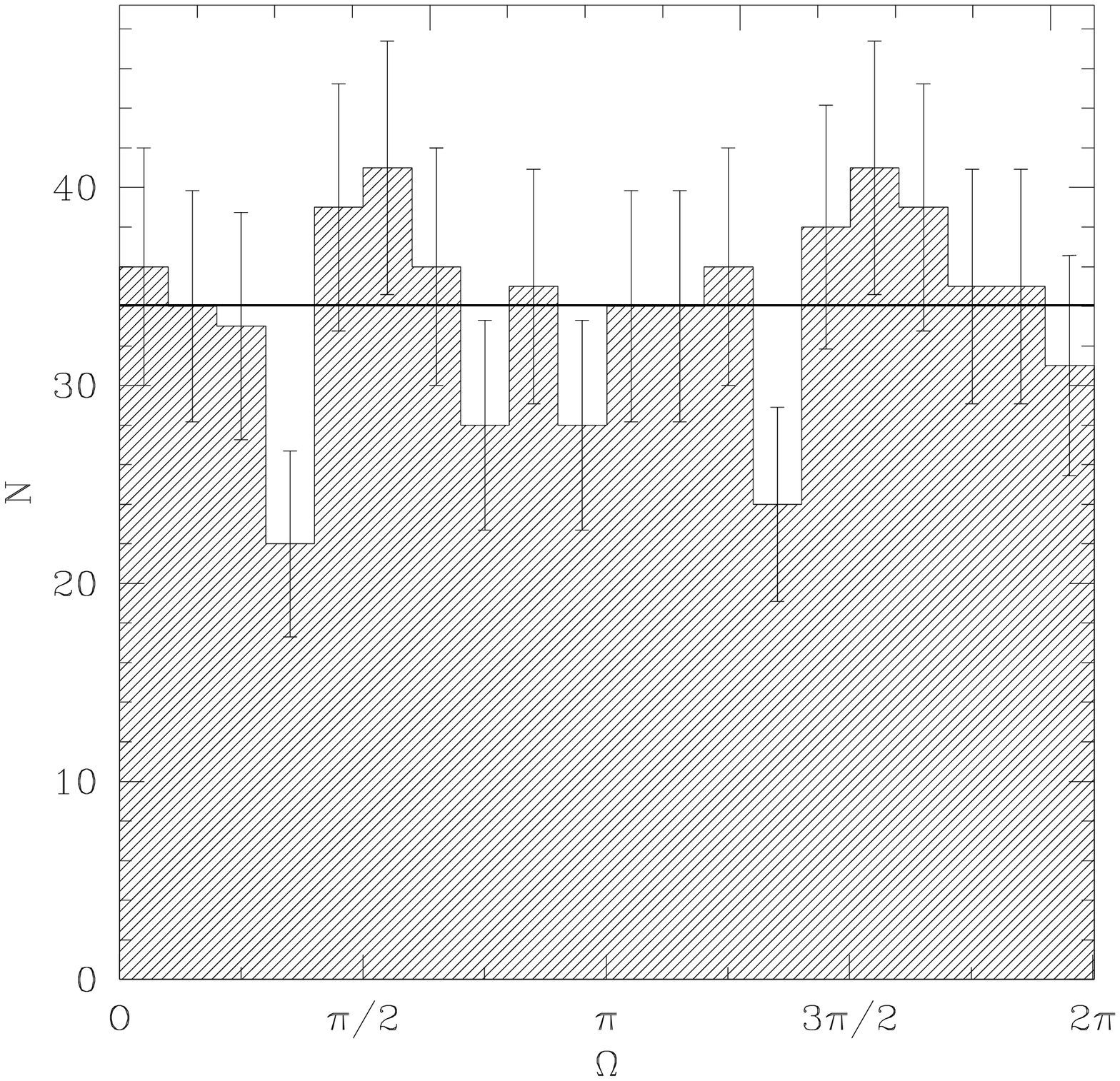,height=3in}
                  \psfig{figure=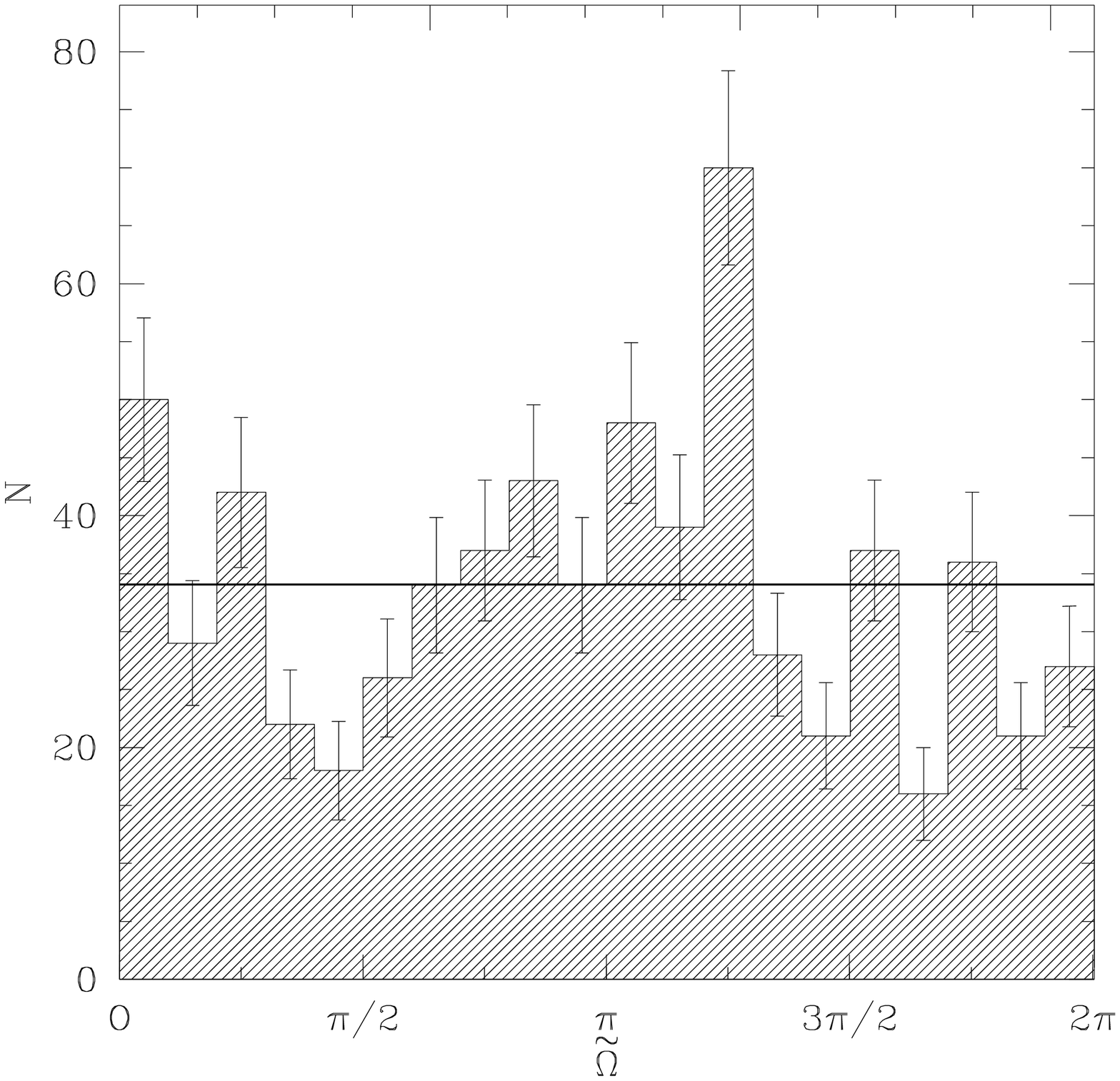,height=3in}}}
\centerline{\hspace*{1.45in}$(a)$\hfill$(b)$\hspace{1.3in}}
\caption{The distribution of the longitude of the ascending node of
the 681 LP comets in (a) ecliptic coordinates, and (b) Galactic coordinates.
Data taken from \protect\cite{marwil93}. }
\label{fi:lasc}
\end{figure}

The distribution of longitude of the ascending node $\Omega$ is
plotted in Fig.~\ref{fi:lasc}. The flat line again indicates a
spherically symmetric distribution.  The Kreutz
sun-grazers are concentrated at $\Omega \sim 0.15$ and $\Og \sim 4$,
and thus are responsible for the highest spike in Fig.~\ref{fi:lasc}.
\nocite{eve67a,eve67b} Everhart (1967a,b) concluded that
$\Omega$-dependent selection effects are likely to be negligible. The
$\chi^2$ and Kolmogorov-Smirnov statistics indicate that the ecliptic
distribution is consistent with a flat distribution at the 80\% and
90\% levels respectively.

\subsection{Argument of perihelion}

\begin{figure}[p]
\centerline{\hbox{\psfig{figure=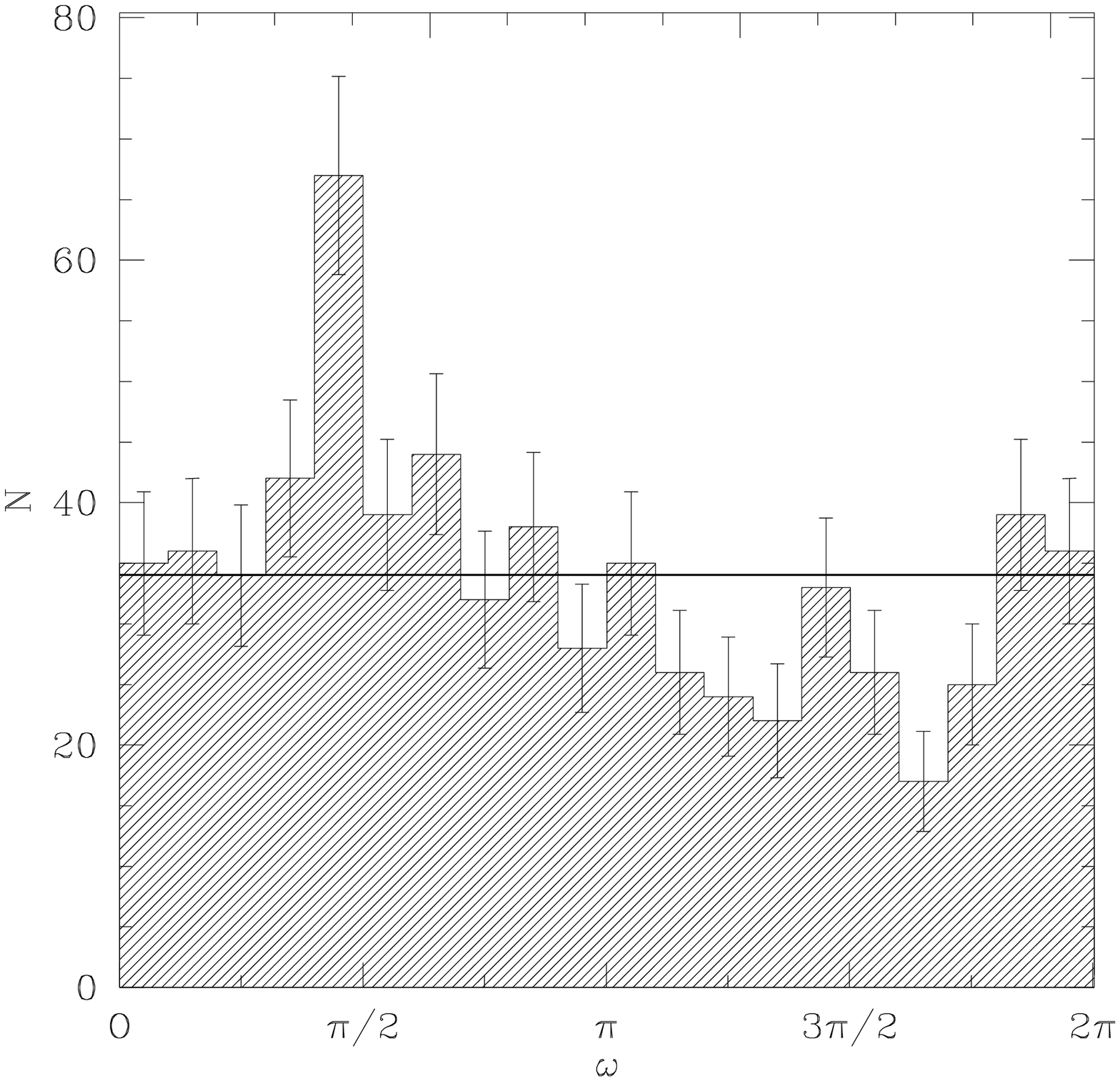,height=3in}
                  \psfig{figure=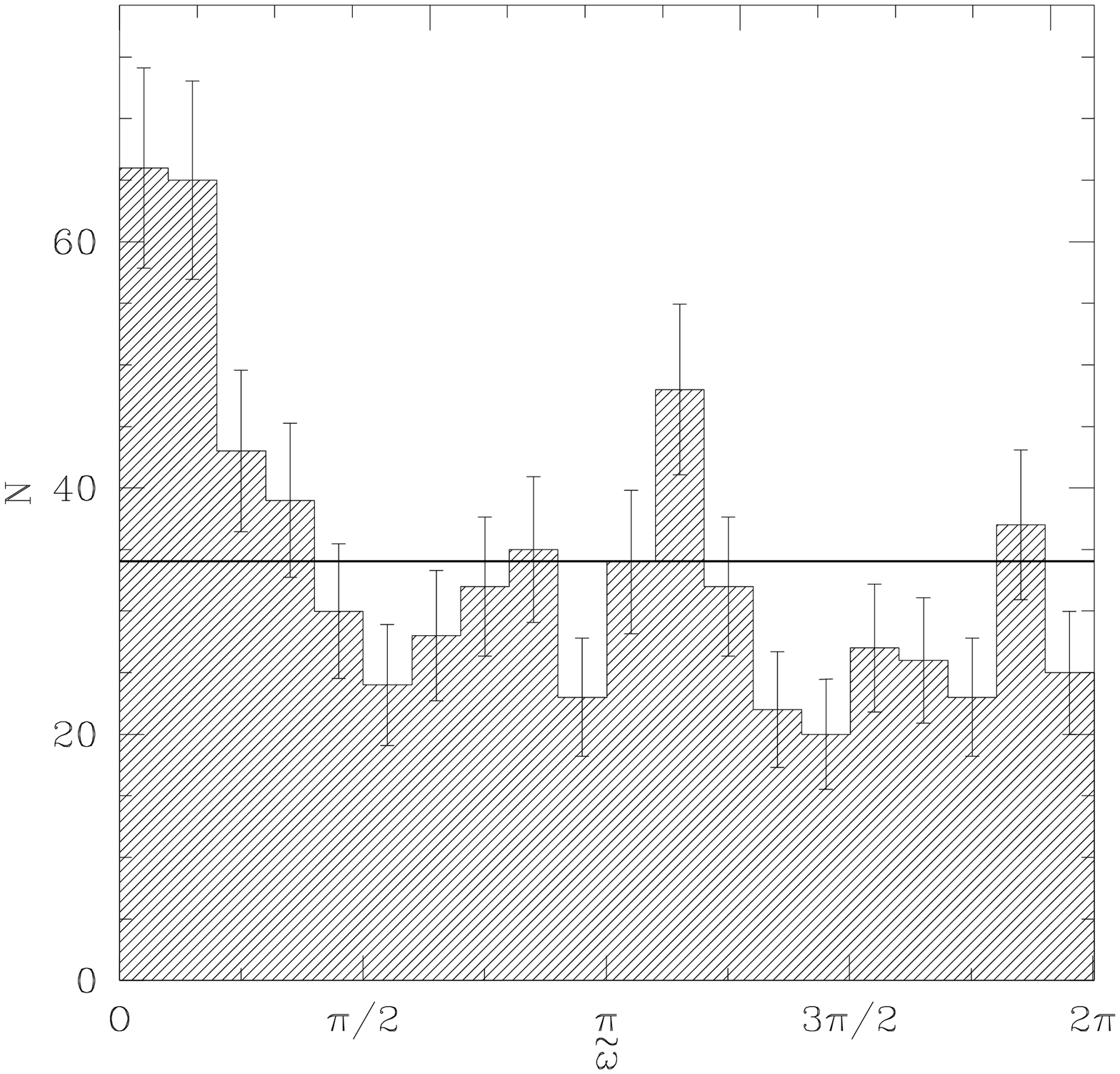,height=3in}}}
\centerline{\hspace*{1.45in}$(a)$\hfill$(b)$\hspace{1.3in}}
\caption{The distribution of the argument of perihelion in $(a)$ the
ecliptic frame, $\omega$, and $(b)$ the Galactic frame, $\og$, for the
681 LP comets.  Data taken from \protect\cite{marwil93}.}
\label{fi:aper}
\end{figure}

Figure~\ref{fi:aper} shows the distribution of the argument of
perihelion $\omega$ for the LP comets. Comets with $0<\omega<\pi$
outnumber those with $\pi<\omega<2\pi$ by a factor of $395/286 = 1.38
\pm 0.11$.  This excess is partly due to the Kreutz
group, which is concentrated at $\omega \sim 1.6$ and $\og \sim
0.15$; but may also be due to observational selection 
\cite[]{eve67b,kre82}: comets with $0 < \omega < \pi$ pass perihelion
above the ecliptic, and are more easily visible to observers in the
northern hemisphere. The lack of observed apparitions with $\omega >
\pi$ reflects the smaller number of comet searchers in the southern
hemisphere until recent times. The distribution in the Galactic frame
has a slight excess of comets with orbits in the range $\sin2 \og > 0$
($399/282 = 0.59 \pm 0.04$).

\subsection{Aphelion direction}

\begin{figure}[p]
\centerline{\vbox{\hbox{\psfig{figure=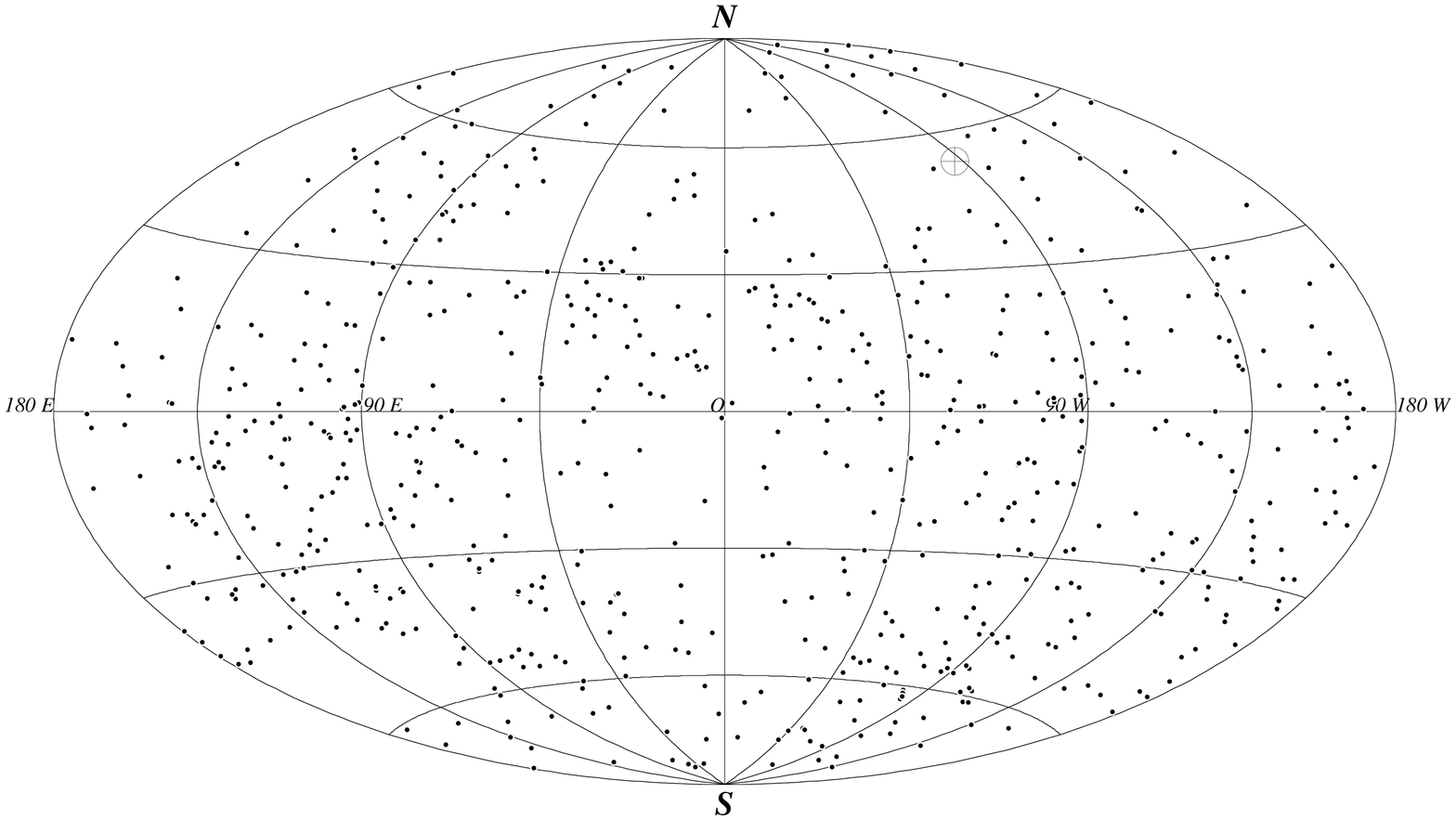,angle=270,height=3in}$(a)$}
                        \vspace{.75cm}
                        \psfig{figure=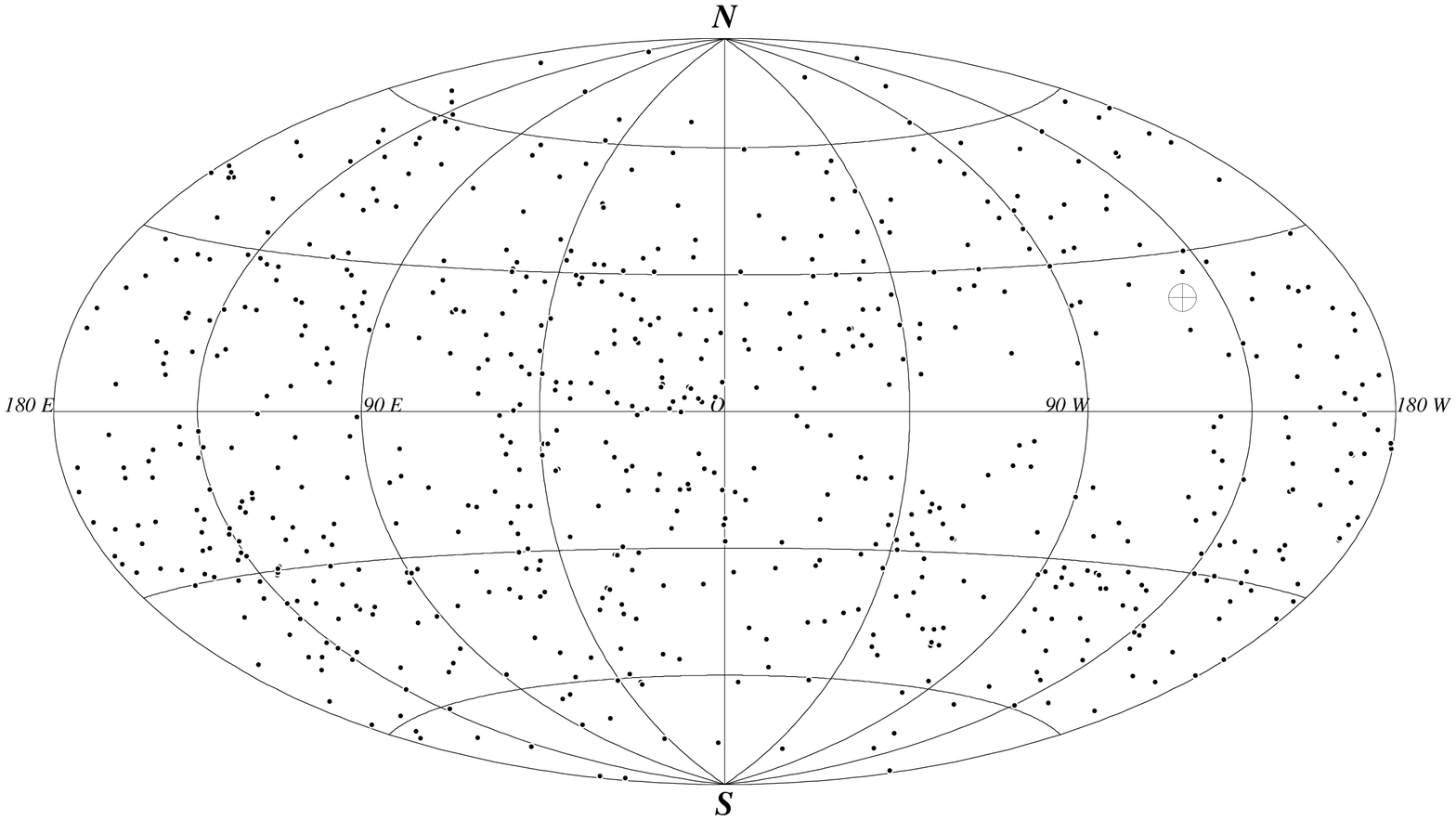,angle=270,height=3in}$(b)$}}
\caption{All 681 long-period comet aphelion directions on ecliptic
$(a)$ and Galactic $(b)$ equal-area maps. More precisely, these are
the antipodes of the perihelion directions. The crossed circle is the
solar apex. Data taken from \protect\cite{marwil93}.}
\label{fi:aphdir}
\end{figure}

Figure~\ref{fi:aphdir} shows the distribution of the aphelion
directions of the LP comets in ecliptic and Galactic coordinates.

Claims have been made for a clustering of aphelion directions around
the solar antapex \cite[\eg]{tyr57,oja75}, but newer analyses with
improved catalogues \cite[\eg]{lus84} have cast doubt on this
hypothesis. The presence of complex selection effects, such as the
uneven coverage of the sky by comet searchers, renders difficult the
task of unambiguously determining whether or not clustering is
present. Attempts to avoid selection effects end up subdividing the
samples into subsamples of such small size as to be of dubious
statistical value.

\cite{whi77} has shown that it is unlikely that there are many
large comet groups \ie comets related through having split from the
same parent body, in the observed sample though the numerous ($\sim$ 20)
observed comet splittings makes the possibility acceptable in
principle. A comet group would likely have spread somewhat in
semimajor axis: the resulting much larger spread in orbital period
$P \propto a^{3/2}$ makes it unlikely that two or more members of
such a split group would have passed the Sun in the 200 years for
which good observational data exist. The Kreutz group of sun-grazing comets
is the only generally accepted exception.

\begin{figure}[p]
\centerline{\hbox{\psfig{figure=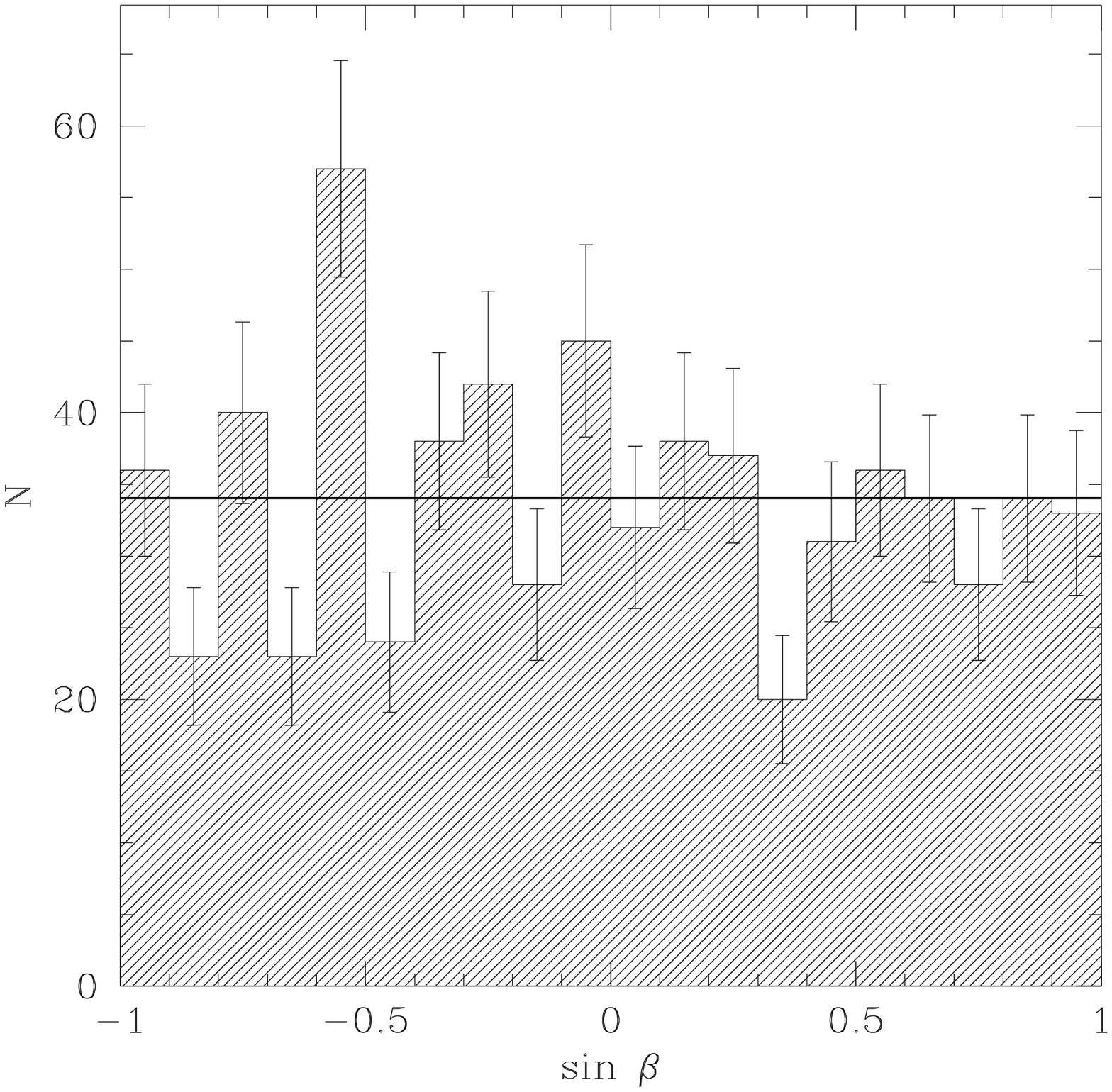,height=3in}
                  \psfig{figure=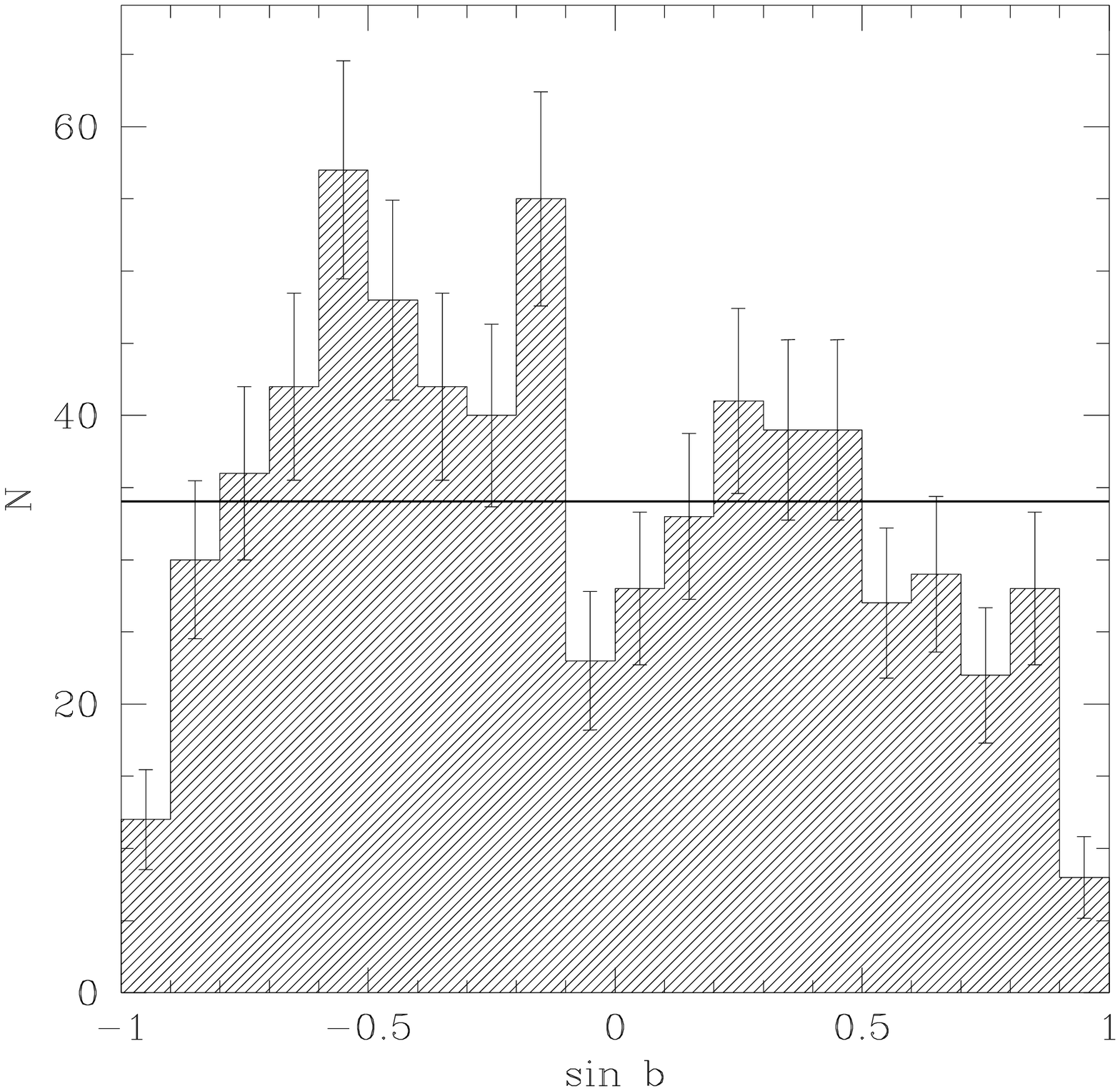,height=3in} }}
\centerline{\hspace*{1.45in}$(a)$\hfill$(b)$\hspace{1.3in}}
\caption{The sine of the aphelion latitudes of the 681 LP comets in
the ecliptic $(a)$ and Galactic $(b)$ reference frames.  The heavy
line indicates a spherically symmetric distribution.  Data taken from
\protect\cite{marwil93}.}
\label{fi:aphlat}
\end{figure}

Figures~\ref{fi:aphlat}a and b show histograms of comet number versus
the sine of the ecliptic latitude $\beta$ and of the Galactic latitude
$b$. The ecliptic latitudes deviate only weakly from a spherically
symmetric distribution and this deviation is likely due to the lack of
southern hemisphere comet searchers.  The Galactic distribution shows
two broad peaks, centred roughly on $\sin b \sim \pm 0.5$.  It will be
shown that these probably reflect the influence of the gravitational
tidal field of the Galaxy (\S~\ref{pa:tide}), which acts most strongly
when the Sun-comet line makes a $45\degree$ angle with the Galactic
polar axis.

\subsection{Orbital elements of new comets} \label{pa:109new}

\begin{figure}[p]
\centerline{\hbox{\psfig{figure=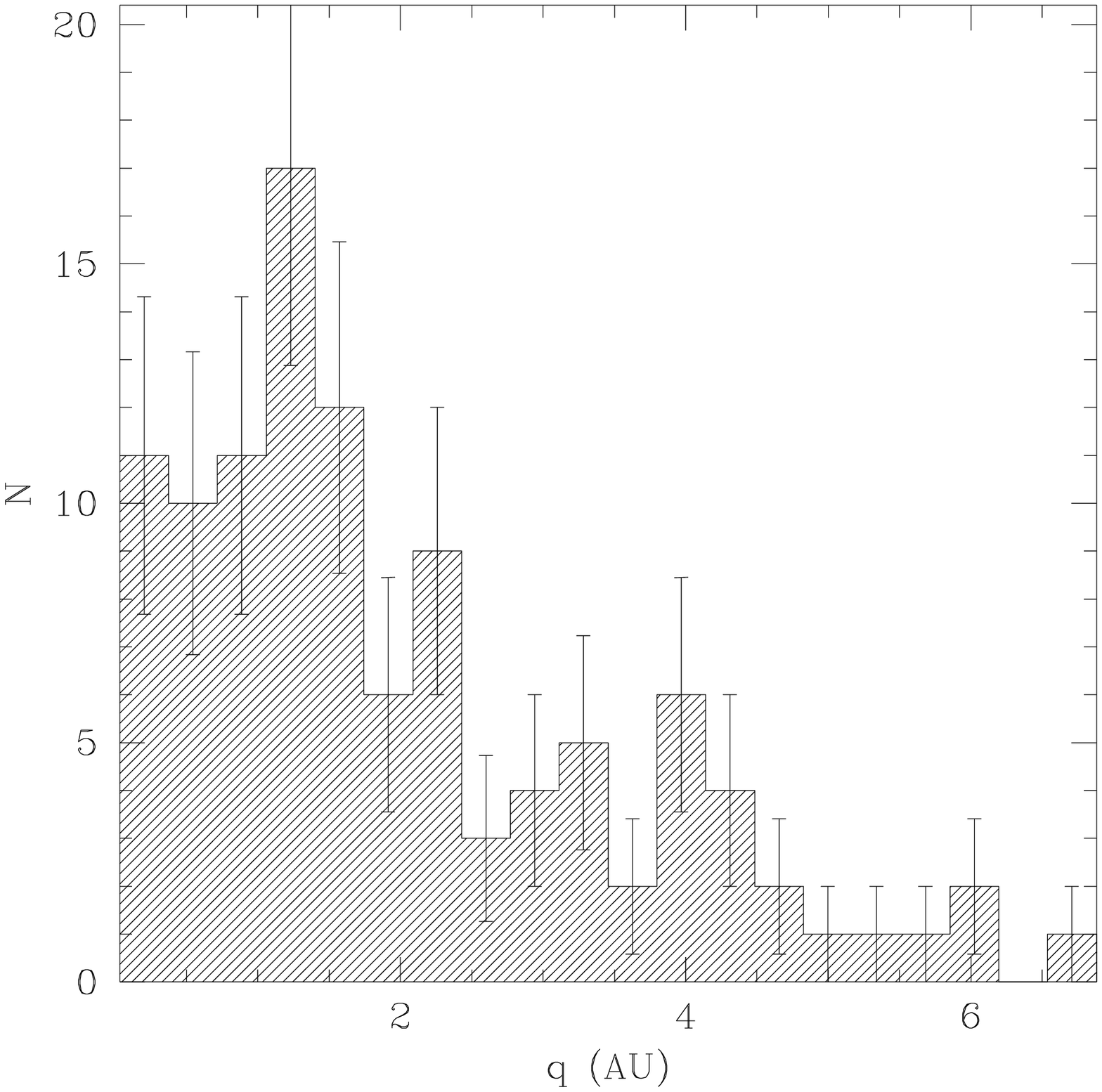,height=3in}
                  \psfig{figure=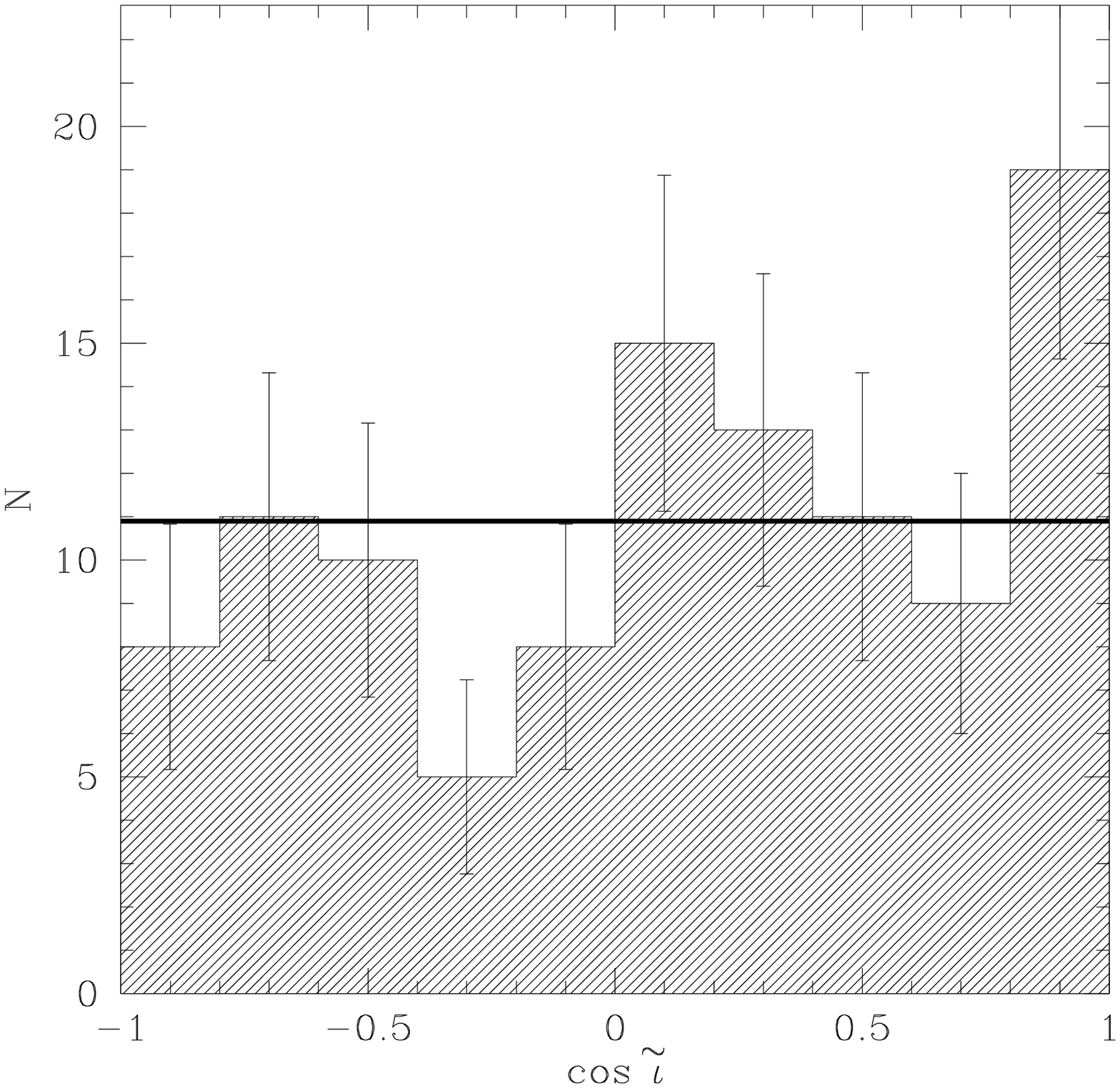,height=3in}}}
\centerline{\hspace*{1.45in}$(a)$\hfill$(b)$\hspace{1.3in}}
\centerline{\hbox{\psfig{figure=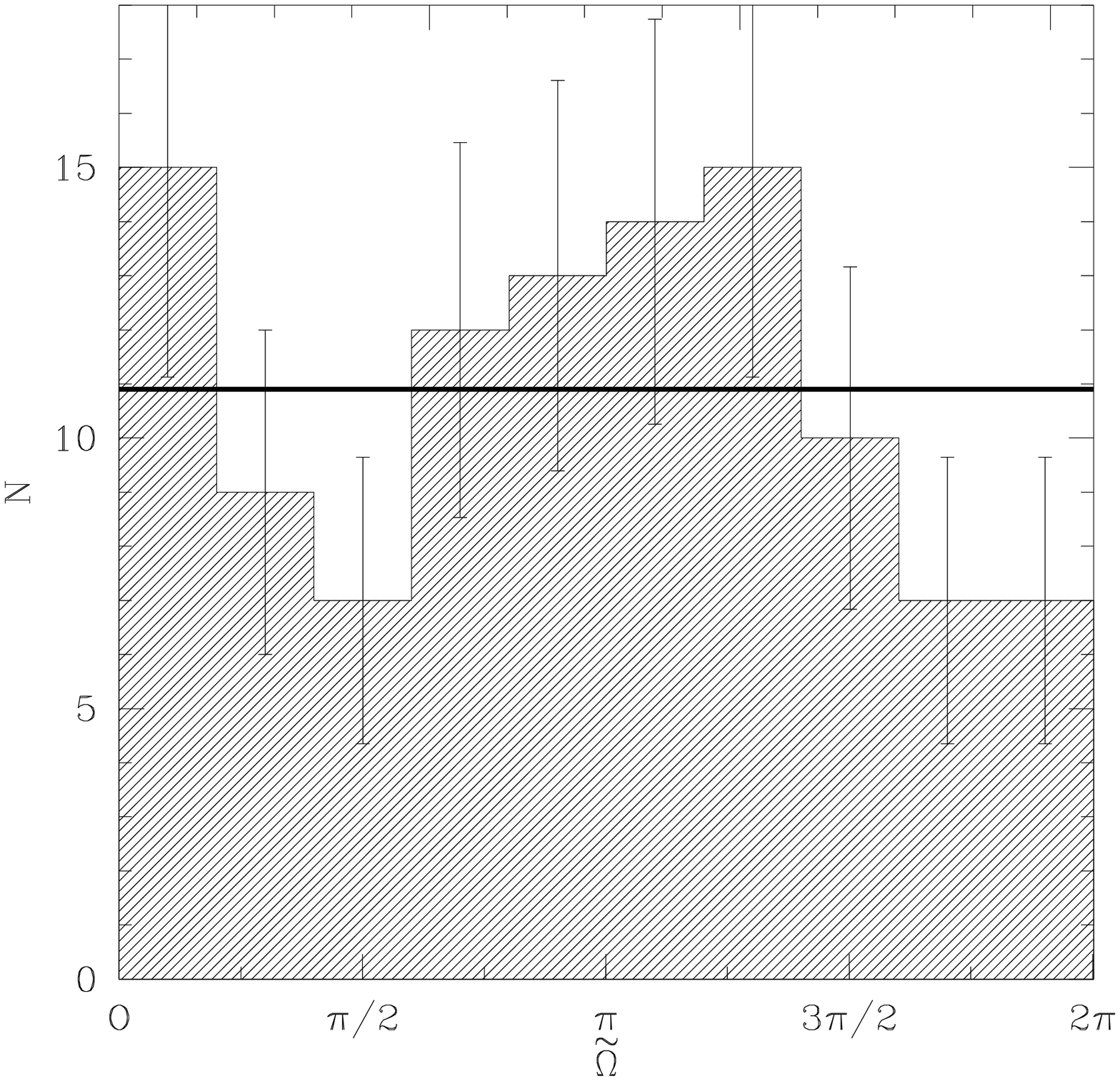,height=3in}
                  \psfig{figure=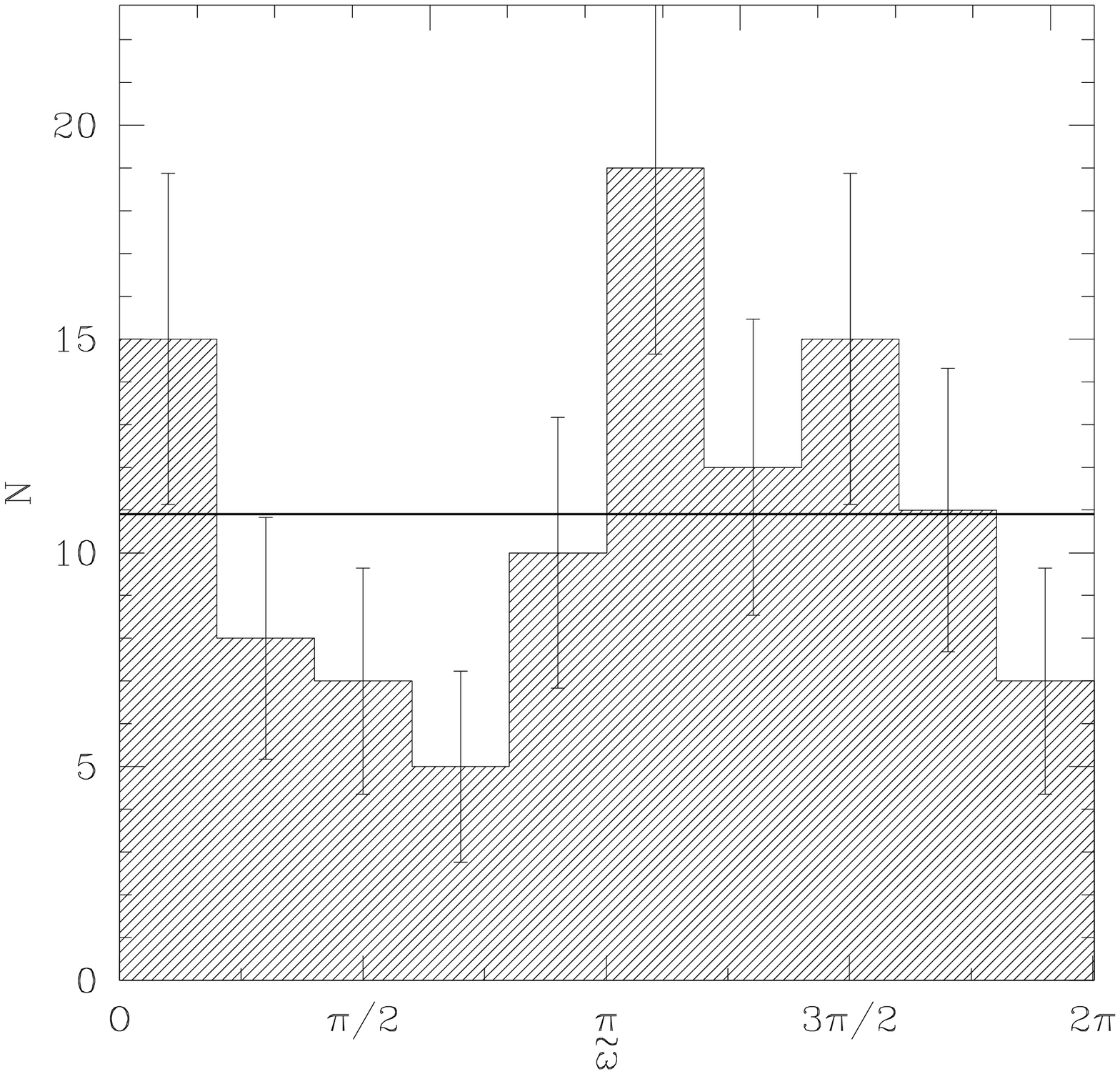,height=3in}}}
\centerline{\hspace*{1.45in}$(c)$\hfill$(d)$\hspace{1.3in}}
\caption{Distribution of orbital elements for the 109 new comets
($1/a<10^{-4}\aui$): (a) perihelion distance; (b) inclination; (c)
longitude of ascending node; (d) argument of perihelion. All angular
elements are measured in the Galactic frame at the aphelion preceding
the first apparition. }
\label{fi:new_i}
\end{figure}

\begin{figure}[p]
\centerline{\vbox{\hbox{
            \psfig{figure=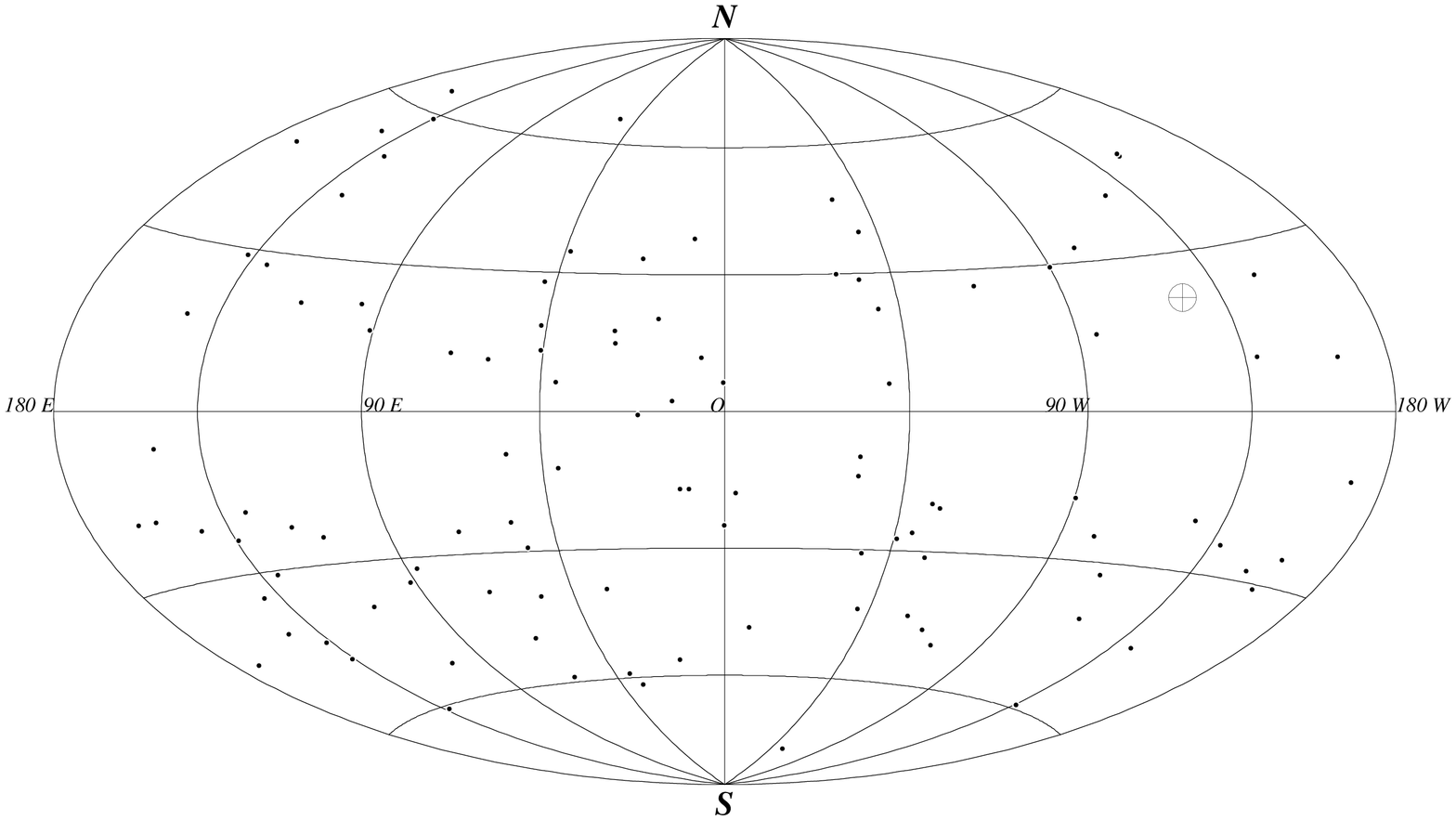,angle=270,height=3in}}}}
\caption{Equal-area plot of the aphelion directions of the 109
 new comets in the Galactic frame.}
\label{fi:new_ap}
\end{figure}

For some purposes it is useful to isolate the distribution of orbital
elements of the 109 new comets whose original semimajor axes lie
in the Oort spike, $x=1/a\le10^{-4}\aui$. The distributions of
perihelion distance, as well as inclination, longitude of the
ascending node, and argument of perihelion in Galactic coordinates are
all shown in Fig.~\ref{fi:new_i}. The distribution of aphelion
directions is shown in Fig.~\ref{fi:new_ap}.

\subsection{Parametrization of the distribution of
elements}\label{sec:psidef} 

For comparison with theoretical models, we shall parametrize the observed
distribution of LP comets by three dimensionless numbers:
\begin{itemize}
\item The ratio of the number of comets in the Oort spike
($1/a<10^{-4}\aui$) to the total number of LP comets is denoted by
$\Psi_1$. This parameter measures the relative strength of the Oort spike.
\item The inverse semimajor axes of LP comets range 
from zero (unbound) to 0.029~{\aui} ($P=200\yr$). Let the ratio of
the number of comets in the inner half of this range
(0.0145 to 0.029~{\aui}) to the total be $\Psi_2$. This parameter measures the
prominence of the ``tail'' of the energy distribution.
\item Let the ratio of the number of prograde comets in the ecliptic frame 
to the total be $\Psi_3$. This parameter measures the isotropy of the LP comet 
distribution. 
\end{itemize}

We estimate these parameters using all LP comets with original orbits
in \cite{marwil93}: 
\begin{eqnarray}
%\Psi_1 & = & 81/246 = 0.329 \pm 0.042,\\
%\Psi_2 & = & 18/246 = 0.073 \pm 0.018,\\
%\Psi_3 & = & 127/246 = 0.516 \pm 0.056.
\Psi_1 & = & 109/289 = 0.377 \pm 0.042,\nonumber\\
\Psi_2 & = & 19/289 = 0.066 \pm 0.016,\\
\Psi_3 & = & 145/289 = 0.501 \pm 0.051.\nonumber
\label{eq:psiobs}
\end{eqnarray}
For consistency, we based our calculation of
$\Psi_3$ on the 289 comets with known original orbits, even though knowledge
of the original orbit is not required since $\Psi_3$ depends only on angular
elements. If we consider all 681 LP comets, we  find $\Psi_3 = 321/681
= 0.471 \pm 0.041$; the two values are consistent within their error bars.

We denote theoretical values of these parameters by $\Psi_i^t$ and compare
theory and observation through the parameters
\begin{equation}
X_i\equiv {\Psi_i^t\over\Psi_i},\qquad i=1,2,3;
\label{eq:xdef}
\end{equation}
which should be unity if theory and observation agree.

\section{Theoretical background} \label{pa:dynamics}

\subsection{The Oort cloud} \label{pa:oortcloud}

The spatial distribution of comets in the Oort cloud 
can be deduced from the assumption that these comets formed in
the outer planetary region and were scattered into the Oort cloud through
the combined perturbations of the tide and planets \cite[]{dunquitre87}. These
calculations suggest---in order of decreasing reliability---that (i) the cloud
is approximately spherical; (ii) the velocity distribution of comets within
the cloud is isotropic; in other words the phase-space distribution is uniform
on the energy hypersurface, except perhaps at very small angular momentum
where the comets are removed by planetary encounters; (iii) the cloud's inner
edge is near 3000~\au, with a space number density of comets roughly
proportional to $r^{-3.5}$ from 3000 to 50~000~\au.

Orbits of comets in the Oort cloud evolve mainly due to torques from
the overall Galactic tidal field, but they are also affected by
encounters with planets, passing stars and molecular clouds.  Comets
are also lost through collisions with the Sun. Through these
mechanisms, between 40\% \cite[]{dunquitre87} and 80\% \cite[]{wei85}
of the original Oort cloud may have been lost over the lifetime of the
Solar System, leaving perhaps $\sim10^{12}$ comets
(cf. eq. \ref{eq:noort}) with mass $\sim50 M_{\oplus}$ \cite[]{wei90}
in the present-day comet cloud.  These numbers are very uncertain.

If the phase-space distribution of comets is uniform on the energy
hypersurface, then the number of comets at a given semimajor axis with
angular momentum less than $J$ should be $\propto J^2$; this in turn implies
that the number of comets with perihelion in the range $[q,q+dq]$ should be
$N(q)dq$, where 
\begin{equation}
N(q)\propto 1 - {q\over a},\qquad q\le a.
\label{eq:qdist}
\end{equation}
This distribution is modified if there are loss mechanisms that depend
strongly on perihelion distance, as we now discuss. 

\subsection{The loss cylinder} \label{pa:lcylinder}

A comet that passes through the planetary system receives a gravitational kick
from the planets. The typical energy kick $\Delta x$ depends strongly on
perihelion distance (and less strongly on inclination): $\Delta
x\approx1\times10^{-3}\au^{-1}$ for $q\lta 6\au$, dropping to
$1\times10^{-4}\au^{-1}$ at $q\simeq10\au$ and $1\times10^{-5}\au^{-1}$ at
$q\simeq20\au$ \cite[]{fer81,dunquitre87}. For comparison, a typical comet in
the Oort spike has $x\lta 10^{-4}\au^{-1}$; since these comets have perihelion
$q \sim 1\au$ they receive an energy kick $\Delta x\gg x$ during perihelion
passage. Depending on the sign of the kick, they will either leave the
planetary system on an unbound orbit, never to return, or be thrown onto a
more tightly bound orbit whose aphelion is much smaller than the size of the
Oort cloud. In either case, the comet is lost from the Oort cloud.

More generally, we can define a critical perihelion distance $q_{\ell}
\sim10\au$ such that comets with $q<q_{\ell}$ suffer a typical energy
kick at perihelion which is larger than the typical energy in the Oort
cloud. Such comets are said to lie in the ``loss cylinder'' in phase
space because they are lost from the Oort cloud within one orbit (the
term ``cylinder'' is used because at a given location 
within the cloud, the constraint $q<q_{\ell}$ is is satisfied in
a cylindrical region in velocity space: for highly eccentric orbits
$q<q_{\ell}$ implies that the angular momentum
$J<J_{\ell}\equiv(2GM_\odot q_{\ell})^{1/2}$, which in turn implies
that the tangential velocity $v_\perp<J_{\ell}/r$). The loss cylinder
is refilled by torques from the Galactic tide and other sources.

The comets in the Oort spike are inside the loss cylinder and hence
must generally be on their first passage through the planetary system
(this is why we designated the 109 comets with $1/a <
10^{-4}$~AU$^{-1}$ as ``new'' in \S~\ref{pa:109new}). The loss
cylinder concept also explains why the energy spread in the Oort spike
is much narrower than the energy spread in the Oort cloud itself:
comets with smaller semimajor axes have a smaller moment arm and
shorter period so their per-orbit angular momentum and perihelion
distance changes are smaller; for $a\lta 2\times10^4\au$ the
perihelion cannot jump the ``Jupiter barrier'' \ie cannot evolve from
$q>q_\ell\sim 10\au$ (large enough to be outside the loss cylinder) to
$q\lta 1\au$ (small enough to be visible) in one orbital period. Thus
the inner edge of the Oort spike is set by the condition that the
typical change in angular momentum per orbit equals the size of the
loss cylinder, and does not reflect the actual size of the Oort cloud
\cite[]{hil81}. The new comets we see come from an outer or active
Oort cloud ($a\gta2\times10^4\au$) in which the typical change in
angular momentum per orbit exceeds the size of the loss
cylinder. Thus, in the outer Oort cloud, losses from planetary
perturbations do not strongly affect the phase-space distribution of
comets near zero angular momentum (the loss cylinder is said to be
``full''), and the equilibrium distribution of perihelion distances
(Eq.~\ref{eq:qdist}) remains approximately valid within the loss
cylinder. The more massive inner Oort cloud ($a\lta2\times10^4\au$)
does not produce visible comets except during a rare comet ``shower''
caused by an unusually close stellar encounter and which perturbs them
sufficiently to jump the Jupiter barrier. In this inner cloud, losses
from planetary perturbations strongly deplete the distribution of
comets at small perihelion distances (the loss cylinder is said to be
``empty'') and thus it does not contribute to the Oort spike.

\subsection{Energy evolution of LP comets}\label{sec:onedran}

Let us examine the motion of an Oort cloud comet after it enters the planetary
system for the first time. The motion of a comet in the field of the giant
planets, the Sun and the Galactic tide is quite complicated, but considerable
analytic insight can be obtained if we make the following approximations: 

\begin{enumerate}

\item Depending on the sign of the energy kick from the planets, the
comet will either be ejected from the Solar System or perturbed onto a
much more tightly bound orbit ($a\sim 10^3\au$). In either case the
Galactic tide plays no further significant role, and can be neglected.

\item The influence of the planets is concentrated near perihelion,
where the moment arm of the comet is small. Thus the angular momentum,
orbit orientation, and perihelion distance are approximately conserved
during perihelion passage; the only significant change is in the
orbital energy. More precisely, the typical changes in angular
elements caused by a planet of mass $M_p$ and semimajor axis $a_p$ to
the orbit of a comet with perihelion $q\lta a_p$ are
\begin{equation}
\Delta i \sim  \Delta \Omega \sim \Delta\omega \sim \Delta q/a_p \sim
M_p/\msun, 
\label{eq:angelems}
\end{equation}
while the fractional change in energy is $\Delta x/x\sim (a M_p/a_p\msun)$,
which is much larger for an Oort cloud comet (by a factor $a/a_p\sim
10^3$--$10^4$). 

\item The orbit of any LP comet looks nearly parabolic when it passes through
the planetary system, so the distribution of energy changes during perihelion
passage is approximately independent of the comet's orbital energy. Therefore
we may define a function $p(\Delta x)d\Delta x$, the probability that the
energy change per perihelion passage due to planetary perturbations is in the
interval $[\Delta x,\Delta x+d\Delta x]$. The function $p(\Delta x)$ is an
implicit function of the inclination, perihelion distance, argument of
perihelion, etc. but as we have seen these change much more slowly than $x$
and so can be considered constants. The properties of $p(\Delta x)$ are
discussed by \cite{eve68}; $p$ is approximately an even function of $\Delta
x$ [the odd component is smaller by O$(M_p/M_\odot)$] and as $|\Delta
x|\to\infty$, $p(\Delta x)\propto |\Delta x|^{-3}$---although despite this
extended tail $p$ is often approximated by a Gaussian. If $q\lta a_p$
then the typical energy change due to a single planet is $\langle|\Delta
x|\rangle\sim (M_p/M_\odot)a_p^{-1}$. Plots of the
second moment of $p(\Delta x)$, averaged over argument of perihelion, as a
function of perihelion distance and inclination are given by \cite{fer81} and
\cite{dunquitre87}. 

\item Comets on escape orbits ($x \leq 0$) are lost from the Solar
System. The appropriate boundary condition at large $x$ is less
clear. Our other approximations fail when $x$ becomes comparable to
the inverse semimajor axes of the planets, which occurs when they
become SP comets at $x=x_{\rm sp}=0.029\au^{-1}$. SP comets will continue
to random walk in energy---some becoming LP comets once again---but
the other orbital elements will also evolve at a comparable rate, so
the approximation of a one-dimensional random walk is no longer
valid. Fortunately we shall find that the fraction of new comets that
survive fading and ejection to become SP comets is small enough that
the details of their evolution are unlikely to affect the overall
distribution of LP comets (cf. Tables~\ref{ta:V1_es} and
\ref{ta:Vm_es}); for most purposes we can simply assume that LP comets
reaching $x_{\rm sp}$ are lost.

\item The orbital periods of most LP comets are sufficiently long that
the orbital phases of the planets and hence the energy kicks that the
comet receives from them are uncorrelated at successive
encounters. Thus the evolution of the comet energy can be regarded as
a Markov process or random walk.\footnote{This argument can be made
more precise \cite[]{chivec86,pet86,sagzas87}. Assume that the energy
kick received by a comet is proportional to the sine of the orbital
phase of Jupiter relative to the comet. Let $x_n$ be the original
energy (in $\au^{-1}$) just before the $n^{\rm th}$ perihelion passage
and let $g_n$ be the orbital phase of Jupiter at that passage. Then
\begin{equation}
x_{n+1}=x_n+\epsilon_1\sin g_n,\qquad g_{n+1}=
g_n+{2\pi\over P_J}x_{n+1}^{-3/2},
\end{equation}
where $P_J$ is Jupiter's orbital period in years and
$\epsilon=2^{-1/2}\epsilon_1$ is the rms energy change for a comet with small
perihelion. Writing $x_n=x_0+\delta x_n$ where $\delta x_n$ is small,
the map becomes
\begin{equation}
I_{n+1}=I_n+\epsilon_1 f\sin g_n,\qquad g_{n+1}=g_n-I_{n+1},
\label{eq:stndrd}
\end{equation}
where $I_n=f\delta x_n$ and $f=3\pi/(P_Jx_0^{5/2})$. This is the standard map,
which exhibits global chaos when $|\epsilon_1 f|\gta 1$; this in turn implies
that the energy kicks received by the comet are effectively random. The
condition $|\epsilon_1 f|\gta1$ can be re-written as
\begin{equation}
a \gta 2^{-1/5}a_J^{3/5}(3\pi\epsilon)^{-2/5}=
20\au\left(5 \times 10^{-4}\au^{-1}
\over\epsilon\right)^{2/5},
\end{equation}
where $\epsilon \approx 5 \times 10^{-4}~\aui$ is a typical energy kick for
comets with perihelia inside Jupiter's orbit
\cite[]{fer81,dunquitre87}.

Thus the orbits of all LP comets with perihelion $q \las a_J$ are expected to
be chaotic.} 

\item A comet may fade as its inventory of volatiles is depleted or
may be disrupted by various mechanisms. We shall use the term
``fading'' to denote any change in the intrinsic properties of the
comet that would cause it to disappear from the observed sample. We
parametrize this process by a function 
$\Phi_m \in [0,1]$, $m=1,2,\ldots$ ($\Phi_1=1$), the
probability that a visible new comet survives fading for at least $m$
perihelion passages.  There are two closely related functions: the
probability that the comet survives fading for precisely $m$
perihelion passages,
\begin{equation}
\phi_m\equiv \Phi_m-\Phi_{m+1};
\label{eq:phimdef}
\end{equation}
and the conditional 
probability that a comet that survives $m$ passages will fade
before the $(m+1)^{\rm st}$ passage, 
\begin{equation}
\psi_m={\phi_m\over\Phi_m}=1-{\Phi_{m+1}\over\Phi_m}.
\end{equation}

\end{enumerate}

With these approximations the evolution of the energy of LP comets can
be treated as a one-dimensional random walk. We assume that visible
new comets arrive directly from the Oort cloud with original energy
$x=0$, at a rate $N$ per year.  Let $f_m(x)dx$ be the number of
visible LP comets per year with original energy in the range
$[x,x+dx]$ which are returning on their $m^{\rm th}$ perihelion
passage (thus $f_1(x)=N\delta(x)$, where $\delta(x)$ is a delta
function). Then in a steady state we must have
\begin{equation}
f_{m+1}(x)=(1-\psi_m)\int_0^\infty f_m(y)p(x-y)dy,\quad m\ge1,\ x\ge0,
\label{eq:fund}
\end{equation}
which can be solved successively for $f_2(x),f_3(x),\ldots$.  The
total number of LP comets with energies in the interval $[x,x+dx]$ is
$f(x)dx$, where $f(x)=\sum_{m=0}^\infty f_m(x)$; theoretical
predictions of $f(x)$ are to be compared with the observed
distribution of LP comets in Fig.  \ref{fi:energy}.

The simplest version of this problem is obtained by assuming that
there is no fading ($\psi_m=0$) and that the energy changes by
discrete steps $\pm\epsilon$ with equal probability
[$p(x)=\half\delta(x+\epsilon)+\half\delta(x-\epsilon)$]. In this case
the possible values of the energy are restricted to a lattice
$x=(j-1)\epsilon$, where $j$ is an integer, and the random walk is
identical to the gambler's ruin problem \cite[]{kan79,fel68}.  The
end-state of ejection ($j=0$) corresponds to bankruptcy; if in
addition we assume that there is an absorbing boundary at $x_{\rm
sp}\equiv (j_{\rm sp}-1)\epsilon$, then evolving to an SP comet
corresponds to breaking the house. Thus, for example, the
probabilities that an LP comet with energy $(j-1)\epsilon$ will be
eventually be ejected or become a short-period comet are respectively
\begin{equation}
p_{\rm ej} = 1-{j\over j_{\rm sp}},\qquad p_{\rm sp} = {j\over
j_{\rm sp}},
\end{equation}
and the mean number of orbits that the comet will survive is
\begin{equation}
\langle m\rangle = j(j_{\rm sp}-j).
\label{eq:mnlf}
\end{equation}
A new comet has $j=1$ and its mean lifetime is therefore
$\langle m\rangle = j_{\rm sp}-1$; the ratio of new to
all LP comets observed in a fixed time interval is
\begin{equation}
\Psi_1^t={1\over\langle m\rangle} ={1\over j_{\rm sp}-1}.
\label{eq:lifegam}
\end{equation}
There are also explicit expressions for the probability that the comet is
ejected or becomes an SP comet at the $m^{\rm th}$ perihelion passage
\cite[]{fel68}. 

The gambler's ruin problem is particularly simple if there is no
boundary condition at large $x$ ($x_{\rm sp}\to\infty$), which is
reasonable since few comets reach short-period orbits anyway
(\S~\ref{pa:sp_comets}). The probability that a new comet will survive
for precisely $m$ orbits is then
\begin{eqnarray}
p_{\rm ej}(m) & = & {1\over 2^m m}\left(m\atop \half m+ \half\right),
\quad\hbox{$m$ odd},\nonumber \\
&=& 0\qquad\qquad \hbox{$m$ even};
\end{eqnarray}
for $m\gg1$, $p_{\rm ej}(m)\to (2/\pi)^{1/2}m^{-3/2}$ for $m$ odd, and zero
otherwise. The mean lifetime $\sum_{m=1}^\infty mp_{\rm ej}(m)$ is
infinite, and the probability that a comet will survive for at least $m$
orbits is $\sim m^{-1/2}$ for large $m$. 

When using the gambler's ruin to model the evolution of LP comets, we
take $\epsilon\simeq5\times10^{-4}\aui$, which is the rms energy
change for comets with perihelion between 5 and $10\au$
\cite[]{fer81,dunquitre87}, and $x_{\rm sp}=0.029\aui$ ($P=200\yr$);
thus $j_{\rm sp}\simeq 60$. Eq.~\ref{eq:lifegam} then predicts
$\Psi_1^t=0.017$; the ratio of the predicted to the observed value for
this parameter (cf. Eq.~\ref{eq:psiobs}) is
\begin{equation}
X_1={\Psi_1^t\over\Psi_1}=0.051\pm0.006.
\label{eq:xonegr}  
\end{equation}
The gambler's ruin model predicts far too few comets in the Oort spike
relative to the total number of LP comets.  

This simple model also makes useful predictions about the inclination
distribution of LP comets. The distribution of new comets is
approximately isotropic, so there are equal numbers of prograde and
retrograde new comets. Since prograde comets have longer
encounter times with the planets, they tend to have larger energy
changes than retrograde comets. Equation~\ref{eq:lifegam} predicts
that the ratio of prograde to retrograde LP comets should be roughly
the ratio of the rms energy change for these two types, $\epsilon_{\rm
retro}/\epsilon_{\rm pro}\simeq$2--3. The fraction of prograde comets
should then be $\Psi_3^t=1/(1+\epsilon_{\rm pro}/\epsilon_{\rm
retro})\simeq0.3$. The ratio of the predicted to the observed value
for this parameter (cf. Eq.~\ref{eq:psiobs}) is
\begin{equation}
X_3={\Psi_3^t\over\Psi_3}=0.58\pm0.06. 
\label{eq:xthrgr} 
\end{equation}
The gambler's ruin model predicts too few prograde comets. 

More accurate investigations of this one-dimensional random walk have
been carried out by many authors. \cite{oor50} approximated $p(\Delta
x)$ by a Gaussian and assumed $\psi_m=k=\hbox{constant}$ and found a
good fit to most of the energy distribution for $k=0.014$; however, he
found that the number of new comets was larger than the model
predicted by a factor of five, and hence was forced to assume that
only one in five new comets survive to the second perihelion
passage---in other words $\psi_1=0.8$, $\psi_m=0.014$ for $m>1$.
\cite{ken61} and \cite{yab79} have analyzed the case $x_{\rm
sp}\to\infty$, $\psi_m=k=\hbox{constant}$,
$p(y)\propto\exp\left(-2^{1/2}|y|/\sigma\right)$, where $\sigma$ is
the rms energy change per perihelion passage. In this case
Eq.~\ref{eq:fund} can be solved analytically to yield\footnote{In
the limit of zero disruption, $k=0$, Eq.~\ref{eq:kendall} yields
$f(x)=N[\delta(x)+2^{1/2}/\sigma]$; in other words, the energy
distribution of observed LP comets should be flat, apart from the Oort
spike.}
\begin{equation}
f(x)=N\delta(x)+{2^{1/2}N\over\sigma}\big(1-k^{1/2}\big)
\exp\big[-(2k)^{1/2}x/\sigma\big];
\label{eq:kendall}
\end{equation}
using this result Kendall derives a reasonable fit to the data if $k=0.04$ and
one in four to six new comets survive to the second perihelion---results
roughly compatible with Oort's.  This model predicts a ratio of new comets to
all LP comets observed in a fixed time interval given by 
\begin{equation}
\Psi_1^t={N\over\int f(x)dx}=k^{1/2}.
\label{eq:kenlife}
\end{equation}
\cite{yab79} gave analytic formulae for $p_{\rm ej}(m)$ for this
model, and showed that the probability that a comet will survive for
at least $m$ orbits is $\sim \exp(-km)/m^{1/2}$ for large
$m$. \cite{whi62} examined survival laws of the form $\phi_m\propto
m^{-\alpha}$ (the proportionality constant is determined by the
condition that $\sum_m\phi_m=1$) and found a good fit to the observed
energy distribution with $\alpha\simeq 1.7$. \cite{eve79} used a
distribution $p(y)$ derived from his numerical experiments and found
$\Phi_m\simeq 0.2$ for all $m>1$; in other words only one in five
comets survived to the second perihelion passage but the fading after
that time was negligible.

 For some purposes the random walk can be approximated as a diffusion
process; in this case the relevant equations and their solutions are
discussed by \cite{yab80}. \cite{bai84} examines solutions of a
diffusion equation in two dimensions (energy and angular momentum) and
includes a fading probability that depends on energy rather than
perihelion number---which is less well-motivated but makes the
equations easier to solve (he justifies his fading function with an
{\it a posteriori} ``thermal shock'' model, in which comets with large
aphelia are more susceptible to disruption because they approach
perihelion with a lower temperature). Bailey finds a good fit to the
observed energy distribution if the fading probability per orbit is
\begin{equation}
\phi(x)=0.3[1+(x/0.004\au)^2]^{-3/2}.
\label{eq:bailfad}
\end{equation}
\cite{emebai96} have modeled the distribution of LP comets using a
Monte Carlo model with $\psi_m=k=\hbox{constant}$ plus an additional
probability per orbit $k^\ast$ that the comet is rejuvenated. Their
preferred values are $k=0.3$ and $k^\ast = 0.0005$.

The most complete model of LP comet evolution based on a random walk in energy
is due to Weissman (1978, 1979, 1980). \nocite{wei78,wei79,wei80} His Monte
Carlo model included the gravitational influence of the planets,
non-gravitational forces, forces from passing stars, tidal disruption by the
Sun, fading and splitting.  In his preferred model, 15\% of the comets have
zero fading probability, and the rest had a fading probability of 0.1 per
orbit. At this cost of this somewhat {\it ad hoc} assumption, Weissman was
able to successfully reproduce the semimajor axis, inclination, and
perihelion distributions.

The one-dimensional random walk is a valuable tool for understanding the
distribution of LP comets. However, some of its assumptions are not
well-justified: (i) Secular changes in perihelion distance, argument of
perihelion, and inclination at each perihelion passage accumulate over many
orbits and can lead to substantial evolution of the orientation and perihelion
\cite[]{quitredun90,baichahah92,thomor96}. 
(ii) Although the probability distribution
of energy changes $p(y)$ is approximately an even function [$\langle
y^2\rangle^{1/2}$ is larger than $\langle y\rangle$ by O$(M_p/M_\odot)$], the
random changes in energy due to the second moment grow only as $m^{1/2}$ where
$m$ is the number of orbits, while the systematic changes due to the first
moment grow as $m$. Thus the small asymmetry in $p(y)$ may have important
consequences.

\subsection{The fading problem} \label{pa:fading}

All the investigations described in the previous subsection reach the
same conclusion: if the LP comets are in a steady state then we cannot
match the observed energy distribution without unexpectedly strong
fading after the first perihelion passage. Therefore either (i) the
comet distribution is not in a steady state, which almost certainly
requires rejecting most of the Oort model\footnote{There are advocates
of this position (see Bailey 1984 for references), but we are not
among them.}, or (ii) we must postulate {\it ad
hoc} fading laws and abandon the use of the energy distribution as a
convincing test of the Oort model. This is the fading problem.

Fading can arise from many possible mechanisms but the most natural
hypothesis is that the comet's brightness fades sharply because its inventory
of volatiles is depleted during the first perihelion passage. \cite{oorsch51}
have argued that this hypothesis is supported by the observation that new
comets have strong continuum spectra due to dust entrained by the gases from a
volatile component, and that the decline of brightness with increasing
heliocentric distance is much slower for new comets. Many authors have looked
for evidence that new comets differ in composition or brightness from older LP
comets, with mixed results; \cite{whi91} summarizes these investigations by
saying that the Oort-Schmidt effect is ``fairly well confirmed''. 

Fading is much slower after the first perihelion passage, as exemplified by
the long history of Halley's comet. \cite{whi92} concludes that there is no
strong evidence that older (i.e. shorter period) LP comets have faded relative
to younger LP comets, consistent with theoretical estimates that
$10^3$--$10^4$ orbits are required for moderate-sized comets to lose their
volatiles \cite[]{wei80} and the lack of strong systematic trends in the
brightness of SP comets. 

Comets may also fade if they disrupt or split. After splitting, the
fragments are fainter and hence less likely to be visible, and in
addition lose their volatiles more rapidly. Moreover, young comets are
more likely to split than old ones: \cite{wei80} gives splitting
probabilities per perihelion passage of $0.10\pm0.04$ for new comets
but only $0.045\pm0.011$ for LP comets in general. The cause of
splitting is not well understood, except in some cases where splitting
is due to tidal forces from a close encounter with a giant planet.

Finally, we note that LP comets are responsible for 10--30\% of the crater
production by impact on Earth \cite[]{sho83}. The observed cratering rate can
therefore---in principle---constrain the total population of LP comets, whether
or not they have faded; however, this constraint is difficult to evaluate, in
part because estimates of comet masses are quite uncertain.

\section{Algorithm} \label{pa:algorithm}

We represent each comet by a massless test particle and neglect interactions
between comets. The orbit of the test particle is followed in the combined
gravitational fields of the Sun, the four giant planets, and the Galactic
tide. We assume that the planets travel around the Sun in circular, coplanar
orbits. We neglect the terrestrial planets, Pluto, the small free inclinations
and eccentricities of the giant planets, and their mutual perturbations as
there is no reason to expect that these play significant roles in the
evolution of LP comets. 

\subsection{Equations of motion}

The equation of motion of the comet can be written as
\begin{equation}
\mten \ddot{{\bf r}} = {\bf F}_{\odot} + {\bf F}_{\rm planets} + 
{\bf F}_{\rm tide} + {\bf F}_{\rm other}, 
\end{equation}
where the terms on the right side represent the force per unit mass from the
Sun, the planets, the Galactic tide, and other sources (e.g. non-gravitational
forces). 

\subsubsection{The planets}

We shall employ two frames of reference: the barycentric frame, whose origin
is the center of mass of the Sun and the four planets, and the heliocentric
frame, whose origin is the Sun. In the barycentric frame,
\begin{equation}
{\bf F}_{\odot}+{\bf F}_{\rm planets}=
-\frac{G \msun}{|{\bf r}-{\bf r}_\odot|^3}({\bf r}-{\bf r}_\odot)
-\sum\limits_p \frac{G M_p}{|{\bf r}-{\bf r}_p|^3}({\bf r}-{\bf r}_p),
\label{eq:baryc}
\end{equation}
where ${\bf r}$, ${\bf r}_\odot$, and ${\bf r}_p$ are 
the positions of the comet,
the Sun, and planet $p$. In the heliocentric frame, the Sun is at the
origin and
\begin{equation}
{\bf F}_{\odot}+{\bf F}_{\rm planets}=
-\frac{G \msun}{|{\bf r}|^3}{\bf r}
-\sum\limits_p \frac{G M_p}{|{\bf r}-{\bf r}_p|^3}({\bf r}-{\bf r}_p)
-\sum\limits_p \frac{GM_p}{|{\bf r}_p|^3}{\bf r}_p;
\label{eq:helioc}
\end{equation}
the last sum is the ``indirect term'' that arises because the heliocentric
frame is not inertial. 

The heliocentric frame is useful for integrating orbits at small radii, $|{\bf
r}|\las |{\bf r}_p|$, because it ensures that the primary force center, the
Sun, is precisely at the origin (see \S\ref{pa:reg}). It is not well-suited
for integrating orbits at large radii, $|{\bf r}|\gas |{\bf r}_p|$, because
the indirect term does not approach zero at large radii, and oscillates with a
period equal to the planetary orbital period---thereby forcing the integrator
to use a very small timestep. In the integrations we switch from heliocentric
to barycentric coordinates when the comet radius $|{\bf r}|$ exceeds a
transition radius $r_c$; tests show
that the integrations are most efficient when $r_c=10\au$.

The code tracks close encounters and collisions between comets and
planets.  A close encounter with a planet is defined to be a passage
through a planet's sphere of influence
\begin{equation}
R_{\sss I} = \left( \frac{M_p}{\msun} \right)^{2/5} a_p, \label{eq:sofi}
\end{equation}
where $a_p$ is the planet's semimajor axis.  Each inward crossing of
the sphere of influence is counted as one encounter, even if there are
multiple pericenter passages while the comet remains within the sphere
of influence. A close encounter with the Sun is defined to be a
passage within 10 solar radii.

\subsubsection{The Galactic tide} \label{pa:tide}

The effects of the Galactic tide on comet orbits are discussed by 
\cite{heitre86}, \cite{mormul86}, \cite{tor86}, and \cite{matwhi89}.
Consider a rotating set of orthonormal vectors $\{{\bf e}_{\xg},{\bf e}_{\yg},
{\bf e}_{\zg} \}$.  Let ${\bf e}_{\xg}$ point away from the Galactic center,
${\bf e}_{\yg}$ in the direction of Galactic rotation, and ${\bf e}_{\zg}$
towards the South Galactic Pole (South is chosen so that the coordinate system
is right-handed).  The force per unit mass from the tide is
\cite[]{heitre86}
\begin{equation}
\mten {\bf F}_{\rm tide} = (A-B)(3A+B)\xg {\bf e}_{\xg} - (A-B)^{2} \yg {\bf
e}_{\yg} - [ 4 \pi G \rho_0 -2(B^{2} - A^{2}) ] \zg {\bf e}_{\zg}, 
\label{eq:totaltide}
\end{equation}
where $\rho_0$ is the mass density in the solar neighborhood, and $A$ and $B$
are the Oort constants. We take $A = 14.4 \pm 1.2 $~km~s$^{-1}~$kpc$^{-1} $
and $B = -12.0 \pm 2.8$~km~s$^{-1}$~kpc$^{-1}$ \cite[]{kerlyn86}.  The local
mass density is less well-known.  Visible matter (stars and gas) contributes
about $0.1$~M$_{\odot}$~pc$^{-3}$, but the amount of dark matter present in the
solar neighbourhood remains controversial.  If the dark matter is distributed
like the visible matter, then the dark/visible mass ratio $P$
is between 0 and 2 \cite[]{oor60,bah84,kuigil89,kui91,bahflygou92}. We adopt
$\rho_0=0.15$ M$_{\odot}$~ pc$^{-3}$ in this paper, corresponding to $P=0.5$.

With these values of $A$, $B$ and $\rho_0$, the $4 \pi G \rho_0$ term of
Eq.~\ref{eq:totaltide} exceeds the others by more than a factor of ten,
and from now on we shall neglect these other terms. The dominant component of
the tidal force arises from a gravitational potential of the
form
\begin{equation}
V_{\rm tide} = 2\pi G \rho_{0} \zg^2. 
\label{eq:tide_potl}
\end{equation}

In practice, of course, the local density $\rho_0$ varies as the Sun
travels up and down, in and out, and through spiral arms during its
orbit around the Galaxy. The amplitude of this variation depends
strongly on the unknown distribution and total amount of disk dark
matter. The maximum-to-minimum density variation could be as large as
3:1 \cite[]{matwhiinn95} but is probably considerably smaller, with a
period around 30 Myr [close to $\half (\pi/G\rho_0)^{1/2}$, the
half-period for oscillations in the potential (\ref{eq:tide_potl})
]. We are justified in neglecting these variations in $\rho_0$,
because the typical lifetime of LP comets after their first apparition
is only 1.4~Myr (see Table~\ref{ta:Vm_es}\ below), which is much shorter.

\subsubsection{Encounters with stars and molecular clouds}\label{sec:passing}

Our model neglects the effects of passing stars on LP comets, for
two main reasons: (i) The delivery rate of Oort cloud comets to the planetary
system due to Galactic tides is higher than the rate due to stellar encounters
by a factor 1.5--2 \cite[]{heitre86,tor86}, except during rare comet showers
caused by an unusually close passage, during which the delivery rate may be
enhanced by a factor of twenty or so \cite[]{hil81,hei90}; we feel justified
in neglecting the possibility of a comet shower because they only last about
2\% of the time \cite[]{hei90}; (ii) The effects of stellar encounters are
highly time-variable whereas the strength of the tide is approximately
constant over the typical lifetime of LP comets; thus by
concentrating on the effects of the tide we focus on a deterministic problem,
whose results are easier to interpret.

The effects of rare encounters with molecular clouds are highly time-variable,
and difficult to estimate reliably because the properties of molecular clouds
are poorly known \cite[]{bai83,drazin84,huttre85,tor86}. Therefore we shall
also assume that the present distribution of LP comets has not been
affected by a recent encounter with a molecular cloud.

\subsubsection{Regularisation} \label{pa:reg}

Integrating the orbits of LP comets is a challenging numerical
problem, because of the wide range of timescales (the orbital period can be
several Myr but perihelion passage occurs over a timescale as short as months)
and because it is important to avoid any secular drift in energy or angular
momentum due to numerical errors.  We have used the Kustaanheimo--Stiefel
(K-S) transformation to convert Cartesian coordinates to regularized
coordinates and have carried out all of our integrations in the regularized
coordinates.  A requirement of K--S regularisation is that the frame origin
must coincide with the primary force centre, which is why we use heliocentric
coordinates at small radii. 

The numerical integrations were carried out using the Bulirsch-Stoer
method, which was checked using a fourth-order Runge-Kutta-Fehlberg algorithm.
All integrations were done in double-precision arithmetic.

\subsection{Non-gravitational forces} \label{pa:nongrava}

The asymmetric sublimation of cometary volatiles results in a net
acceleration of the nucleus. These
non-gravitational\footnote{Traditionally, the term ``non-gravitational
forces'' has been reserved for the reaction forces resulting from the
uneven sublimation of cometary volatiles, and it will be used here in
that manner. Other factors of a non-gravitational nature have been
considered, including radiation and solar wind pressure, drag from the
interplanetary/interstellar medium, and the heliopause, but were found to be
negligible in comparison to the outgassing forces \cite[]{wie96}.}
(NG) forces are limited to times of significant outgassing (\ie coma
production), and remain small even then.

Non-gravitational forces are difficult to model. Their strength obviously
depends on the comet's distance from the Sun, but displays less
regular variability as well:  gas production may vary by a factor of 2
or more between the pre- and post-perihelion legs of the orbit
\cite[]{sek64,fes86}, and jets and streamers are observed to evolve on
time scales of less than a day \cite[]{fesricwes93b}, suggesting that
NG forces change on similar time scales.  Further complications arise
from the rotation of the nucleus, which is difficult to measure
through the coma, and which may be complicated by precession
\cite[]{wil87}. 

The NG acceleration ${\bf F}_{\rm jet}$ is written as
\begin{equation}
{\bf F}_{\rm jet} = F_1 {\bf e}_1 + F_2 {\bf e}_2 + F_3 {\bf e}_3,
\end{equation}
where ${\bf e}_1$ points radially outward from the Sun, ${\bf e}_2$ lies
in the orbital plane, pointing in the direction of orbital motion and normal
to ${\bf e}_1$, and ${\bf e}_3={\bf e}_1 \times {\bf e}_2$. 
A naive model of NG accelerations, which is all the data allows, assumes that
the short timescale components are uncorrelated and cancel out, leaving only
fairly regular, longer timescale components as dynamically important. We shall
use the  Style~II model of \cite{marsekyeo73}, which assumes that
accelerations are symmetric about perihelion, and can be represented
by
\begin{equation}
 F_1(r) = A_{1} g(r), \hspace{1cm}
 F_2(r) = A_{2} g(r), \hspace{1cm}
 F_3(r) = A_{3} g(r). 
\end{equation}
Here $\{A_1,A_2,A_3 \}$ are independent constants, and $g(r)$ is a
non-negative function describing the dependence on the comet-Sun distance $r$.
The form of $g(r)$ is based on an empirical water sublimation curve by
\cite{delmil71},
\begin{equation}
 g(r) = \alpha \left( \frac{r}{r_{0}} \right)^{-m} \left[ 1 + \left(
\frac{r}{r_{0}} \right)^{n} \right]^{-k}, \label{eq:gr}
\end{equation}
where $m=2.15$, $k=4.6142$, $n=5.093$, $r_{0} = 2.808$~\au \ and $\alpha$ is
chosen to be 0.1113 so that $g(1~$\au$)=1$. Note that $g(r)$ is roughly
proportional to $r^{-m} \approx r^{-2}$ for $r \ll r_{0}$.  At $r \gg r_{0}$,
$g(r)$ drops much faster than the simple inverse square that describes the
incident solar flux.

The constants $A_i$ are determined by fitting individual comet orbits 
\cite[]{marsekyeo73}; the value of $A_1$ is typically $10^{-7}$ to $10^{-9}$
\au\ day$^{-2}$,  $|A_2|$ is typically only 10\% of $|A_1|$, and $A_3$ is
consistent with zero.

\subsection{Initial conditions}

\label{sec:init}

\subsubsection{Initial phase-space distribution}\label{sec:oor}

The distribution of comets in the Oort cloud is only poorly known, although it
is plausible to assume that the cloud is roughly spherical and that the comets
are uniformly distributed on the energy hypersurface in phase space, except
possibly at very small angular momenta (cf. \S \ref{pa:oortcloud}). Then the
phase-space density is a function only of $L\equiv(GM_\odot a)^{1/2}$, which
we assume to be
\begin{equation}
f(L) =\left\{ 
\begin{array}{cc}
 0, & L < L_{-}=(GM_\odot a_-)^{1/2},\\
 f_0L^{2\alpha+3}, & L_- \le L  \le L_+,\\
 0, & L>L_+=(GM_\odot a_+)^{1/2},
\end{array} \right. 
\label{eq:ooouuu}
\end{equation}
where $f_0$ and $\alpha$ are constants, and $a_-$ and $a_+$ are the inner and
outer edges of the Oort cloud, respectively. We show below (footnote
7) that the total number of Oort cloud comets with semimajor axes in
the range specified by $[L,L+dL]$ is $(2\pi)^3 f(L)L^2dL$; this in turn
implies that the number density of comets is $\propto r^\alpha$ for
$a_-\ll r\ll a_+$.

Simulations of the formation of the Oort cloud by \cite{dunquitre87} suggest
that the number density of Oort cloud comets is $\propto r^{-(3.5\pm0.5)}$
between 3000 and 50~000~{\au}. Thus we set $\alpha=-3.5$, $a_-=10~000~\au$ and
$a_+=50~000$~\au. The inner edge of the cloud was placed at ~10~000~{\au}
instead of 3000~{\au} because comets with $a<10~000\au$ cannot become visible
except in occasional comet showers, yet would consume most of the computer
time in our simulation.

If the comets are uniformly distributed on the energy hypersurface,
the fraction of cloud comets with perihelion less than $q\ll a$ is
$J^2(q)/L^2=2q/a=0.003(q/40\au)(25,000\au/a)$ (which is consistent with $\int
N(q)dq$ as given by Eq.~\ref{eq:qdist}).  Since the effects of the
planets decline rapidly to zero when $q\gas 40\au$, only a small
fraction of cloud comets are influenced by planetary perturbations.
Therefore to avoid wasting computer time we analyze the motion of
comets with larger perihelion distance analytically, as we now
describe.

\subsubsection{Orbit-averaged evolution}

For comets in the Oort cloud, the tidal potential (\ref{eq:tide_potl}) is much
smaller than the Kepler Hamiltonian $H_{\rm Kep} = -\half GM_\odot/a$. Thus the
evolution of the comet under the Hamiltonian $H_{\rm Kep}+V_{\rm tide}$ can be
approximately described by averaging $V_{\rm tide}$ over one period of a
Kepler orbit to obtain the orbit-averaged Hamiltonian \cite[]{heitre86}
\begin{equation}
H_{\rm av} = -\frac{G \msun}{2a} + \pi G \rho_0 \, a^2 \sin^2 \! \ig \: (1-e^2
+ 5 e^2 \sin^2 \og); 
\label{eq:ham_avg}
\end{equation}
here $\ig$ and $\og$ are the inclination and argument of perihelion measured
in the Galactic frame. It is useful to introduce canonical momenta
\begin{equation}
L\equiv(G \msun a)^{1/2},\quad J = [G \msun a(1-e^2)]^{1/2},\quad 
J_{\zg} = J \cos \ig
\end{equation}
and their conjugate coordinates
\begin{equation}
f,\qquad \og, \qquad \Og.
\end{equation}
Here $J$ is the usual angular momentum per unit mass, $J_{\zg}$ is its
component normal to the Galactic plane, $f$ is the true anomaly and $\Og$ is
the longitude of the ascending node on the Galactic plane\footnote{ At this
point we may prove a result mentioned in \S\ref{sec:oor}: if the phase-space
density is $f=f(L)$ then the total number of comets in the range $[L,L+dL]$ is
$dN=f(L)dL\int_0^L dJ\int_{-J}^J dJ_{\zg}\int_0^{2\pi}d\Og\int_0^{2\pi}
d\og\int_0^{2\pi}df=(2\pi)^3f(L)L^2dL.$}. In terms of the canonical
coordinates and momenta the orbit-averaged Hamiltonian is
\begin{equation}
H_{\rm av}  =  - \frac{ (G \msun)^2 }{2 L^2} + \frac{\pi \rho_0}{G \msun^2}
\frac{L^2}{J^2} (J^2 - J_{\zg}^2) \left[ J^2 +  5(L^2 - J^2) \sin^2 \og
\right].  
\label{eq:ham}
\end{equation}
The canonical variables $f$ and $\Og$ are absent from
Eq.~\ref{eq:ham}, so the conjugate momenta $L$ and $J_{\zg}$ are
conserved. The conservation of $L$ implies that semimajor axis is
conserved as well.  The solution of the equations of motion
(\ref{eq:ham}) is discussed by \cite{heitre86} and \cite{matwhi89} but
is not needed for our purposes.

The rate of change of angular momentum is given by
\begin{eqnarray}
\dot{J} &=& - \frac{\partial H_{\rm av}}{\partial \og}, \\
 &=& - \frac{5 \pi \rho_0}{G \msun^{2}} \frac{L^{2}}{J^{2}} \: 
(J^2-J_{\zg}^{2}) (L^{2} - J^{2})\: \sin  2\og, \label{eq:jdot1} \\
 & = & -\frac{5 \pi \rho_0}{G \msun^{2}} e^2 L^4 \sin^2 \ig \sin 2 \og, 
\label{eq:jdot2} 
\end{eqnarray}

We now define the ``entrance surface'' to be the boundary of the region of
phase space with $J\le J_E(a)\equiv[2GM_\odot q_E(a)]^{1/2}$. We shall follow
cometary orbits only after they cross the entrance surface. We choose
$q_E=\max(q_{1},q_{2})$ where $q_{1}$ and $q_{2}$ reflect two criteria
that must be satisfied by the entrance surface: (1) Planetary perturbations
must be negligible outside the entrance surface; we take $q_{1}=60\au$ since
outside this perihelion distance the rms fractional energy change per orbit
caused by the planets is $\las 0.1\%$ for a typical Oort cloud comet. (2) The
orbit-averaged approximation for the effects of the Galactic tide must be
reasonably accurate outside the entrance surface; thus we demand that $J_E$
must exceed $\eta>1$ times the maximum change in angular momentum per orbit,
which in turn requires 
\begin{equation}
J_E(L) \ge \eta\frac{10 \pi^2 \rho_0}{G^3 \msun^4} L^7,\quad\hbox{or}\quad 
q_{2}= \eta^2\frac{50 \pi^4 \rho_0^2}{\msun^2} a^7,
\label{eq:es_J}
\end{equation}
where we have assumed $e\sim1$. In this paper we
take $\eta=3$.

The semimajor axis $a_{1,2}$ where $q_{1}=q_{2}$ is 
\begin{eqnarray}
a_{1,2} & = &\left( \frac{\msun^2 \: q_{1} }{50 \pi^4 \eta^2 
  \rho_0^2} \right)^{1/7} \mten \; , \label{eq:a_cross}\\
 &= & 2.41\times10^4  \mbox{\au}\Bigg( \frac{\eta}{3} \Bigg)^{-2/7} \Bigg( 
  \frac{q_{1}}{\mbox{60 \au}} \Bigg)^{1/7} \left( \frac{\rho_0}
  {\mbox{0.15 M$_{\odot}$ pc$^{-3}$}} \right)^{-2/7}.
  \label{eq:es_qmin}
\end{eqnarray}
Thus
\begin{equation}
\qe =\left\{ 
\begin{array}{cc}
 \mbox{60 \au} &\mbox{where } a \le a_{1,2}\\
 {\displaystyle \mbox{60 \au} \left( \frac{a}{a_{1,2}} \right)^7} & \mbox{where } 
a > a_{1,2} \end{array} \right. \label{eq:es2}
\end{equation}

\subsubsection{The flux of comets into the entrance surface} 
\label{pa:cometflux}

We have assumed in \S\ref{sec:oor} that the phase-space density is a
function only of energy or semimajor axis, $f=f(L)$. This assumption is not
in general correct for small angular momentum, where the comets are removed by
planetary encounters. However, all we require is the flux into the entrance
surface, most of which arises from comets whose angular momentum is steadily
decreasing under the influence of the Galactic tide. Such comets are
unaffected by the planets until after they cross the entrance surface, and
hence the assumption that $f=f(L)$ should be approximately correct.

Let $\Phi(L,J_E,J_{\zg}, \Og, \og, f)\, dL \, dJ_{\zg} \, d\Og \, d\og
\, df$ be the current of Oort cloud comets crossing into the entrance
surface $J_E$ at a given point. Then from Eq.~\ref{eq:jdot1}
\begin{eqnarray}
\Phi(L,J_E,J_{\zg}, \Og, \og, f) & = & \left\{\begin{array}{cc}
- \dot J\,f(L), &\mbox{where $\dot J<0$} \\ 0&\mbox{otherwise}
\end{array}\right. \nonumber \\
 & = & \left\{\begin{array}{cc}
- \frac{\displaystyle 5 \pi \rho_0}{\displaystyle G \msun^2} \frac{ 
\displaystyle L^2}{\displaystyle J_E^2}f(L) 
(J_E^2 - J_{\zg}^2)(L^2 - J_E^2) \sin 2 \og&\mbox{where 
$\sin 2 \og > 0$, } \\ 0& \mbox{otherwise}.
\end{array}\right. \\
\label{eq:phase_dens}
\end{eqnarray}

In our simulations, the initial orbital elements of the comets are drawn from
the distribution described by $\Phi$, using the energy distribution
(Eq.~\ref{eq:ooouuu}).

\subsection{End-states} \label{pa:endstates}

End-states may represent the loss or destruction of a
comet or simply an intermediate stopping point, from which the
simulation can subsequently be restarted. The possible end-states are:

\begin{description}
\item[Collision] The distance between the comet and the Sun or one of
the giant planets is less than that object's physical radius. To
ensure that we detect collisions, when a comet is close to a Solar
System body we interpolate between timesteps using a Keplerian orbit
around that body.

\item[Ejection] The comet is either (i) leaving the
Solar System on an orbit which is unbound \ie parabolic or hyperbolic with
respect to the Solar System's barycentre, or (ii) has ventured beyond the last
closed Hill surface around the Sun, and is thus considered stripped from the
Solar System by the action of passing stars, molecular clouds, \etc \ In
either case, the simulation is not terminated until the comet is at
least $10^5$~{\au} from the Sun, to allow for the possibility that subsequent
perturbations will result in the comet losing energy and returning to a
``bound'' state.

\item[Exceeded age of Solar System] The elapsed time has
exceeded the age of the Solar System, $5 \times
10^9$~yr.

\item[Exceeded orbit limit] The comet has completed more than 
5000 orbits without reaching one of the other end-states.  The integration
is terminated and the orbital elements are saved for later examination.
This is a safeguard to prevent extremely long-lived
comets from consuming excessive computer time.

\item[Faded] The comet is considered to have faded through loss of volatiles,
splitting or other mechanisms, and is no longer bright enough to be observed,
even if its orbit should carry it close to the Sun or Earth. We shall
investigate various empirical models for fading. The fading end-state is not
activated in any simulations unless explicitly mentioned in the accompanying
text.

\item[Perihelion too large] The comet's perihelion $q$ has evolved beyond
some limit, usually taken to be 40~\au, and is moving outwards under
the influence of the tide. Such a comet is
unlikely to become visible in the near future. 

\item[Short-Period] The comet's orbital period has decreased below 200~yr: it
has become a short-period comet. Continued planetary perturbations may cause
short-period comets to evolve back into LP comets, but we shall see that the
fraction of comets that reach this end-state is very small (at most a few
percent; see Tables~\ref{ta:V1_es} and \ref{ta:Vm_es}), so former short-period
comets are not a significant contaminant.

\item[Visible] The comet passes within 3~{\au} of the Sun for the first time,
an event we shall call the first apparition. Such comets continue to evolve,
but the first apparition provides a useful intermediate stopping point for the
simulations.

\end{description}

\section{Results} \label{pa:results}

We follow the trajectories of our sample comets from the time they cross the
entrance surface until they reach one of the end-states in \S
\ref{pa:endstates}. We divide the evolution into two stages: the
pre-visibility stage, which lasts until the comet first becomes visible, that
is, until their first passage within $3\au$ of the Sun (the first apparition;
cf. \S \ref{pa:perih}); and the post-visibility stage, which lasts from the
first apparition until the comet reaches one of the other end-states.

We call the set of LP comets at their first apparition the $V_1$ comets.
Similarly, those making their $m^{\rm th}$ apparition are called the $V_m$
comets. The union of the sets of orbital elements $V_1,V_2,\ldots$ is
called the $V_\infty$ comets.

We intend to compare the distribution of elements of the $V_1$ comets to the
observed distribution of elements of new comets, and the $V_\infty$ comets to
the visible LP comets. Note that the $V_\infty$ comets represent all
apparitions of a set of Oort cloud comets that first crossed the entrance
cylinder in a given time interval, while the observations yield all the comets
passing perihelion in a given time interval---one is a fixed interval of
origin and the other is a fixed interval of observation. However, in a steady
state these two distributions must be the same except for normalization.

For some purposes it is useful to estimate this normalization, i.e., to
estimate the time interval to which our simulation corresponds. To do this, we
first estimate the number of perihelion passages per year of new comets with
$q<q_{\rm v}=3\au$, which we call $\Phi_{\rm new}$. \cite{krepit78} find that
the rate of long-period comets passing within Jupiter's orbit ($5.2\au$) is
25~yr$^{-1}$. Taking a round number of 10~yr$^{-1}$ passing within 3~{\au} and
assuming one in three of these is new \cite[]{fesricwes93b}, we find
$\Phi_{\rm new}\simeq 3$~yr$^{-1}$. The number of $V_1$ comets produced in our
simulation (see below) is 1368; hence our simulation corresponds to a time
interval
\begin{equation}
t_s = 450\hbox{ yr}\left(3 \hbox{ yr}^{-1}\over\Phi_{\rm new}\right).
\label{eq:simlength}
\end{equation}
The total number of comets crossing the entrance surface in our
simulation is 125~495. Using our assumed form for the semimajor axis
distribution of comets in the Oort cloud (Eq.~\ref{eq:ooouuu}, with
$\alpha=-3.5$), and our formula for the flux through the entrance
cylinder (Eq.~\ref{eq:phase_dens}), we may deduce that the
normalization constant in Eq.~\ref{eq:ooouuu} is
$f_0=2.3\times10^{12}(\Phi_{\rm new}/3\hbox{ yr}^{-1})$ and the total
population of the Oort cloud is
\begin{equation}
N_{\rm Oort}=5.1\times10^{11}(\Phi_{\rm new}/3\hbox{ yr}^{-1}),
\label{eq:noort}
\end{equation} 
from 10,000~AU to 50,000~AU. Extrapolating in to 3000~AU yields a
population 5\% higher. For comparison, \cite{hei90} found that 0.2 new
comets per year with perihelion $<2$~AU are expected per $10^{11}$
Oort cloud objects outside 3000~AU; this corresponds to an Oort cloud
population that is a factor of two higher than the one given in
Eq. \ref{eq:noort}.  Of course, these estimates depend strongly on
uncertain assumptions about the extent of the inner Oort cloud.

\subsection{Pre-visibility evolution} \label{pa:V1}

The dynamically new or $V_1$ comets can be used as a starting point
for any investigation of phenomena that only affect the comet after
its first apparition (non-gravitational forces, fading, etc.). The
elements of the $V_1$ comets are measured in the barycentric frame
200~AU from the Sun.

The simulations reported here followed the evolution of 125~495 Oort cloud
comets that crossed the entrance surface. The orbital elements at the entrance
surface were determined as described in \S\ref{sec:init}.  Of the comets
crossing the entrance surface, 84\% had minimum perihelion distances
(determined from contours of the averaged Hamiltonian in Eq.~\ref{eq:ham})
greater than 40~{\au}, too far outside the planetary system to suffer
significant ($\gas 1\%$) perturbations in semimajor axis from the planets.
These comets were transferred to the {\sc Perihelion too large}
end-state. The orbits of the remaining 20~286 comets were followed in the
field of the Galactic tide and the Sun and planets.  Table~\ref{ta:V1_es}
shows the distribution of these comets among the various end-states; 1368 or
6.7\% became $V_1$ comets. Only 57 comets triggered the {\sc Exceeded Orbit
Limit} flag (see \S~\ref{pa:endstates}), set at 5000 revolutions; these are
discussed further in \S~\ref{pa:overmax}.  These computations consumed eight
weeks of CPU time on a 200 MHz workstation.

\begin{table}[p]
\centerline{
\begin{tabular}{|l|rrrrrr|r|} \hline
End-state & Ejection  &  Exc. Limit & Large $q$ & Short Pd. & Visible &  Total \\ \hline \hline
%                      ej    overmax   lq     sp    vis   total
%
Number         & 3807 & 57    & 15023 & 32    & 1368   & 20286  \\
Fraction       &0.1877& 0.0028& 0.7406&0.0015 & 0.0674 & 1.0000 \\  
Minimum $t_x$  & 6.80 & 17.2  & 7.46  & 11.7  & 7.14   & 6.80   \\
Median $t_x$   & 28.7 & 152   & 35.2  & 29.3  & 26.8   & 33.3 \\
Maximum $t_x$  & 342  & 480   & 1182  & 72.4  & 147    & 1182 \\
Minimum $m_x$  & 1    & 5000  &  1    & 6     & 1      & 1    \\
Median $m_x$   & 8    & 5000  &  5    & 387   & 5      & 6     \\  
Maximum $m_x$  & 4799 & 5000  &  4872 & 3432  & 2937   & 5000 \\ \hline
\end{tabular}}
\caption{The distribution of end-states of the 20~286 Oort cloud
comets with minimum perihelia $<40\au$. The minimum, median and
maximum lifetimes $m_x$ and $t_x$ are shown in orbital periods and
Myr, respectively.  No comets suffered collisions with the planets or the
Sun, or survived for the lifetime of the Solar System.} 
\label{ta:V1_es}
\end{table}

During the pre-visibility stage there were 729 close encounters
(Eq.~\ref{eq:sofi}) with the giant planets by 343 individual
comets, distributed as shown in Table~\ref{ta:ce}.

\begin{table}[p]
\centerline{
\begin{tabular}{|l|rrrr|r|} \hline
Planet & Jupiter & Saturn & Uranus & Neptune & Total \\ \hline \hline
Number of comets             & 60  & 145 & 71  & 67  & 343   \\
Number of encounters         & 210 & 317 & 109 & 93  & 729   \\
Encounters/comet             & 3.5 & 2.19& 1.53& 1.39& 2.13  \\ 
Collisions                   & 0   & 0   & 0   & 0   & 0     \\
Captures                     & 0   & 0   & 0   & 0   & 0     \\
Min. distance ($R_{\sss I}$) &0.023&0.043&0.074&0.049& 0.023 \\ 
Min. distance ($R_p$)        & 16.0& 38.7& 150 & 167 & 16.0  \\ 
$R_{\sss I} (R_p)$           & 674 & 907 & 2030& 3510& ---   \\
Outer satellite ($R_p$)      & 326 & 216 & 23  & 222 & ---   \\ \hline  
\end{tabular}}
\caption{Planetary encounter data during the pre-visibility stage for
the 20~286 Oort cloud comets with minimum perihelia
$<40\au$. Encounters for the 57 comets in the {\sc Exceeded Time
Limit} end-state are included only up to their 5000$^{th}$
orbit. ``Captures'' are considered to occur when the comet has a
planetocentric eccentricity less than unity at planetocentric
pericentre. The radius of the planet's sphere of influence $R_{\sss
I}$ (Eq.~\protect\ref{eq:sofi}) and the semimajor axis of its
outermost satellite are also given, in units of the planetary radius
$R_p$.}
\label{ta:ce}
\end{table}

\begin{figure}[p]
\centerline{\hbox{\psfig{figure=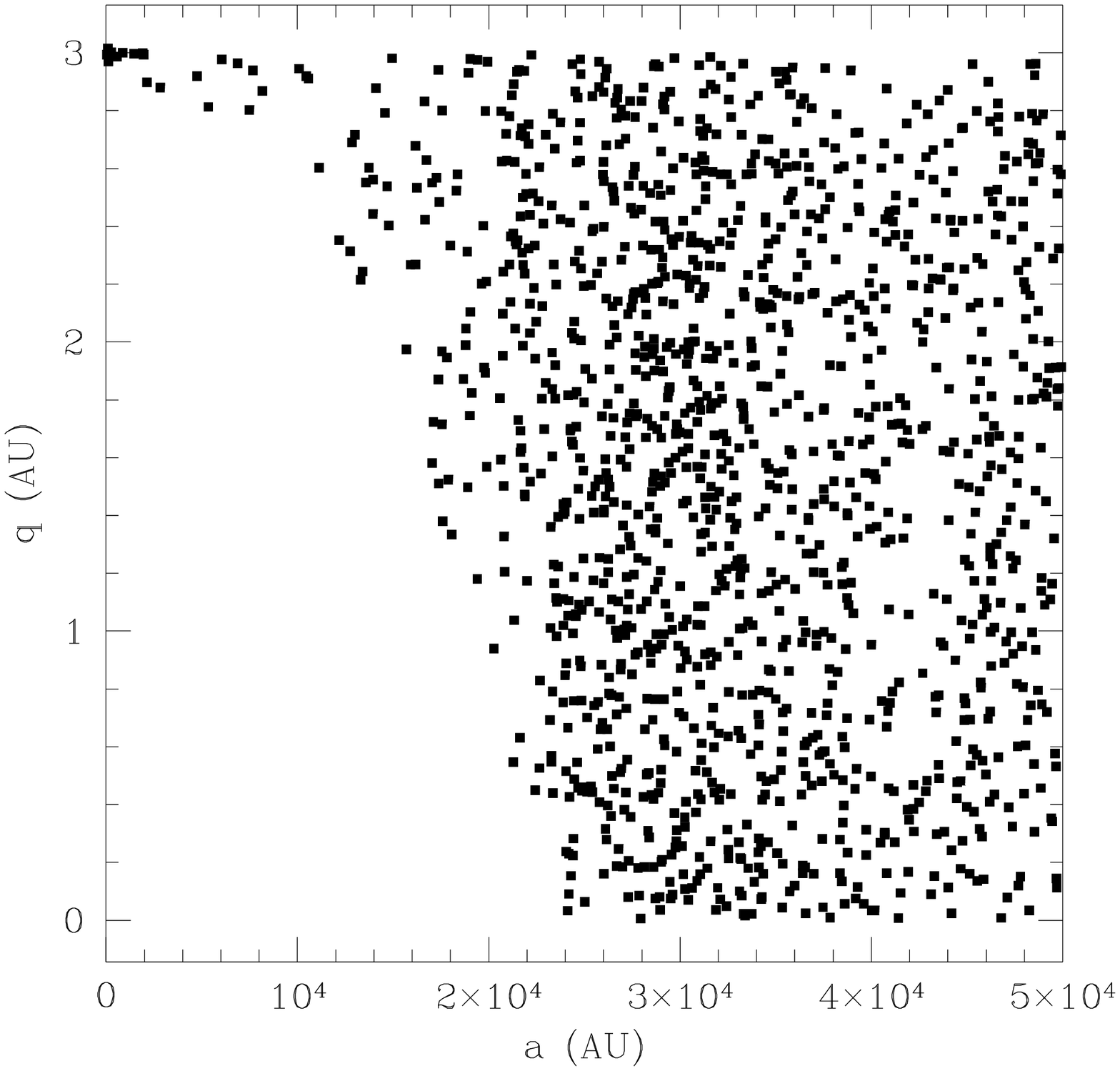,height=3in}
	          \psfig{figure=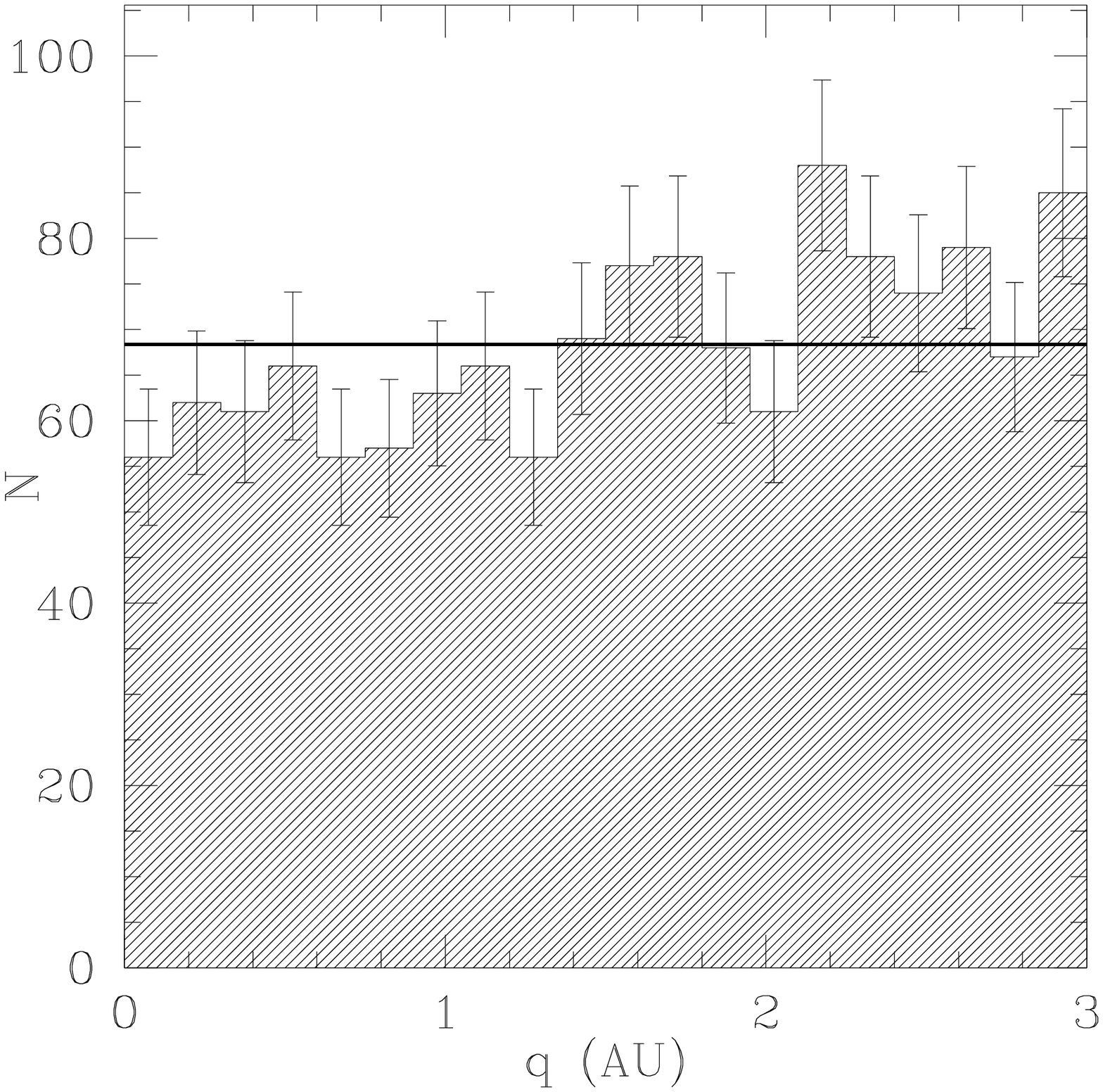,height=3in}}}
\centerline{\hspace*{1.45in}$(a)$\hfill$(b)$\hspace{1.3in}}
\caption{The $V_1$ or new comets: (a) perihelion distance $q$ versus original
semimajor axis $a$; (b) number as a function of perihelion
distance.}
\label{fi:V1_q}
\end{figure}

A scatter plot of perihelion distance versus original semimajor axis
for the $V_1$ comets is shown in Fig.~\ref{fi:V1_q}a. There is a sharp
lower bound to the distribution of semimajor axes for comets with
perihelion $q\lta 1\au$, which is due to the Jupiter barrier (\S
\ref{pa:lcylinder}). This lower bound shifts to smaller semimajor axes
at larger perihelion distances, since the
angular momentum ``hop'' over the Jupiter barrier is smaller. As a
result the number of $V_1$ comets as a function of perihelion distance
(Fig.~\ref{fi:V1_q}b) is approximately flat, as predicted by
Eq.~\ref{eq:qdist}, but slightly larger for large perihelion
distance. In comparison, the distribution of perihelion distances for
the observed new comets (Fig.~\ref{fi:new_i}a) is not flat, but this
is probably a result of the strong selection effects acting against
comets with large perihelion.

\begin{figure}[p]
\centerline{\hbox{\psfig{figure=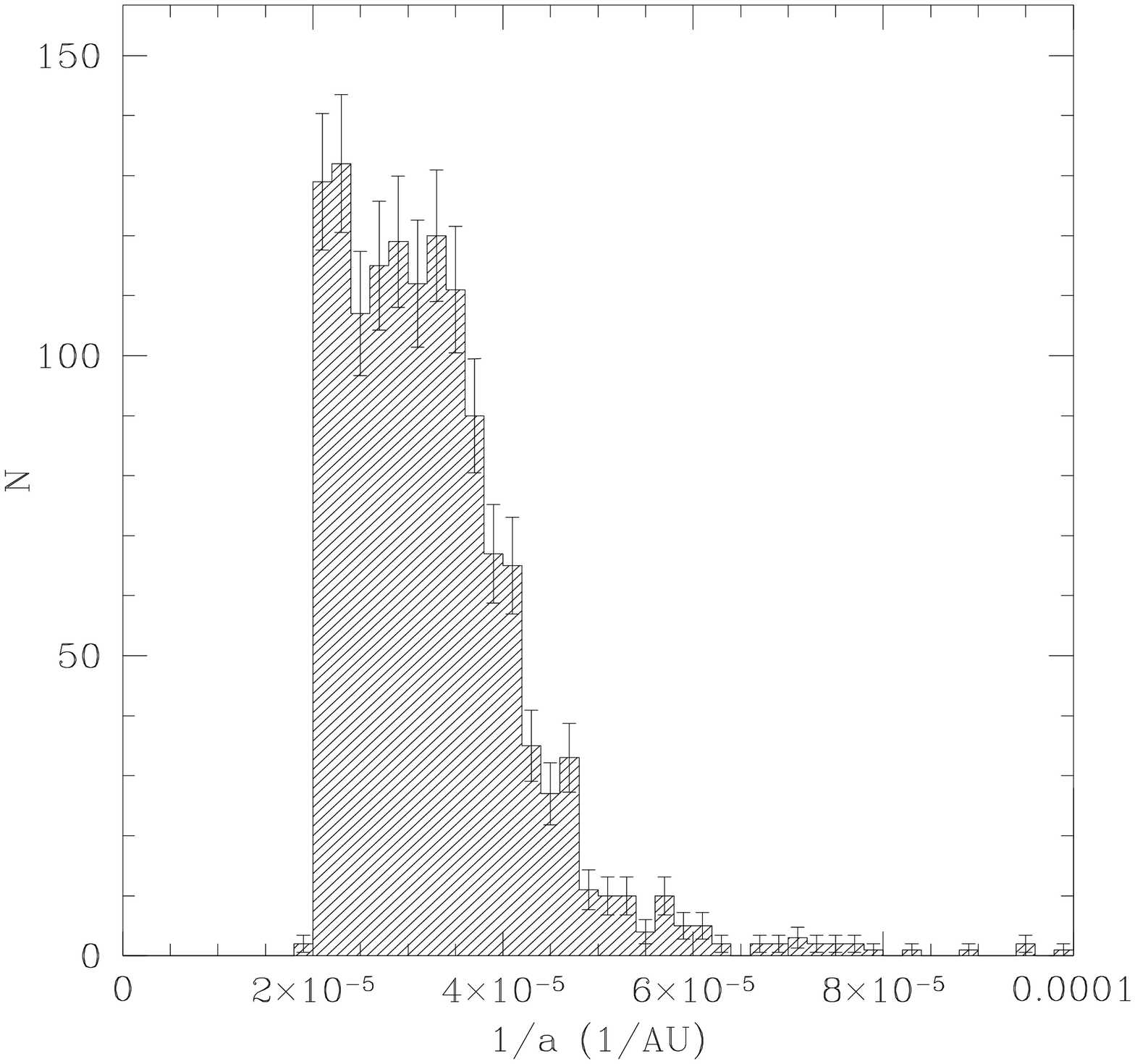,height=3in}
                  \psfig{figure=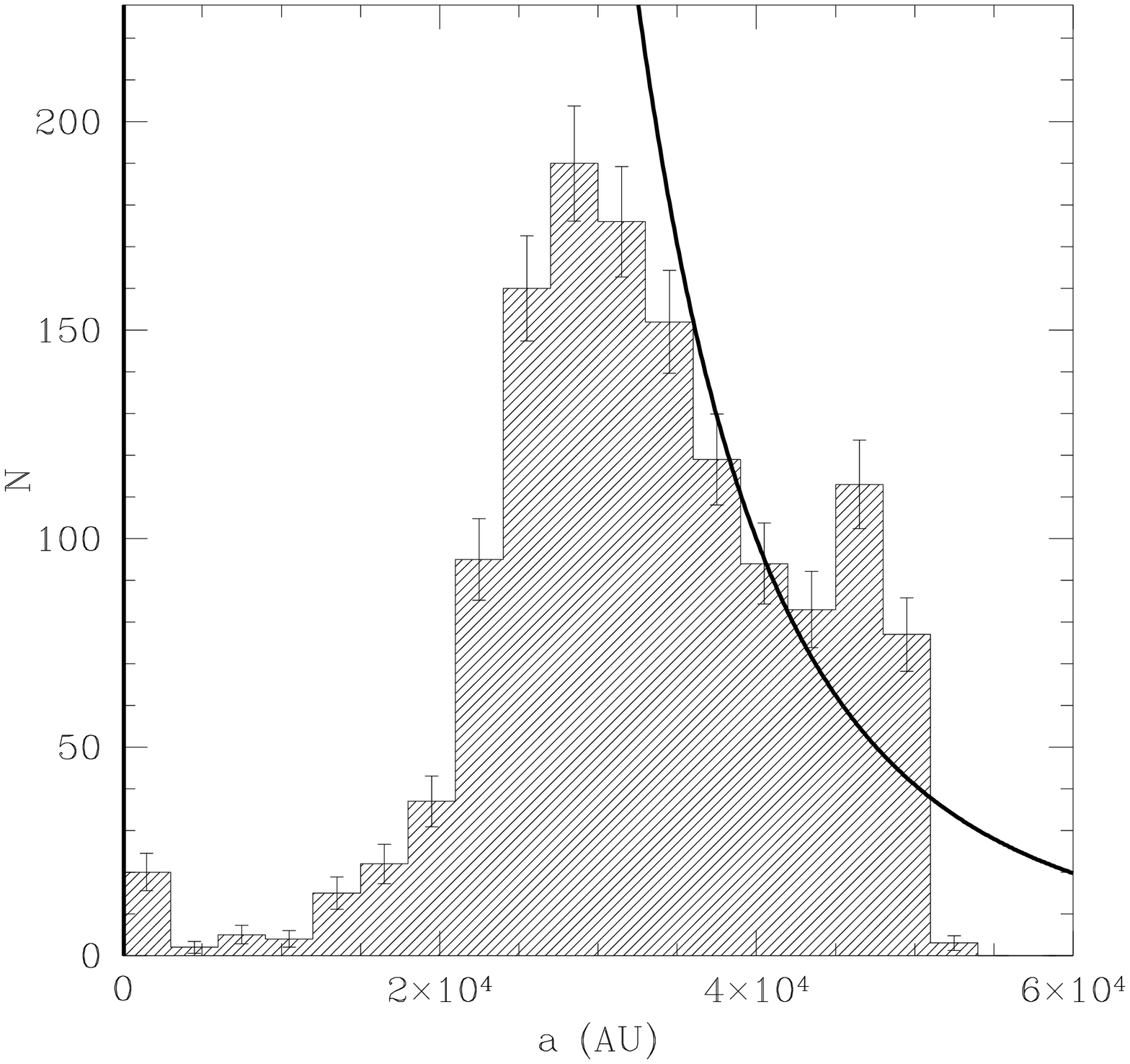,height=3in}}}
\centerline{\hspace*{1.45in}$(a)$\hfill$(b)$\hspace{1.3in}}
\caption{Distribution of original energies $x=1/a$ and semimajor axes
$a$ for the $V_1$ comets. An additional 28 comets, 2\% of the total,
have $x>10^{-4}\aui$. The curve in (b) is an analytical approximation
to the expected distribution when the loss cylinder is full, derived
in \protect\cite{wie96}. }
\label{fi:V1_b}
\end{figure}

The distribution of original semimajor axes of the $V_1$ comets is
shown in Fig.~\ref{fi:V1_b}a.  The cutoff at $1/a=2 \times
10^{-5}$~{\aui} or $a=50~000\au$ is an artifact of our choice of a
sharp outer boundary for the Oort cloud at this point
(\S\ref{sec:oor}). All but 2\% of the new comets have original
energies in the range $0<x<10^{-4}\aui$, consistent with our
assumption that observed comets in this range are new comets.  The
mean energy of the $V_1$ comets is $\langle1/a\rangle=3.3 \pm 1 \times
10^{-5}$~{\aui}, in good agreement with Heisler's (1990)
\nocite{hei90} estimate of $3.55 \times 10^{-5}~\aui$ outside of
showers. Heisler's Monte Carlo simulations included both the Galactic
tide and passing stars; the agreement confirms that our omission of
stellar perturbers does not strongly bias the distribution of new
comets.

The curve in Fig.~\ref{fi:V1_b}b shows an analytical approximation to
the expected flux of new comets when the loss cylinder is full
\cite[]{wie96}. The agreement between the analytical curve and the
distribution of $V_1$ comets for $a\gta 30~000\au$ confirms that the
inner edge of the distribution of new comets is caused by the emptying
of the loss cylinder as the semimajor axis decreases. The source of
the smaller peak at 47~000~{\au} is unclear: if the sample is split
into two parts, it appears only in one half, and thus may be a
statistical fluke even though the deviation from the analytical curve
is several times the error bars. In any event it is unlikely to play a
significant role in determining the overall distribution of LP comets
for two reasons: first, only a few percent of the $V_1$ comets are
involved in the peak; and second the subsequent planet-dominated
evolution of the $V_1$ comets is relatively insensitive to the comets'
original semimajor axes.

\begin{figure}[p]
\centerline{\hbox{\psfig{figure=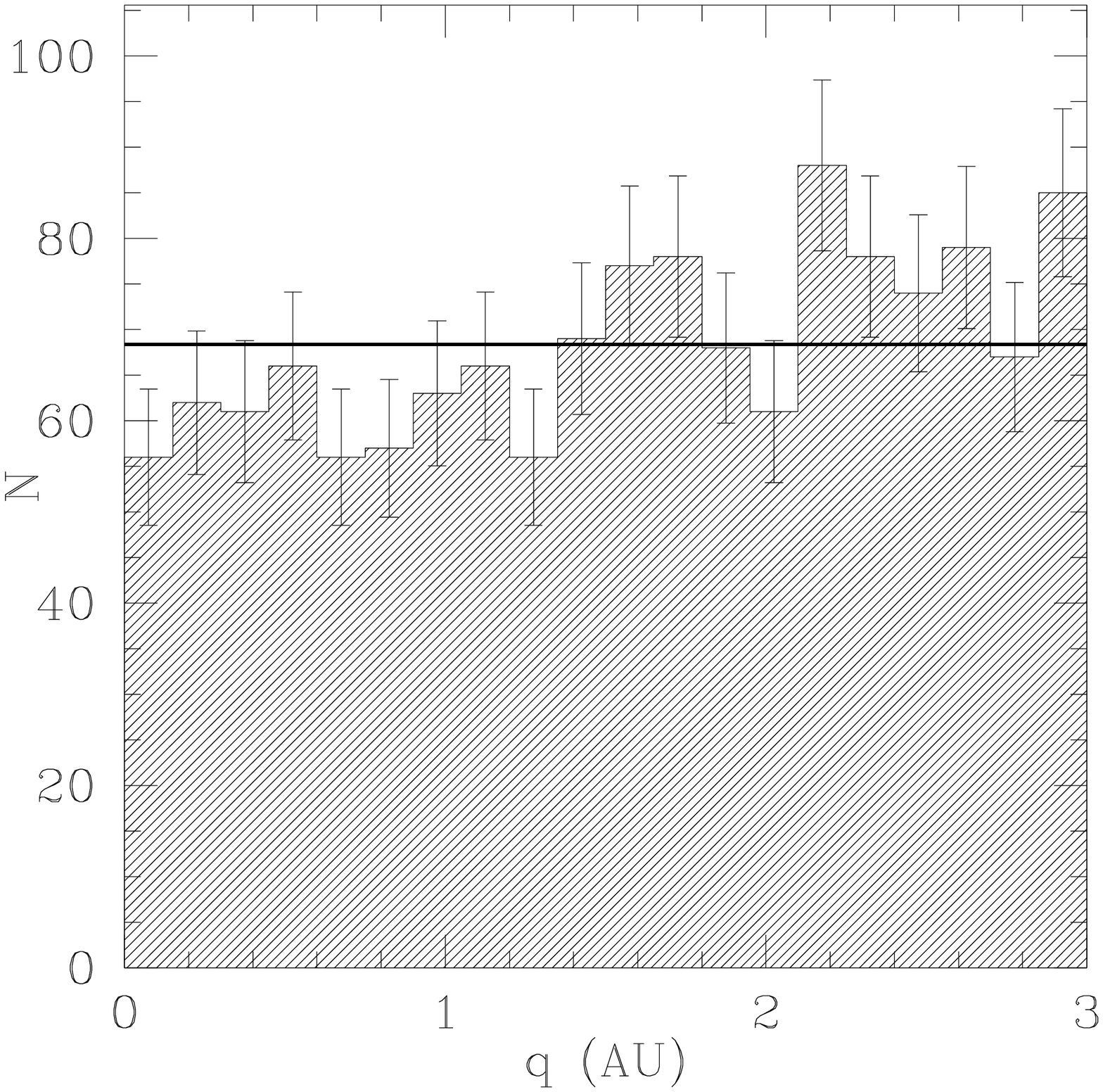,height=3in}
                  \psfig{figure=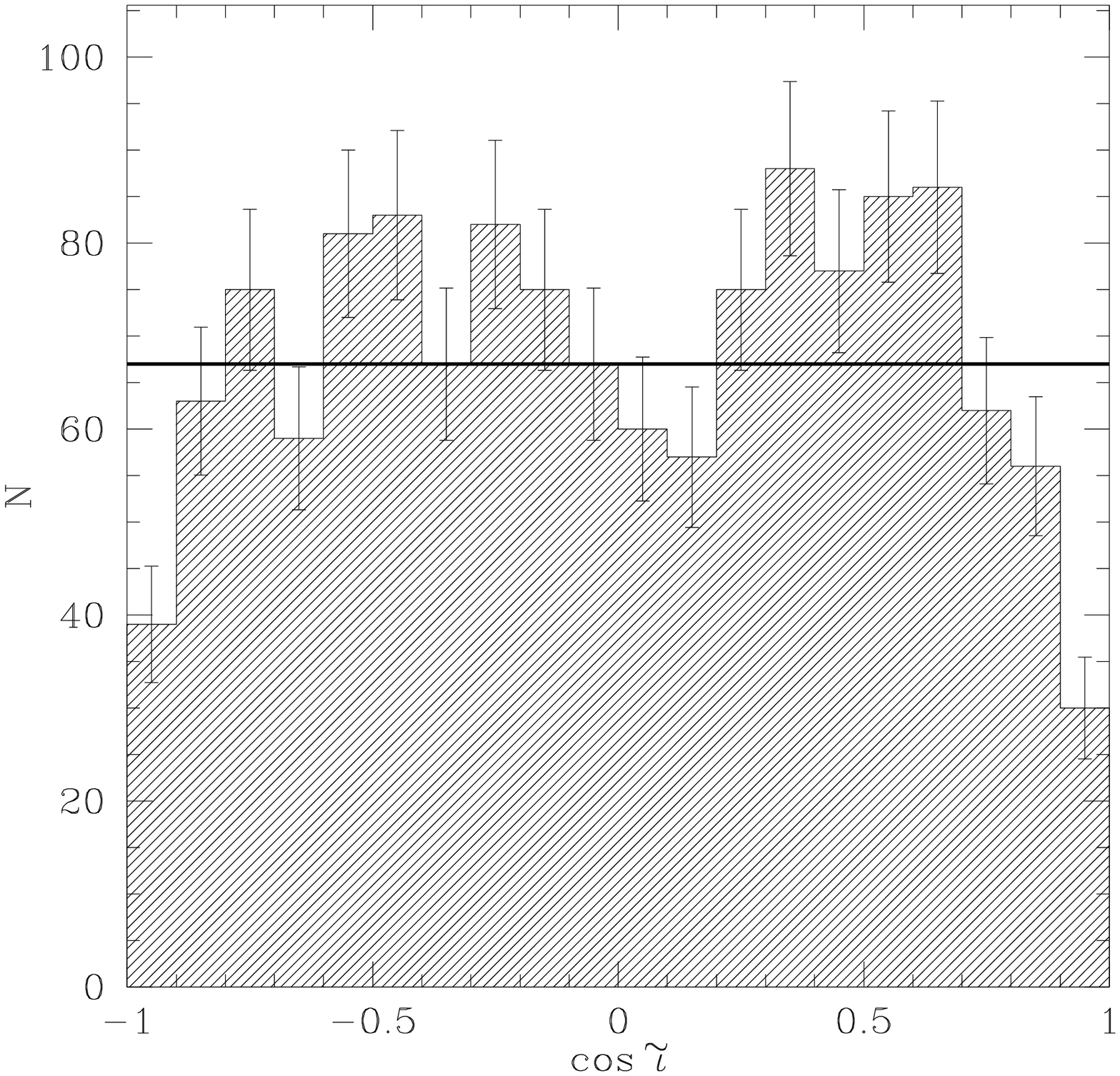,height=3in}}}
\centerline{\hspace*{1.45in}$(a)$\hfill$(b)$\hspace{1.3in}}
\centerline{\hbox{\psfig{figure=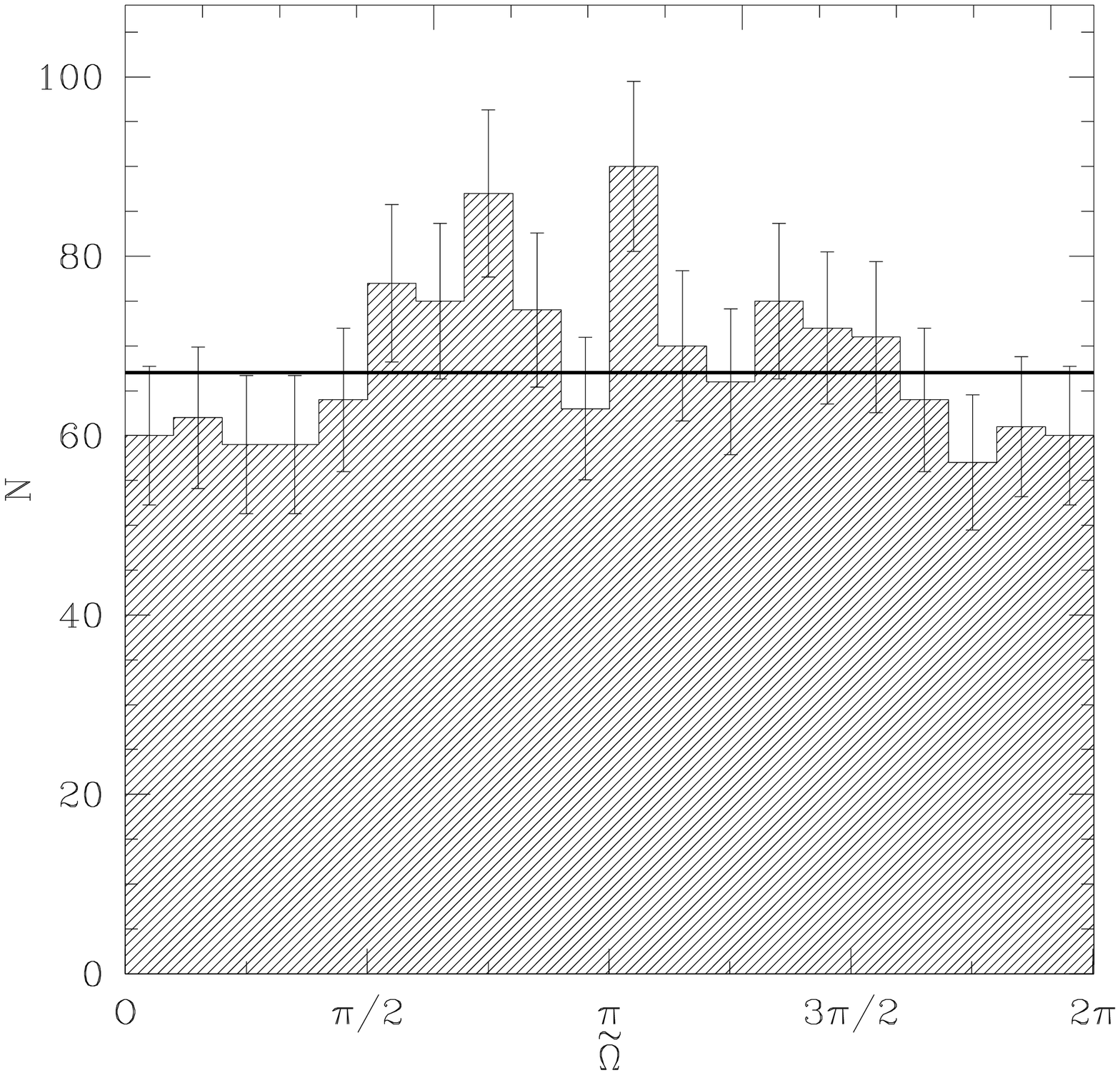,height=3in}
                  \psfig{figure=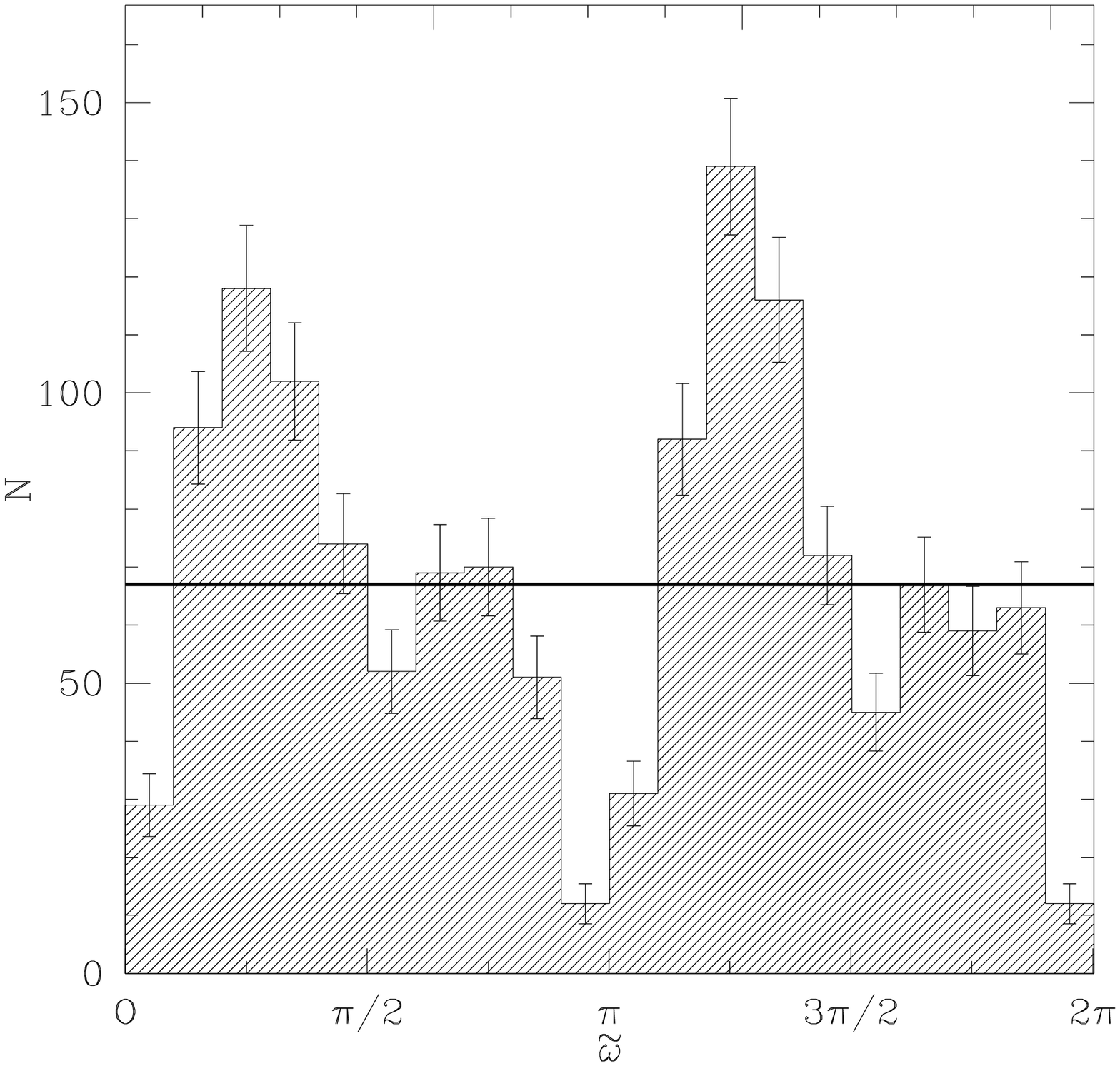,height=3in}}}
\centerline{\hspace*{1.45in}$(c)$\hfill$(d)$\hspace{1.3in}}
\caption[]{Distribution of orbital elements for the $V_1$ comets: (a)
perihelion distance; (b) inclination; (c) longitude of ascending node;
(d) argument of perihelion. All angular elements are measured in the
Galactic frame when the comet passes 200~AU on its inbound leg.}
\label{fi:V1_i}
\end{figure}

\begin{figure}[p]
\centerline{\vbox{\hbox{
            \psfig{figure=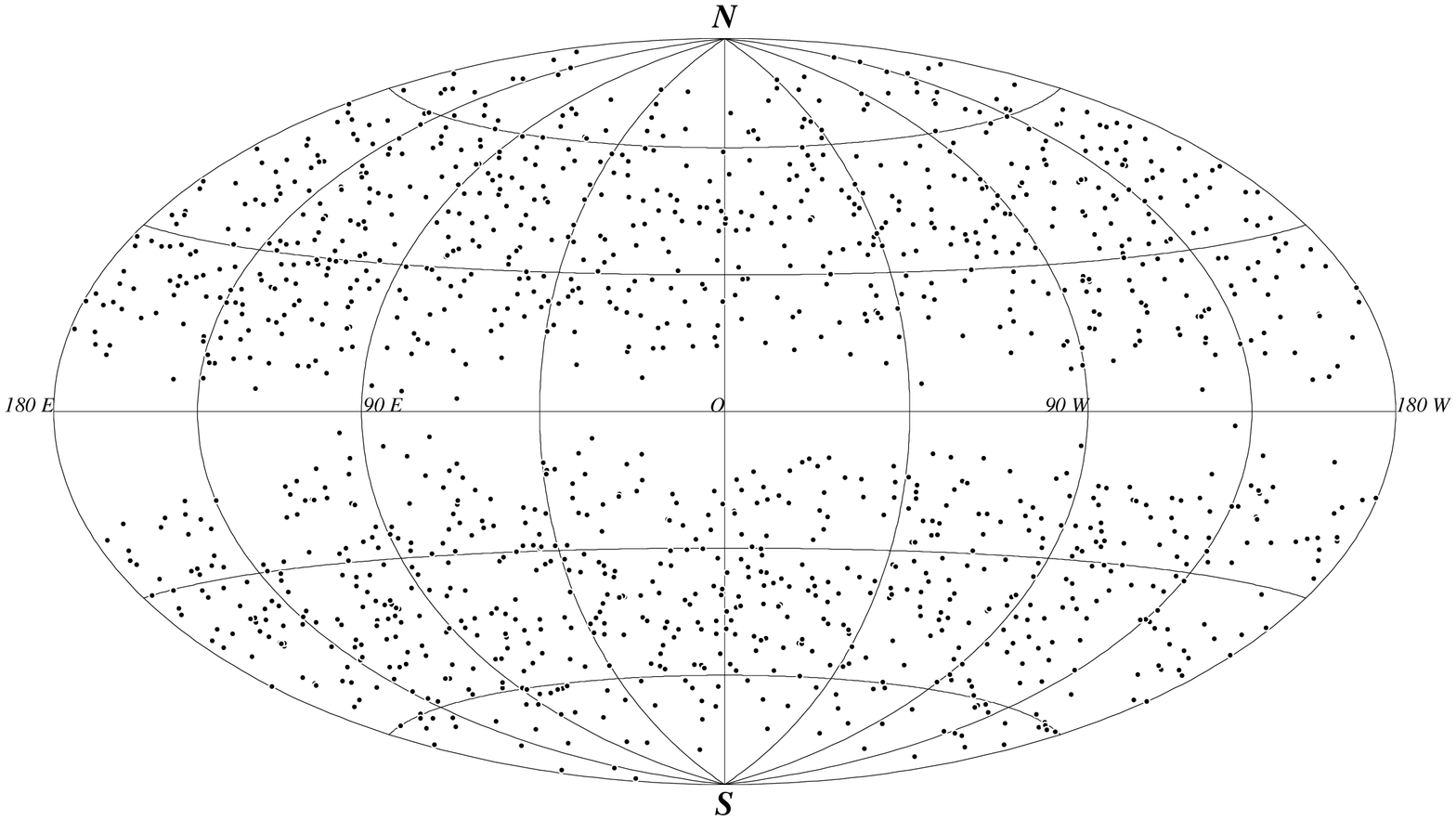,angle=270,height=3in}$(b)$}}}
\caption{Equal-area plot of the aphelion directions of the $V_1$
comets in the Galactic frame. More precisely, we have plotted the
antipode of the perihelion direction, since this is what is
observable.}
\label{fi:V1_ap}
\end{figure}

The distributions of perihelion and angular orbital elements for the
$V_1$ comets are shown in Fig.~\ref{fi:V1_i}, which can be compared to
the observed distributions in Fig. \ref{fi:new_i}.  The observed
perihelion distribution is strongly affected by selection effects, so
no comparison is practical there. The angular element distributions
are reasonably consistent between the two figures; in particular the
$\og$ distributions both show peaks in the regions where $\sin 2 \og
>0$, reflecting the role of the Galactic tide in creating new comets.

The aphelion directions of the $V_1$ comets are shown in
Fig.~\ref{fi:V1_ap}, which can be compared to the observed
distribution in Fig.~\ref{fi:new_ap}.  The most striking feature in
Fig.~\ref{fi:V1_ap} is the concentration towards mid-Galactic
latitudes, again pointing to the importance of the Galactic tide as a
LP comet injector.  The real distribution of aphelion directions is
expected to be much clumpier, due to the injection of comets by
passing stars; however, the number of new comets in
Fig.~\ref{fi:new_ap} is too small for any reliable comparisons to be
made.

\subsubsection{The longest-lived comets} \label{pa:overmax}

\begin{figure}[p]
\centerline{\hbox{\psfig{figure=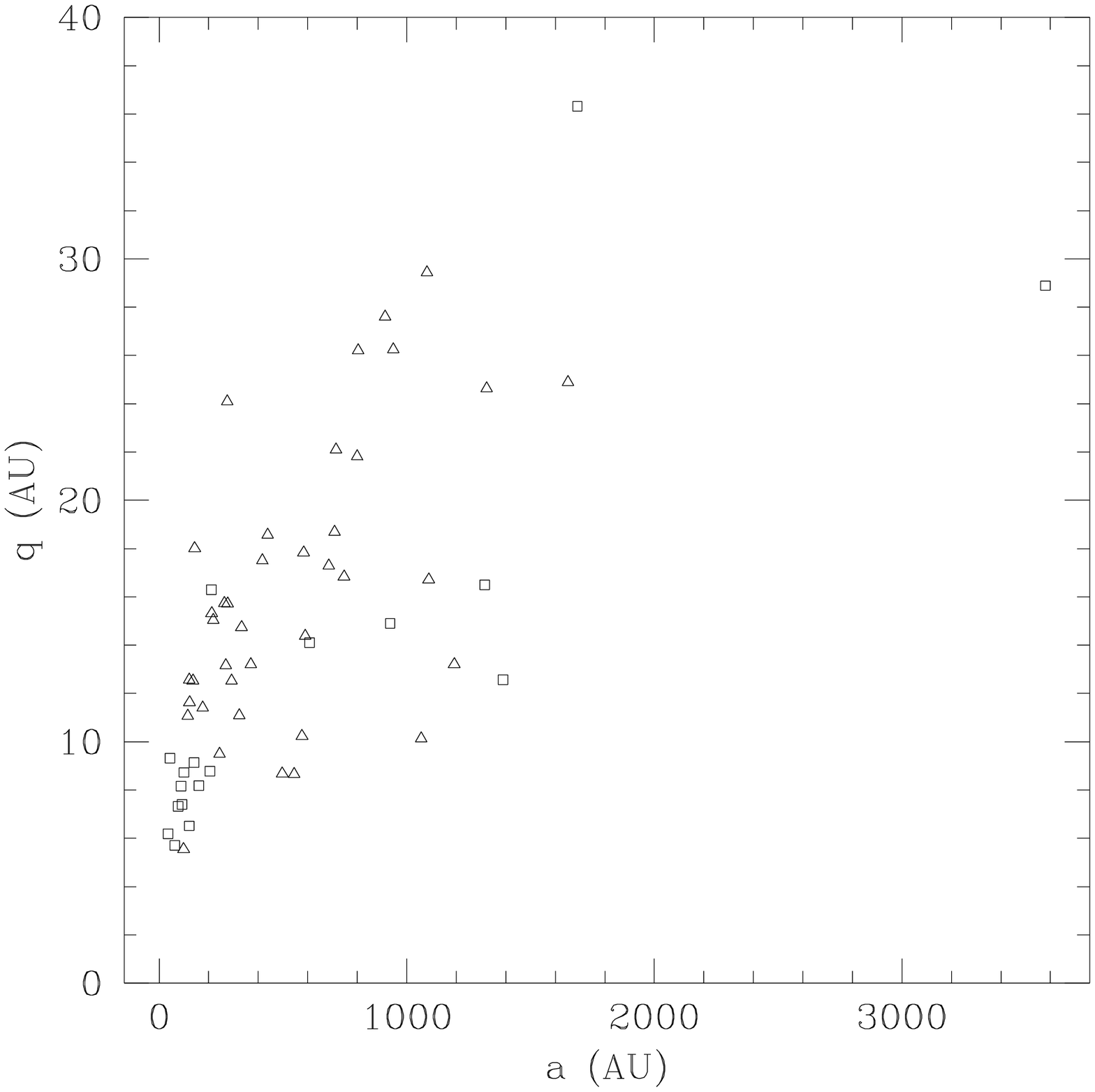,height=3in}
                  \psfig{figure=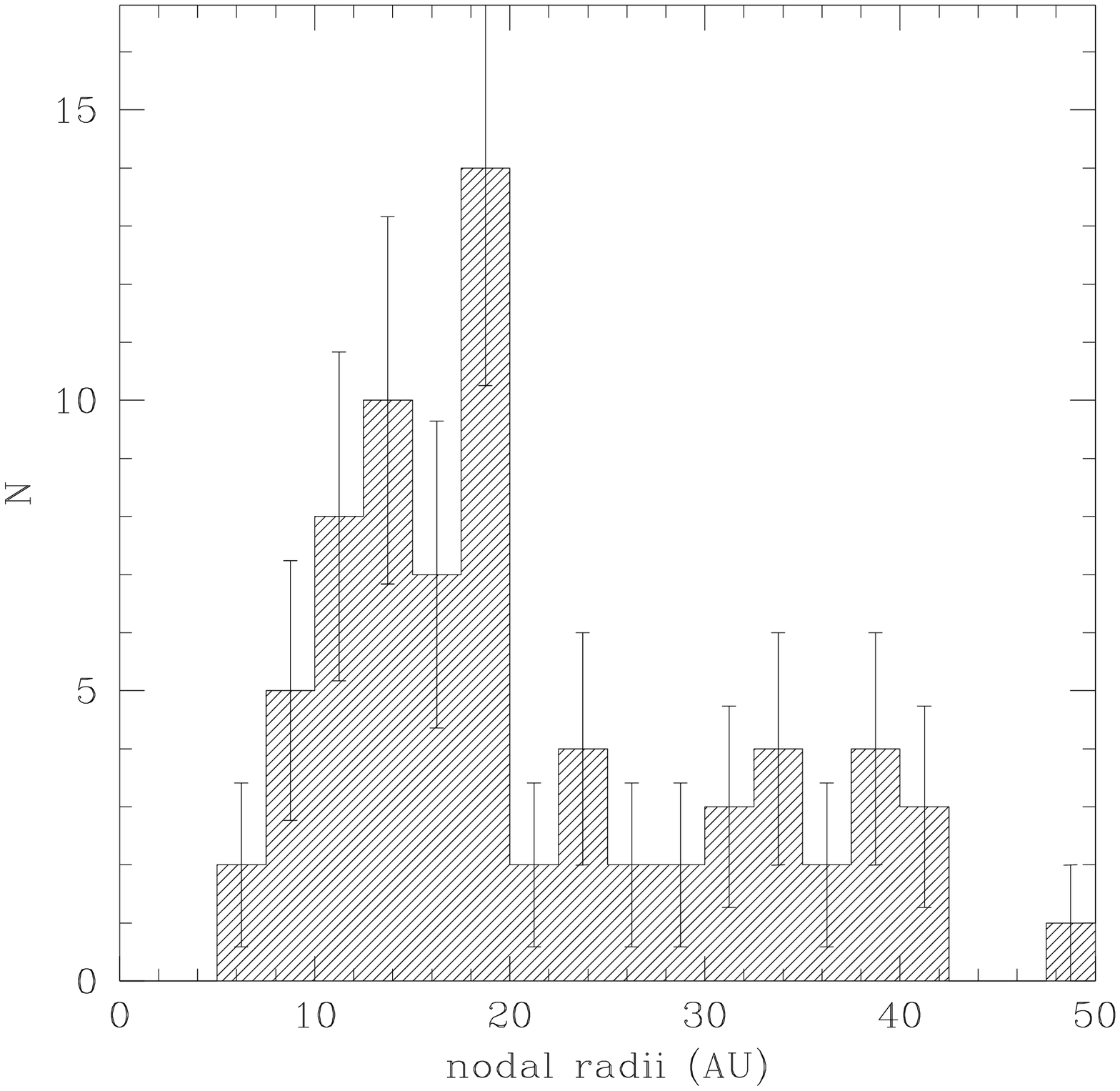,height=3in}}}
\caption{For the 57 comets that survived 5000 orbits, ($a$) their perihelion distance $q$ versus semimajor axis $a$, and ($b$) the distances of their nodes.  In ($a$), triangles are prograde comets, squares, retrograde.}
\label{fi:Cm_elems}
\end{figure}

Although most comets reach one of the end-states within a few orbits
(see Table \ref{ta:V1_es}) a small fraction survive for much longer
times: 57 of the 20~286 initial comets in our simulation triggered the
{\sc Exceeded Orbit Limit} flag after 5000 orbits. The population of
these comets decays only very slowly and their fate cannot be
determined without prohibitive expenditures of CPU time.  The
perihelion distances and semimajor axes of these comets on their
5000$^{th}$ orbit are indicated in Fig.~\ref{fi:Cm_elems}. Also
shown is the distance at which they cross the ecliptic. Most have nodes and 
perihelia outside Saturn's orbit, where the energy perturbations are
relatively small.

\subsection{Post-visibility evolution: the standard model} \label{pa:Vm}

We now follow the orbits of the $V_1$ comets forward in time until they reach
one of the end-states (obviously, the {\sc visible} end-state is disabled in
these simulations). Each time one of these comets makes an apparition its
orbital elements are added to the set of $V_\infty$ comets. The $V_\infty$
comets are to be compared to the observed distribution of LP comets.

The errors in the distribution of elements of the $V_\infty$ comets are not
Poisson, as a single comet may contribute hundreds or thousands of
apparitions.  The errors that we quote and show in the figures are determined
instead by bootstrap estimation \cite[]{efr82,nr2}.

The ``standard model'' simulation of post-visibility evolution has no fading,
and no perturbers except the giant planets and the Galactic tide.

The distribution of end-states for the standard model is shown in
Table~\ref{ta:Vm_es}. The {\sc Exceeded orbit limit} end-state
(\S~\ref{pa:endstates}) is invoked after 10~000 orbits for these simulations,
but no comets reach this end-state. The mean lifetime is 45.3 orbits, compared
to 60 predicted by the gambler's ruin model (Eq.~\ref{eq:mnlf}).  Ejection by
the giant planets is by far the most common end-state (89\% of $V_1$
comets). Most of the remaining comets (about 8\% of the total) move back out
to large perihelion distances. Their median energy when they reach this
end-state is given by $1/a=4 \times 10^{-5}$~{\aui} ($a = 25~000$~\au); in
other words these comets have suffered relatively small energy perturbations
and remain in the outer Oort cloud.

The distribution of orbital elements of the $V_\infty$ comets may be
parametrized by the dimensionless ratios $X_i$ defined in
Eq.~\ref{eq:xdef}: the ratio of theoretical parameters $\Psi_i^t$ for
the standard model to the observed parameters (Eq.~\ref{eq:psiobs}) is
\begin{equation}
X_1 = {\Psi_1^t\over\Psi_1}=0.075 \pm 0.011,\qquad X_2={\Psi_2^t\over\Psi_2}=4.4 \pm 1.2,\qquad X_3 = {\Psi_3^t\over\Psi_3}= 0.61 \pm 0.13.
\label{eq:xstd}  
\end{equation}
The standard model agrees much better with the predictions of the simple
gambler's ruin model ($X_1=0.05$, $X_3=0.58$, see eqs. \ref{eq:xonegr}\ and
\ref{eq:xthrgr}) than it does with the observations ($X_i=1$).

\begin{table}[p]
\centerline{
\begin{tabular}{|l|rrr|r|} \hline
End-state & Ejection &  Large $q$ & Short pd. & Total \\ \hline \hline
%                    ej     lq     sp   
%
Number        & 1223  & 109   & 36    & 1368 \\
Fraction      & 0.894 & 0.080 & 0.026 & 1.000 \\ 
Minimum $t_x$ & 0.296 & 2.61  & 0.014 & 0.014 \\ 
Median $t_x$  & 1.33  & 4.62  & 0.67  & 1.40 \\ 
Maximum $t_x$ & 31.7  & 71.0  & 7.94  & 71.0 \\ 
Minimum $m_x$ & 1     & 1     & 13    & 1 \\ 
Median $m_x$  & 1     & 2     & 330   & 1 \\ 
Maximum $m_x$ & 5832  & 2158  & 4277  & 5832 \\ \hline
\end{tabular}}
\caption{The distribution of end-states of the $V_1$ comets in the
standard model. The minimum, median and maximum lifetimes $t_x$ of
these comets are measured from their first apparition in Myr. 
 No comets suffer collisions with the planets or Sun, or
survive for the age of the Solar System.}
\label{ta:Vm_es}
\end{table}

\begin{figure}[p]
\centerline{\hbox{\psfig{figure=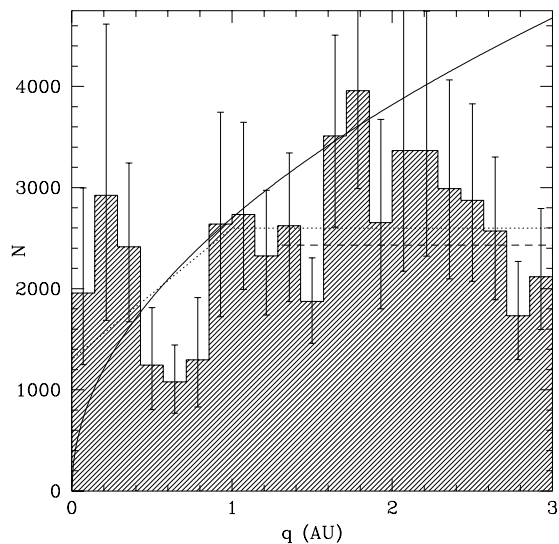,height=3in}}}
\caption{Distribution of perihelion distances $q$ for the $V_\infty$
comets in the standard model.  Error bars are determined from
bootstrap estimators and represent one standard deviation. 
The curves are Everhart's
(1967a, dotted line), Kres\'ak and Pittich's (1982, solid line) and
Shoemaker and Wolfe's (1982, dashed line) estimates of the intrinsic
perihelion distribution.  The correct normalizations are unclear, and
have been made somewhat arbitrarily.}
\nocite{eve67a,kre82,showol82}
\label{fi:Vm_q}
\end{figure}

The perihelion distribution of the $V_\infty$ comets in the standard model is
shown in Fig.~\ref{fi:Vm_q}. Although the figure represents 52~303
apparitions, the error bars---as determined by bootstrap---remain large,
reflecting strong contributions from a few long-lived comets: over 45\% of the
apparitions are due to the 12 comets that survive for 1000 or more orbits
after their first apparition.  This figure can be compared to the observed
perihelion distribution (Fig.~\ref{fi:lp_q}), which however reflects the
strong selection effects favouring objects near the Sun or the Earth.  We note
that not all perihelion passages made by comets after their first apparition
are visible: in addition to the 52~303 apparitions made by the $V_\infty$
comets, there were 9561 perihelion passages with $q > 3\au$.

Let the total number of comets with perihelia in the range $[q,q+dq]$
be $N(q)dq$. A linear least-squares fit to Fig.~\ref{fi:Vm_q} yields
$N(q)$ roughly proportional to $1 + (q/1\au)$, similar to Everhart's
(1967a) \nocite{eve67a} earlier estimate of the intrinsic perihelion
distribution. The perihelion distribution is not flat, as would be
expected if the distribution were uniform on the energy hypersurface
(Eq.~\ref{eq:qdist}). The simulations are noisy enough to be
consistent with any number of slowly varying functions of perihelion
over $0 < q < 3$~\au, possibly including $N(q)\propto q^{1/2}$, as
proposed by \cite{krepit78}. The estimates of the intrinsic perihelion
distribution of LP comets published by
Everhart, by Kres\'ak and Pittich, and by Shoemaker and Wolfe are indicated
on Fig.~\ref{fi:Vm_q}.

\begin{figure}[p]
\centerline{\hbox{\psfig{figure=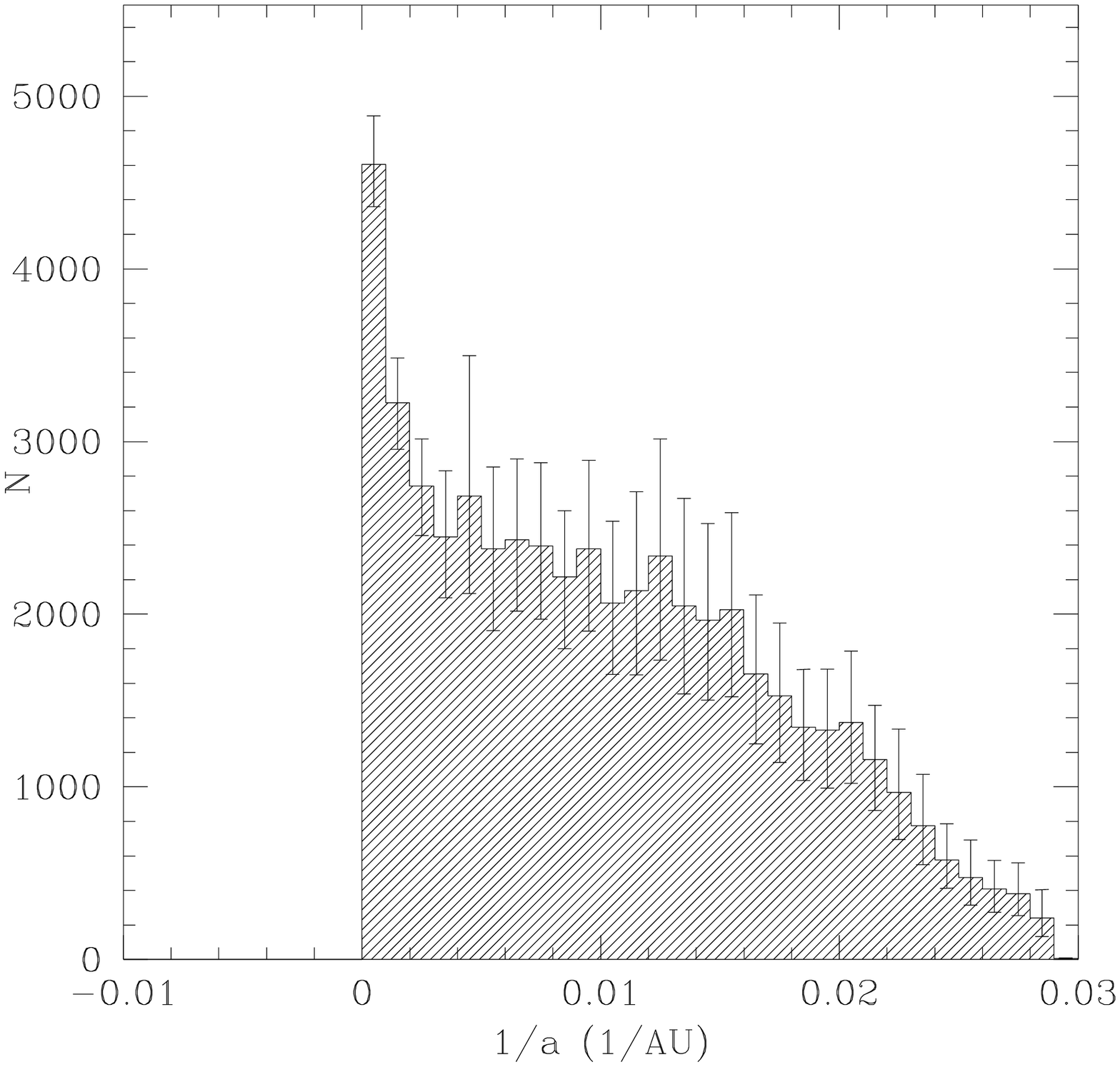,height=3in}
                  \psfig{figure=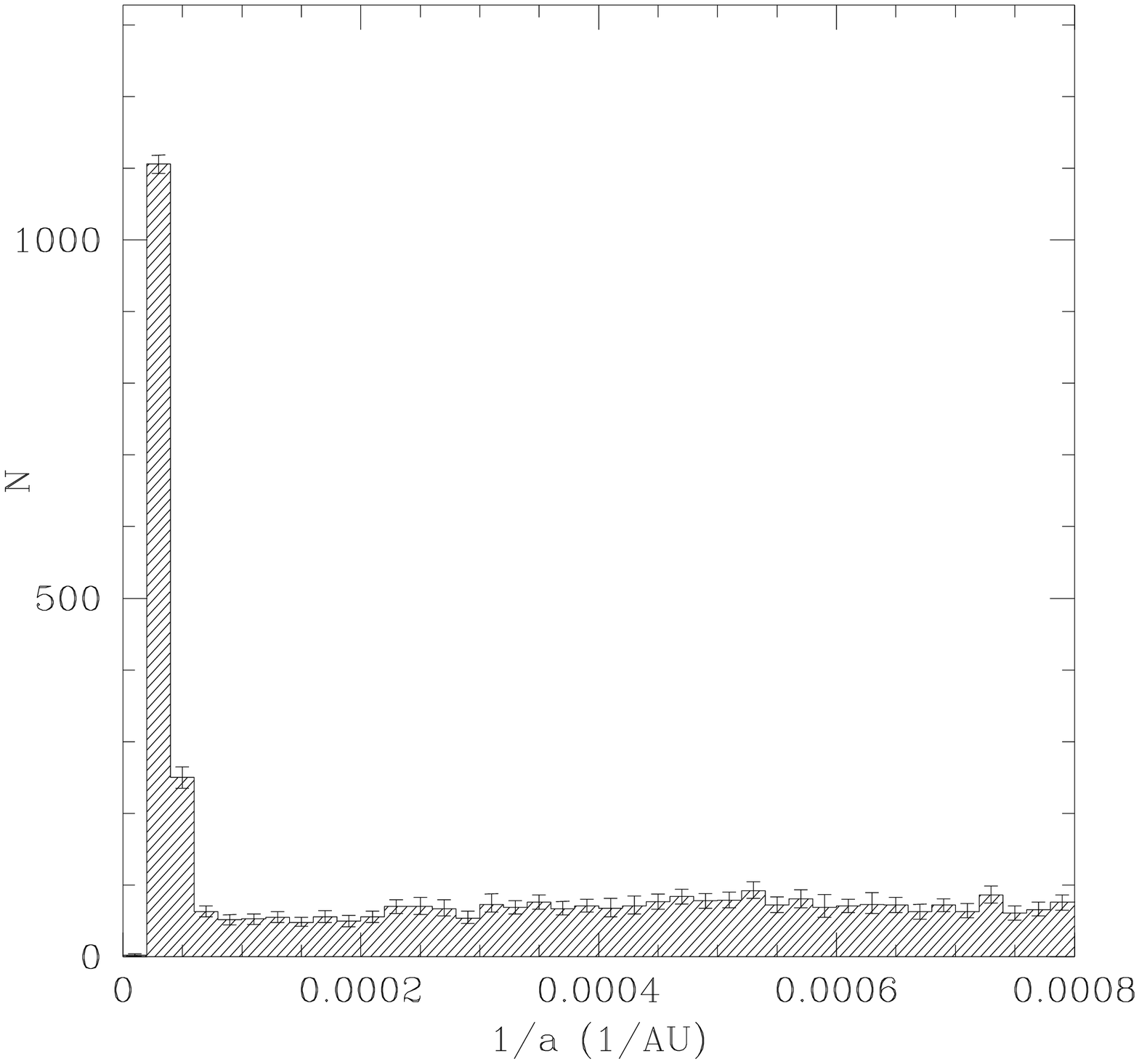,height=3in}}}
\centerline{\hspace*{1.45in}$(a)$\hfill$(b)$\hspace{1.3in}}
\caption{Distribution of original energies for the $V_\infty$ comets
in the standard model for all 52~303 apparitions ($q < 3$~\au).}
\label{fi:Vm_b}
\end{figure}

The original energy distribution of the $V_\infty$ comets in the
standard model is shown in Fig.~\ref{fi:Vm_b}, at two different
magnifications, for all 52~303 apparitions.  These figures should be
compared with the observations shown in Fig. \ref{fi:energy}. As
already indicated by the statistic $X_1$ (Eq.~\ref{eq:xstd}), the
standard model has far too many LP comets relative to the number of
comets in the Oort spike: the simulation produces 35 visible LP comets
for each comet in the spike, whereas in the observed sample the ratio
is 3:1.  This disagreement is at the heart of the fading problem: how
can the loss of over 90\% of the older LP comets be explained?

These simulations allow us to estimate the contamination of the Oort spike by
dynamically older comets. There are 1368 $V_1$ comets, of which 1340 have $1/a
< 10^{-4}\aui$, but a total of 1475 apparitions are made in this energy range
in the standard model. Thus roughly 7\% of comets in the Oort spike are not
dynamically new. Of course, this estimate neglects fading, which would further
decrease the contamination of the Oort spike by older comets.

\begin{figure}[p]
\centerline{\hbox{\psfig{figure=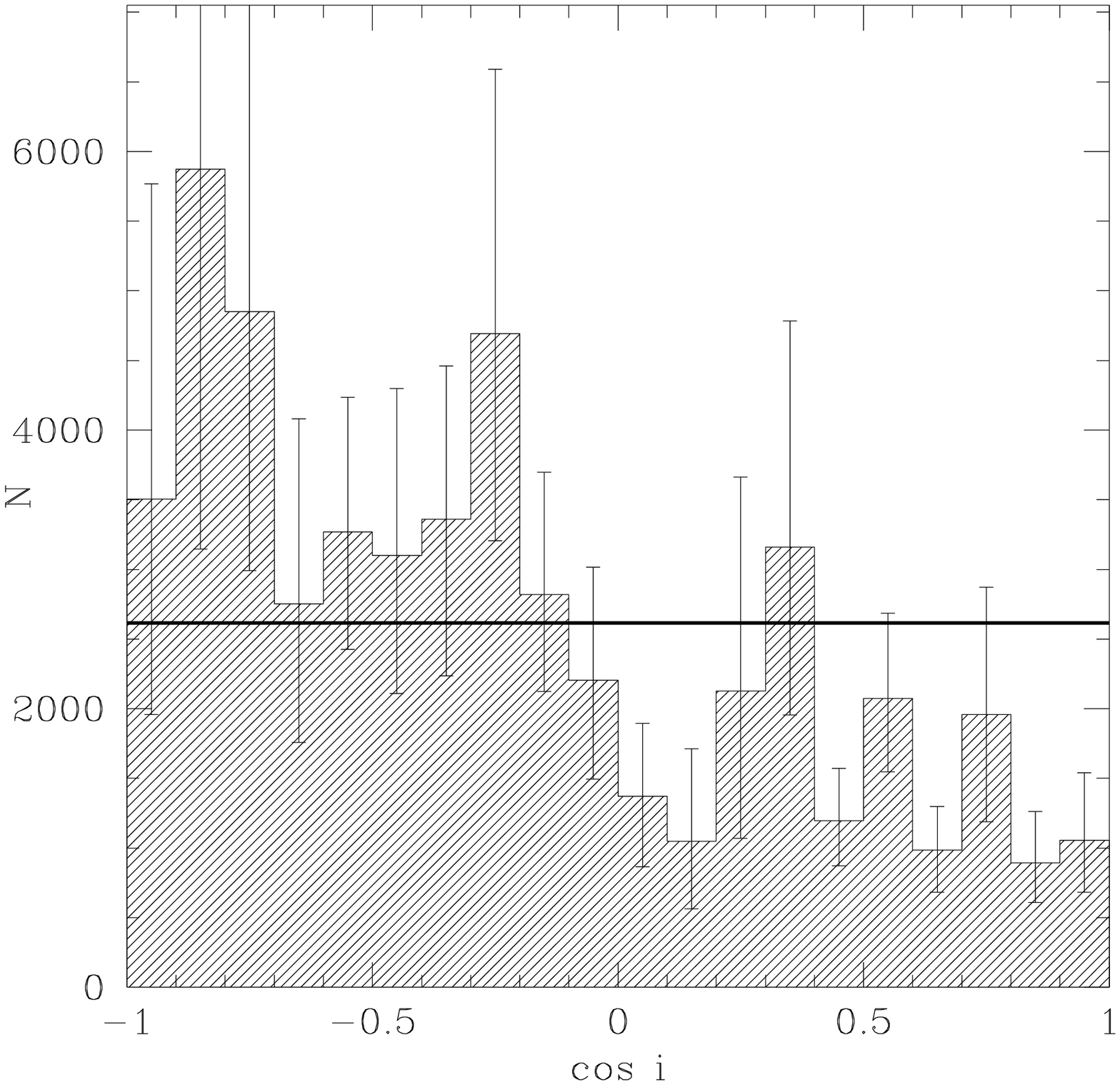,height=3in}
                  \psfig{figure=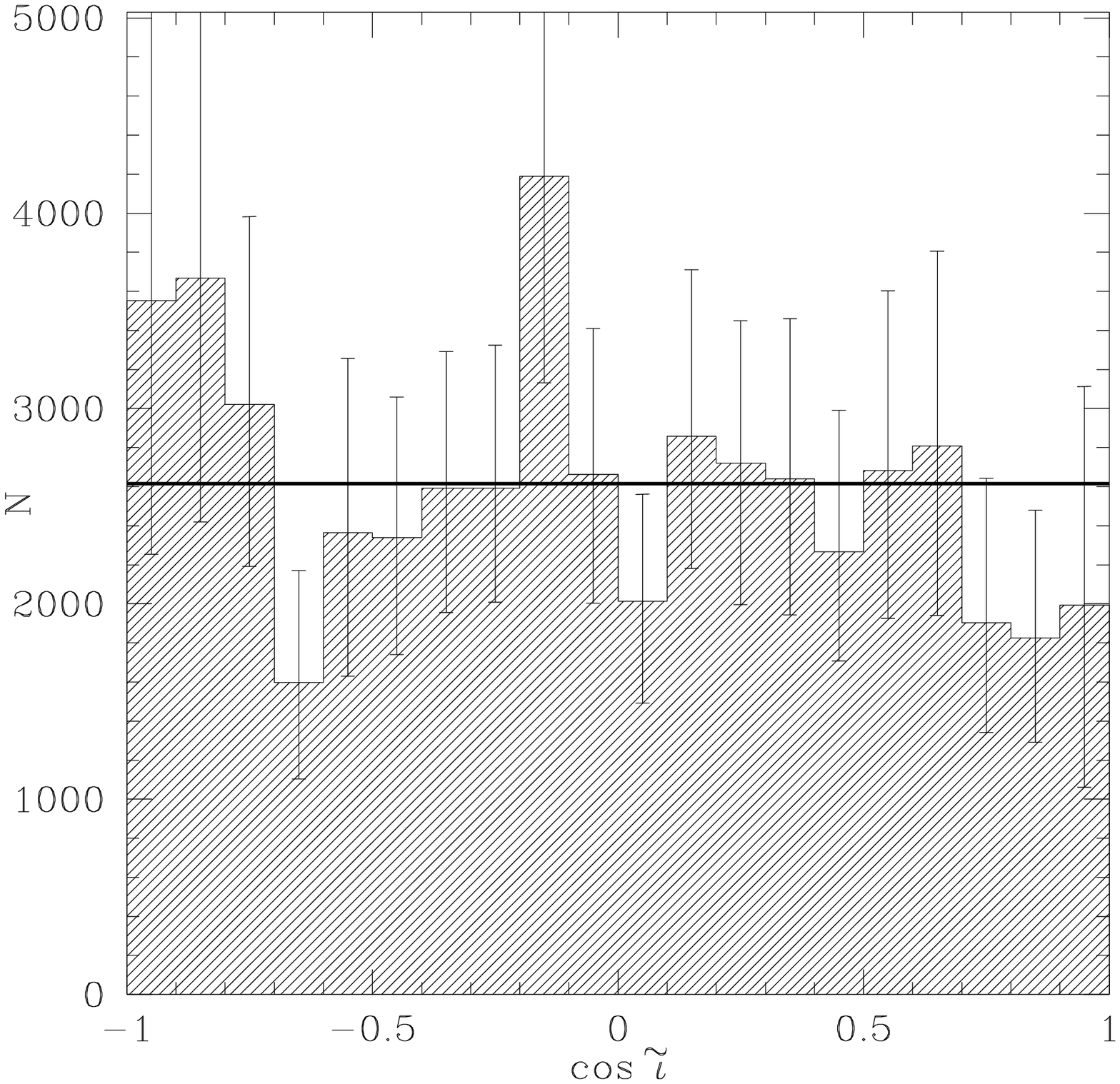,height=3in}}}
\centerline{\hspace*{1.45in}$(a)$\hfill$(b)$\hspace{1.3in}}
\caption{Distribution of the cosine of the inclination for the
$V_\infty$ comets in the standard model, (a) at perihelion in the
ecliptic frame, and (b) at 200~{\au} on the inbound leg in the
Galactic frame. The heavy line indicates a uniform distribution.}
\label{fi:Vm_i}
\end{figure}

Figure~\ref{fi:Vm_i} shows the inclination distribution of the $V_\infty$
comets in the standard model. There is a noticeable excess of comets in
ecliptic retrograde orbits: the fraction on prograde orbits is $15875/52303
\approx 0.3$. This is inconsistent with observations, which show an isotropic
distribution (Fig.~\ref{fi:cosi}a), but consistent with the predictions of
the gambler's ruin model (Eq.~\ref{eq:xthrgr}).

\begin{figure}[p]
\centerline{\hbox{\psfig{figure=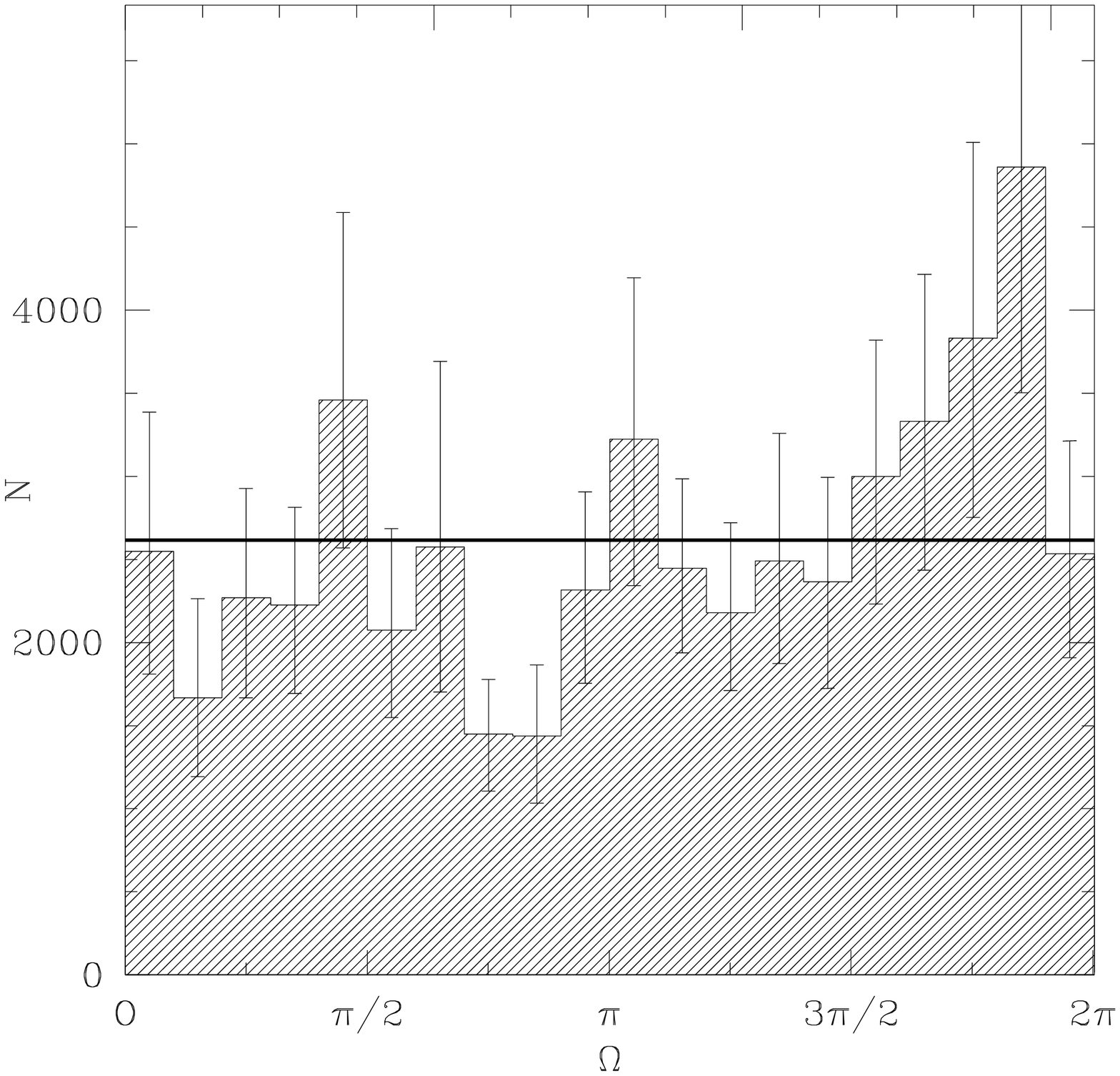,height=3in}
                  \psfig{figure=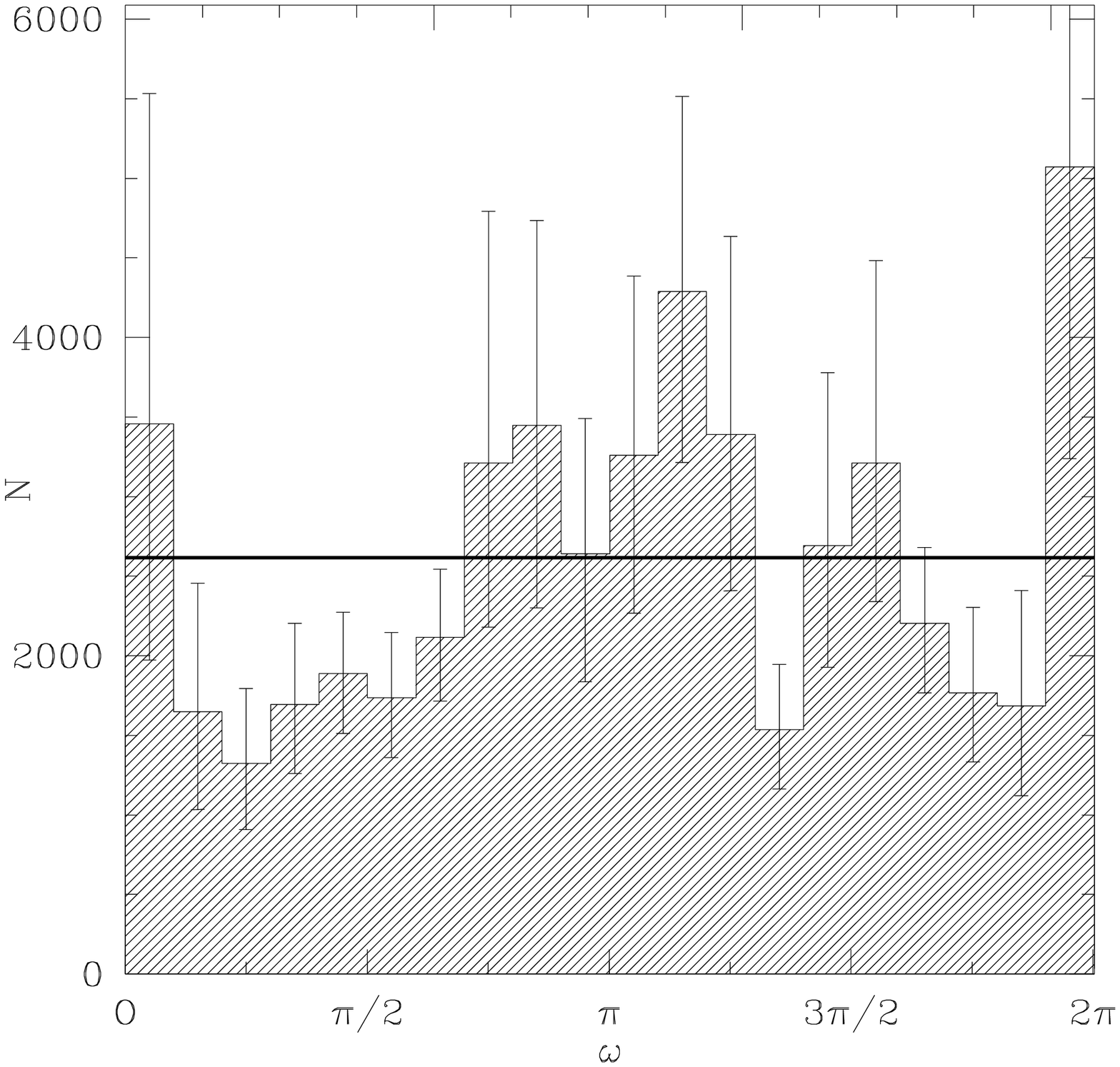,height=3in}}}
\centerline{\hspace*{1.45in}$(a)$\hfill$(b)$\hspace{1.3in}}
\caption{Distribution of the longitude of the
ascending node and the argument of perihelion for the $V_\infty$ comets in the
standard model. The elements are measured at perihelion in the ecliptic frame.
The heavy line indicates a uniform distribution.}
\label{fi:Vm_lasc}
\end{figure}

Figure~\ref{fi:Vm_lasc} shows the distribution of the longitude of the
ascending node and the argument of perihelion, in the ecliptic
frame. The large error bars suggest that the structure in these
figures is probably not statistically significant.

The principal conclusion from this analysis is that the standard model
provides a poor fit to the observed distribution of LP comets. The standard
model agrees much better with models based on a one-dimensional random walk,
suggesting that the basic assumptions of the analytic random-walk models in
\S\ref{sec:onedran}\ are valid. In \S\S \ref{sec:nongrav}--\ref{sec:nonparm}
we shall explore whether variants of the standard model can provide a better
match to the observations.

\subsubsection{Short-period comets from the Oort cloud} \label{pa:sp_comets}

\begin{figure}[p]
\centerline{\vbox{\hbox{\psfig{figure=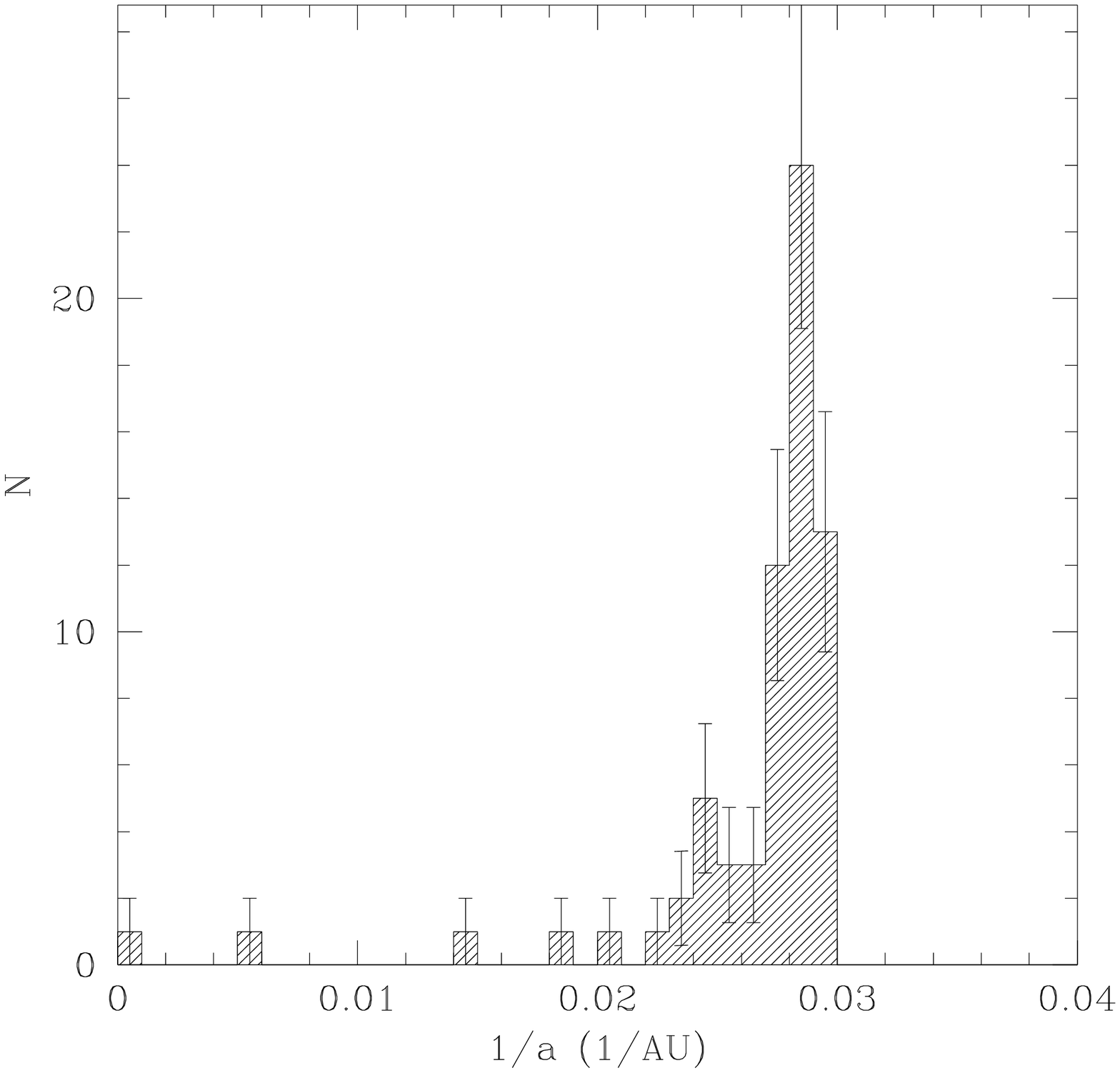,height=3in}
                        \psfig{figure=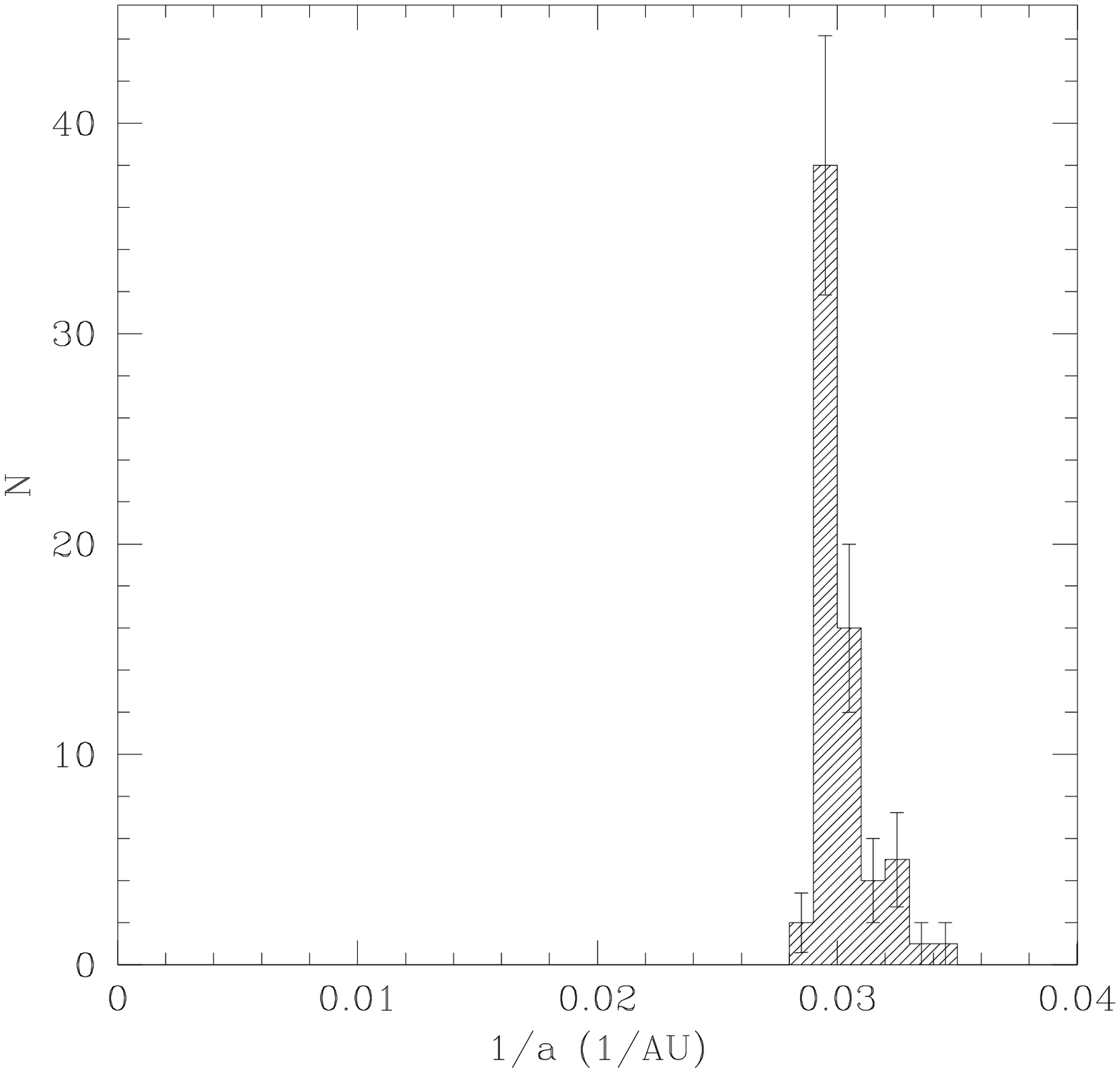,height=3in}}
                  \hbox{\psfig{figure=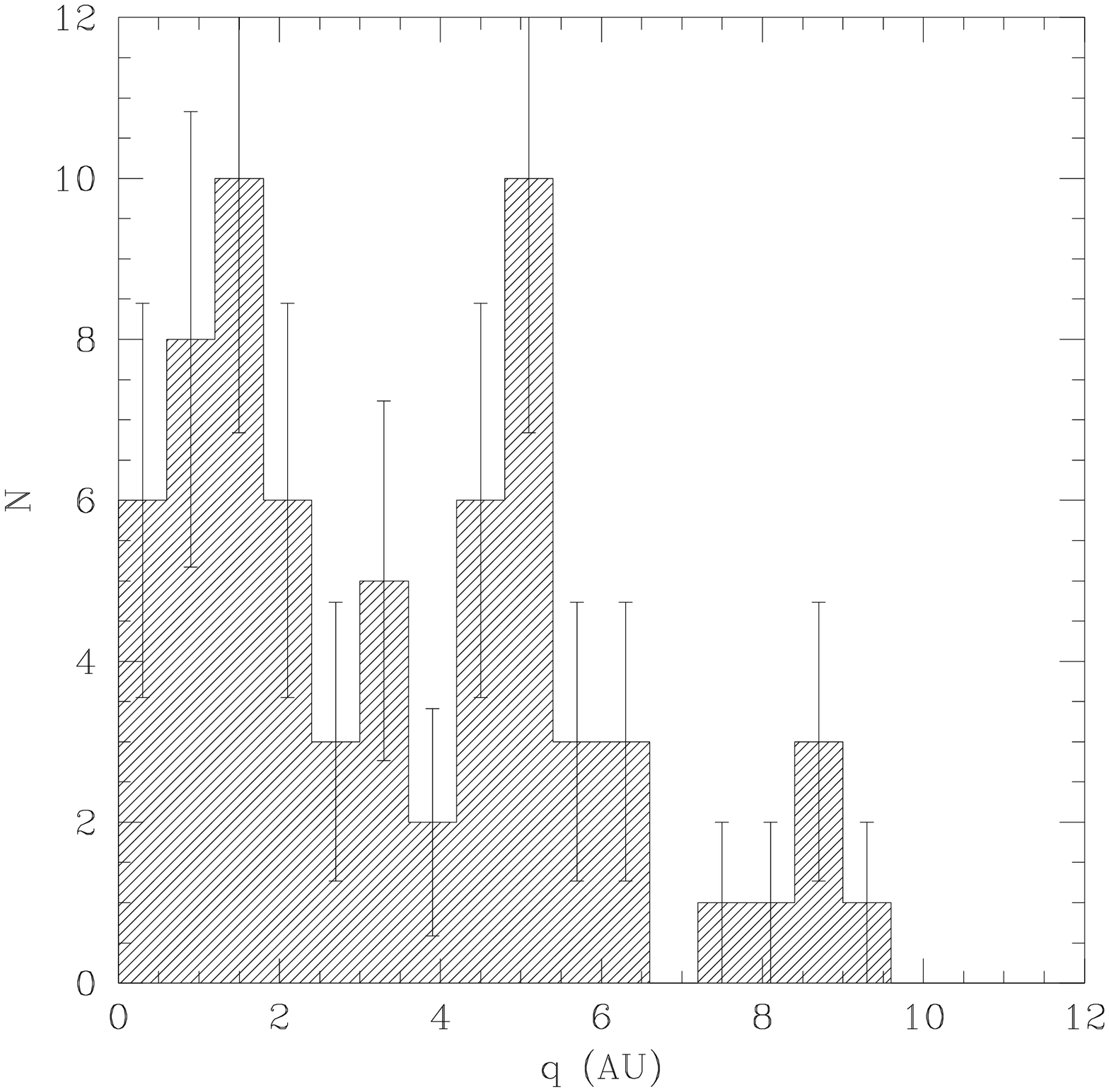,height=3in}
                        \psfig{figure=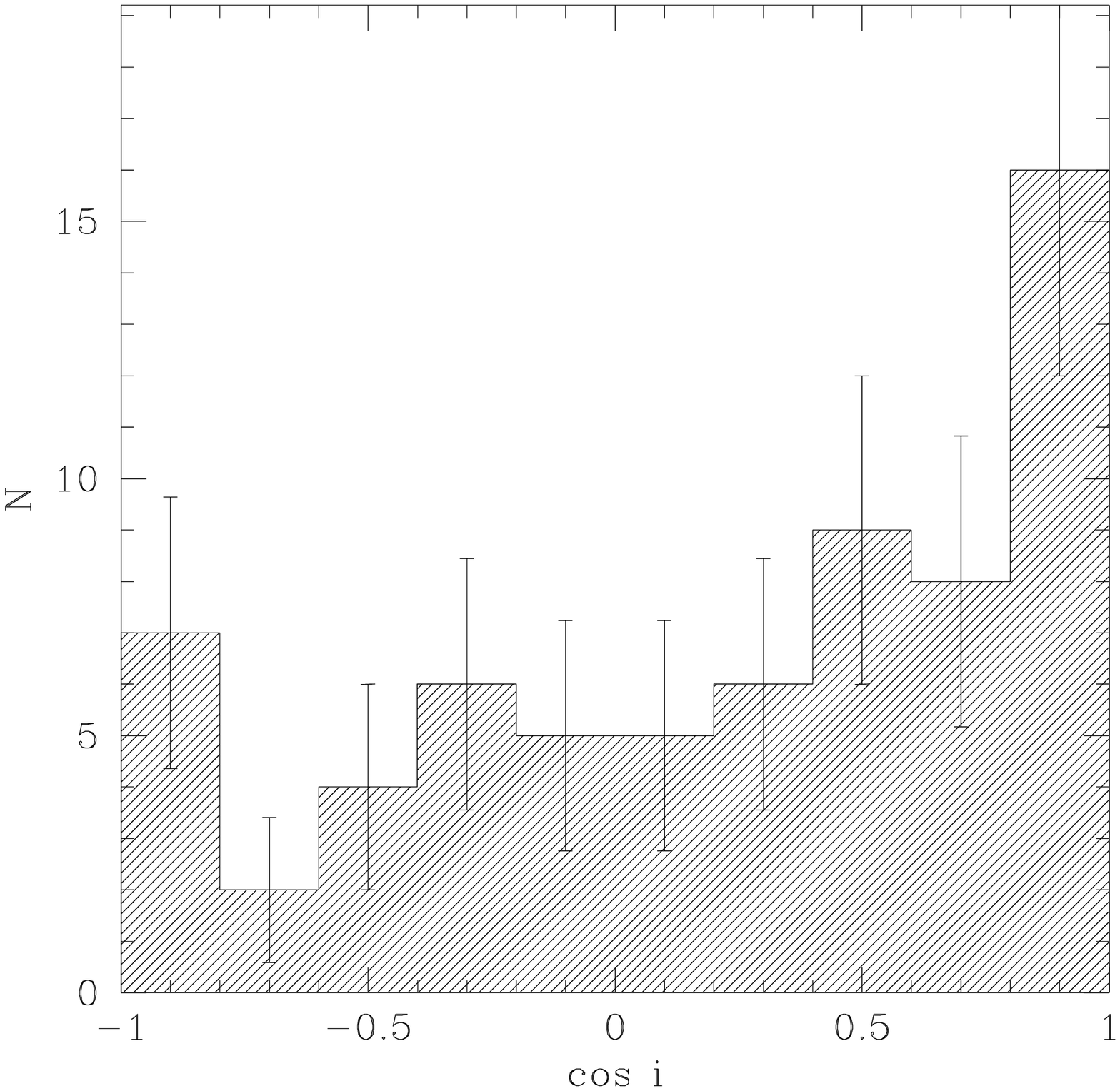,height=3in}}}}
\caption{The distribution of the inverse semimajor axis $1/a$,
perihelion distance $q$ and cosine of the ecliptic inclination $i$ for
the short-period comets originating in the Oort cloud. The
distribution of $1/a$ on the left is measured at the aphelion previous
to, and the other distributions measured at, the initial perihelion
passage as a short-period comet.}
\label{fi:sp_elems}
\end{figure}

During our simulations only 68 Oort cloud comets eventually become
short-period comets, 36 of them after having made one or more
apparitions as LP comets.  The distributions of inverse semimajor
axis, perihelion distance and inclination for these comets are shown
in Fig.~\ref{fi:sp_elems}. In no case is an Oort cloud comet
converted to a short-period comet in a single perihelion passage: the
largest orbit at the previous aphelion has a semimajor axis of only
1850~{\au}. There is a distinct concentration of orbits near zero
ecliptic inclination, as expected from studies of captures by Jupiter
\cite[]{eve72}, but the concentration is much less than that of
short-period comets in our Solar System.  The prograde fraction is
$44/68 \simeq 0.65$.

Our simulation corresponds to approximately 450 years of real time
(Eq.~\ref{eq:simlength}).  Thus we deduce that $68/450 \simeq 0.15$
short-period comets per year arrive (indirectly) from the Oort cloud
(in the absence of fading). For comparison, on average five new
short-period comets are discovered each year
\cite[]{fesricwes93a}; we conclude that the Oort cloud contributes
less than 3\% of the population of short-period comets, and another
source, such as the Kuiper belt, is required. 
Only about 10\% of the known SP comet apparitions are Halley-family,
and thus the Oort cloud may contribute a significant fraction of
these objects, though the picture is clouded by the multiple
apparitions by individual comets in this sample.

\subsubsection{Planetary encounter rates} \label{pa:plan_enc}

Close encounters of the $V_\infty$ comets with the giant planets are
described in Table~\ref{ta:ce_Vm}. Note the high frequency of multiple
encounters between a giant planet and a single comet, though this does
not indicate capture by the planet in the traditional sense
(\ie planetocentric eccentricity less than unity).

\begin{table}[p]
\centerline{
\begin{tabular}{|l|rrrrr|r|} \hline
Planet & Sun & Jupiter & Saturn & Uranus & Neptune & Total \\ \hline \hline
Number of comets            &  7  & 28  & 12  & 2    & 3    & 52  \\
Number of encounters        & 16  & 43  & 16  & 4    & 3    & 82  \\
Encounters/comet            & 2.3 & 1.5 & 1.3 & 2.0  & 1.0  & 1.6 \\ 
Collisions                  & 0   & 0   & 0   & 0    & 0    & 0 \\
Captures                    & --- & 0   & 0   & 0    & 0    & 0  \\
Min. distance ($R_{\sss I}$)& --- &0.018&0.086& 0.17 & 0.16 & 0.018\\ 
Min. distance ($R_p$)       & 1.61& 12.5& 77.9& 335  & 553  & 12.5 \\ 
Outer satellite ($R_p$)     & --- & 326 & 216 & 23   & 222  & ---  \\ \hline  
\end{tabular}}
\caption{Planetary and solar encounter data for the $V_\infty$ comets in
the standard model (post-visibility stage). The
distance to each planet's outermost satellite is given in the last
row.}
\label{ta:ce_Vm}
\end{table}

Since our simulation corresponds to roughly 450 years of real time
(Eq.~\ref{eq:simlength}) we can calculate the rate of close encounters
between the LP comets and the giant planets. During the combined pre-
and post-visibility phases of the comets' evolution,
a total of 253 encounters were recorded for Jupiter,
333 for Saturn, 111 for Uranus and 96 for Neptune. These numbers
translate to total currents ${\cal J}_p$ of 0.56, 0.74, 0.25 and 0.21
comets per year passing through the spheres of influence
(Eq.~\ref{eq:sofi}) of Jupiter through Neptune respectively.

If we assume that these currents ${\cal J}_p$ reflect a uniform flux
of LP comets across the sphere of influence of each planet, then the
rate of impacts between LP comets and the giant planets can be deduced
to be
\begin{equation}
n_p = {\cal J}_p \left( \frac{M_p}{\msun} \right)^{-4/5} \left(\frac{R_p}{a_p}
\right)^2\left(1+{2\over3}{M_p\over \msun}{a_p\over R_p}\right),
\end{equation}
where $a_p$ and $R_p$ are the planets' semimajor axis and radius,
$M_p$ is the planetary mass, and the second term is a crude correction
for gravitational focusing, assuming the comets are on nearly
parabolic orbits. The resulting collision rates are
$1.0\times10^{-5}$, $5.0\times10^{-6}$, $2.4 \times 10^{-7}$, $1.3
\times 10^{-7}$ per year for Jupiter through Neptune respectively. It
should be noted that Comet Shoemaker-Levy 9, which collided with
Jupiter in July of 1994, was not a LP comet but rather a
Jupiter-family comet \cite[]{benmck95}.
  
\subsection{Post-visibility evolution: the effect of 
non-gravitational forces}\label{sec:nongrav}

Asymmetric sublimation of volatiles leads to significant non-gravitational (NG)
forces on comets. As described in \S\ref{pa:nongrava}, we specify
NG forces using two parameters $A_1$ and $A_2$.  The parameter
$A_1$ is proportional to the strength of the radial NG force,
and is always positive, as outgassing accelerates the comet away from
the Sun. The parameter $A_2$ is proportional to the strength of the tangential
force, is generally less than $A_1$, and may have either sign depending on
the comet's rotation. Comet nuclei are likely to have randomly
oriented axes of rotation, with a corresponding random value of $A_2$. Rather
than make a complete exploration of the available parameter space for $A_1$
and $A_2$, we shall investigate a few representative cases.

We assume that $|A_2|=0.1A_1$, and consider two distributions for the sign of $A_2$:
\begin{enumerate}
\item Half the comets have positive values of $A_2$, half negative, and the
sign of $A_2$ is constant throughout a comet's lifetime---as if the axis of
rotation of the nucleus remained steady throughout the comet's dynamical
lifetime.
\item The sign of $A_2$ is chosen at random after each perihelion passage---as
if the axis of rotation changed rapidly and chaotically.
\end{enumerate}

We examined four values of $A_1$: $10^{-8},10^{-7},10^{-6},10^{-5}\au$
day$^{-2}$. The first two of these are reasonably consistent with the
NG forces observed in LP comets \cite[]{marsekyeo73}. The two remaining
values for $A_1$ are probably unrealistically large.

\begin{figure}[p]
\centerline{\vbox{\hbox{\psfig{figure=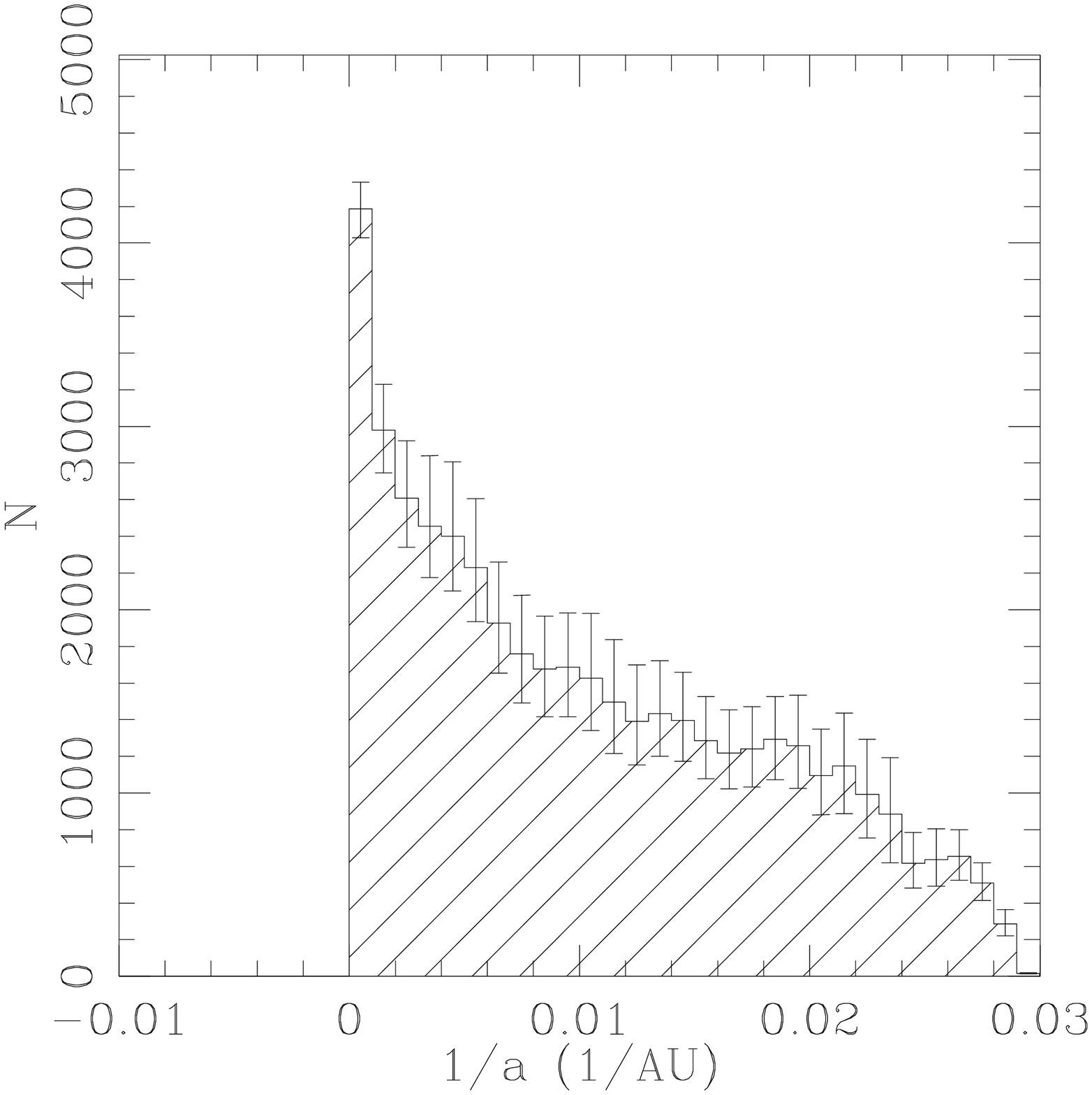,height=1.5in}
                        \psfig{figure=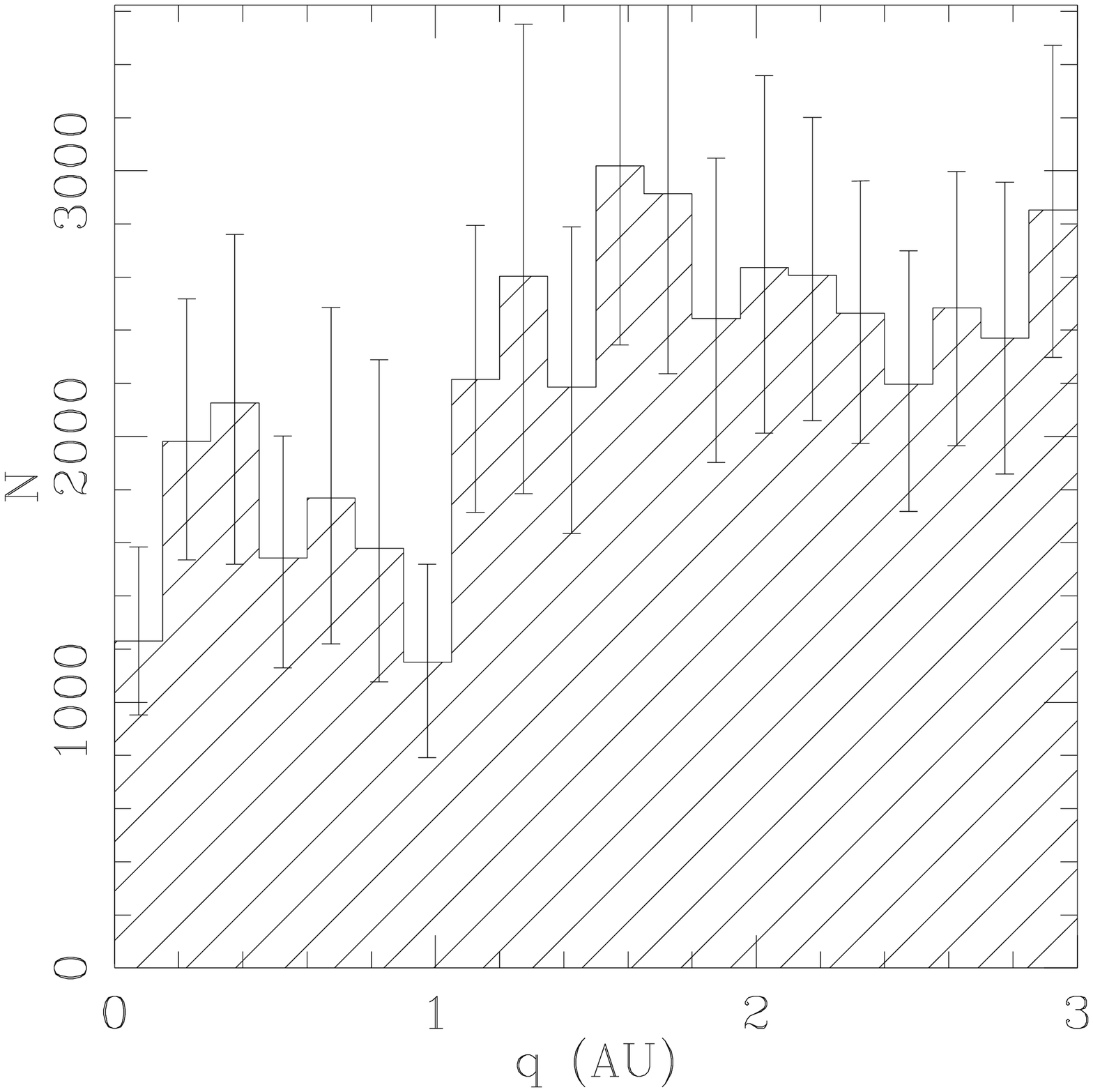,height=1.5in}
                        \psfig{figure=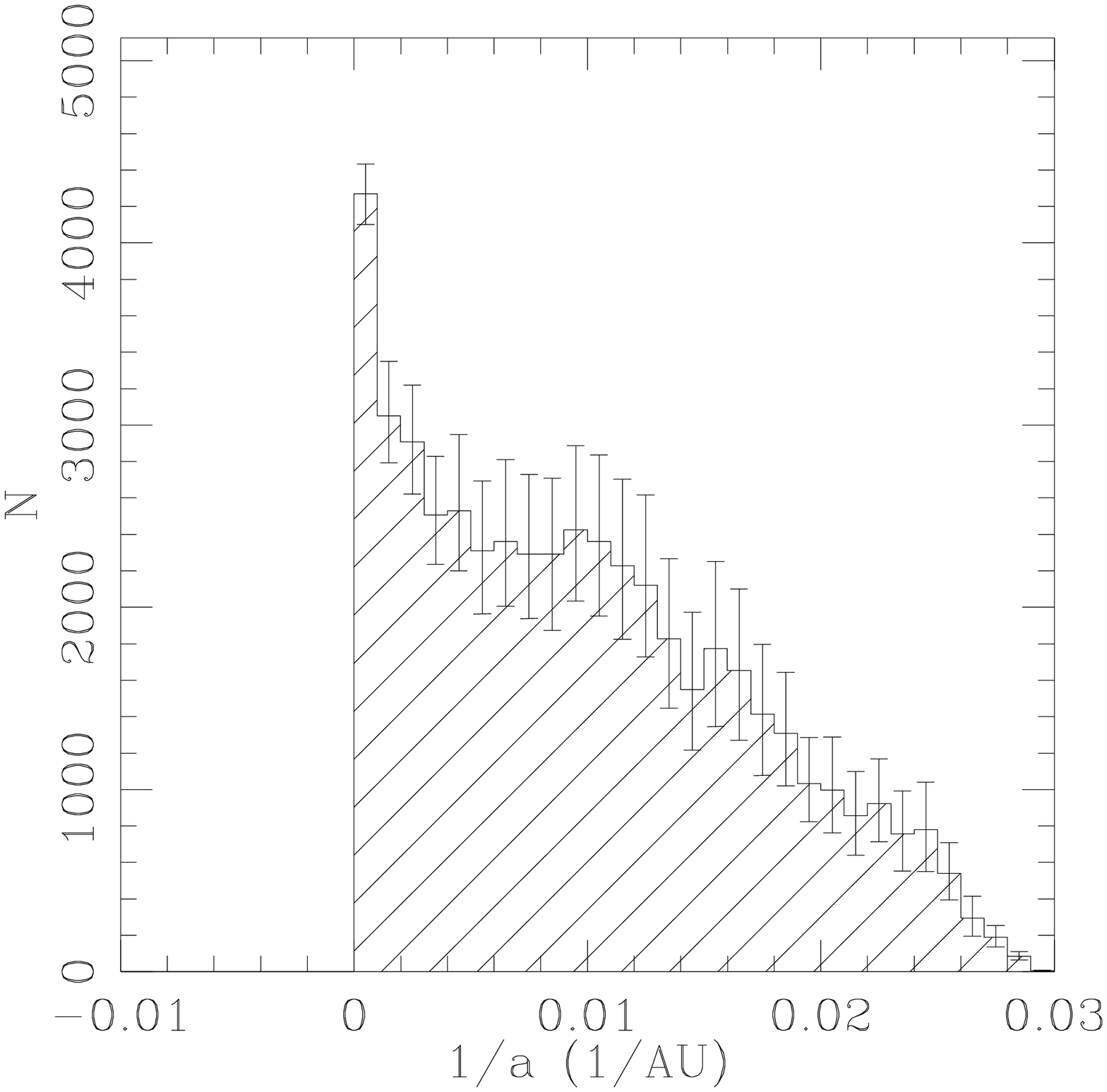,height=1.5in}
                        \psfig{figure=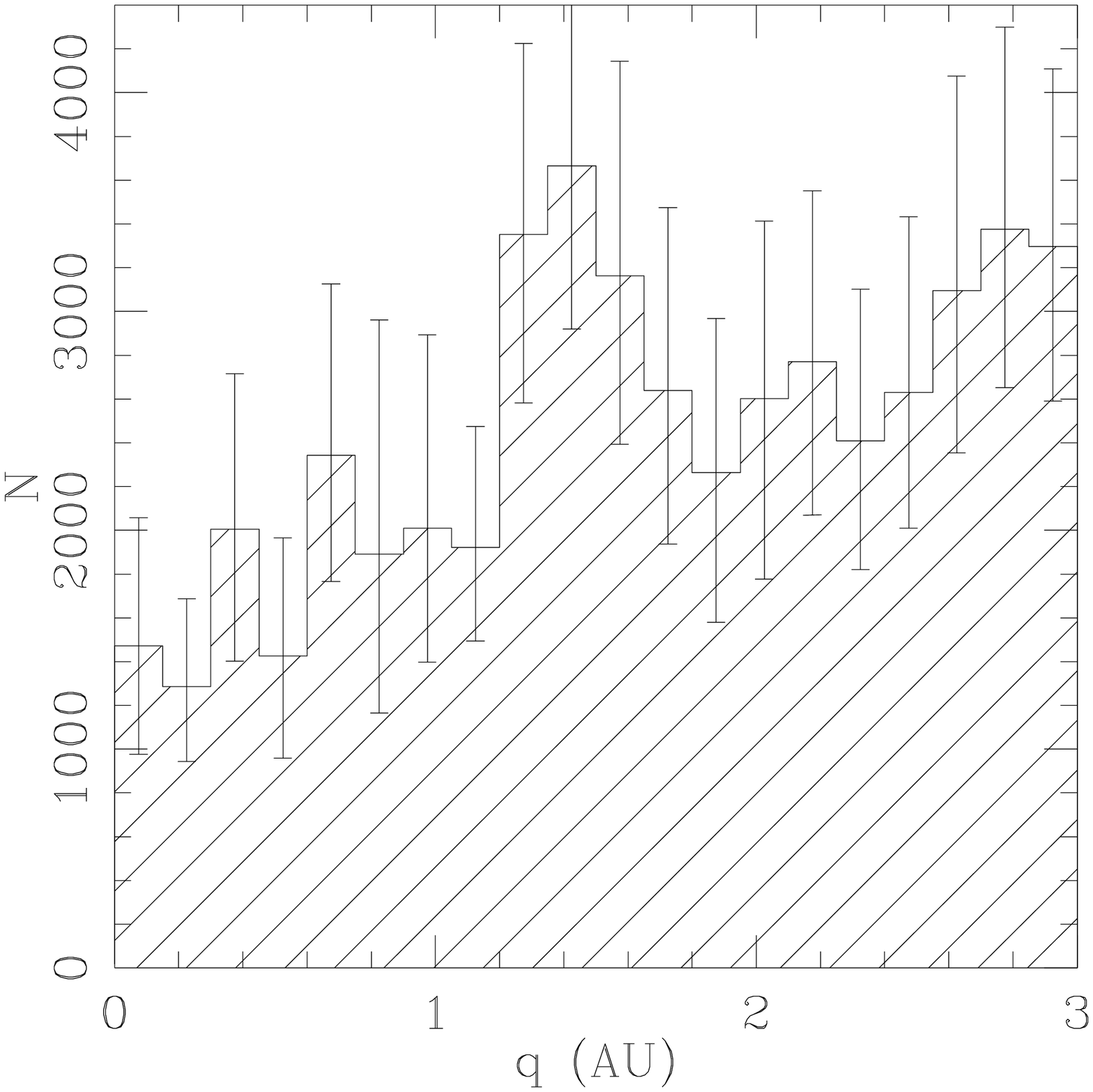,height=1.5in}}
                  \hbox{\psfig{figure=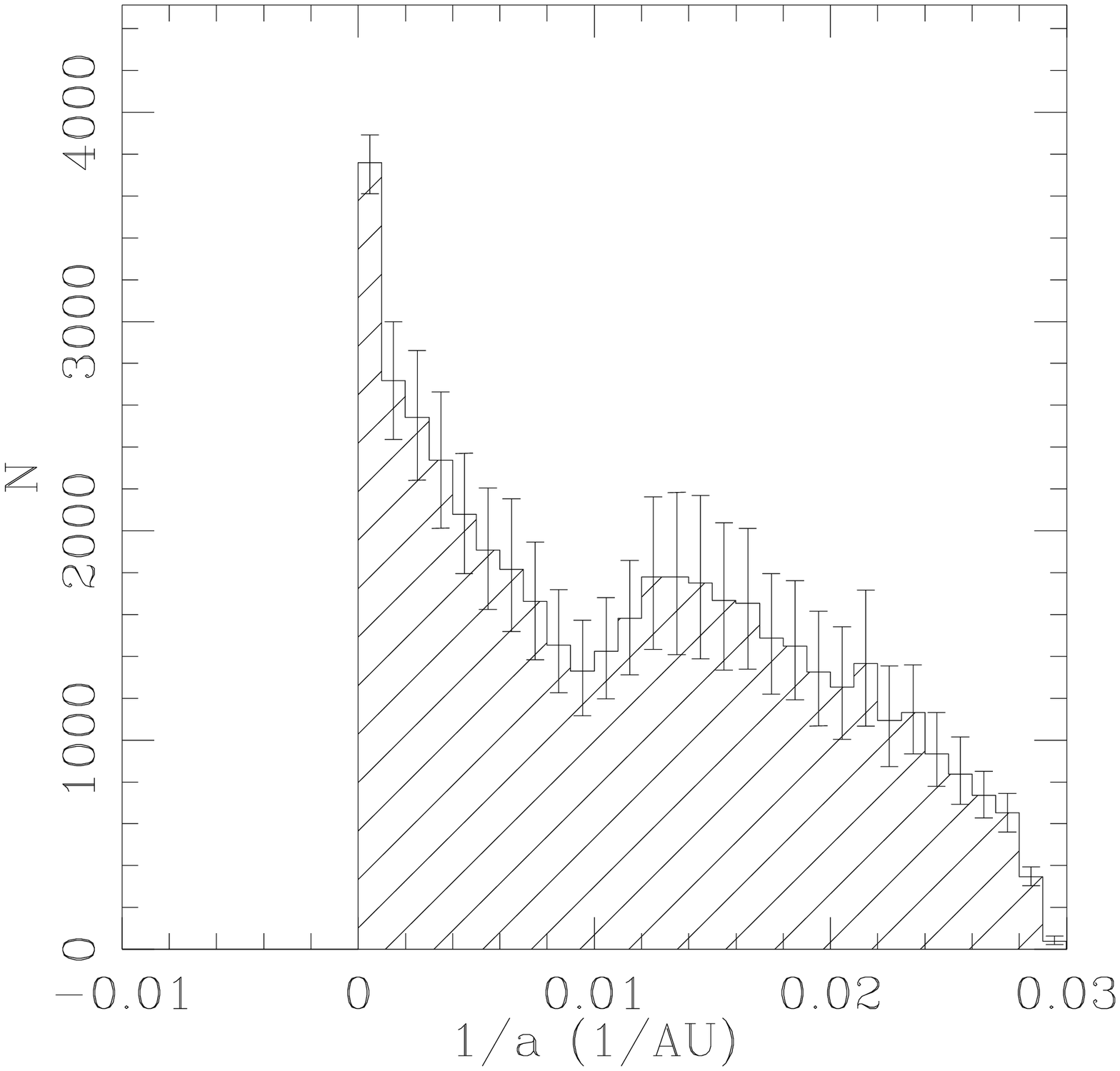,height=1.5in}
                        \psfig{figure=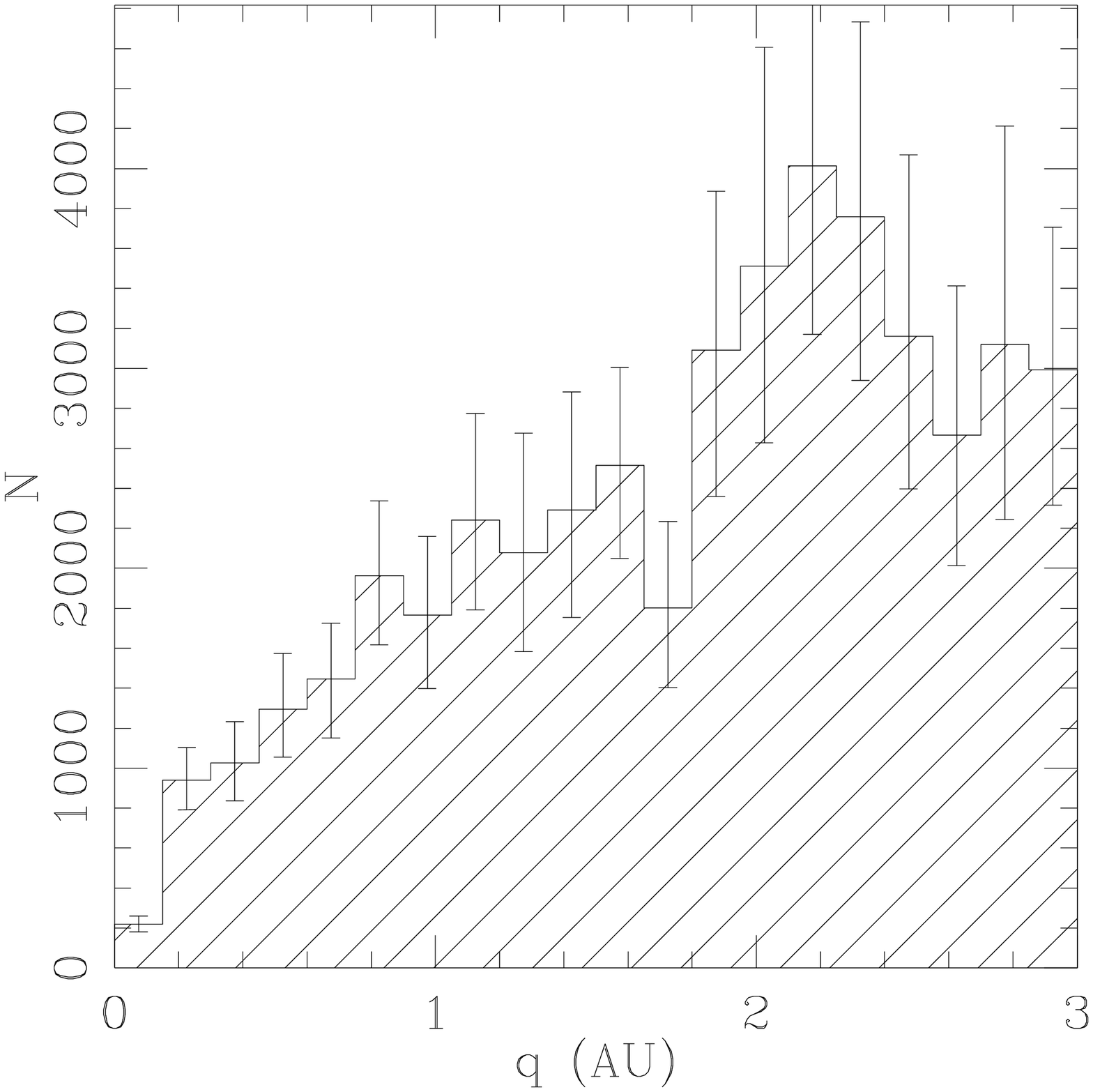,height=1.5in}
                        \psfig{figure=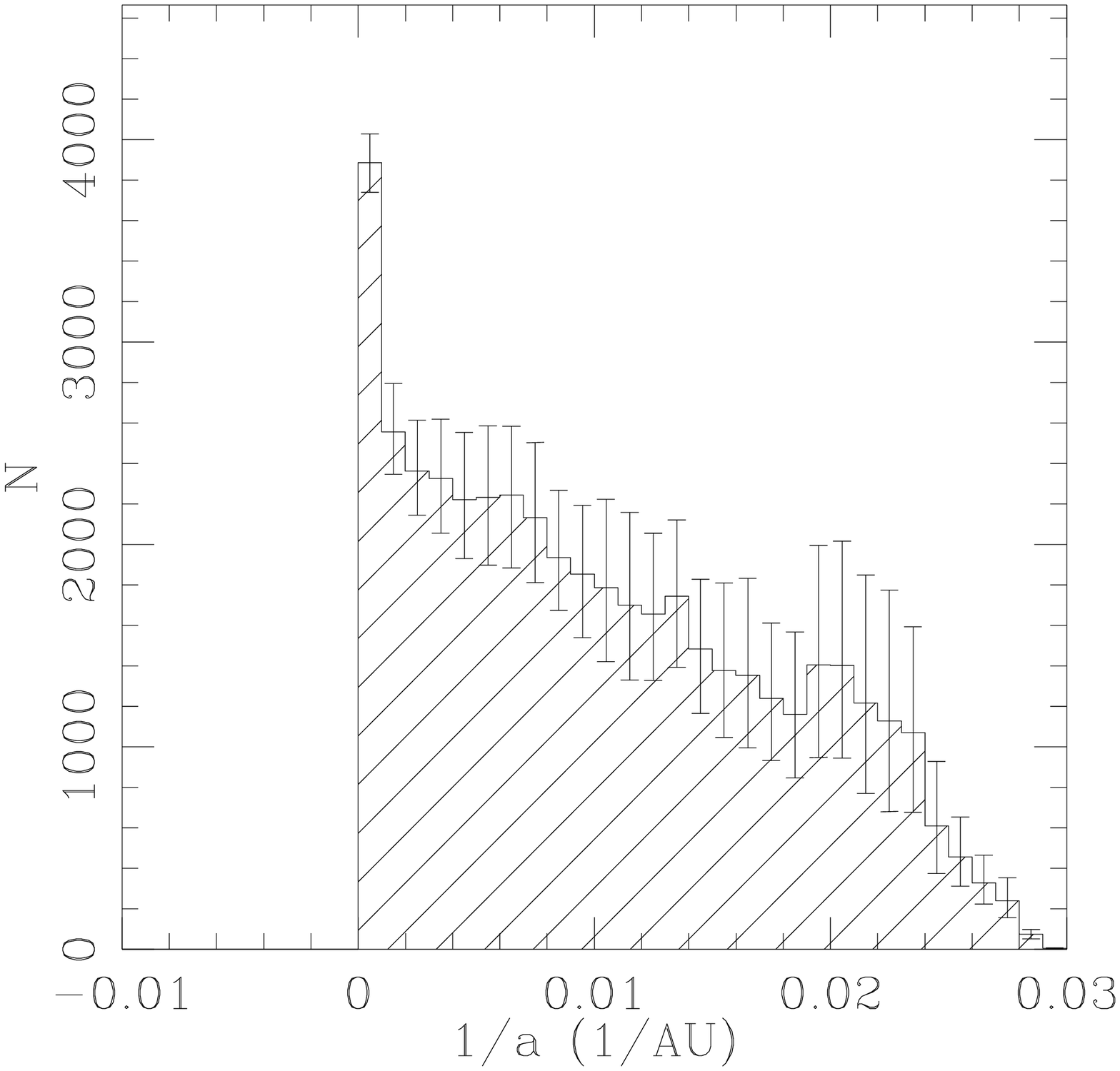,height=1.5in}
                        \psfig{figure=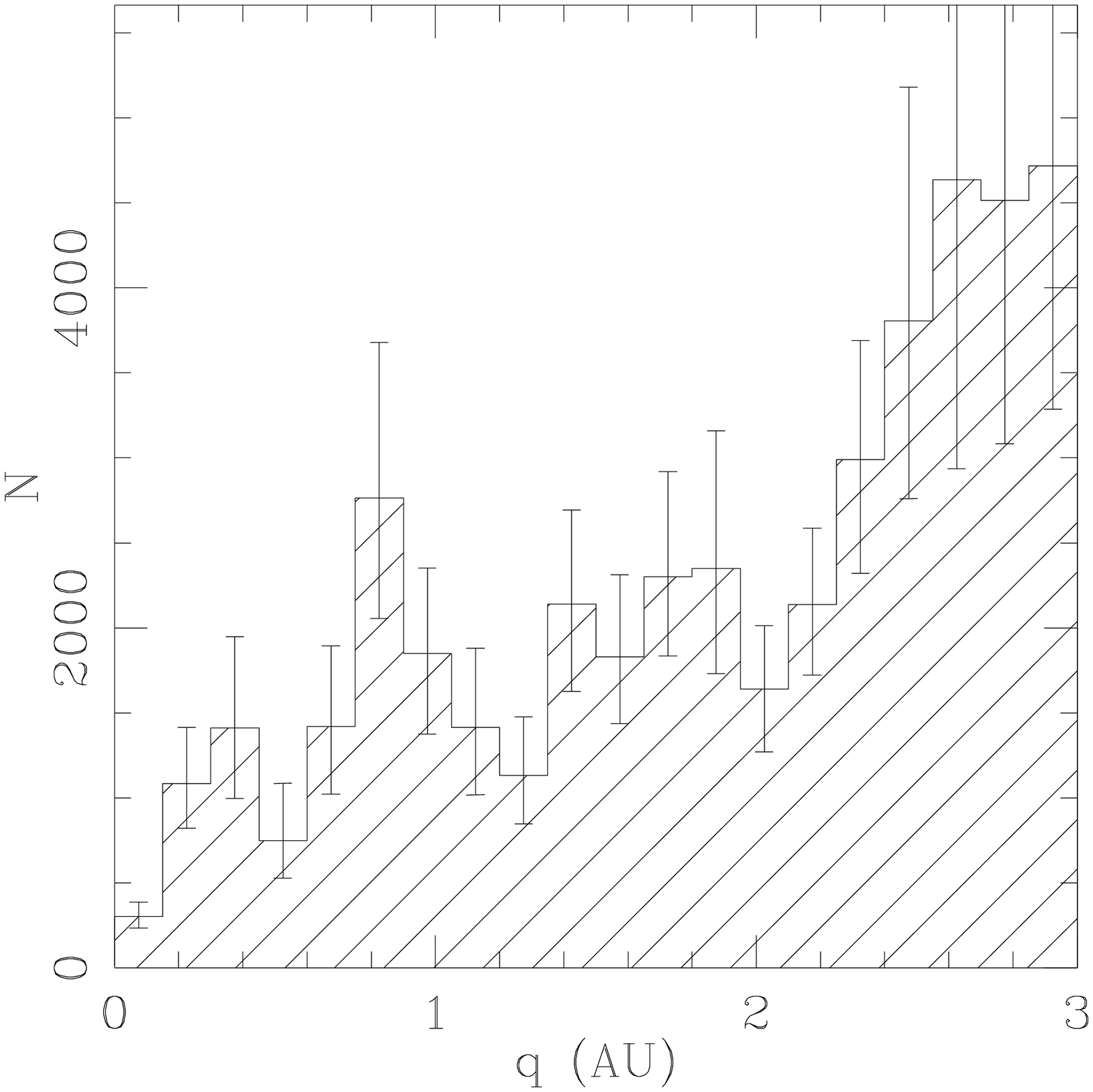,height=1.5in}}
                  \hbox{\psfig{figure=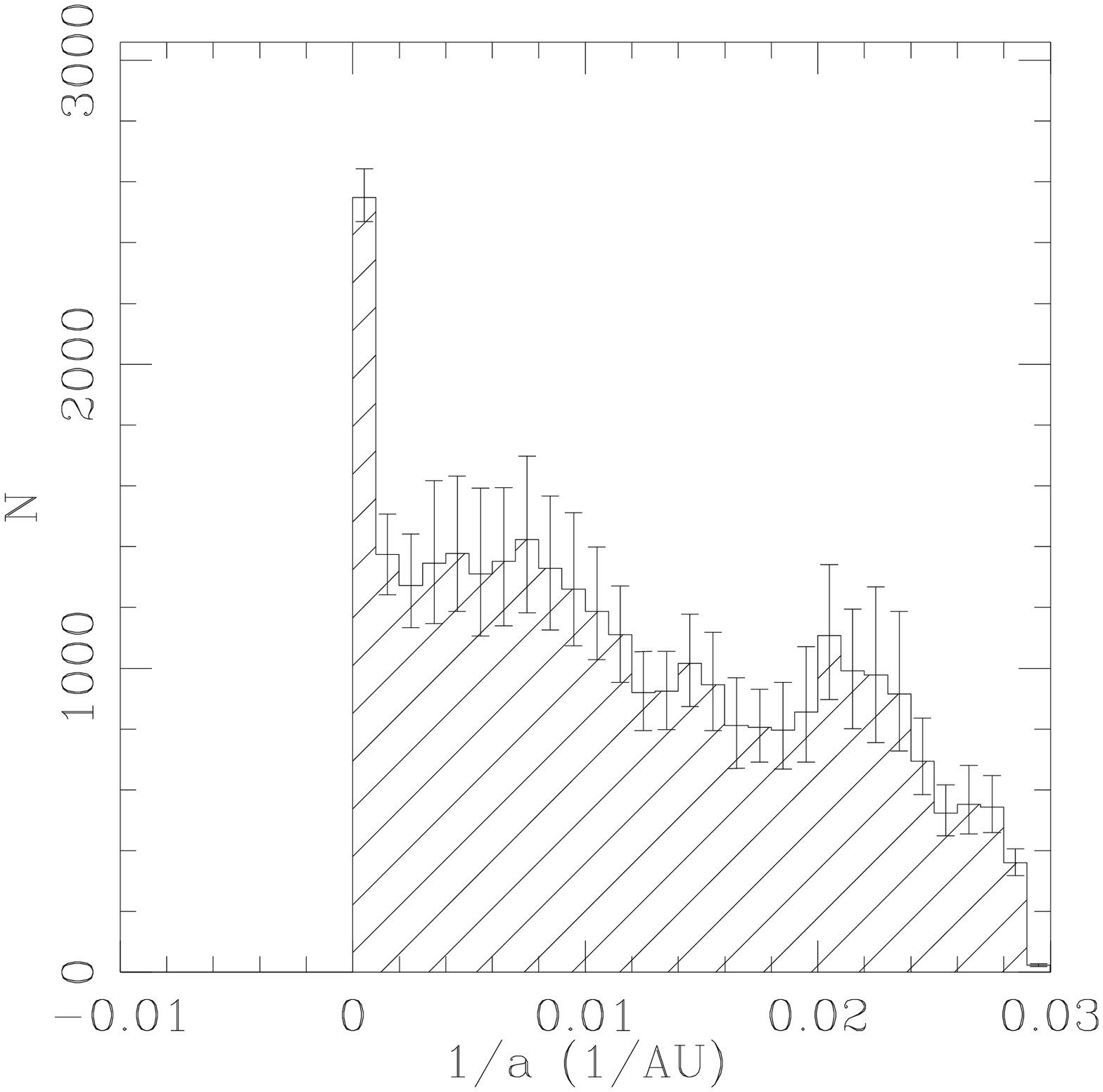,height=1.5in}
                        \psfig{figure=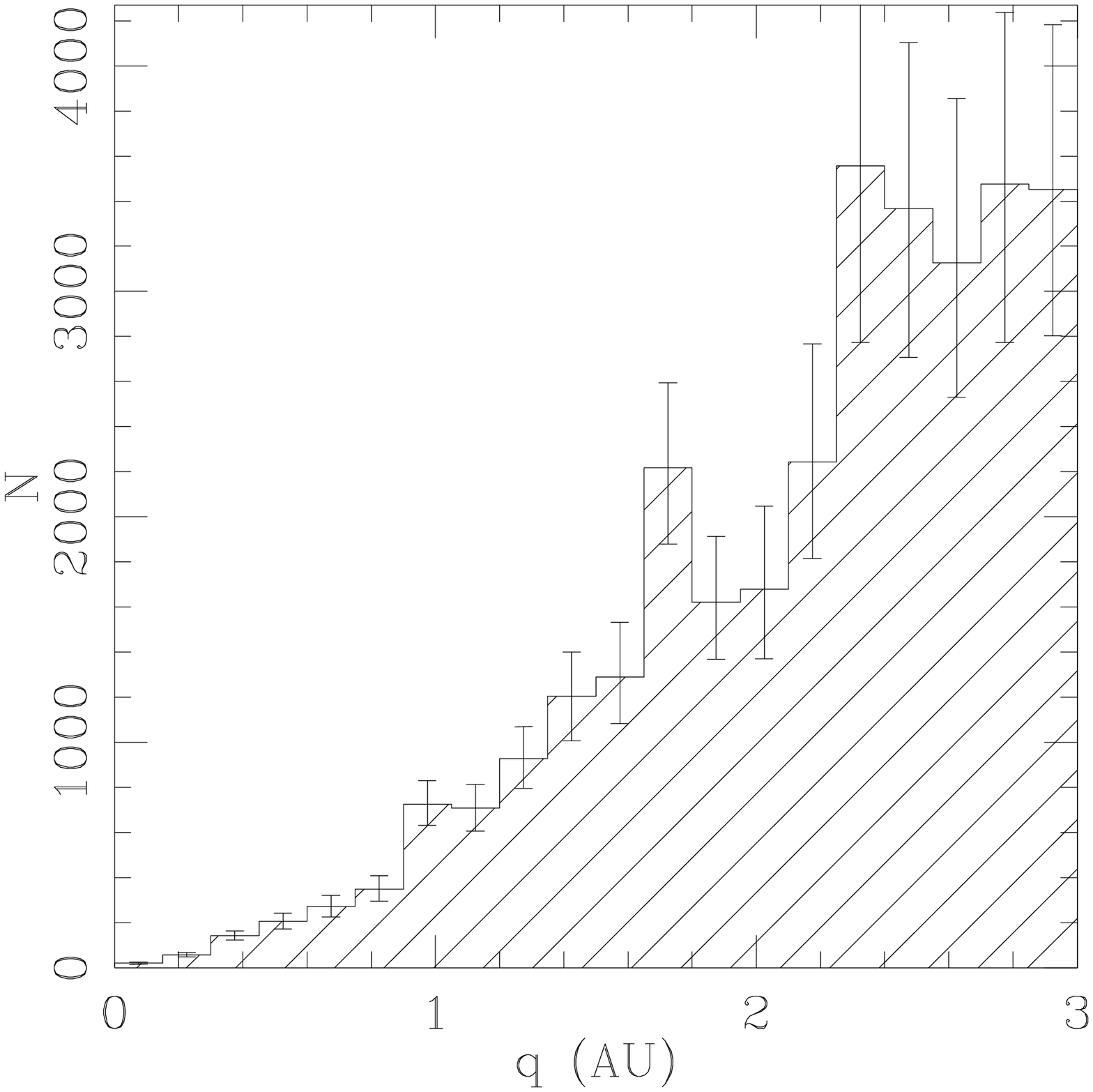,height=1.5in}
                        \psfig{figure=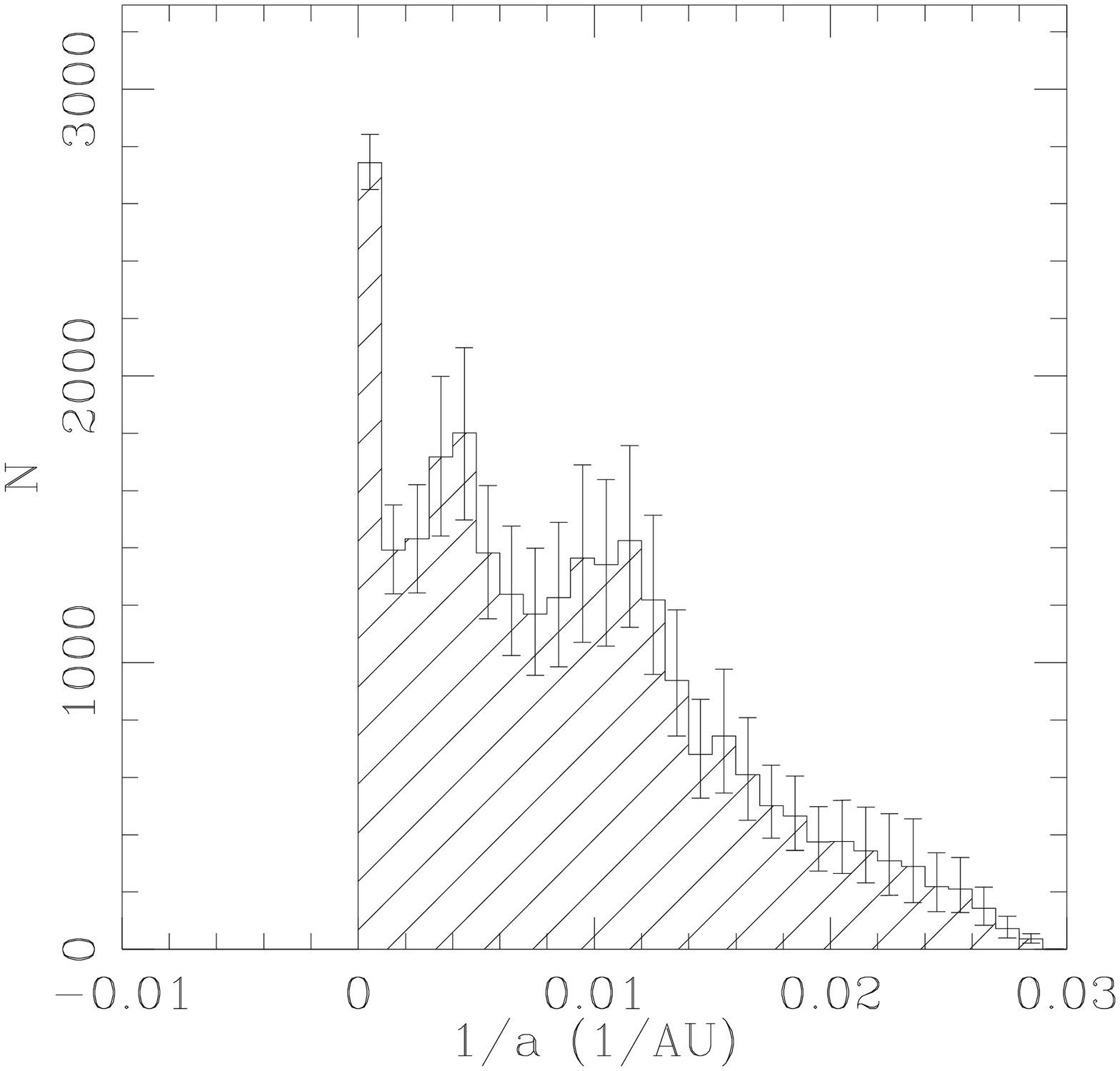,height=1.5in}
                        \psfig{figure=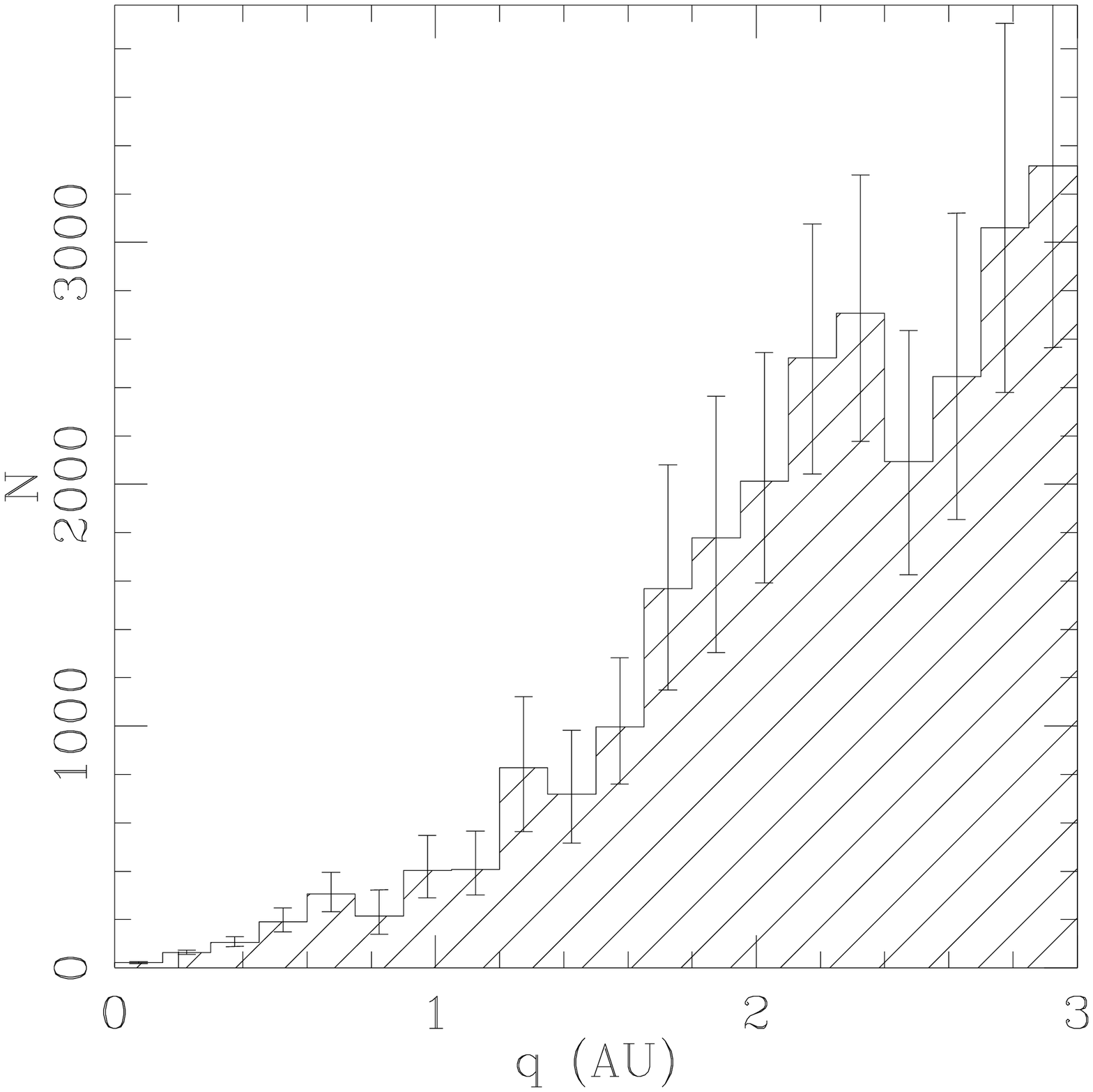,height=1.5in}}
                  \hbox{\psfig{figure=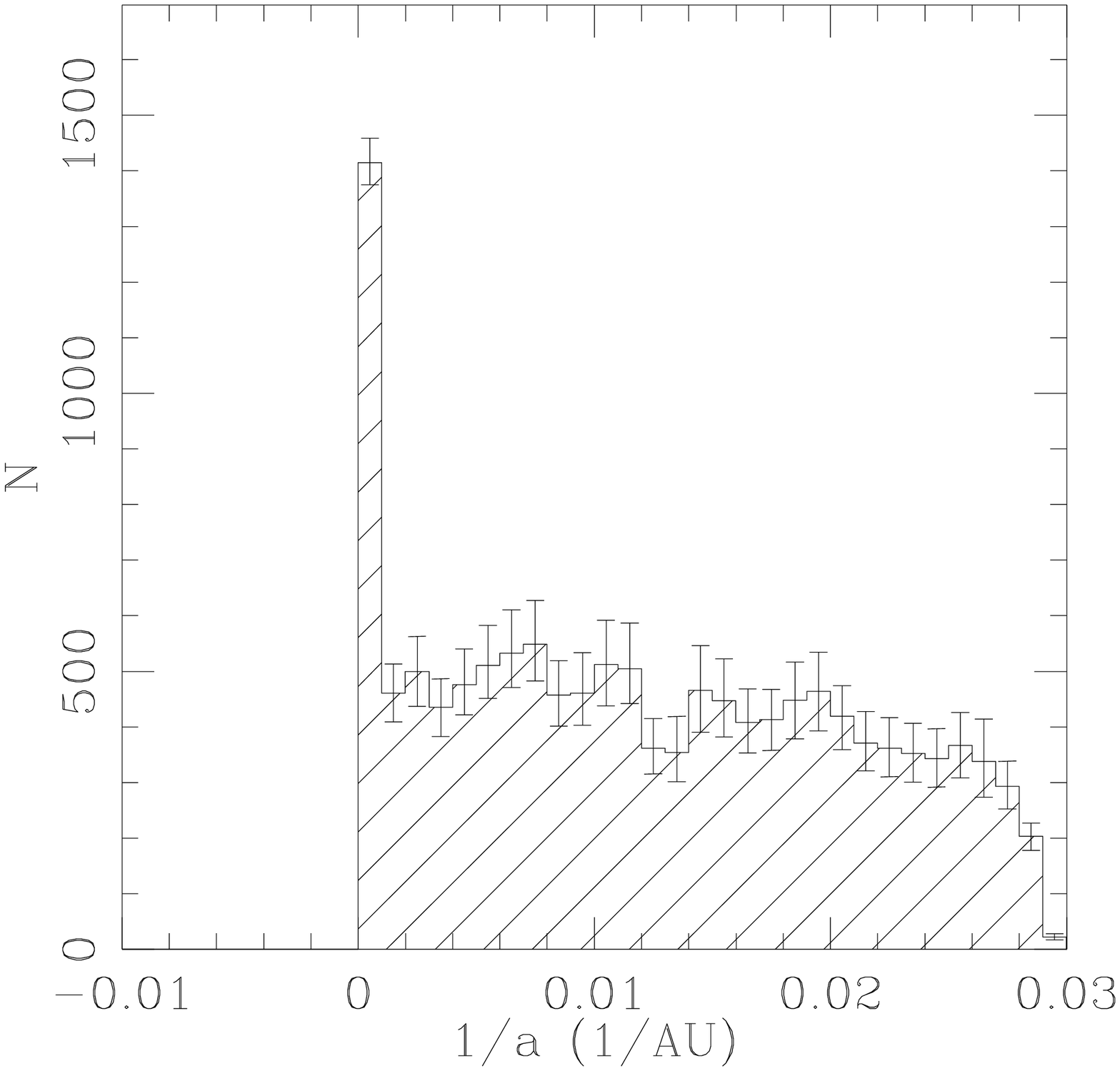,height=1.5in}
                        \psfig{figure=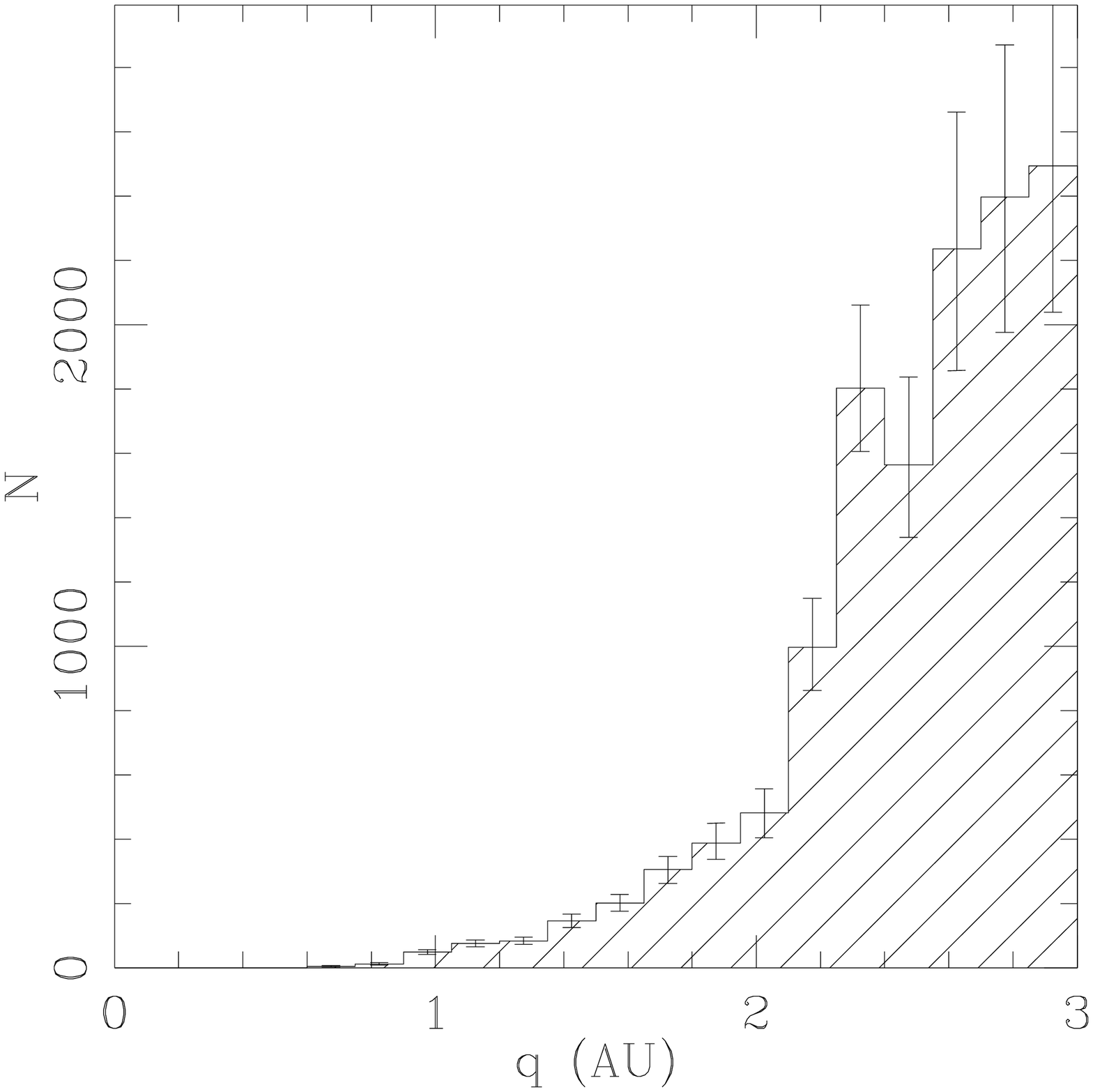,height=1.5in}
                        \psfig{figure=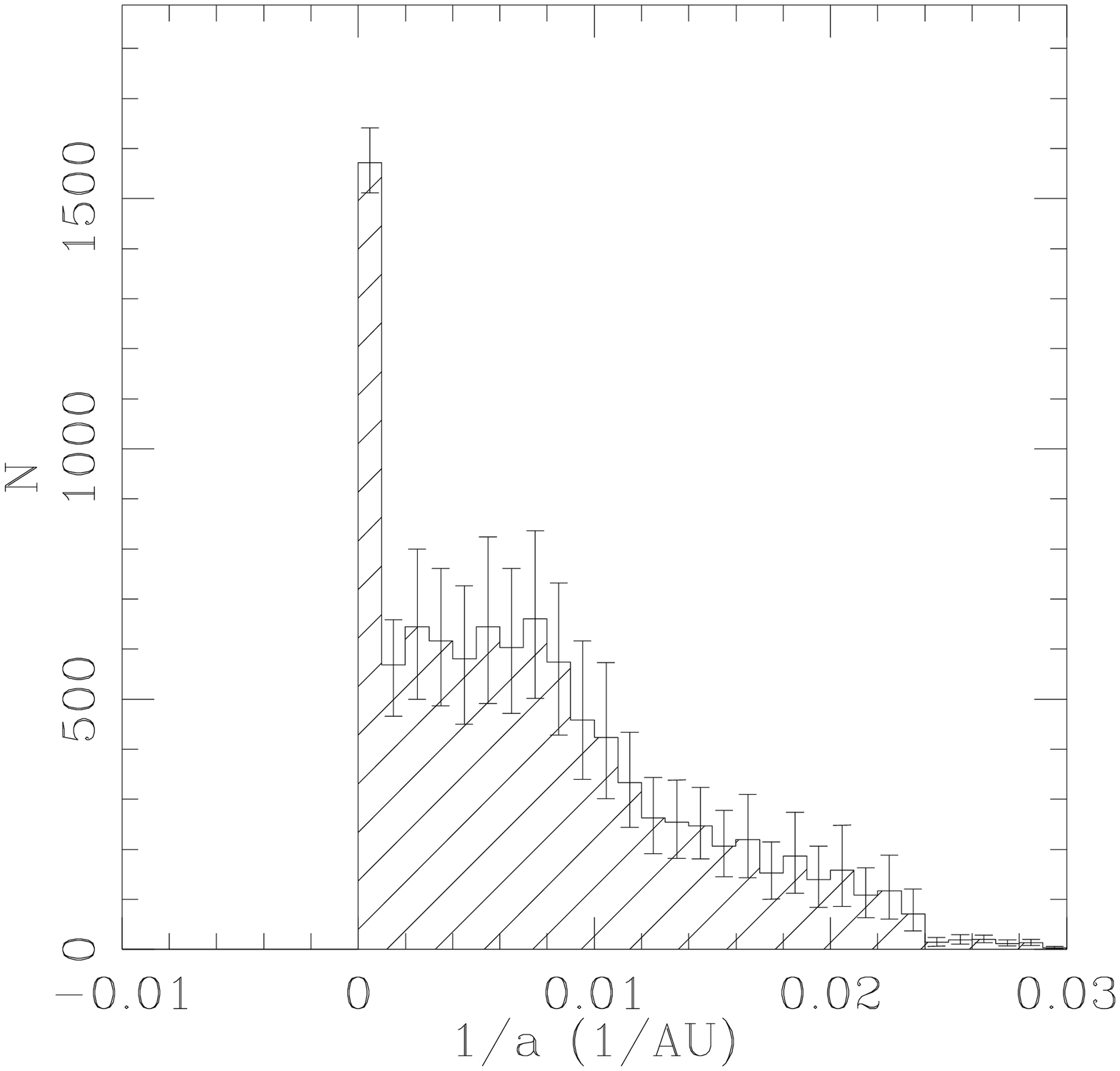,height=1.5in}
                        \psfig{figure=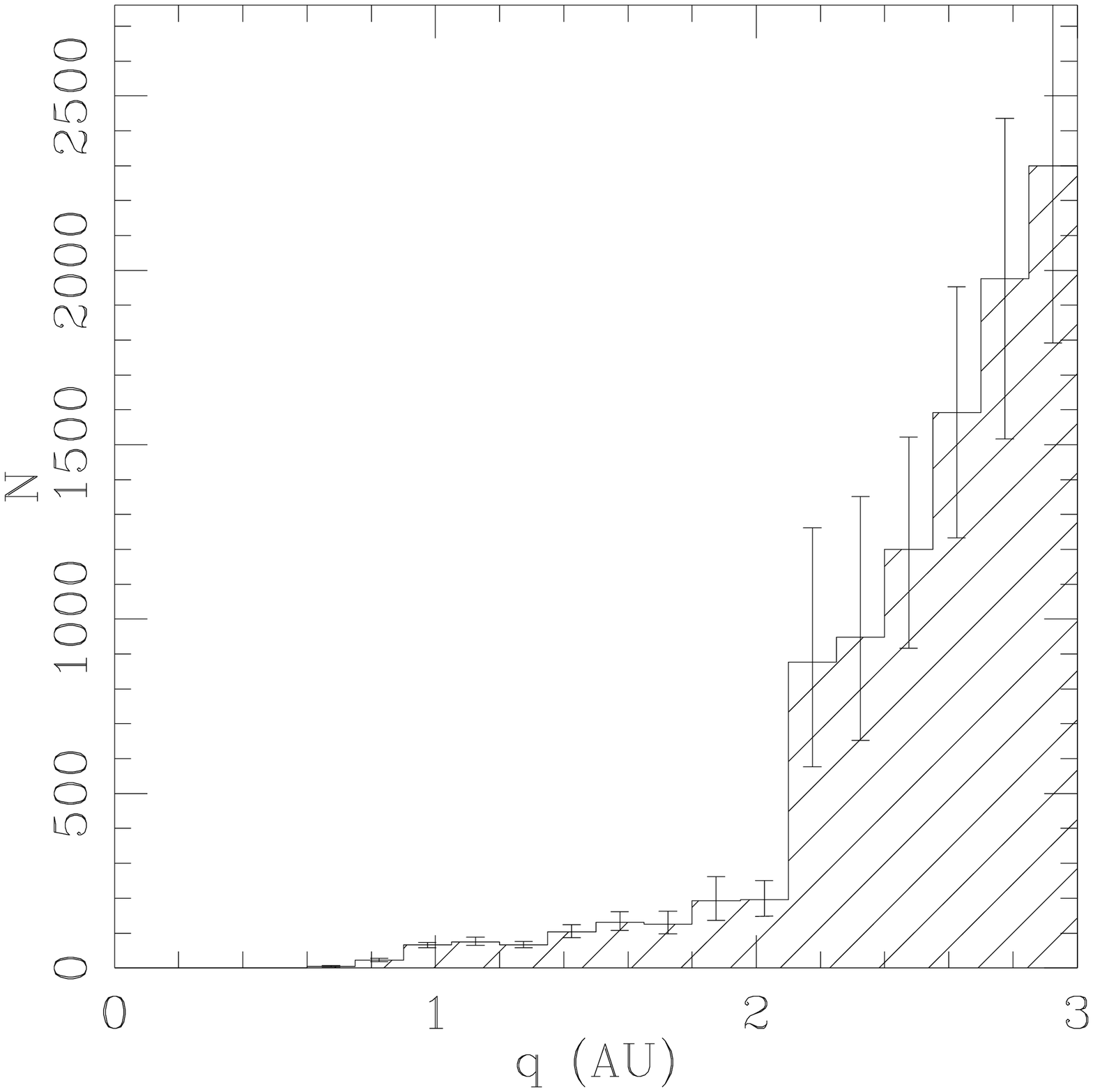,height=1.5in}}
		  \rule[2mm]{6.2in}{0.5mm}
                  \hbox{\psfig{figure=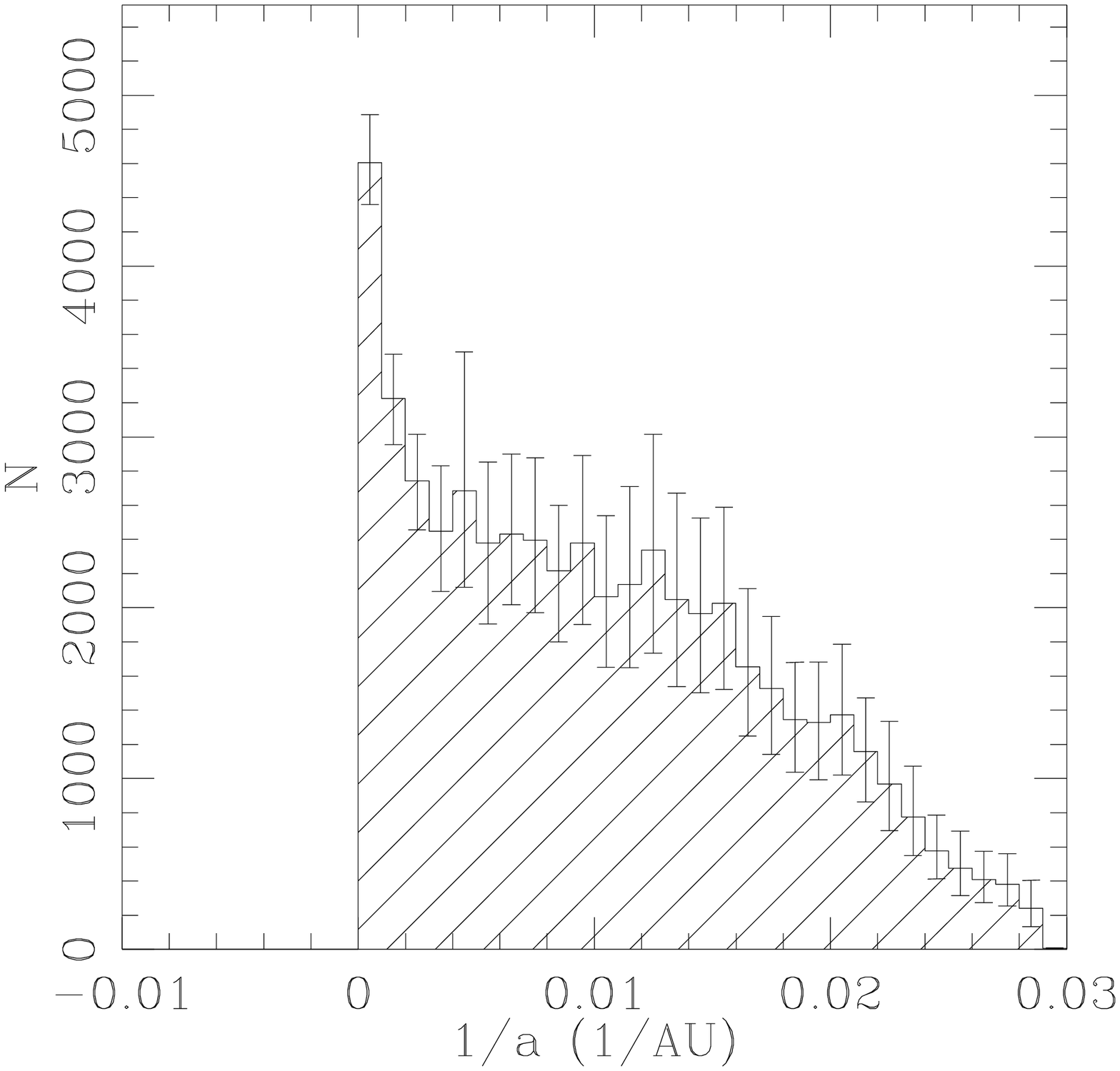,height=1.5in}
                        \psfig{figure=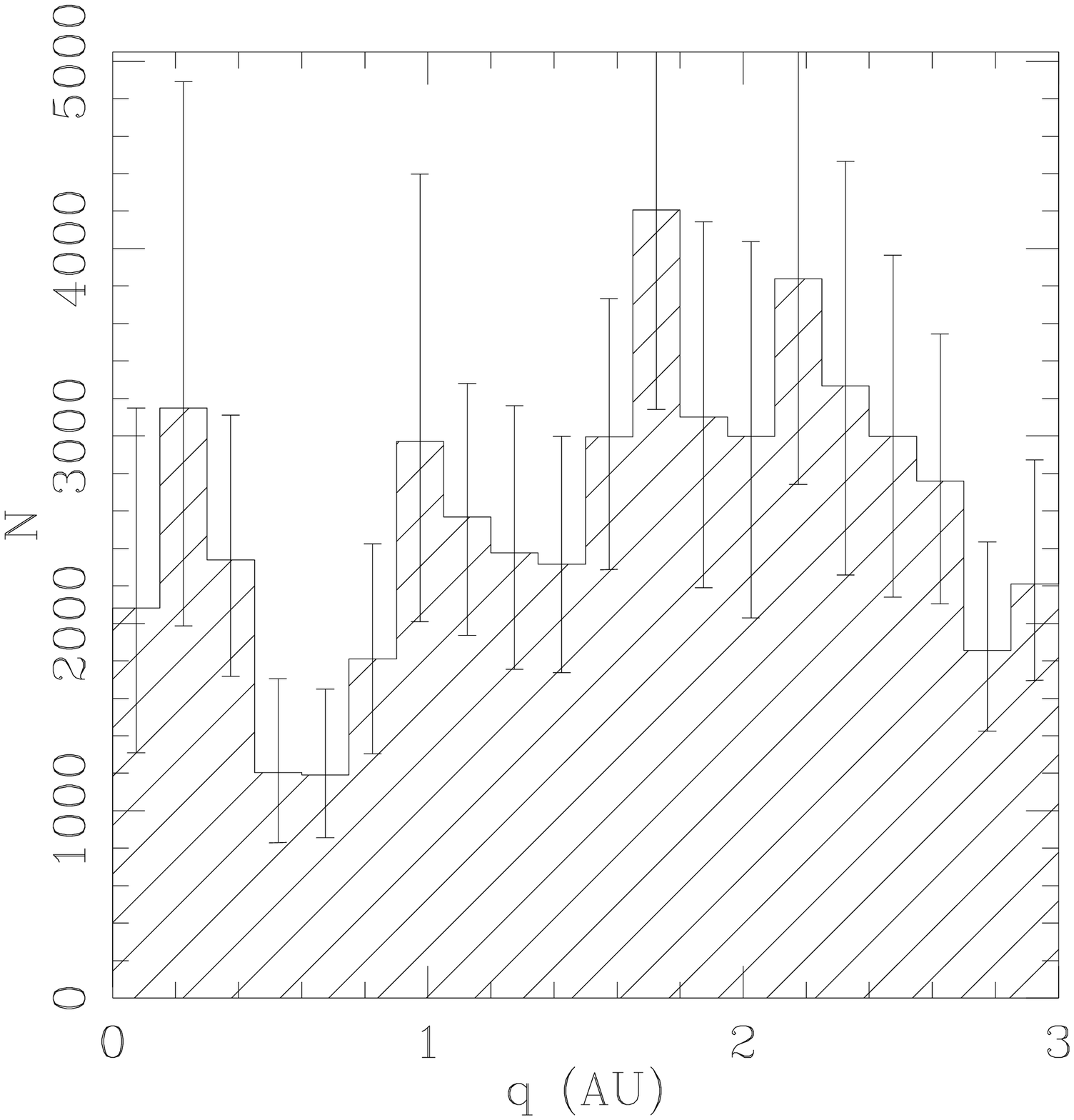,height=1.5in}
                        \psfig{figure=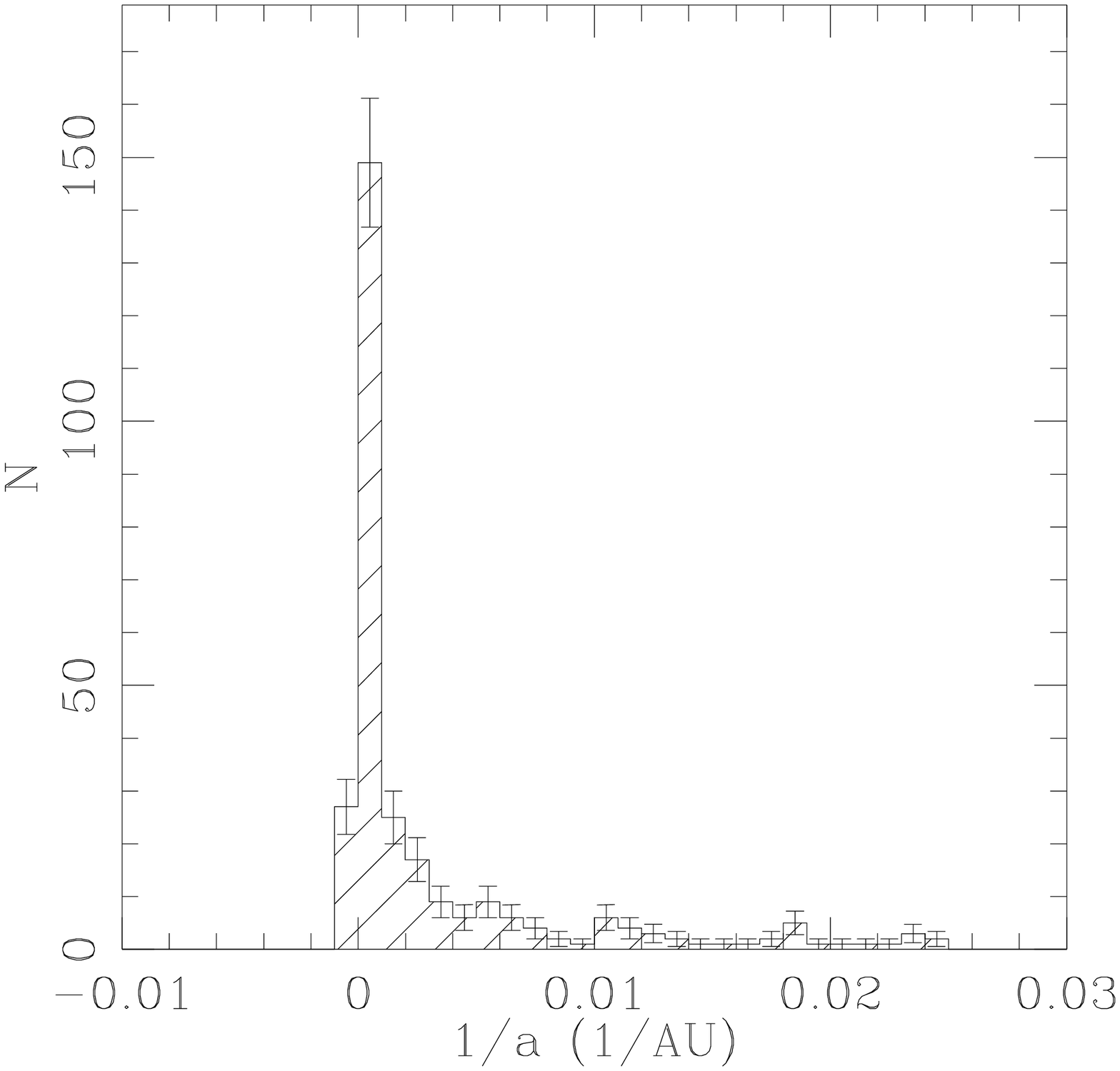,height=1.5in}
                        \psfig{figure=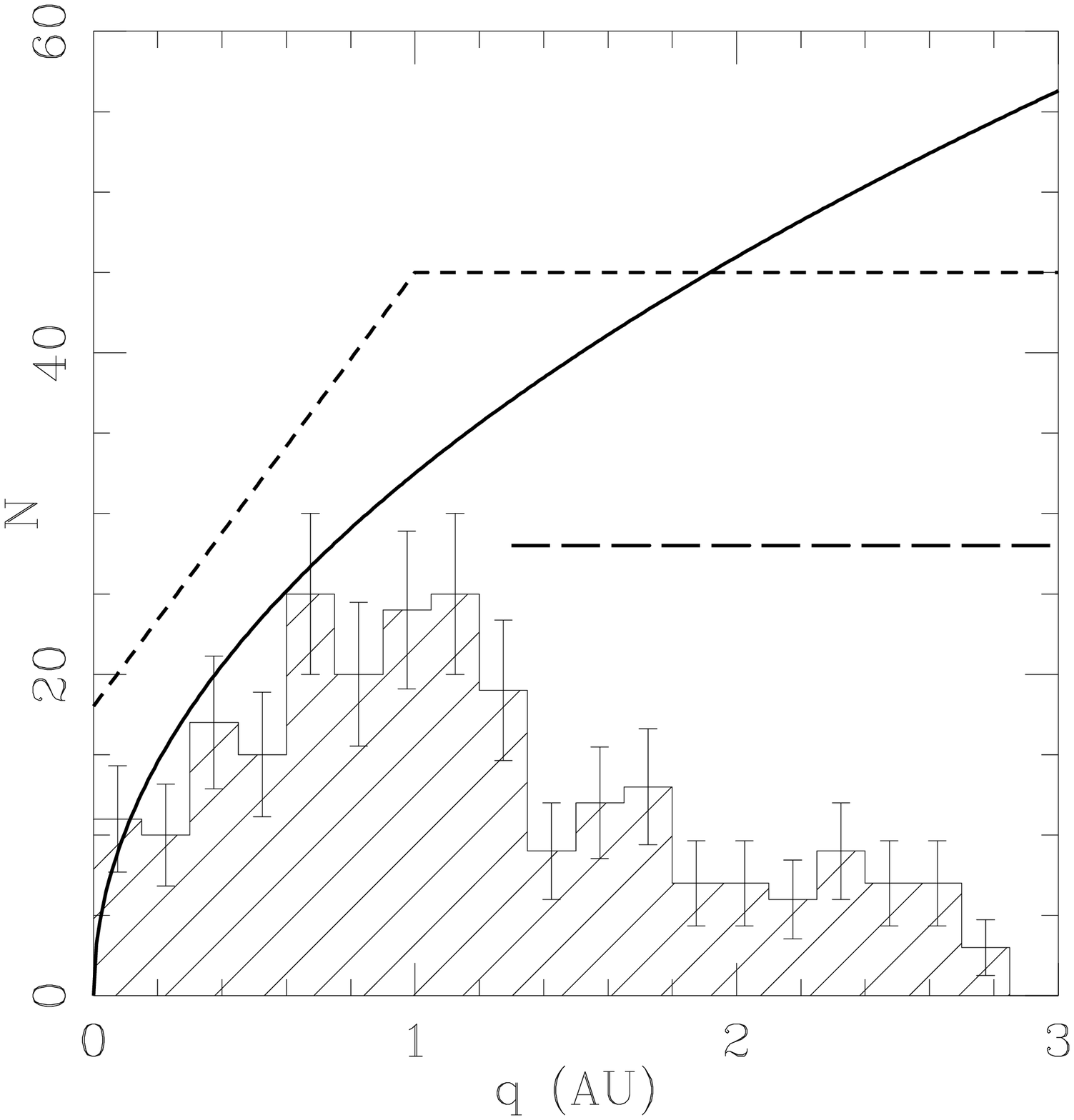,height=1.5in}}}}
\caption{\small Distribution of the inverse semimajor axis $1/a$ and
perihelion distance $q$ for $V_\infty$ comets subjected to
non-gravitational forces. Left panels: constant values for $A_2$, half
positive, half negative. Right panels: the sign of $A_2$ is randomised
for each perihelion passage. From the top down, $A_1=10^{-8}$,
$10^{-7}$, $10^{-6}$ and $10^{-5}$~\aud2, with $|A_2|=0.1A_1$. The
bottom line of panels is for comparison, and includes the standard model
(left side) and the observations (right side). The observed perihelion
distribution includes curves indicating the estimated intrinsic
distribution (see Fig. \ref{fi:Vm_q} for details).}
\label{fi:ng3}
\end{figure}

Figure~\ref{fi:ng3} and Table~\ref{ta:ng_parms} illustrate the effects of NG
forces on the energy and perihelion distributions, and on the parameters $X_i$
defined in Eq.~\ref{eq:xdef}, which should be unity if the simulated
and observed element distributions agree. The figure shows that NG forces do
decrease the number of dynamically older comets relative to the number of new
comets and hence improve agreement with the observations (\ie increasing
$X_1$, decreasing $X_2$); however, the same forces erode the population of
comets at small perihelion distances, thereby worsening the agreement with the
observed perihelion distribution. Even unrealistically large NG forces cannot
bring the distribution of inverse semimajor axes into line with observations,
and these produce an extremely unrealistic depletion of comets at small
perihelia.

The effects of NG forces can be summarized as follows:
\begin{itemize}
\item The semimajor axis perturbation due to radial NG forces averages to zero
over a full orbit (assuming that the radial force is symmetric about
perihelion, as in the model discussed in \S\ref{pa:nongrava}). Thus
radial forces have little or no long-term effect on the orbital distribution. 
\item Positive values of the tangential acceleration $A_2$ reduce the
tail of the population, resulting in an increase in $X_1$ towards
unity and improving the match with observations, but erode the
population at small perihelia, a depletion
which is not seen in the observed sample.
\item Negative values of $A_2$ preserve a reasonable perihelion distribution,
but increase the number of comets in the tail of the energy
distribution, thus reducing $X_1$ so that the disagreement between the
observed and simulated energy distribution becomes even worse. 
\end{itemize}

We have also conducted simulations with a more realistic model for
observational selection effects (Eq.~\ref{eq:discoverprob2}) but
this does not alter our conclusions.

Although we have not exhaustively explored the effects of NG forces on the LP
comet distribution, we are confident that conventional models of NG forces
cannot by themselves resolve the discrepancy between the observed and predicted
LP comet distribution. 

\nocite{krepit78,eve67a,showol82}

\begin{table}[p]
\centerline{
\begin{tabular}{|cc|cccccccc|} \hline
$A_1$ & $A_2$ & Total & Spike & Tail & Prograde &$X_1$&$X_2$&$X_3$ &
$\langle m \rangle$ \\  \hline \hline
 0.0 & 0.0       &52303 &1473 &15004 & 15875  & 0.07 & 4.37  & 0.61 & 45.4 \\
 1.0 & 0.1       &35370 &1457 &7368  & 12381  & 0.11 & 3.17  & 0.70 & 36.1 \\
 1.0 & $-0.1$    &57819 &1462 &19364 & 21110  & 0.07 & 5.10  & 0.73 & 51.0 \\
 1.0 &$\pm 0.1^a$&44383 &1461 &13705 & 19021  & 0.09 & 4.70  & 0.86 & 38.4 \\
 10  &$\pm 1^a$  &45899 &1425 &16628 & 18504  & 0.08 & 4.42  & 0.80 & 42.5 \\
 100 &$\pm 10^a$ &30660 &1341 &11296 & 11012  & 0.12 & 5.61  & 0.72 & 33.1 \\
 1000&$\pm 100^a$&13248 &995  &5432  & 5872   & 0.20 & 6.24  & 0.88 & 14.4 \\
 1.0 &$\pm 0.1^b$&49642 &1450 &13203 & 16387  & 0.08 & 4.05  & 0.66 & 46.7 \\
 10  &$\pm 1^b$  &45202 &1448 &13631 & 17311  & 0.08 & 4.59  & 0.76 & 41.4 \\
 100 &$\pm 10^b$ &25774 &1364 &4969  & 11452  & 0.14 & 2.93  & 0.88 & 27.7 \\
 1000&$\pm 100^b$&9878  &1035 &1536  & 5042   & 0.28 & 2.37  & 1.02 & 13.2 \\
\hline
\end{tabular}}
\caption{Parameters of the distribution of $V_\infty$ comets subjected
to non-gravitational forces. The superscript $^a$ indicates that half
the sample have positive $A_2$, half negative; $^b$ indicates that
$A_2$ has a randomly chosen sign for each perihelion
passage. ``Total'' is the total number of apparitions (\ie perihelion
passages with $q<3$~\au), ``Spike'' is the number of these with
original inverse semimajor axes $1/a<10^{-4}\aui$, ``Tail'' is the
number with $0.0145\aui<1/a<0.029\aui$, and ``Prograde'' the number
with ecliptic inclination less than $90\degree$. The parameters $X_i$
are defined in Eq.~\ref{eq:xdef}. The mean lifetime in orbits
$\langle m \rangle$ includes all perihelion passages, whether visible
or not, after the initial apparition. The units
of $A_1$ and $A_2$ are AU~day$^{-2}$.}
\label{ta:ng_parms}
\end{table}

\subsection{Post-visibility evolution: the effect of a
solar companion or disk} \label{pa:other_scen}

In this section we investigate the influence of two hypothetical components of
the Solar System on the evolution of LP comets:
\begin{enumerate}

\item A massive circumsolar disk extending to hundreds of $\au$ or even
further. Such a disk might be an extension of the Kuiper belt
\cite[]{jewluuche96} or related to the gas and dust disks that have been
detected around stars (especially $\beta$ Pictoris) and young stellar objects
\nocite{fervid94}(Ferlet and Vidal-Madjar 1994).  
Residuals in fits to the orbit of Halley's comet imply
that the maximum allowed mass for a disk of radius $r$ is roughly
\cite[]{hogquitre91}
\begin{equation}
M_{\rm max}\simeq 10 M_\oplus \left(r\over 100\au\right)^3.
\label{eq:maxmassd}
\end{equation}
Current estimates of the mass in the Kuiper belt are much smaller, typically
$\sim0.1M_\oplus$ from direct detection of 100~km objects \cite[]{jewluuche96}
or from models of diffuse infrared emission \cite[]{bacdasste95}, but these
are based on the uncertain assumption that most of the belt mass is in the
range 30--50$\au$. The disk around $\beta$~Pic is detected in the infrared to
radii exceeding 1000~{\au} \cite[]{smiter87}; the dust mass is probably less
than $1M_\oplus$ \cite[]{art94}, but there may be more mass in condensed
objects.

\item A solar companion, perhaps a massive planet or brown dwarf, orbiting at
hundreds of $\au$.  Residuals in fits to the orbits of the outer planets imply
that the maximum allowed mass for a companion at radius $r$ is roughly
\cite[]{tre90,hogquitre91}
\begin{equation}
M_{\rm max}\simeq 100 M_\oplus \left(r\over 100\au\right)^3.
\label{eq:maxmassc}
\end{equation}
There are also significant but model-dependent constraints on the
characteristics of a solar companion from the IRAS infrared all-sky survey
\cite[]{hogquitre91}. 

\end{enumerate}

To reduce computational costs, we used the $V_1$ comets as a starting point
for these investigations; that is, the effect of the disk or companion is
ignored before the comet's first apparition (more precisely, we started the
integration at the aphelion preceding the comet's initial apparition, in order
to correctly calculate any perturbations occurring on the inbound
leg). Starting at this point is an undesirable oversimplification, but one
that should not compromise our conclusions.

\subsubsection{Circumsolar disk}

The circumsolar disk is represented by a Miyamoto-Nagai
potential \cite[\eg]{bintre87},
\begin{equation}
V_{\rm disk}(x,y,z)= \frac{-G M_d}{ \Big[ x^2 + y^2 + \left( a_{d} + 
\sqrt{z^2 + b_{d}^2} \, \right)^2 \Big]^{1/2} }. \label{eq:disk_potl} 
\end{equation}
Here $M_{d}$ is the disk mass, and $a_{d}$ and $b_{d}$ are parameters
describing the disk's characteristic radius and thickness. We assume that the
disk is centered on the Solar System barycenter and coplanar with the
ecliptic. We considered disk masses $M_d$ of 0.1, 1, and 10 Jupiter masses,
disk radii $a_d$ of $100$ and $1000\au$, and a fixed axis ratio
$b_d/a_d=0.1$. The two more massive disks with $a_d=100\au$ are unrealistic
because they strongly violate the constraint (\ref{eq:maxmassd}), but we
examine their effects in order to explore as wide a range of parameters as
possible.

Comets arriving from the Oort cloud have fallen through the disk
potential and hence are subjected to a shift in their original inverse
semimajor axis. This offset can be as large as $2 \times
10^{-4}$~{\aui} for a 10 Jupiter-mass disk with radius 100~{\au}, but
is much smaller for disks that do not already violate the
observational constraint (\ref{eq:maxmassd}).  This shift is not shown
in the Figures below, for which the semimajor axis is measured at
aphelion. 

As usual, the {\sc Perihelion Too Large} end-state
(\S~\ref{pa:endstates}) was entered if $q>40\au$ and $\sin 2\og >
0$. The assumption that such comets are unlikely to become visible in
the future is only correct if the torque is dominated by the Galactic
tide, and this may not be the case when a disk is present. However,
there is no significant difference in the numbers or semimajor axes of
the comets reaching this end-state in simulations with and without a
circumsolar disk, suggesting that evolution to this end-state is
indeed dominated by the Galaxy.

\begin{figure}[p]
\centerline{\vbox{\hbox{\psfig{figure=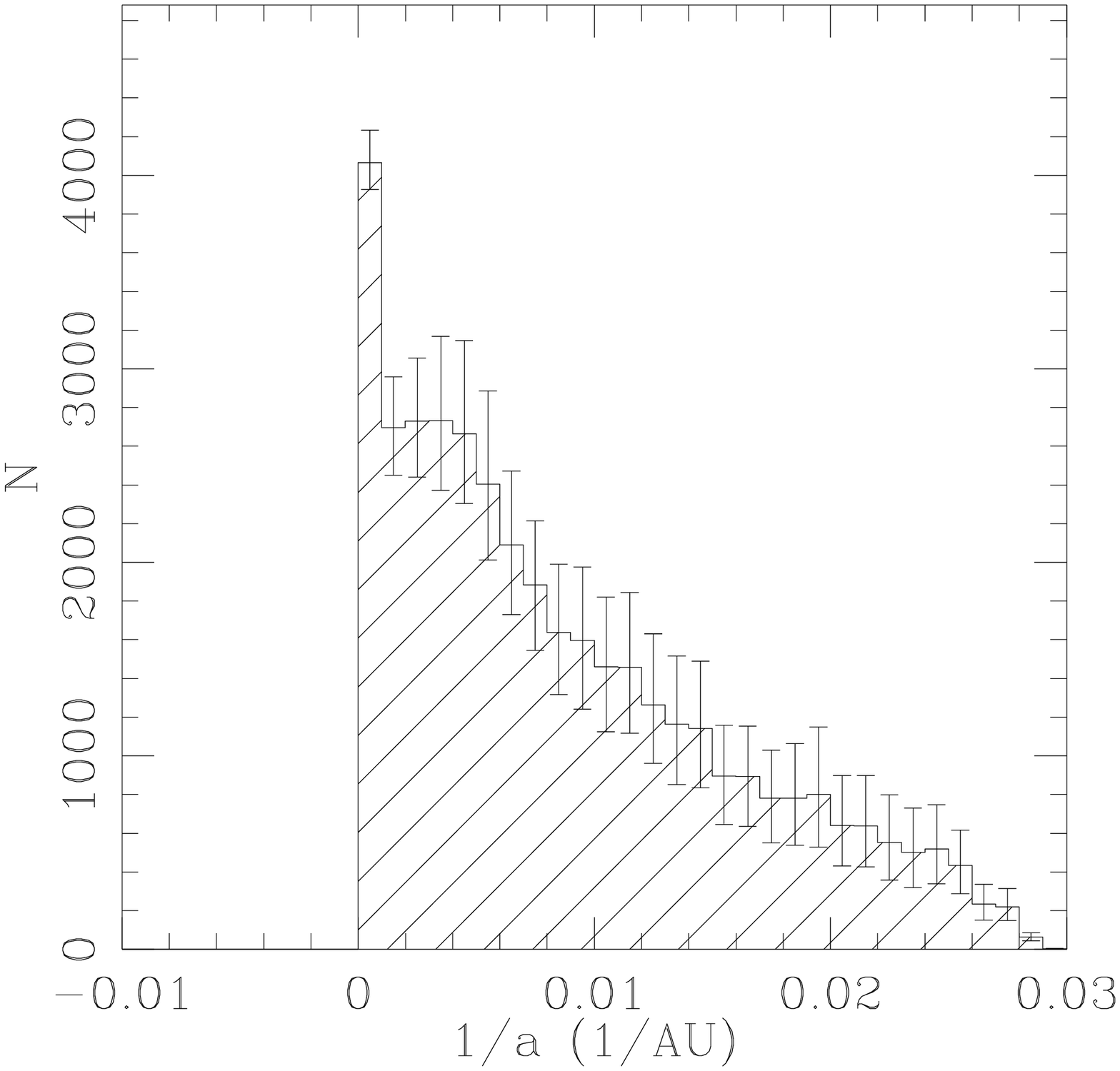,height=1.5in}
                        \psfig{figure=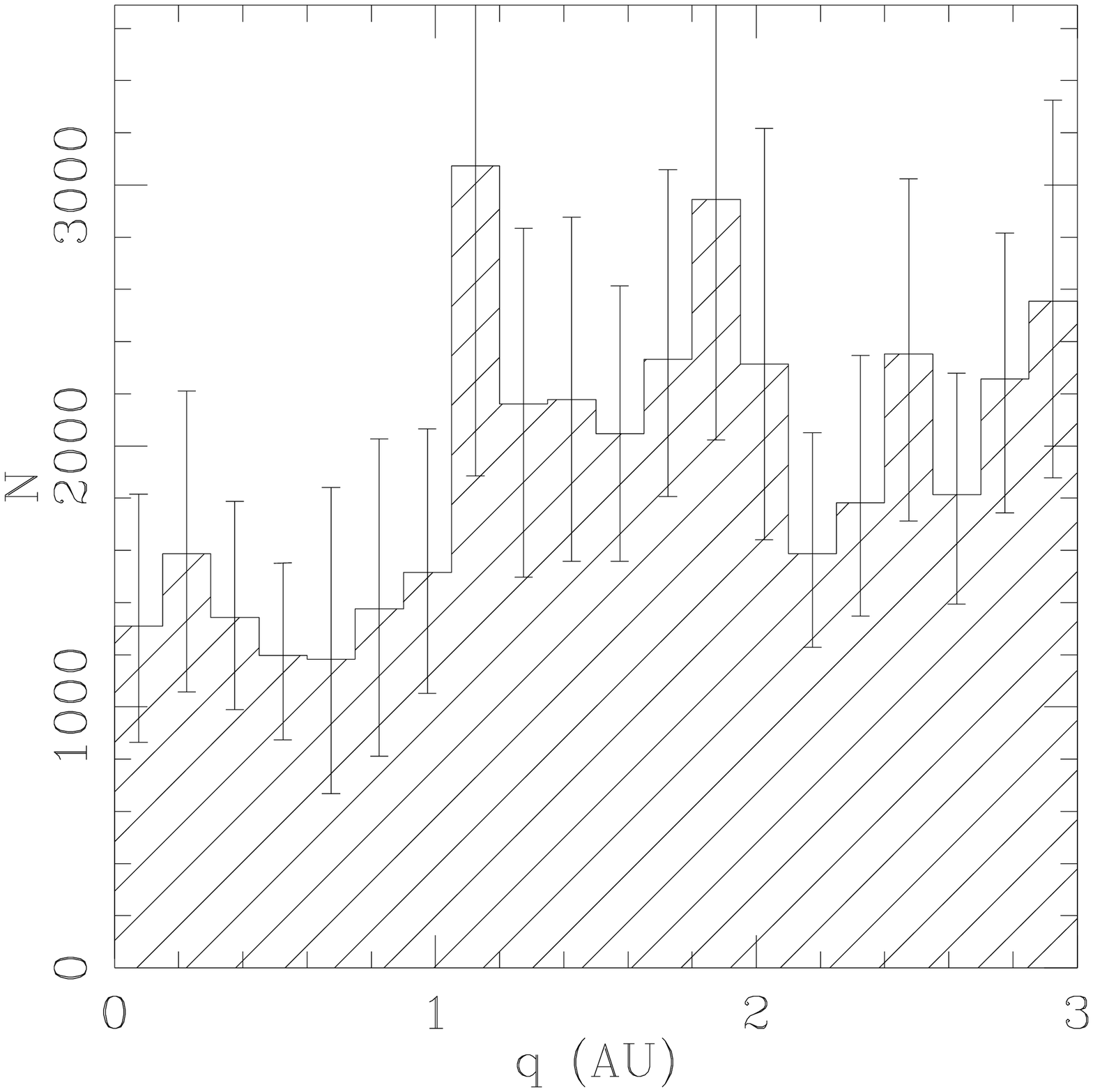,height=1.5in}
                        \psfig{figure=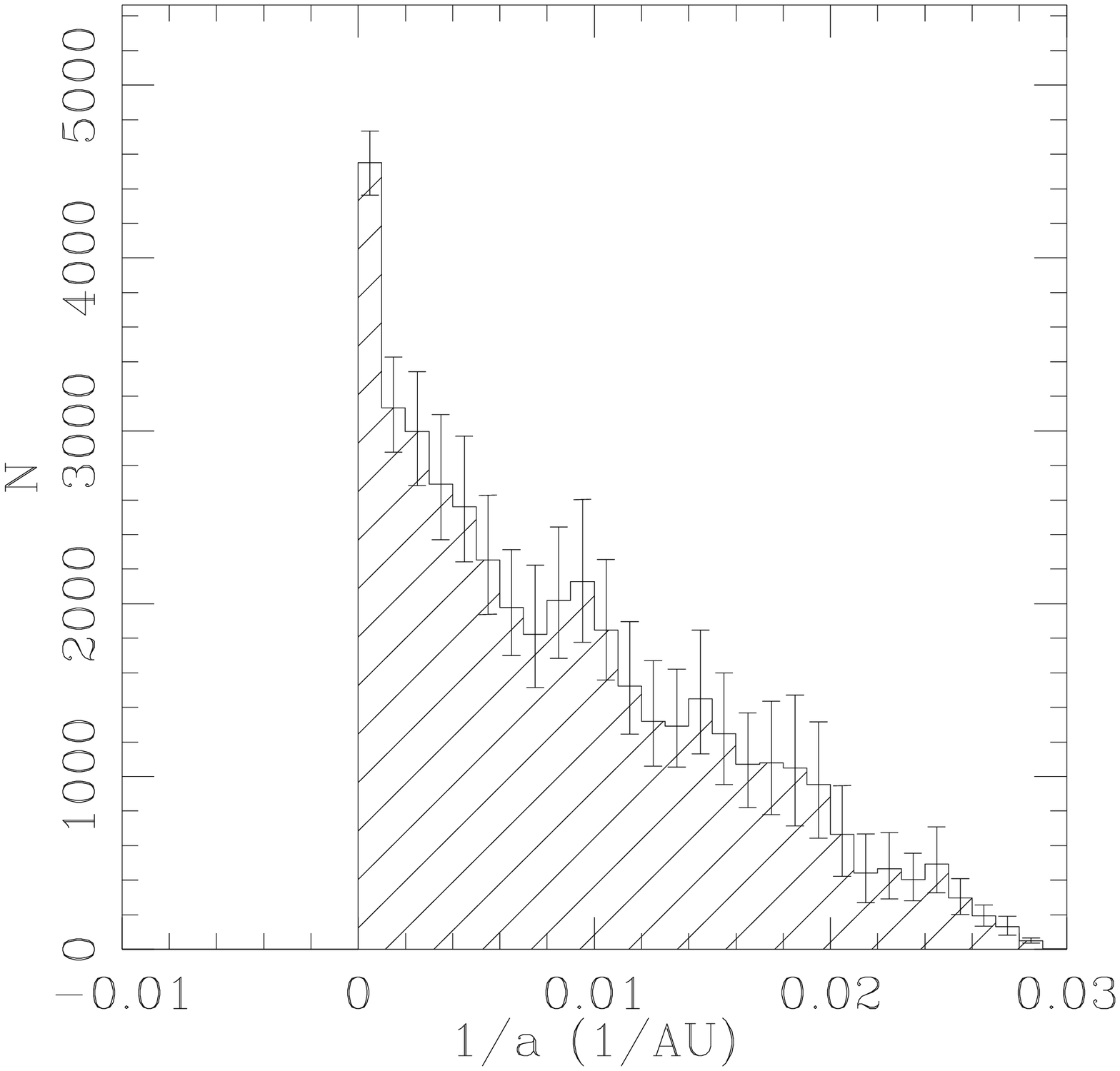,height=1.5in}
                        \psfig{figure=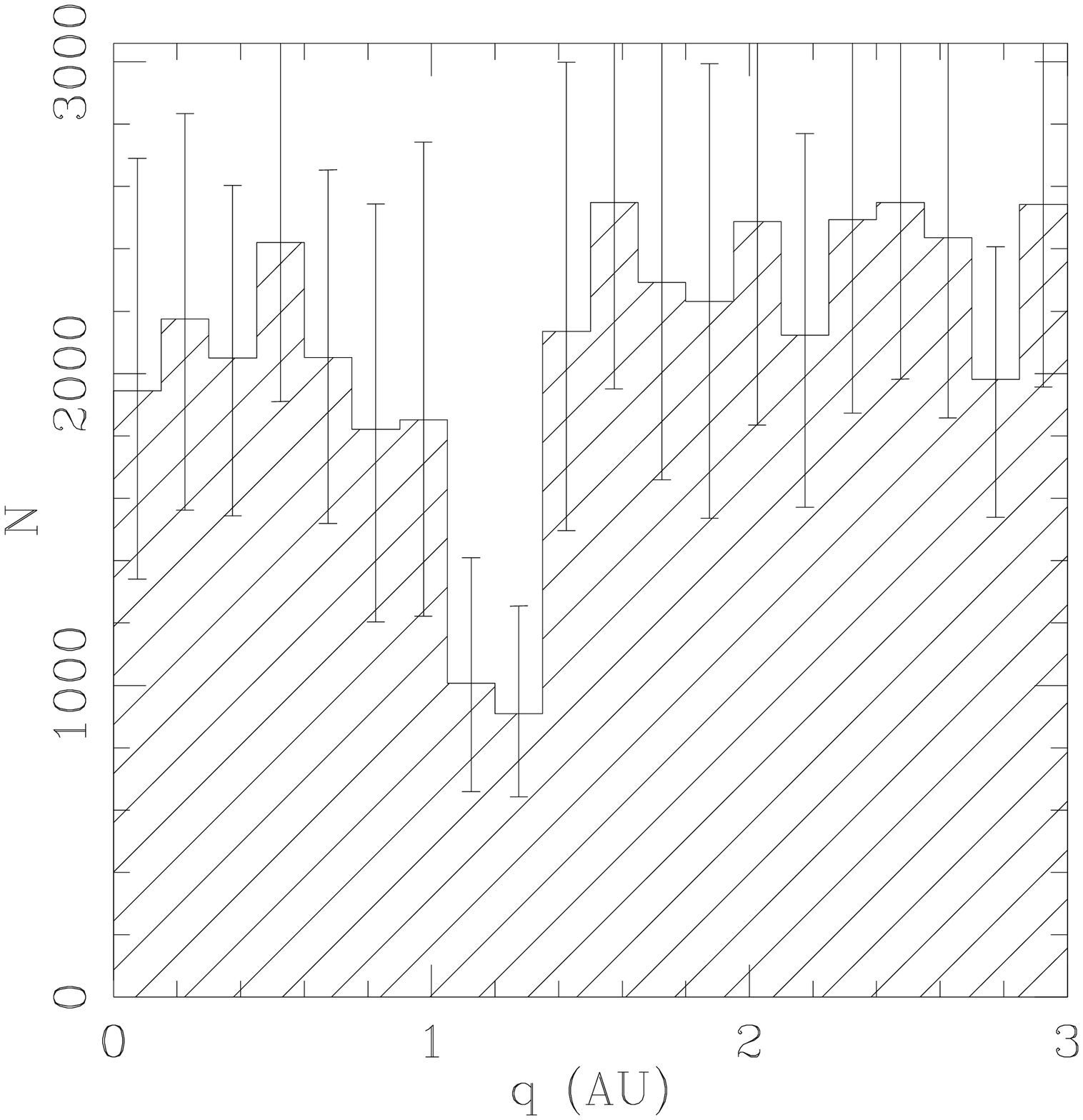,height=1.5in}}
                        \hbox{\psfig{figure=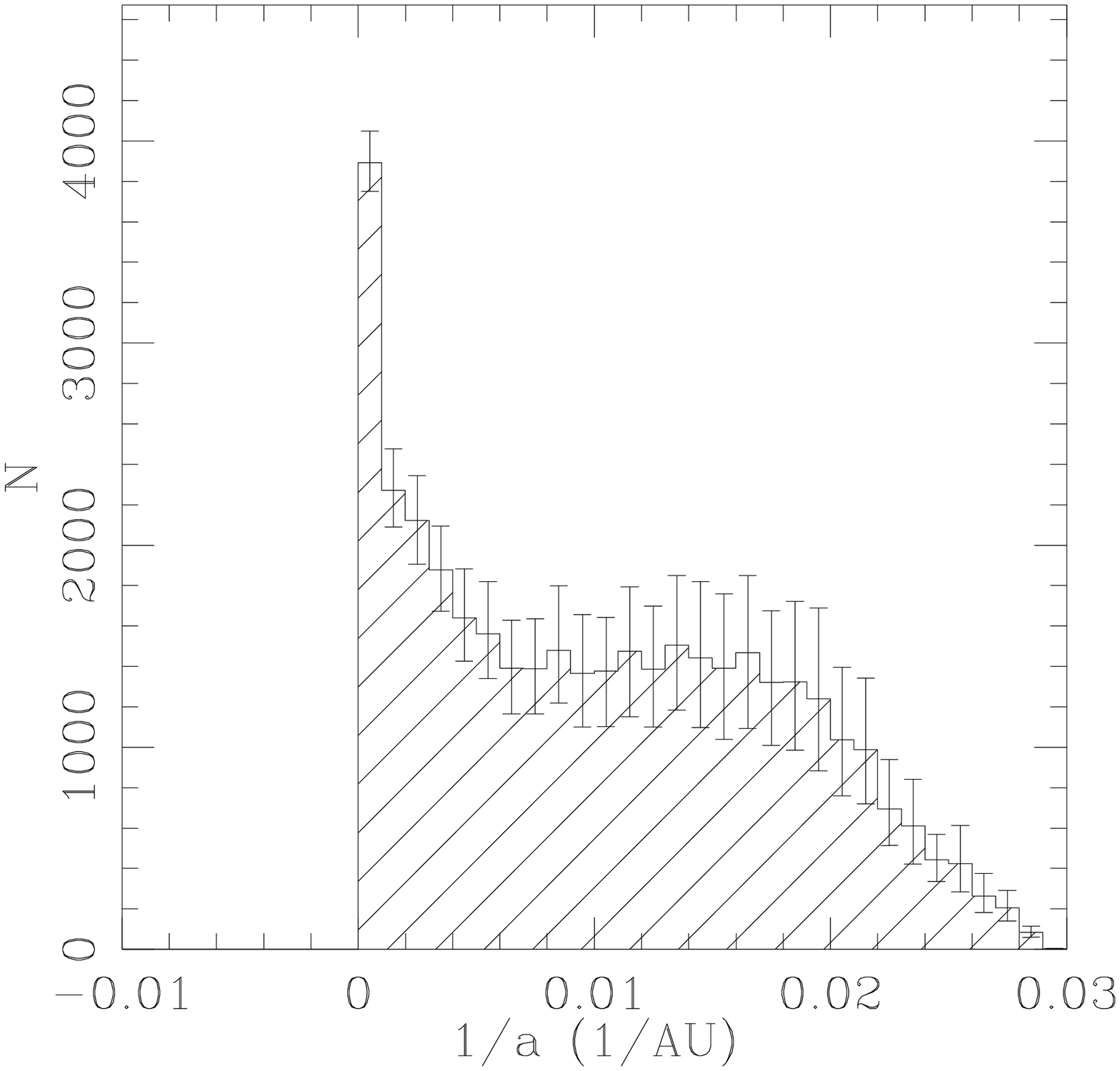,height=1.5in}
                        \psfig{figure=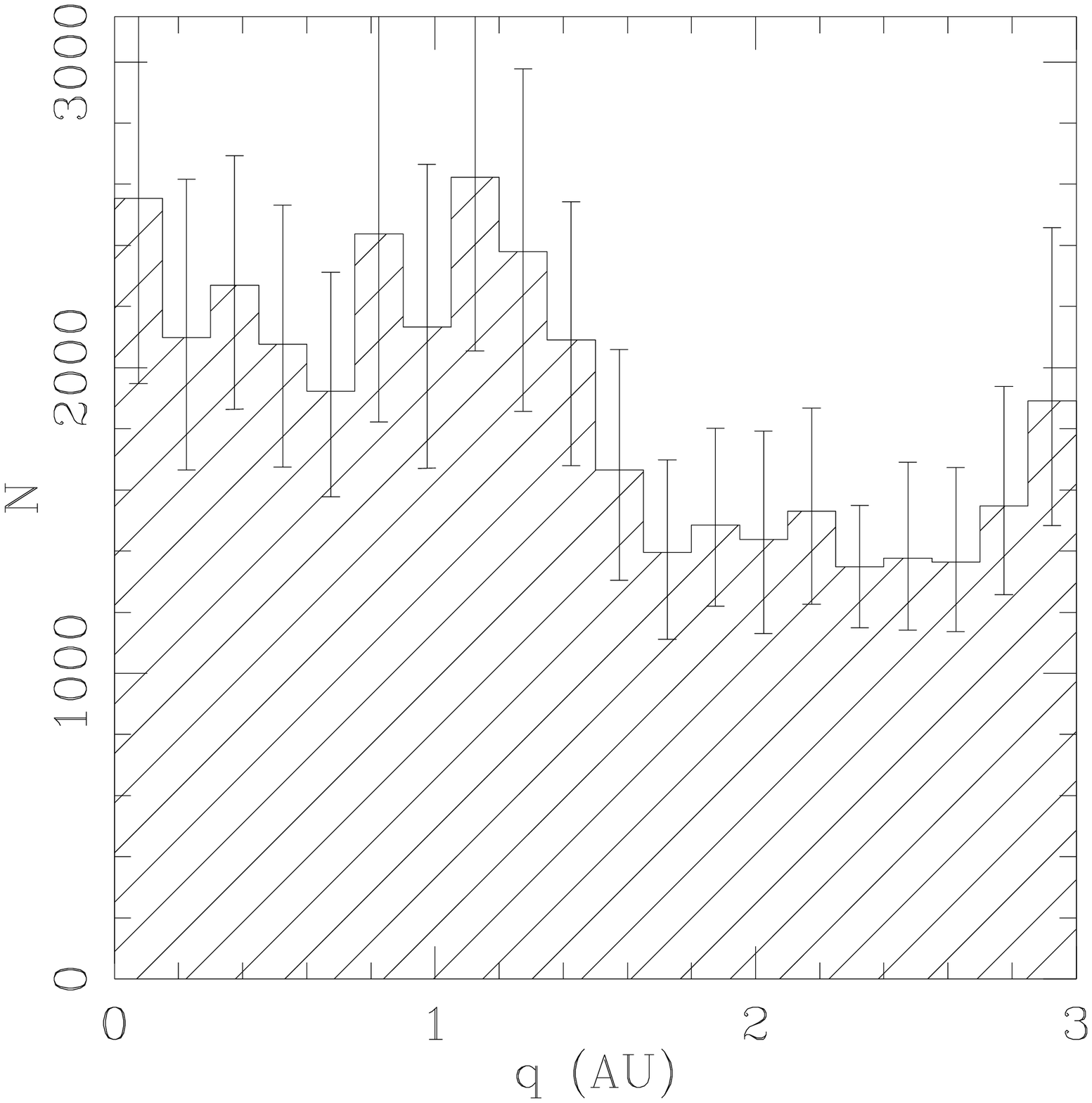,height=1.5in}
                        \psfig{figure=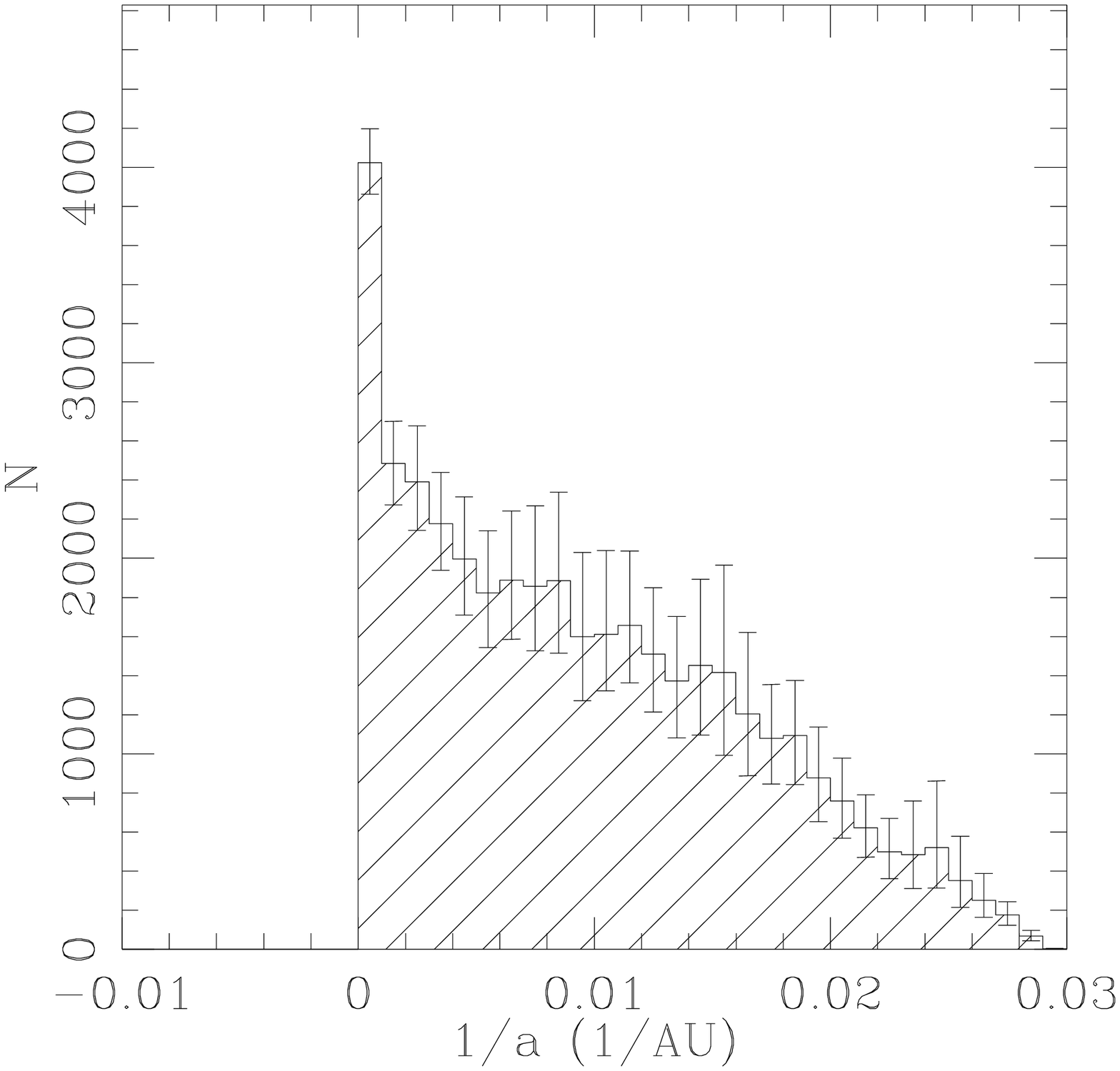,height=1.5in}
                        \psfig{figure=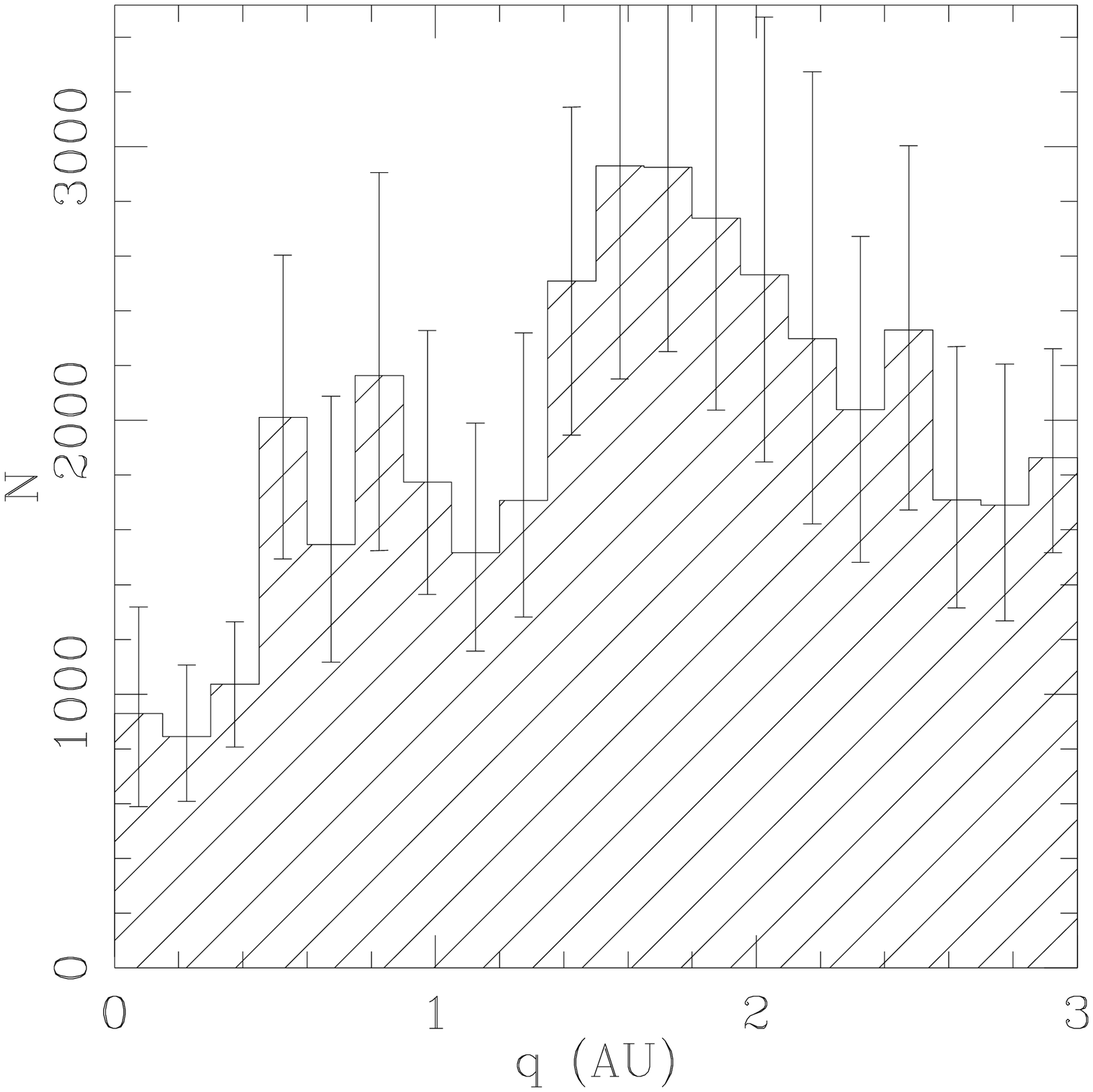,height=1.5in}}
                        \hbox{\psfig{figure=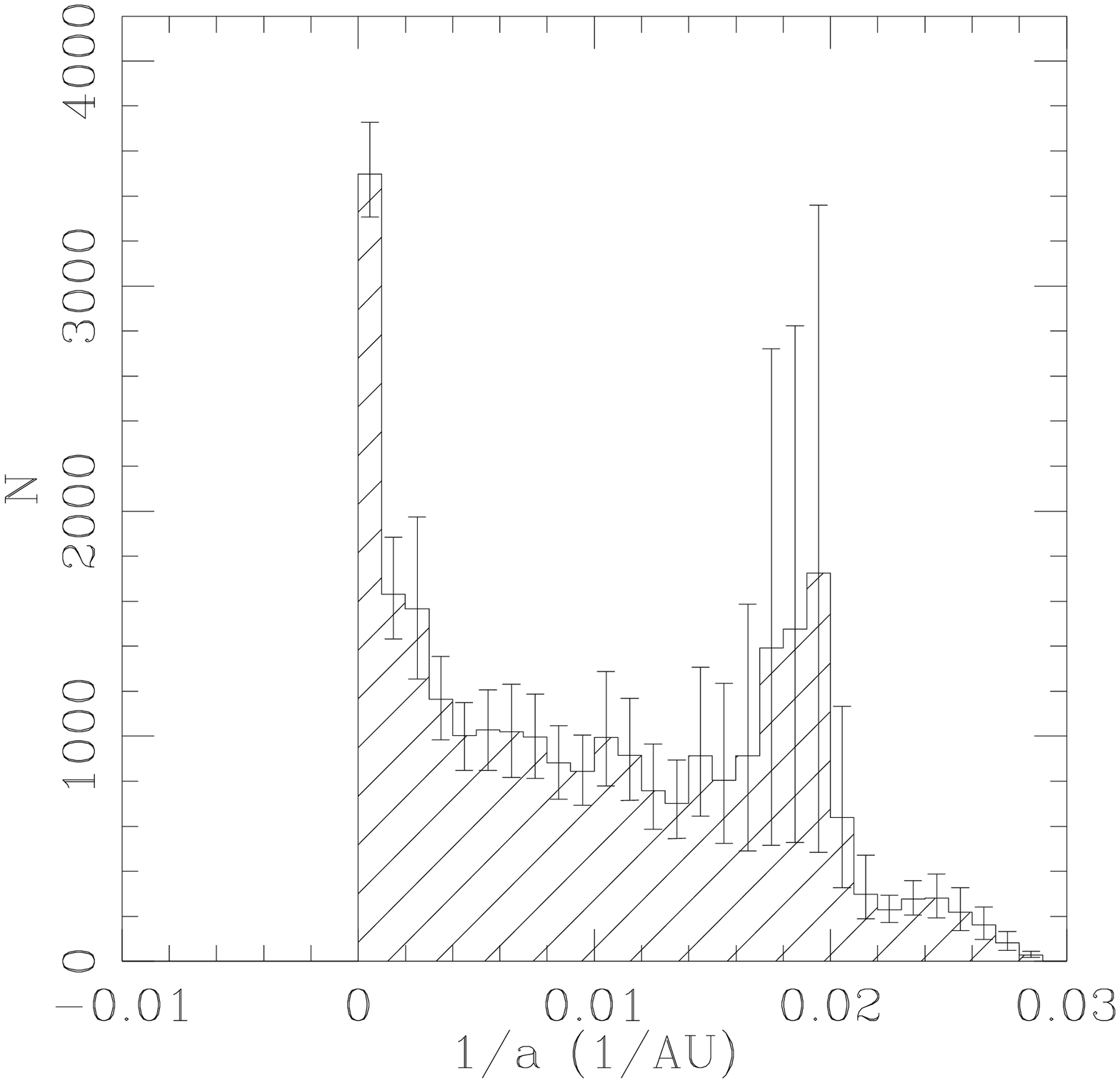,height=1.5in}
                        \psfig{figure=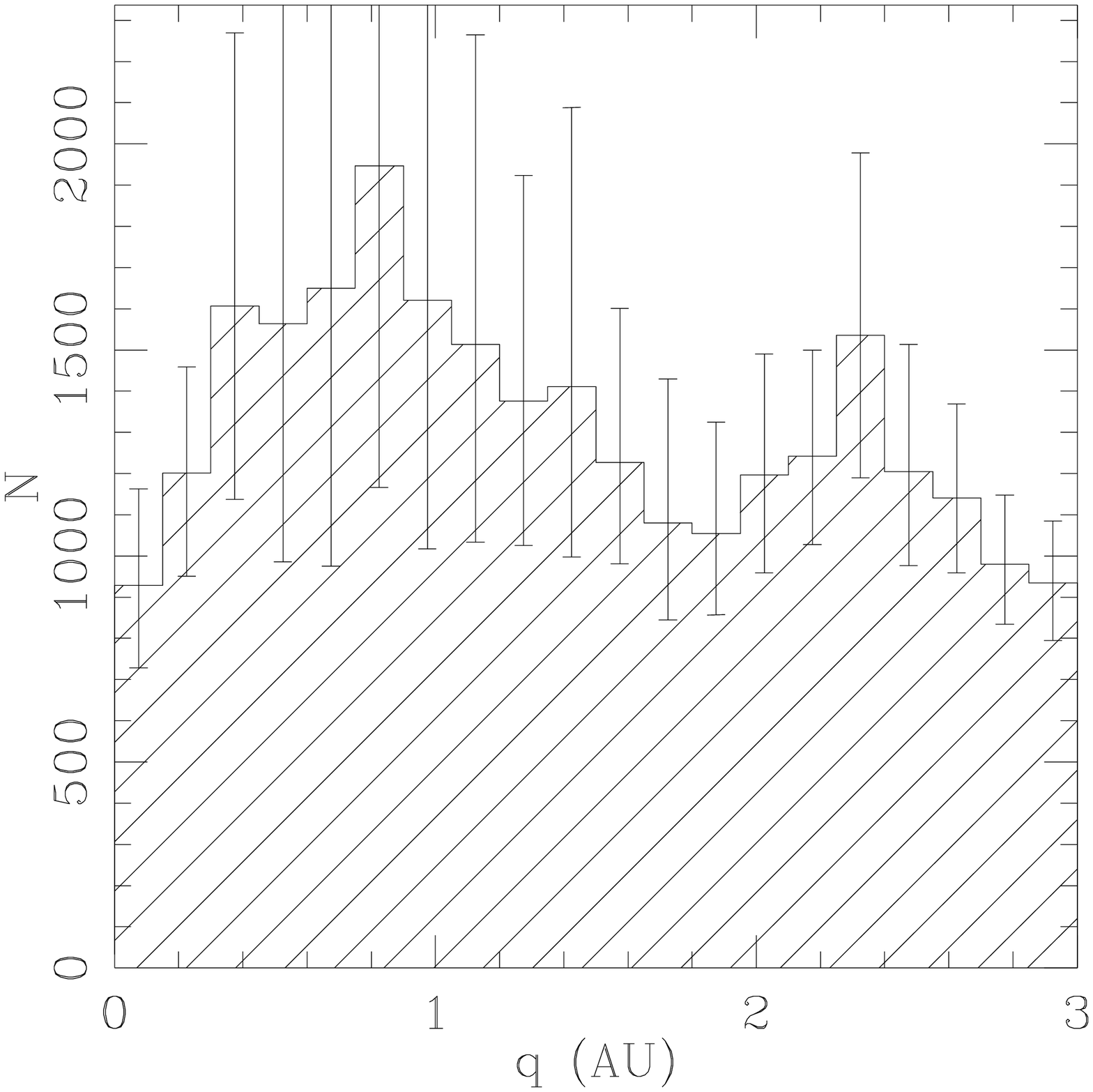,height=1.5in}
                        \psfig{figure=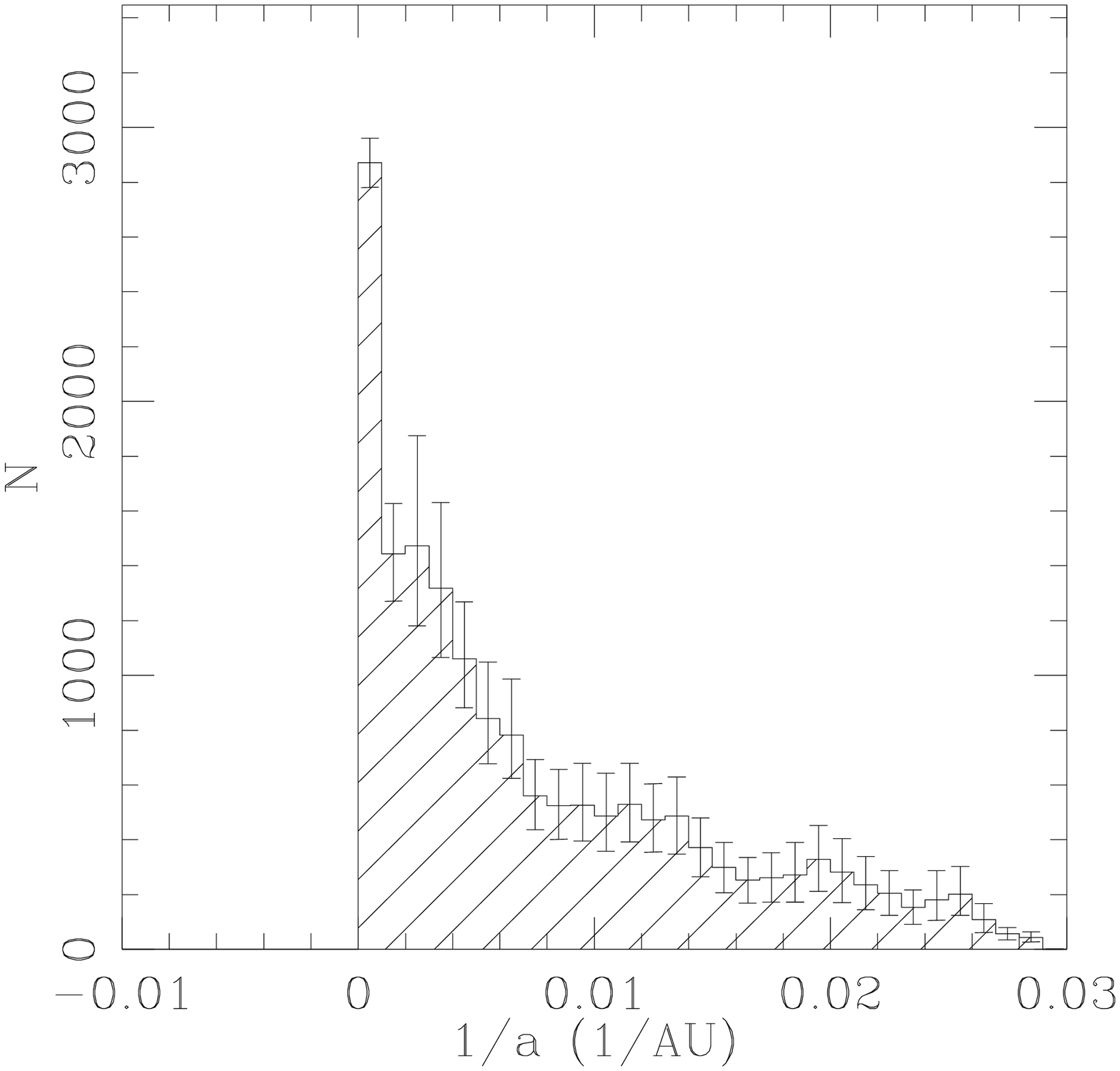,height=1.5in}
                        \psfig{figure=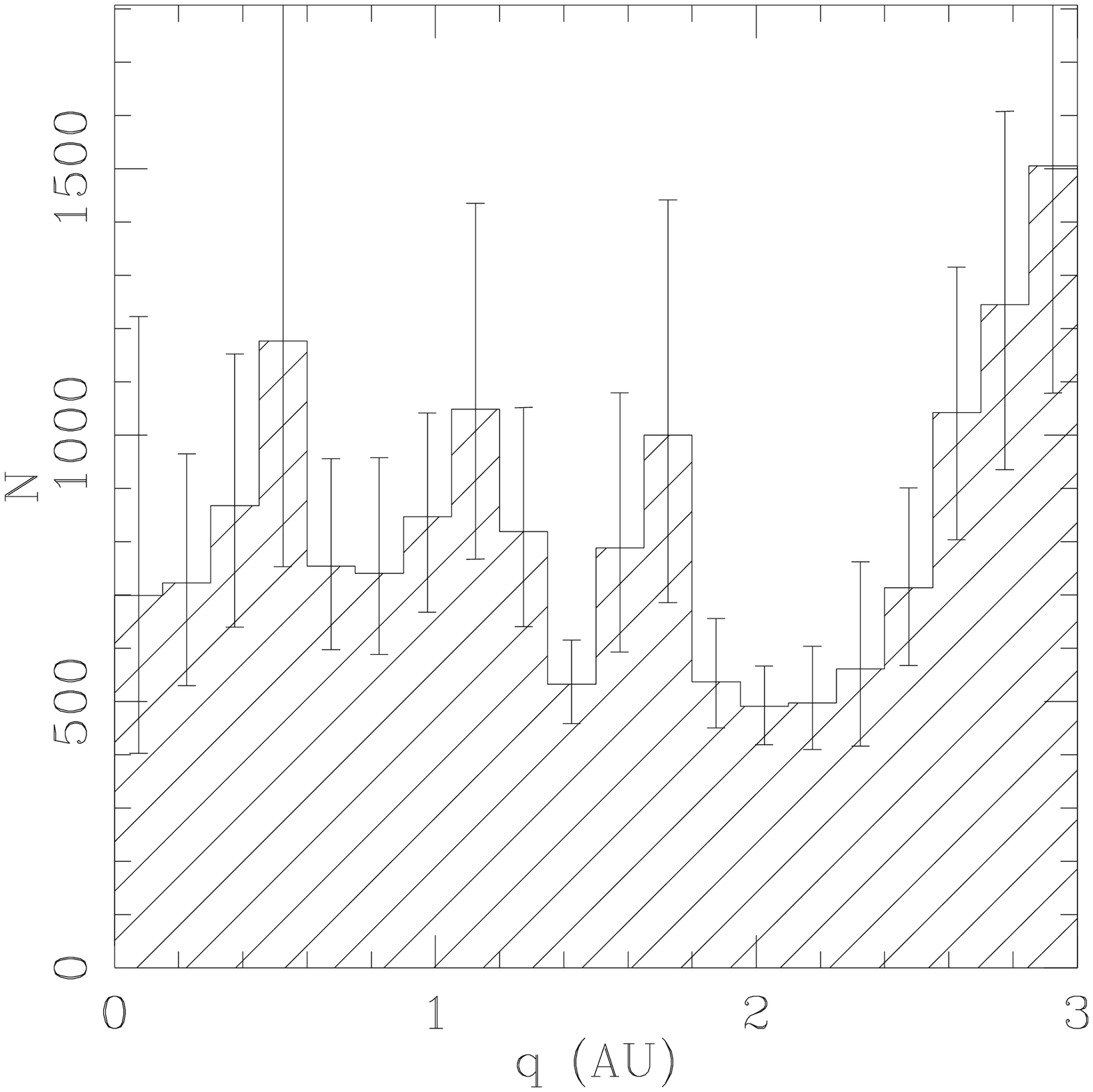,height=1.5in}}
                        \rule[2mm]{6.2in}{0.5mm}
                        \hbox{\psfig{figure=tvp_b.eps,height=1.5in}
                        \psfig{figure=tvp_q.eps,height=1.5in}
                        \psfig{figure=tmars_b.eps,height=1.5in}
                        \psfig{figure=lp_qlines_new.eps,height=1.5in}}}}
\caption{Distribution of the inverse semimajor axis $1/a$ and
perihelion distance $q$ for the $V_\infty$ comets, when the Solar
System contains a massive circumsolar disk.  Left panels:
characteristic disk radius $a_d =100$~\au. Right panels: disk radius
$a_d =1000$~\au. From the top down, the disk masses are 0.1, 1 and 10
Jupiter masses. The bottom line of panels is for comparison, and includes the
standard model (left side) and the observations (right side). The
observed perihelion distribution includes curves indicating the
estimated intrinsic distribution (see Fig. \ref{fi:Vm_q} for
details).}
\label{fi:disk1}
\end{figure}

The results from simulations including a circumsolar disk are displayed in
Fig.~\ref{fi:disk1} and Table~\ref{ta:disk_parms}.  One plot of the energy
distribution in Fig.~\ref{fi:disk1} shows a strong peak near $1/a=0.02\aui$;
as the large error bars suggest, this peak is caused by a single comet and has
little statistical significance.

The principal effect of the disk is to exert an additional torque on the
comets, resulting in oscillations of the comet's perihelion distance.  This
effect normally increased the comet's lifetime, as the risk of ejection is
greatly reduced when the comet is outside Saturn's orbit. The perihelion
oscillations also enhance the probability of collision with the Sun
(Table~\ref{ta:disk_parms}).

The perihelion distribution of visible comets is not strongly affected
by the disk. The presence of a massive disk reduces the number of
dynamically old comets (because their perihelia are no longer
nearly constant, only a fraction of them are
visible at any given time), but not enough so that the energy
distribution is consistent with the observations. This conclusion is
confirmed by examining the $X$ parameters in
Table~\ref{ta:disk_parms}, which should be unity if the simulated
element distribution agrees with the observations
(cf. Eq.~\ref{eq:xdef}). The values of $X_1$, which measures the ratio
of number of comets in the spike to the total number, are far smaller
than unity even for the most massive disks.  Increasing the disk mass
tends to improve $X_2$ and $X_3$ for the 1000~{\au} disk, but degrades
the fit for the $100\au$ disk. There is no set of disk parameters than
comes close to producing a match with observations. Using a more
elaborate model for selection effects (Eq.~\ref{eq:discoverprob2})
does not alter this conclusion (see bottom half of Table
~\ref{ta:disk_parms}).

We conclude that a circumsolar disk cannot by itself resolve the discrepancy
between the observed and predicted LP comet distribution.

\begin{table}[p]
\centerline{
\begin{tabular}{|cc|ccccccccc|} \hline
$M_{d} $ & $a_d$  & Total & Spike & Tail & Prograde & $X_1$ & $X_2$ & $X_3$ &
$\langle m \rangle$ & $R_{\odot}$ \\  \hline \hline
0  & ---   & 52303 & 1473 & 15004& 15875 & 0.07 & 4.37 & 0.61 & 45.4&0\\
0.1& 100   & 38947 & 1486 & 8382 & 15178 & 0.10 & 3.28 & 0.78 & 60.4&0\\
0.1& 1000  & 42106 & 1496 & 9122 & 16957 & 0.09 & 3.30 & 0.80 & 33.7&1\\
1  & 100   & 37676 & 1459 & 12027& 11888 & 0.10 & 4.86 & 0.63 & 60.8&2\\
1  & 1000  & 39138 & 1458 & 9944 & 16141 & 0.10 & 3.87 & 0.82 & 44.7&1\\
10 & 100   & 26445 & 1416 & 8881 & 6813  & 0.14 & 5.11 & 0.51 & 62.6&5\\
10 & 1000  & 16636 & 1324 & 3020 & 7555  & 0.21 & 2.76 & 0.91 & 66.9&3\\
\hline
$0^d$  & --- & 33449 & 957  & 10190& 14308 & 0.09 & 4.17 & 0.83 & 45.5&0\\
$0.1^d$&100  & 24535 & 968  & 6086 & 8589  & 0.10 & 3.76 & 0.70 & 60.4&0\\
$0.1^d$&1000 & 26335 & 969  & 5261 & 11950 & 0.10 & 3.04 & 0.91 & 33.7&1\\
$1^d$  &100  & 27655 & 947  & 9514 & 8712  & 0.09 & 5.24 & 0.63 & 60.8&2\\
$1^d$  &1000 & 25200 & 947  & 7070 & 9881  & 0.10 & 4.27 & 0.78 & 44.7&1\\
$10^d$ &100  & 18769 & 939  & 7104 & 4103  & 0.13 & 5.76 & 0.44 & 62.6&5\\
$10^d$ &1000 & 10600 & 910  & 1541 & 4650  & 0.23 & 2.21 & 0.86 & 66.9&3\\
\hline
\end{tabular}}
\caption{Parameters of the distribution of $V_\infty$ comets, when the
Solar System contains a circumsolar disk. The disk mass $M_d$ is
measured in Jupiter masses and the disk radius $a_d$ is measured in
$\au$. The rightmost column indicates the number of comets that
collided with the Sun. The superscript $^d$ indicates that the
discovery probability from Eq.~\ref{eq:discoverprob2} has been
applied. The definitions of the other columns are the same as in
Table~\ref{ta:ng_parms}.}
\label{ta:disk_parms}
\end{table}

\subsubsection{Solar companion} \label{pa:planetx}

For simplicity, we shall assume that the solar companion has a circular orbit
in the ecliptic (the orientation and eccentricity of the companion orbit
should not strongly affect its influence on the LP comets since the comets are
on isotropic, highly eccentric robits).

We examined companion masses $M_X$ of 0.1, 1 and 10 Jupiter masses and
orbital radii of 100 and 1000~{\au}.  The most massive companion at
$100\au$ is unrealistic because it strongly violates the constraint
(\ref{eq:maxmassc}).  As in the previous subsection, the original
semimajor axes of the comets are measured at aphelion, and thus do not
include the energy offset caused by their fall through the companion's
gravitational potential. 

\begin{figure}[p]
\centerline{\vbox{\hbox{\psfig{figure=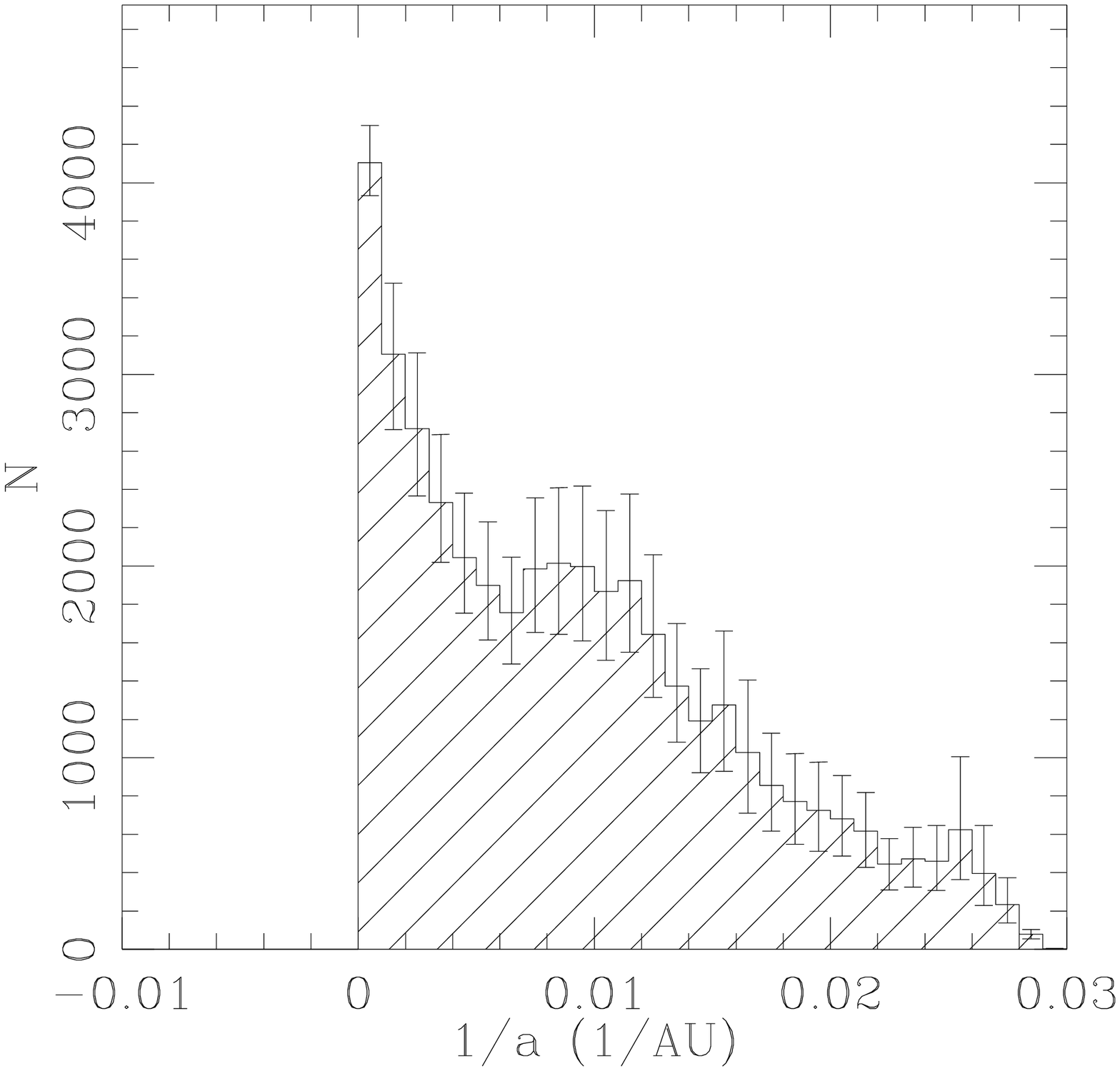,height=1.5in}
                        \psfig{figure=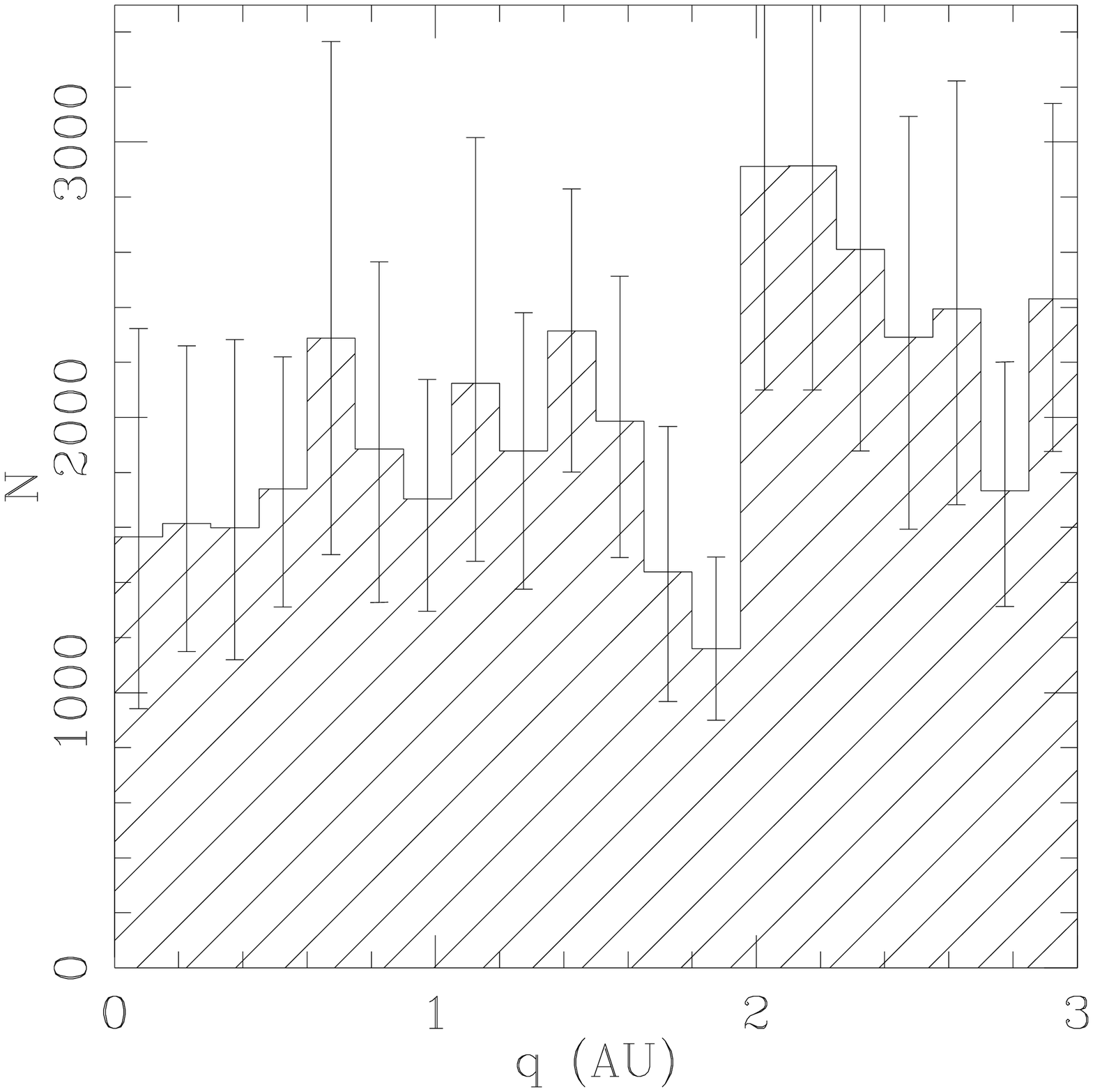,height=1.5in}
                        \psfig{figure=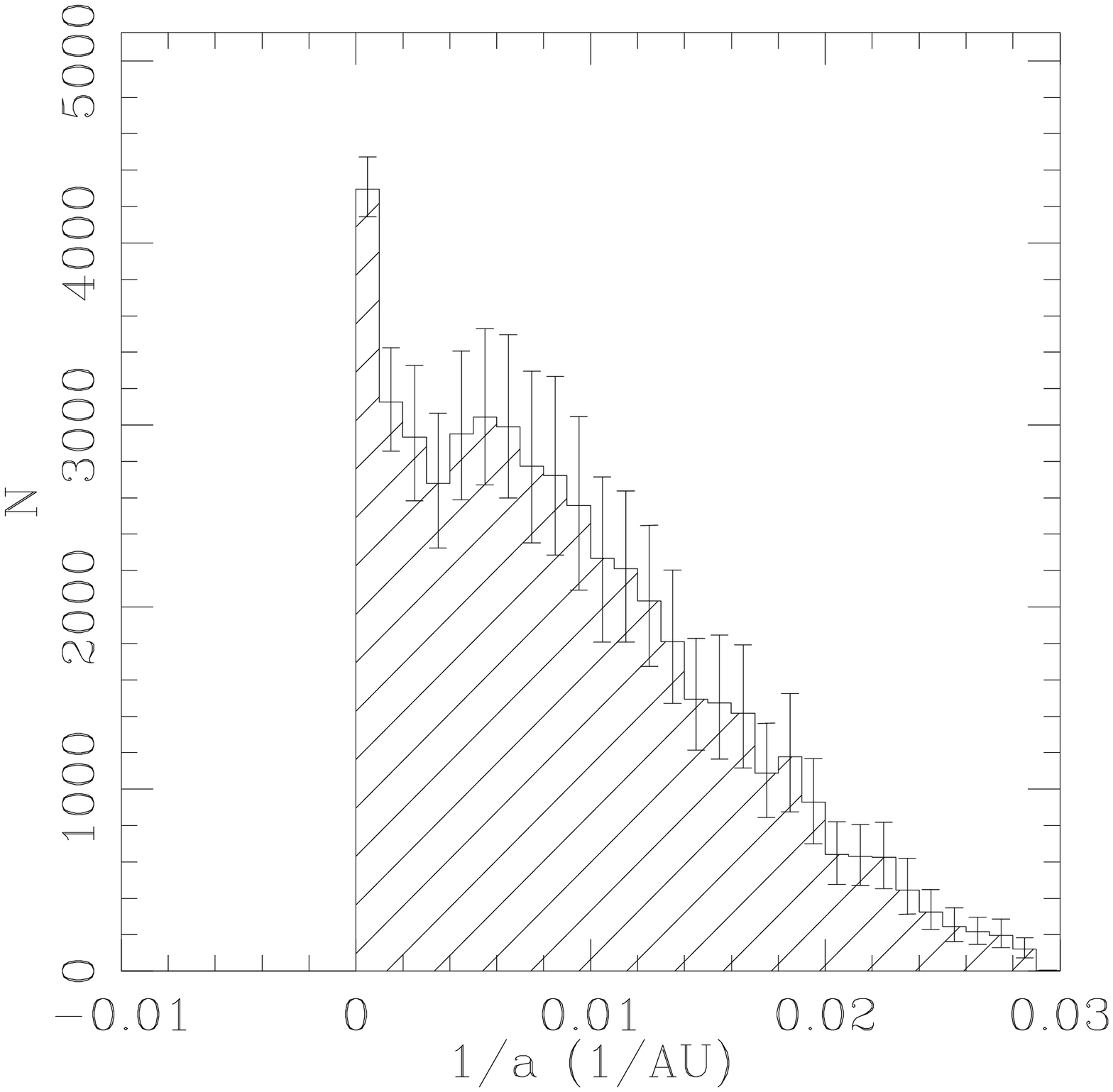,height=1.5in}
                        \psfig{figure=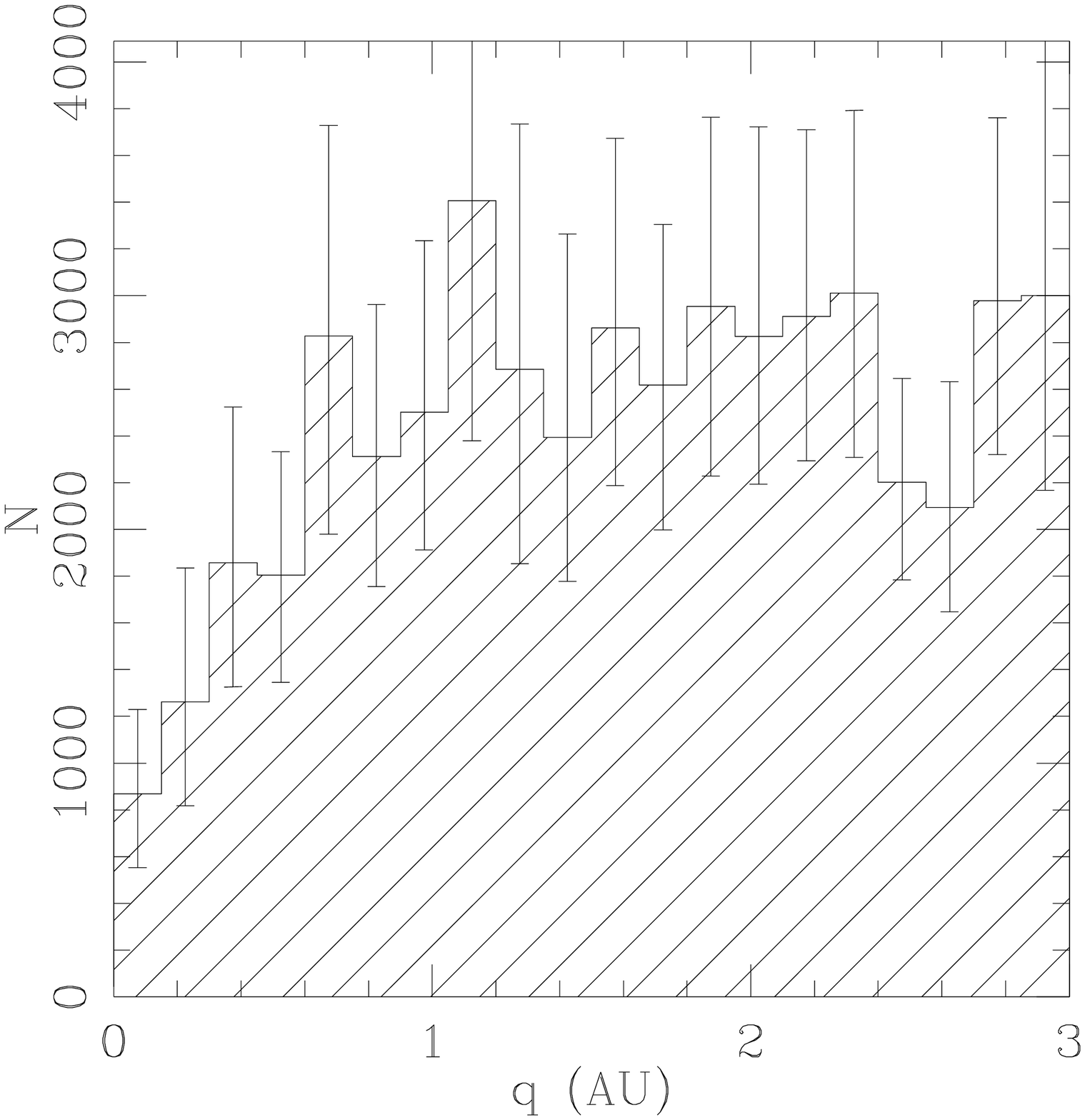,height=1.5in}}
                  \hbox{\psfig{figure=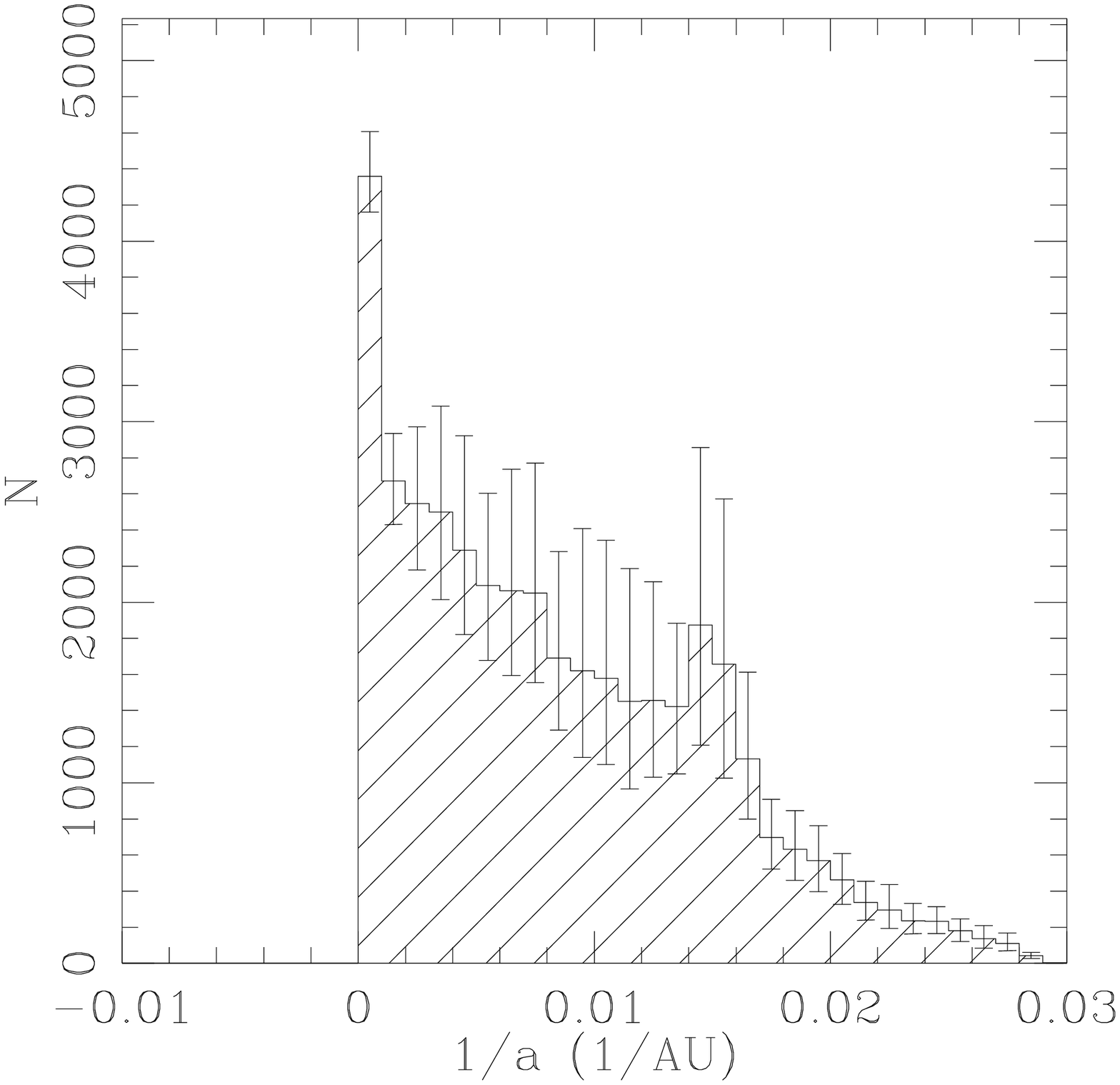,height=1.5in}
                        \psfig{figure=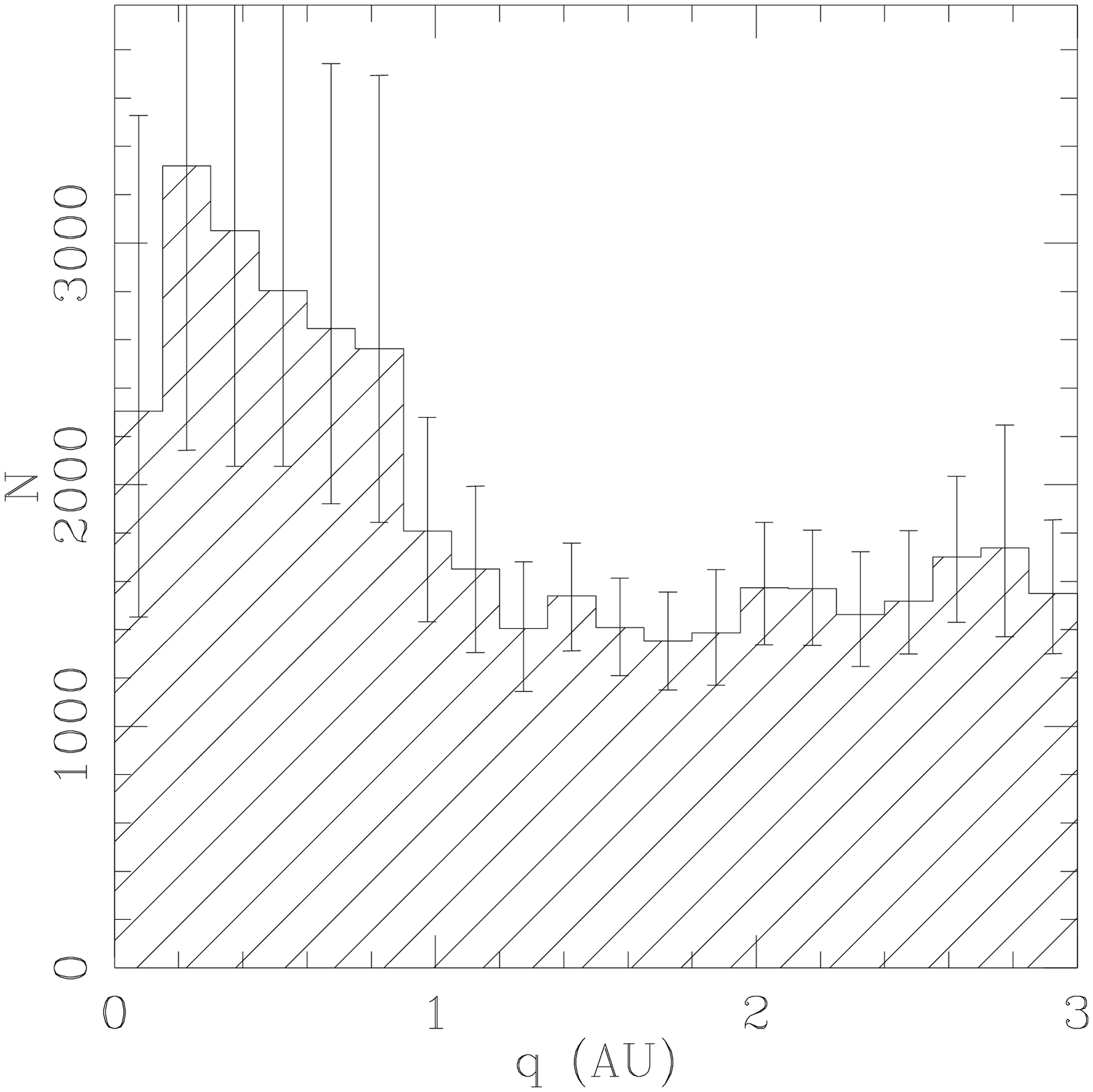,height=1.5in}
                        \psfig{figure=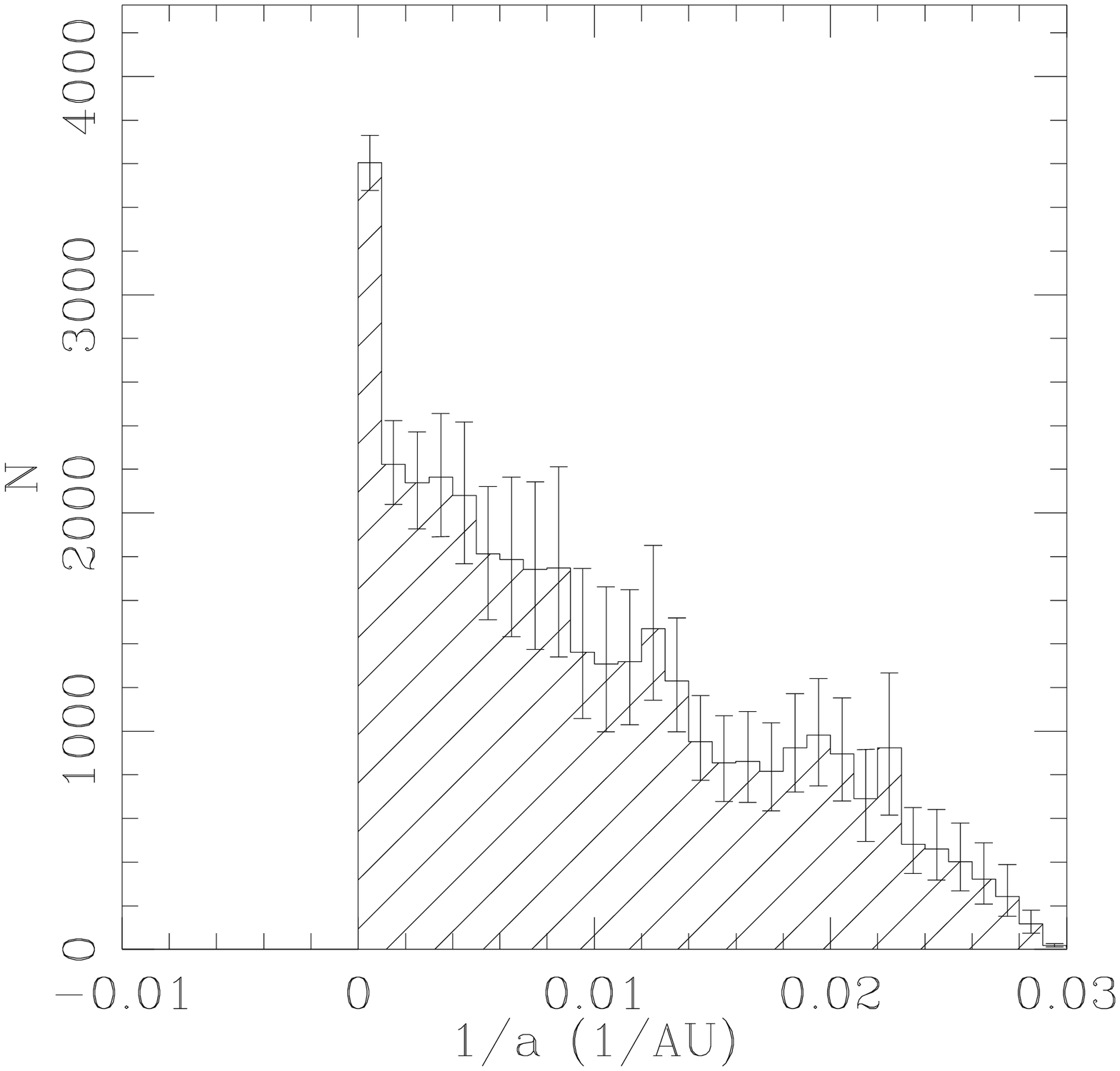,height=1.5in}
                        \psfig{figure=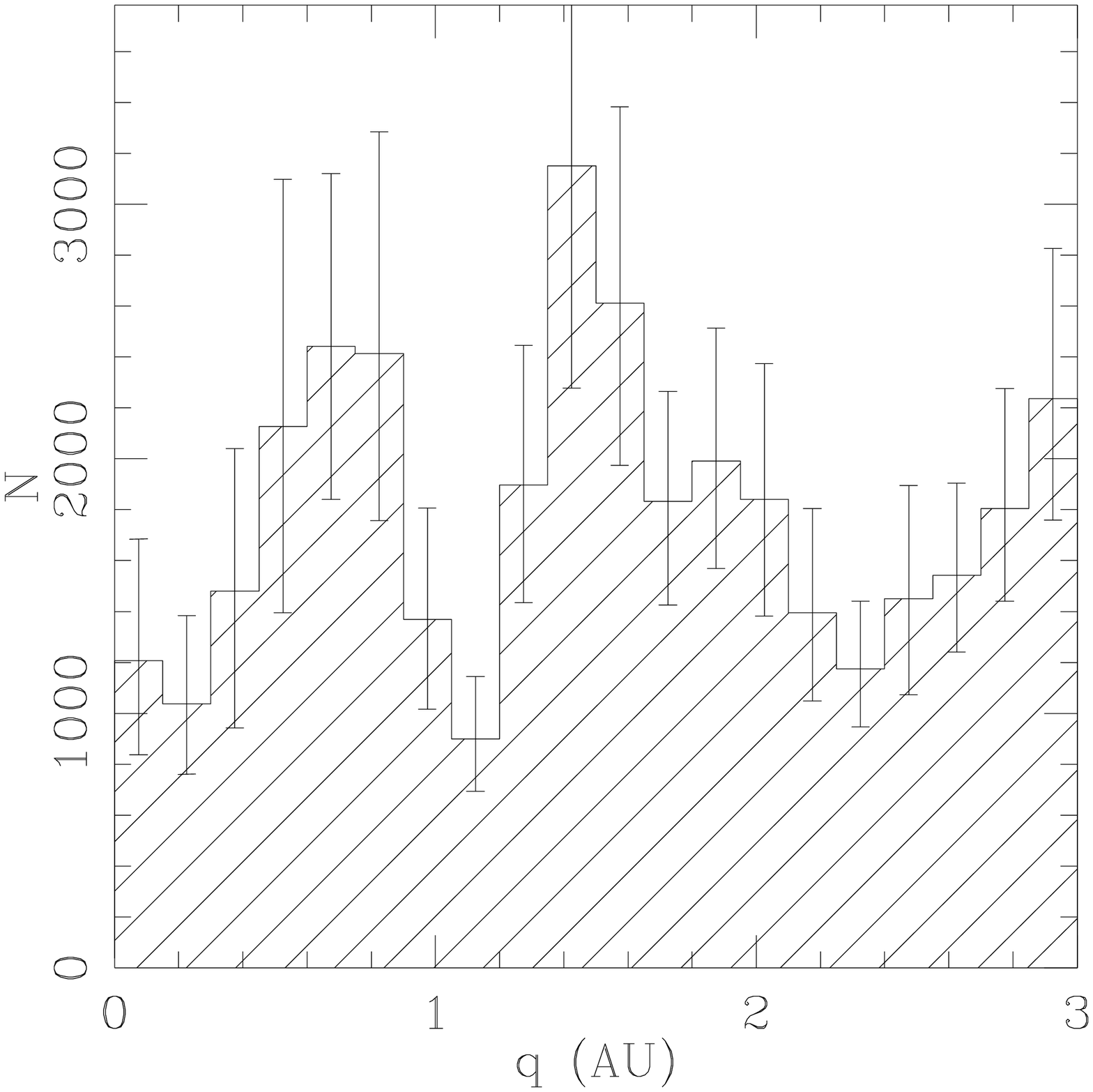,height=1.5in}}
                  \hbox{\psfig{figure=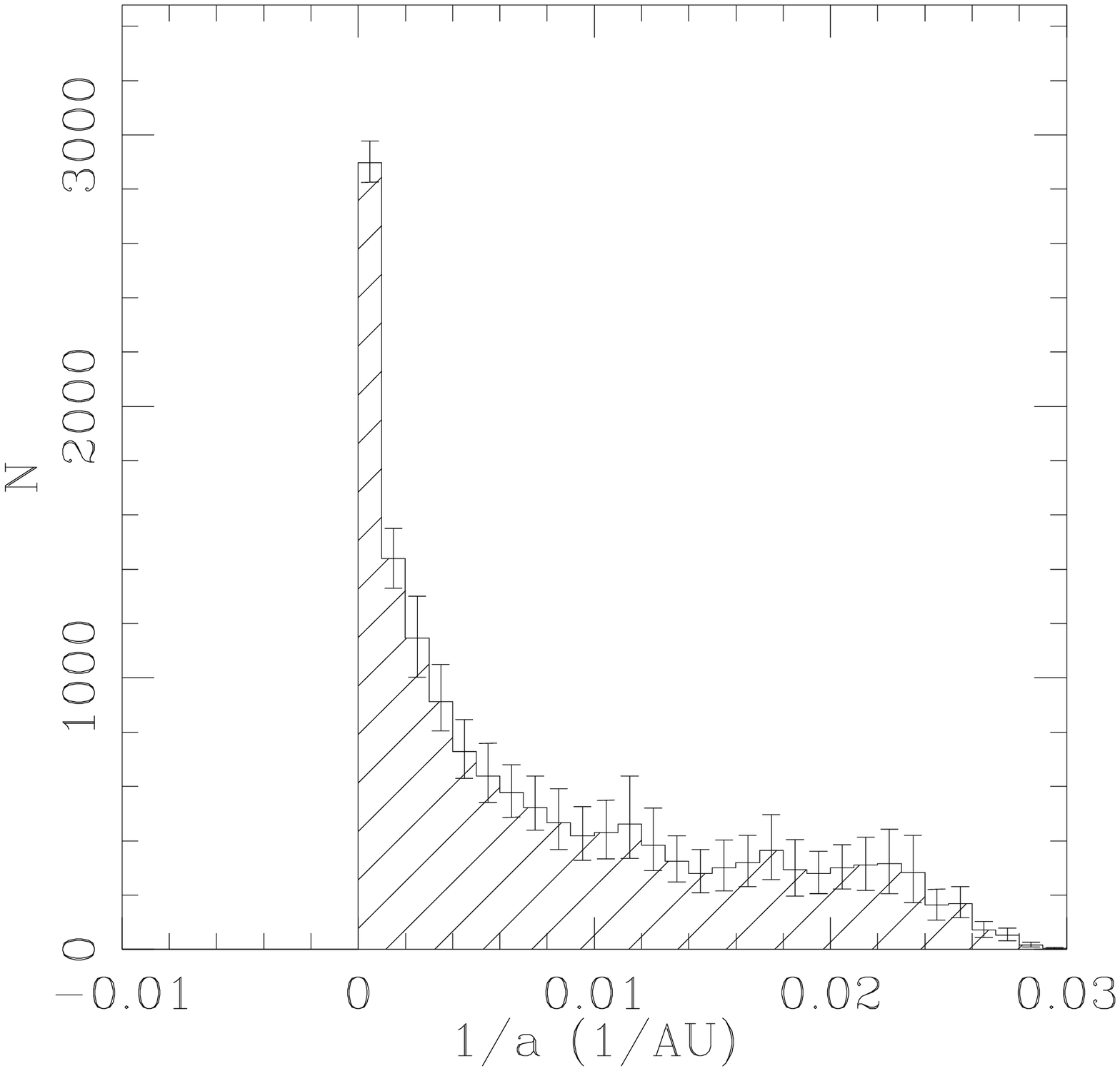,height=1.5in}
                        \psfig{figure=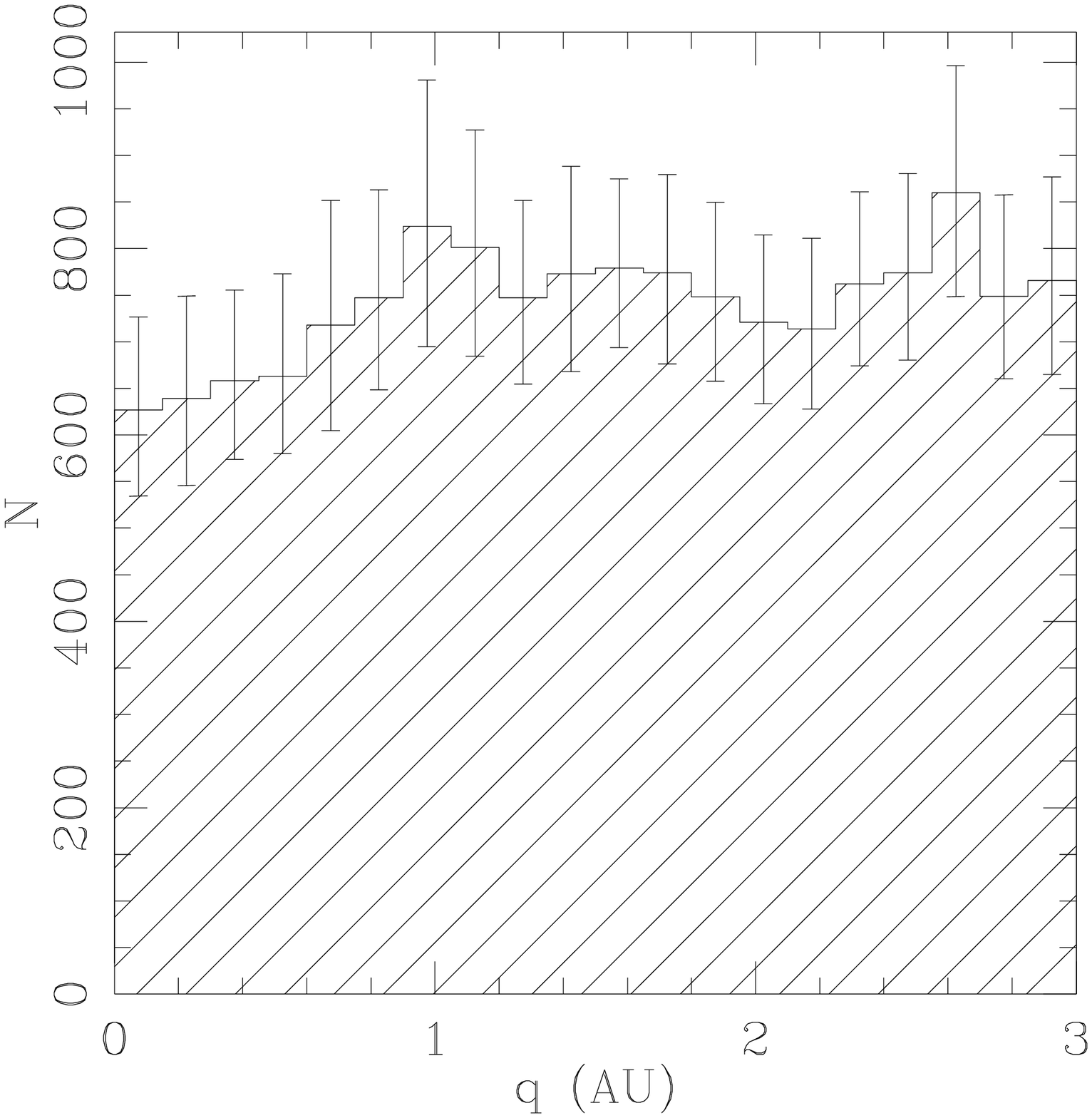,height=1.5in}
                        \psfig{figure=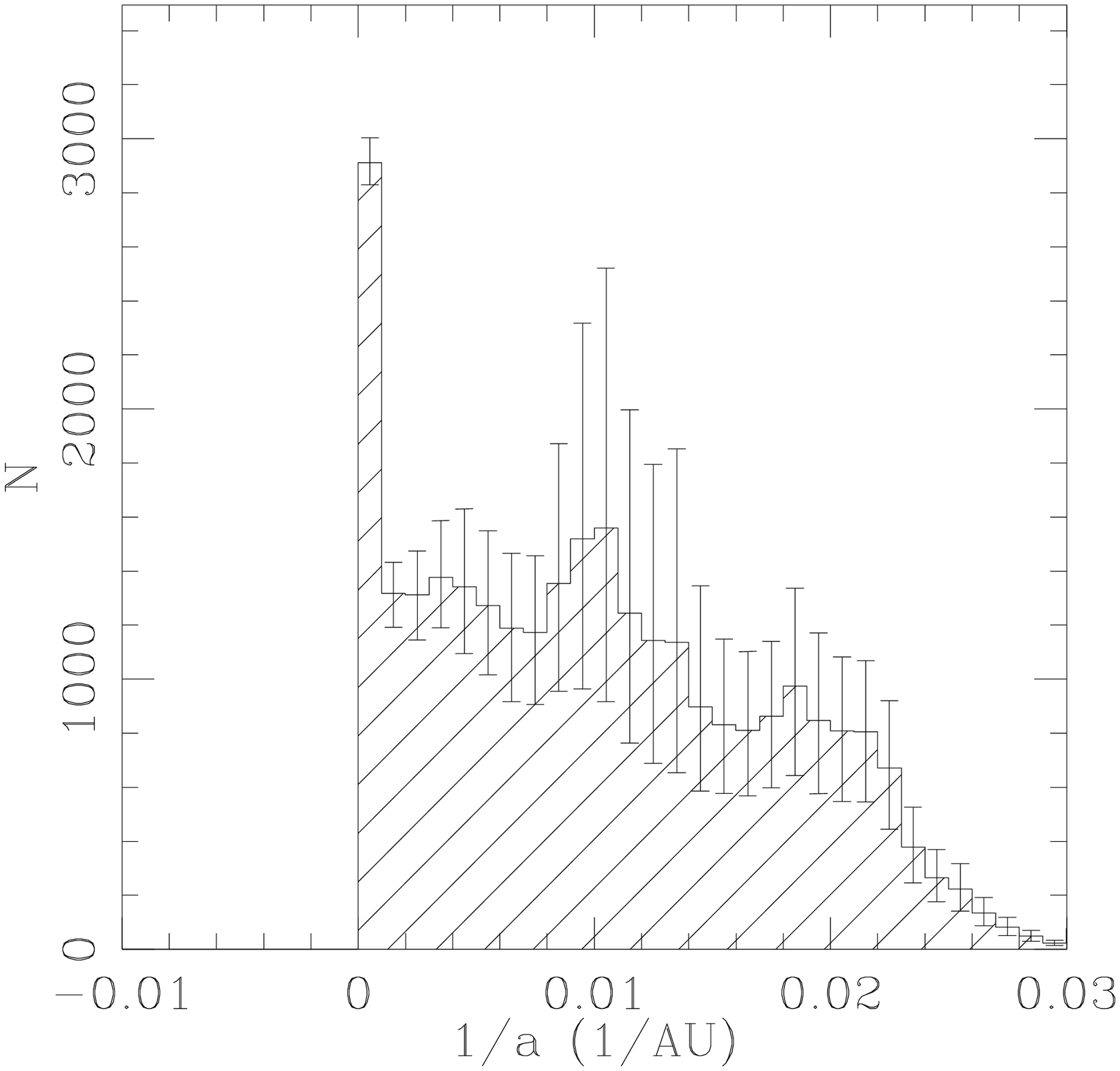,height=1.5in}
                        \psfig{figure=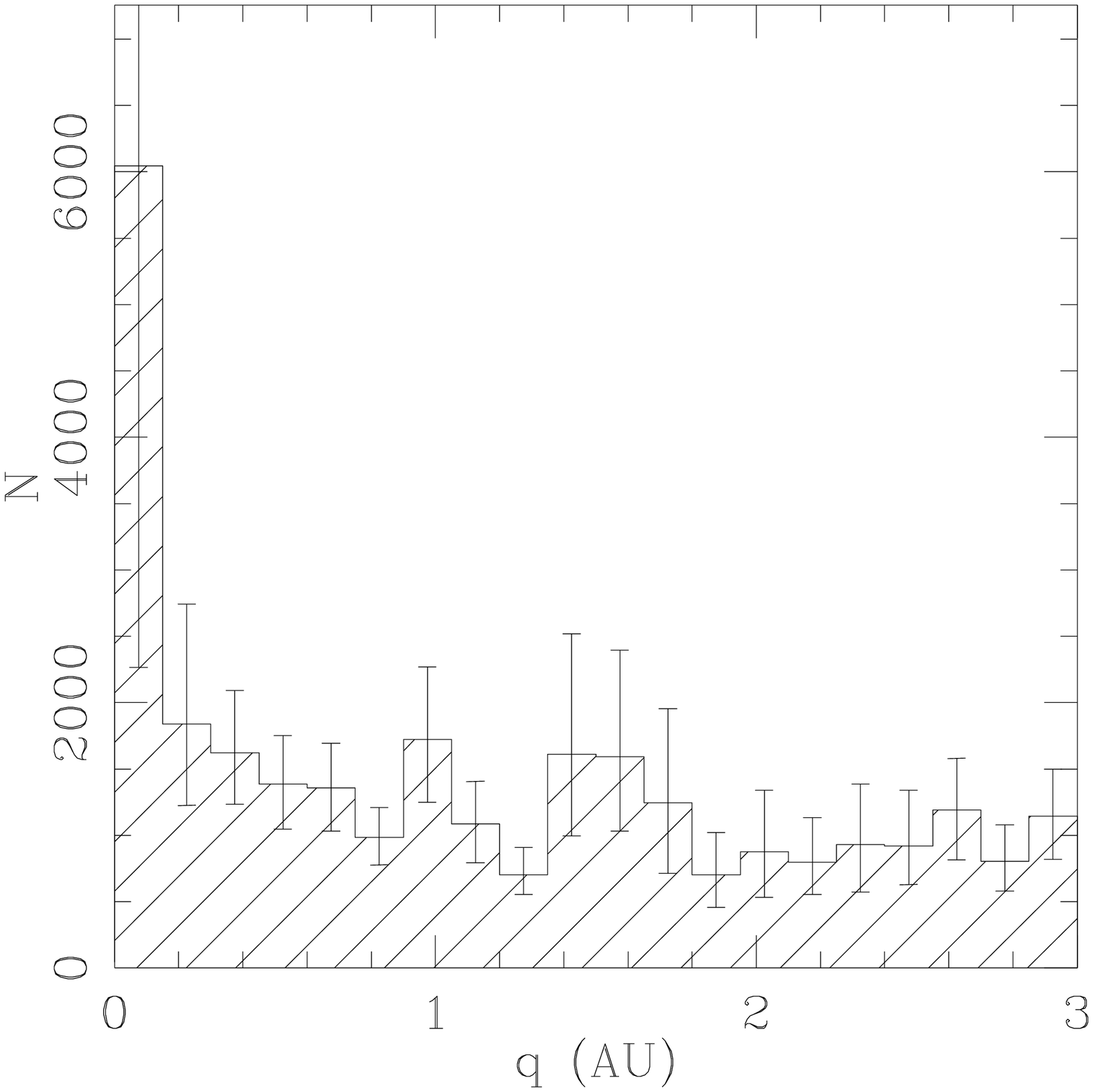,height=1.5in}}
		  \rule[2mm]{6.2in}{0.5mm}
                  \hbox{\psfig{figure=tvp_b.eps,height=1.5in}
                        \psfig{figure=tvp_q.eps,height=1.5in}
                        \psfig{figure=tmars_b.eps,height=1.5in}
                        \psfig{figure=lp_qlines_new.eps,height=1.5in}}}}
\caption{Distribution of the inverse semimajor axis $1/a$ and
perihelion distance $q$ for the $V_\infty$ comets, when the Solar
System contains a massive solar companion.  Left panels: companion
orbital radius of $100$~\au. Right panel: orbital radius $1000\au$. From
the top down, the companion masses are 0.1, 1 and 10 Jupiter
masses. The bottom line of panels is for comparison, and includes the
standard model (left side) and the observations (right side). The
observed perihelion distribution includes curves indicating the
estimated intrinsic distribution (see Fig. \ref{fi:Vm_q} for
details).}
\label{fi:px1}
\end{figure}

The results are presented in Fig.~\ref{fi:px1} and Table~\ref{ta:px_parms}.
The $X$ parameters are listed in Table~\ref{ta:px_parms}.  As the companion
mass is increased, the fraction of prograde to total comets ($X_3$) improves.
However, $X_1$ and $X_2$ remain far from unity. There is no evidence that a
solar companion can significantly improve the agreement between the observed
and predicted LP comet distribution. 

\begin{table}[p]
\centerline{
\begin{tabular}{|cc|ccccccccc|} \hline
$M_{X} $ & $a_{X}$ & Total & Spike & Tail  & Prograde & $X_1$ & $X_2$ & $X_3$
& $\langle m \rangle$ & $R_{\odot}$ \\  \hline \hline
0.1    & 100  & 40662 & 1451 & 9111 & 14074& 0.11 & 3.07 & 0.67 & 43.1 & 1 \\
0.1    & 1000 & 49420 & 1490 & 10057& 13550& 0.09 & 2.79 & 0.53 & 44.4 & 1 \\
1      & 100  & 38397 & 1473 & 7465 & 9379 & 0.12 & 2.66 & 0.47 & 85.4 & 4 \\
1      & 1000 & 35940 & 1438 & 9338 & 13544& 0.12 & 3.56 & 0.73 & 68.1 & 1 \\
10     & 100  & 14877 & 1379 & 3365 & 5846 & 0.28 & 3.10 & 0.76 & 66.0 & 4 \\
10     & 1000 & 28600 & 1400 & 8183 & 15489& 0.15 & 3.92 & 1.05 & 146.3& 2\\
\hline
0.1$^d$& 100  & 25300 & 944  & 6762 & 8893 & 0.11 & 3.66 & 0.68 & 43.1 & 1 \\
0.1$^d$& 1000 & 31376 & 975  & 6206 & 8623 & 0.09 & 2.71 & 0.53 & 44.4 & 1 \\
1$^d$  & 100  & 27918 & 963  & 4764 & 6047 & 0.10 & 2.34 & 0.42 & 85.4 & 4 \\
1$^d$  & 1000 & 24740 & 943  & 6713 & 8281 & 0.12 & 3.72 & 0.65 & 68.1 & 1 \\
10$^d$ & 100  & 9749  & 928  & 2197 & 4059 & 0.29 & 3.09 & 0.81 & 66.0 & 4 \\
10$^d$ & 1000 & 22177 & 1030 & 6052 & 12649& 0.14 & 3.74 & 1.11 & 146.3&2\\
\hline
\end{tabular}}
\caption{Parameters of the distribution of $V_\infty$ comets, when the
Solar System contains a massive solar companion. The companion mass
$M_X$ is in Jupiter masses, and its orbital radius $a_X$ is measured
in $\au$. The rightmost column indicates the number of comets that
collided with the Sun. The superscript $^d$ indicates that the
discovery probability from Eq.~\protect\ref{eq:discoverprob2} has
been applied. The definitions of the other columns are the same as in
Table~\ref{ta:ng_parms}.}
\label{ta:px_parms}
\end{table}

\subsection{Post-visibility evolution: fading}\label{sec:nonparm}

The concept of fading was introduced in \S\S \ref{sec:onedran} and
\ref{pa:fading}. We use the term ``fading'' to denote any change in
the intrinsic properties of the comet that would cause it to disappear
from the observed sample.  Our focus is on modeling the fading process
empirically, rather than attempting to elucidate the physical processes
involved.  The distributions of inverse semimajor axis and 
ecliptic inclination will serve as our primary fading benchmarks,
through the values of the parameters $X_1$, $X_2$ and $X_3$
(Eq.~\ref{eq:xdef}).

We shall generally assume that fading depends only on the number of
apparitions (perihelion passages with $q<3\au$). We parametrize the fading
process by a function $\Phi_m$ (cf. Eq.~\ref{eq:phimdef}), the probability
that a visible new comet survives fading for at least $m$ apparitions (thus
$\Phi_1=1$).

We shall conduct simulations with and without plausible non-gravitational (NG)
forces (\S\S \ref{pa:nongrava}, \ref{sec:nongrav}). When NG forces are
included, we shall use the parameters  $A_1 =
10^{-7}$~\aud2, $A_2 = \pm 10^{-8}$~\aud2, $A_3 = 0$, with a random sign for
$A_2$ at each perihelion passage (henceforth the ``standard NG model''). 

The most direct way to determine the fading function $\Phi_m$ would be to
break down the simulated data set into individual distributions, one for each
perihelion passage \ie $\{ V_1, V_2, V_3, \ldots\}$, and then fit the observed
distribution of orbital elements to the parameters $\Phi_1,\Phi_2,\ldots$
where $\Phi_{m+1}\le\Phi_m$.  Unfortunately, this problem is poorly
conditioned. Instead, we shall experiment with a few simple parametrized
fading functions.

\subsubsection{One-parameter fading functions}

The fading functions we shall examine include:
\begin{description}
\item[a) Constant lifetime] Each comet is assigned a fixed lifetime,
measured in apparitions. Thus
\begin{equation}
\Phi_m=1,\quad m\le m_v,\qquad\qquad \Phi_m=0,\quad m>m_v.
\label{eq:fadea}
\end{equation}
\item[b) Constant fading probability] Comets are assigned a fixed probability
$\lambda$ of fading, per apparition. Thus
\begin{equation}
\Phi_m=(1-\lambda)^{m-1}.
\label{eq:fadeb}\end{equation}
\item[c) Power-law] The fraction of comets remaining is 
\begin{equation}
\Phi_m=m^{-\kappa},
\label{eq:fadec}
\end{equation}
where $\kappa$ is a positive constant.
\end{description}

We have also investigated fading functions in which $\Phi$ depends on
the elapsed time $t$ since the first apparition. Such laws are less
physically plausible than fading functions based on the number of
apparitions, since by far the harshest environment for comets occurs
as they pass perihelion; and in fact the functions $\Phi(t)$ that we
investigated all produced relatively poor matches to the
observations. Also, fading functions in which $\Phi$ depends on the
number of perihelion passages produce results very similar to laws
based on the number of apparitions.

\begin{figure}[p]
\centerline{\vbox{\hbox{\psfig{figure=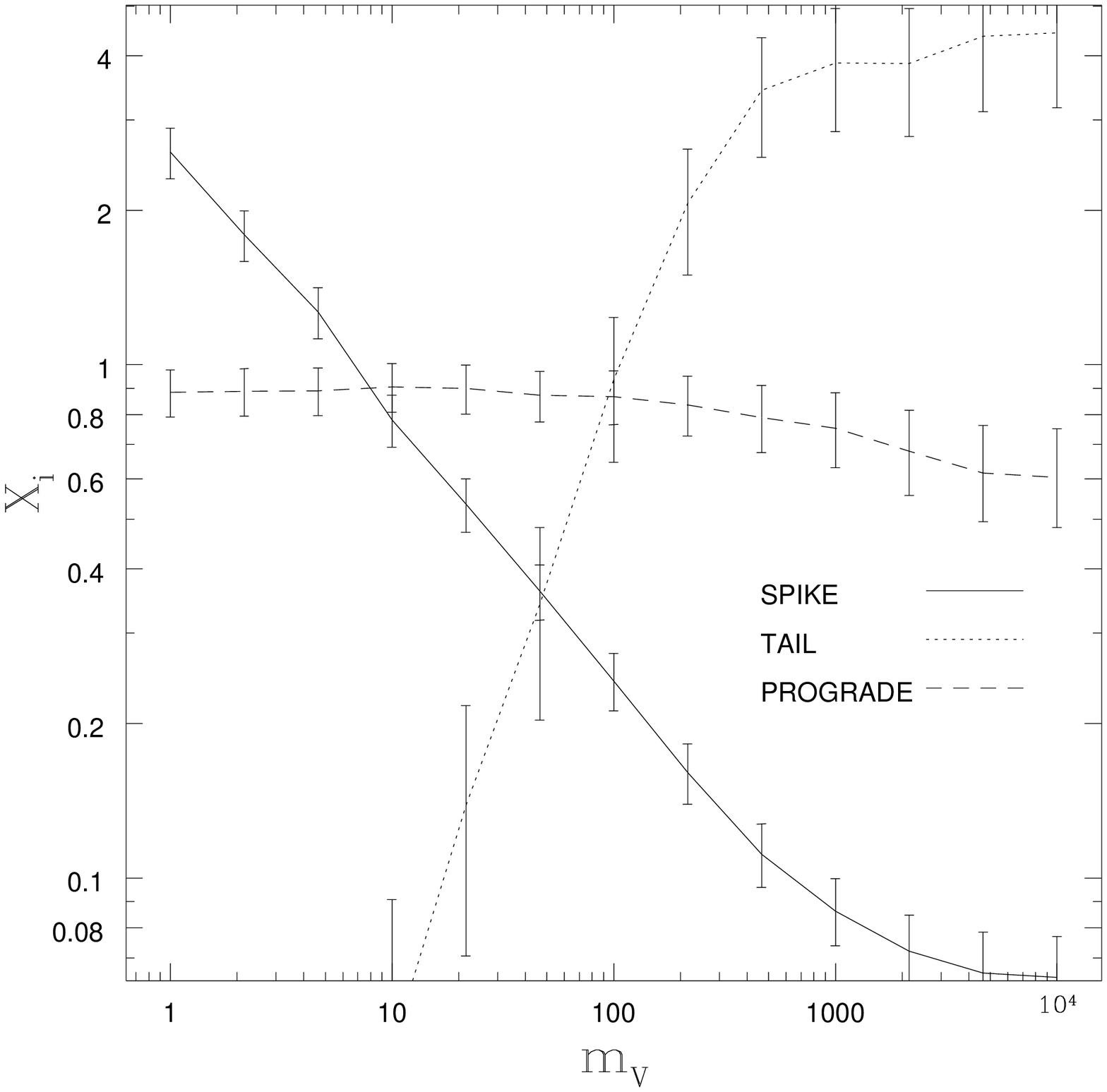,height=3in}
                        \psfig{figure=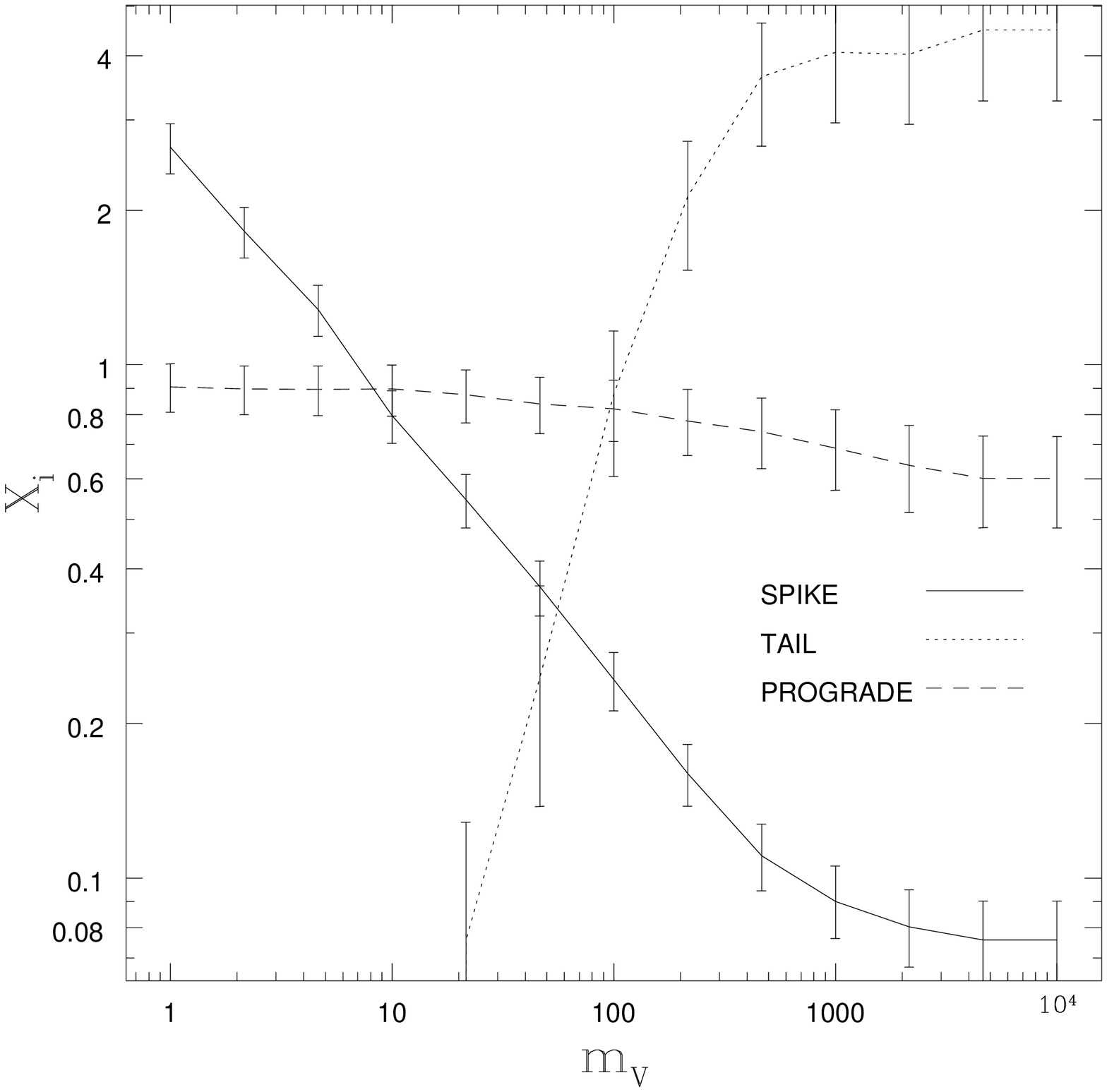,height=3in}}
                  \hbox{\psfig{figure=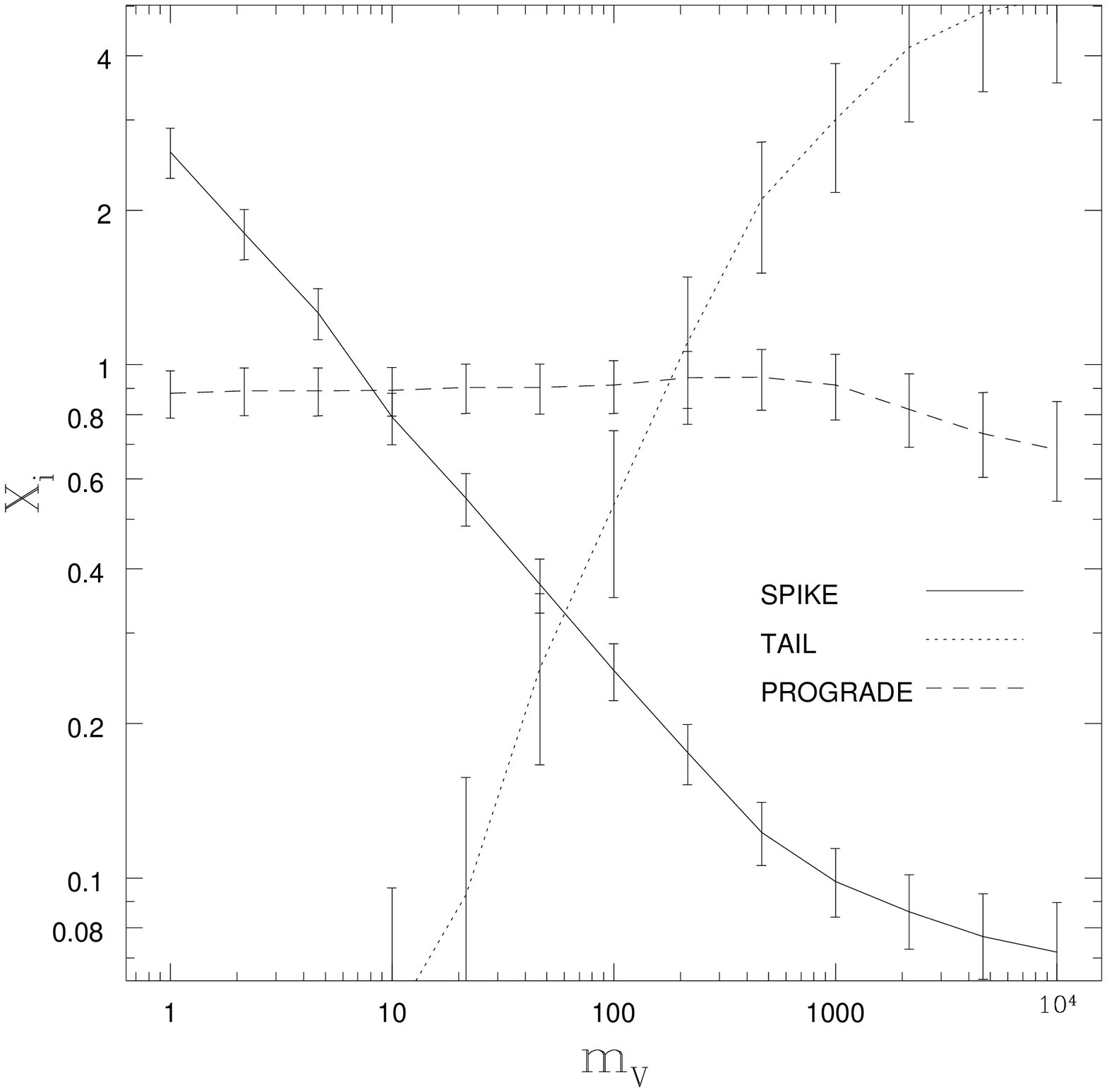,height=3in}
                        \psfig{figure=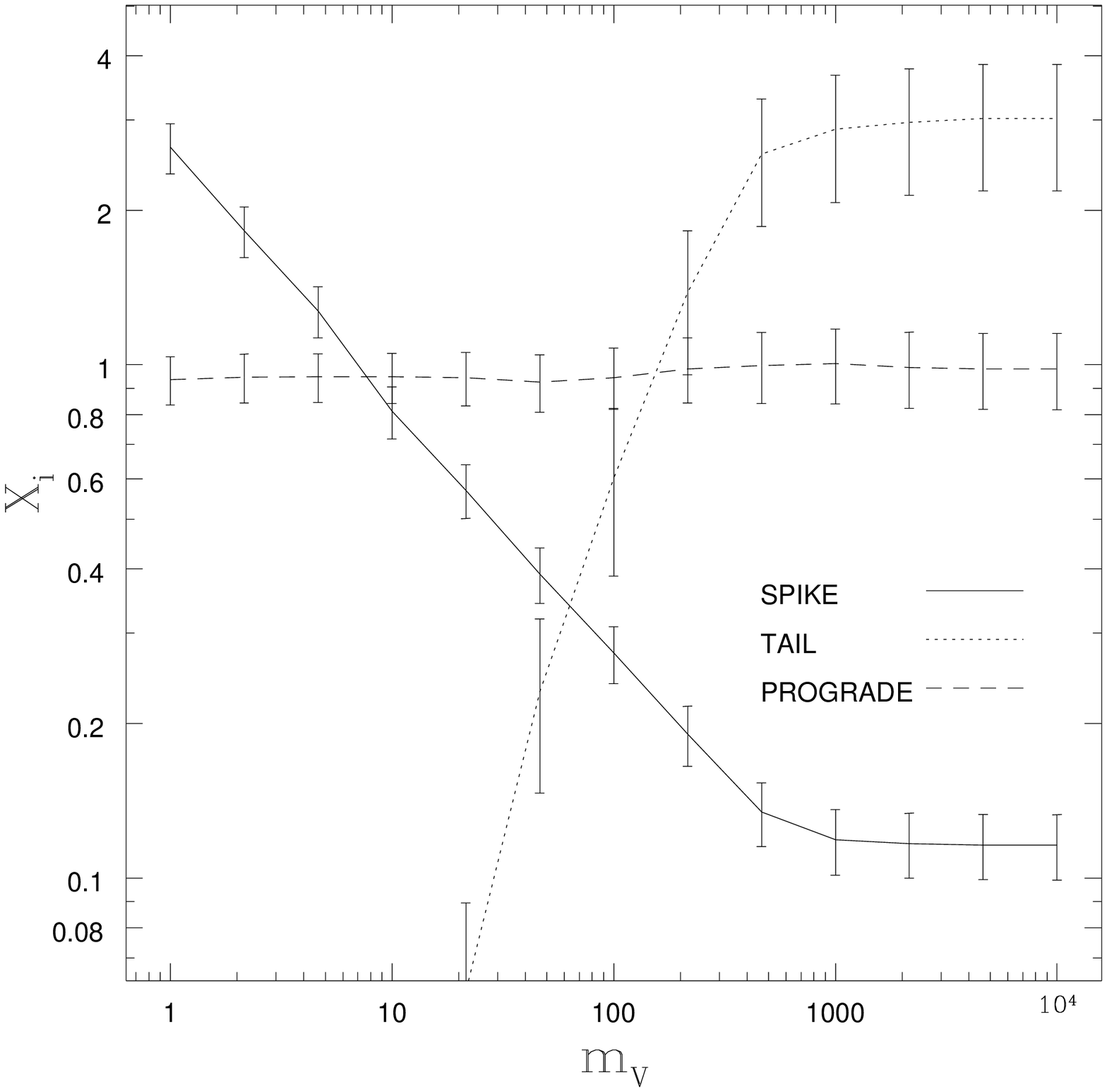,height=3in}}}}
\caption{ The values of the parameters $X_i$ for a fading function
with a fixed lifetime of $m_v$ apparitions (model [a],
Eq.~\ref{eq:fadea}). If the simulation agrees with the observations
then $X_i=1$, $i=1,2,3$. The parameter $X_1$ is based on the fraction
of LP comets in the Oort spike (solid curve); $X_2$ is based on the
fraction of comets in the energy tail, $x>0.0145\aui$ (dotted curve);
$X_3$ is based on the fraction of prograde comets (dashed curve)
(cf. \S\ref{sec:psidef}).  The panels on the left are based on the
visibility criterion $q<3\au$, and those on the right are based on the
visibility probability (Eq.~\ref{eq:discoverprob2}). The upper panels are
based on the standard model with no NG forces, and the lower panels
are based on the standard NG model.}
\label{fi:const_mv}
\end{figure}

\begin{figure}[p]
\centerline{\vbox{\hbox{\psfig{figure=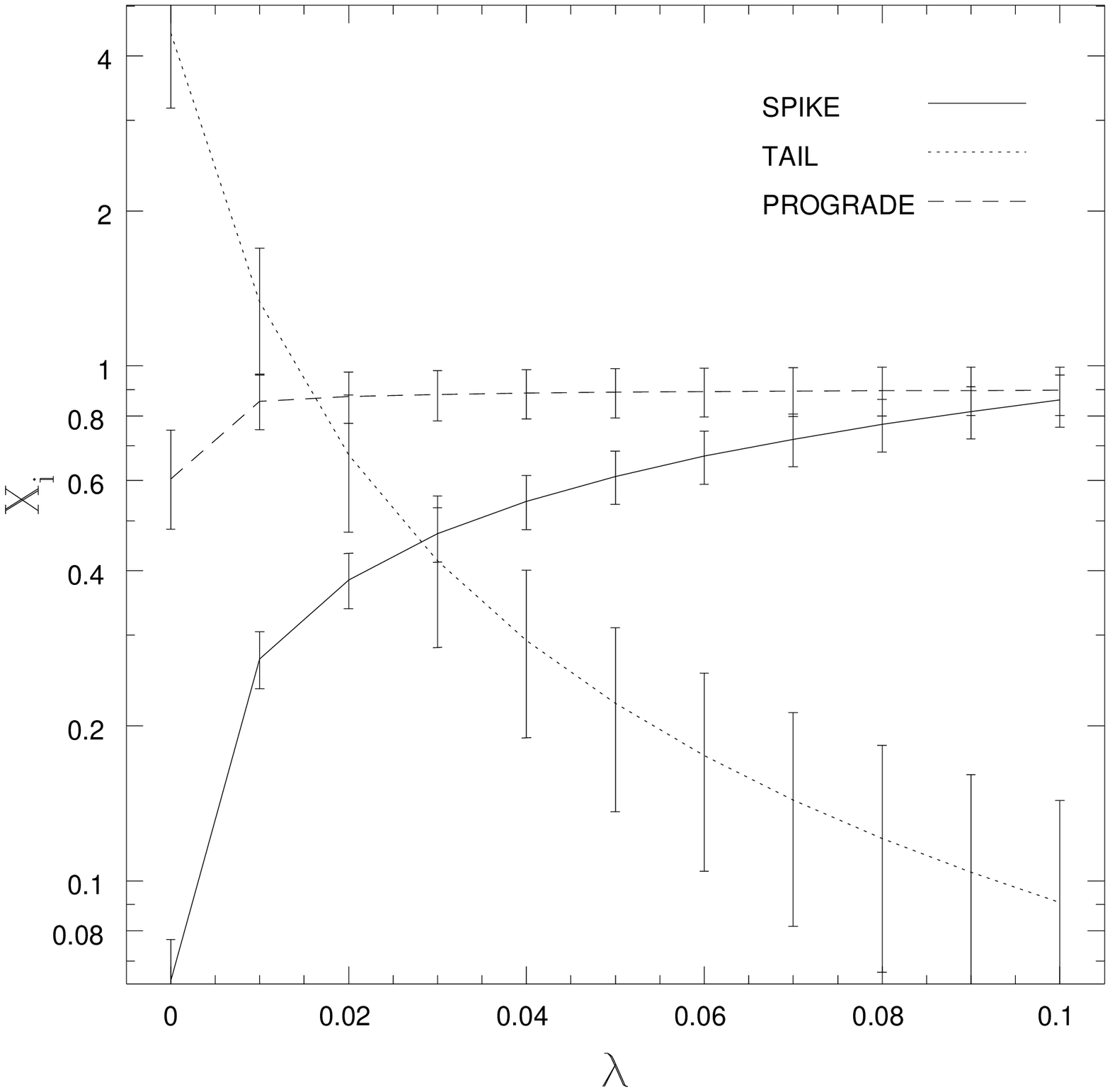,height=3in}
                        \psfig{figure=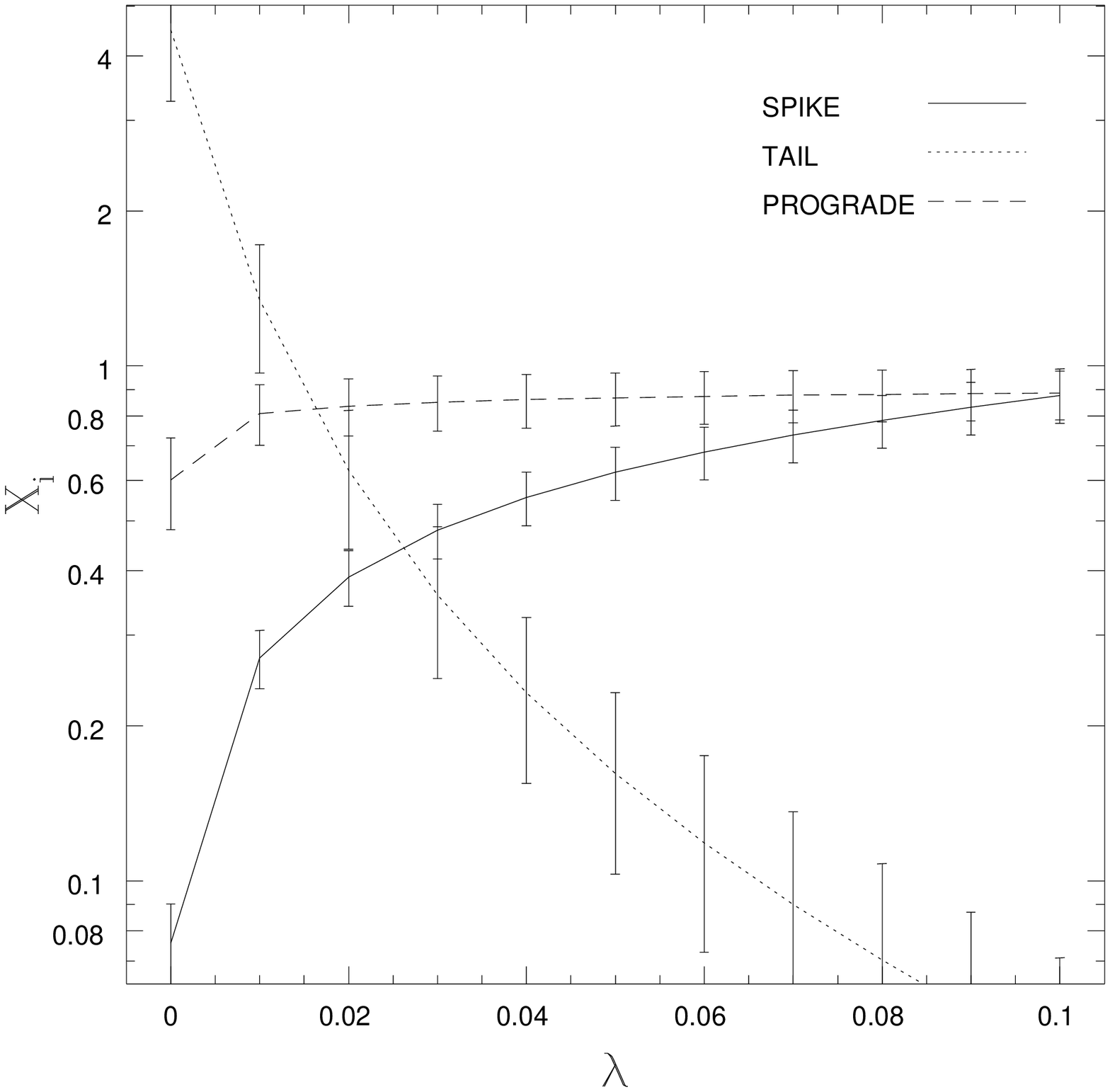,height=3in}}
                  \hbox{\psfig{figure=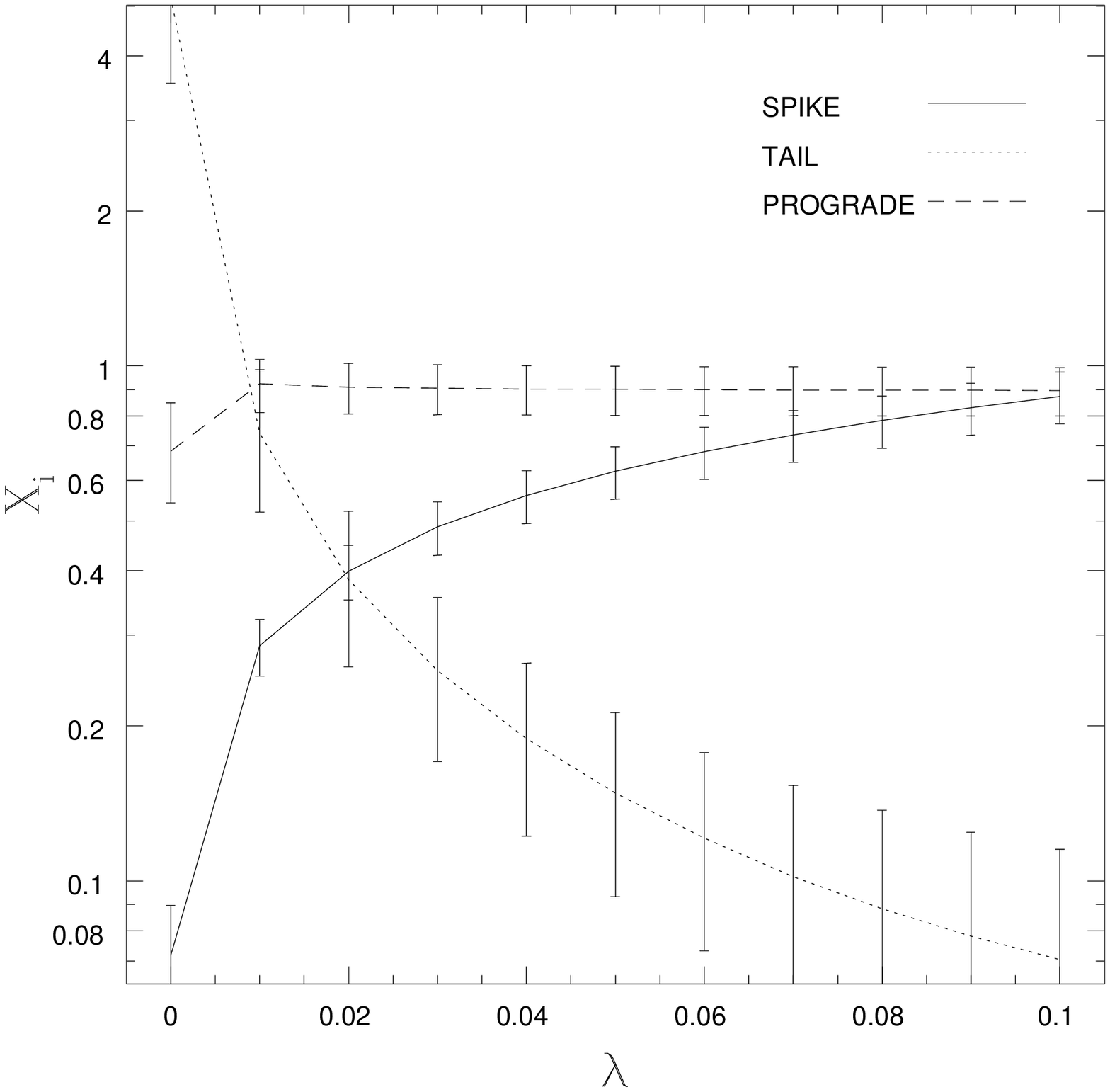,height=3in}
                        \psfig{figure=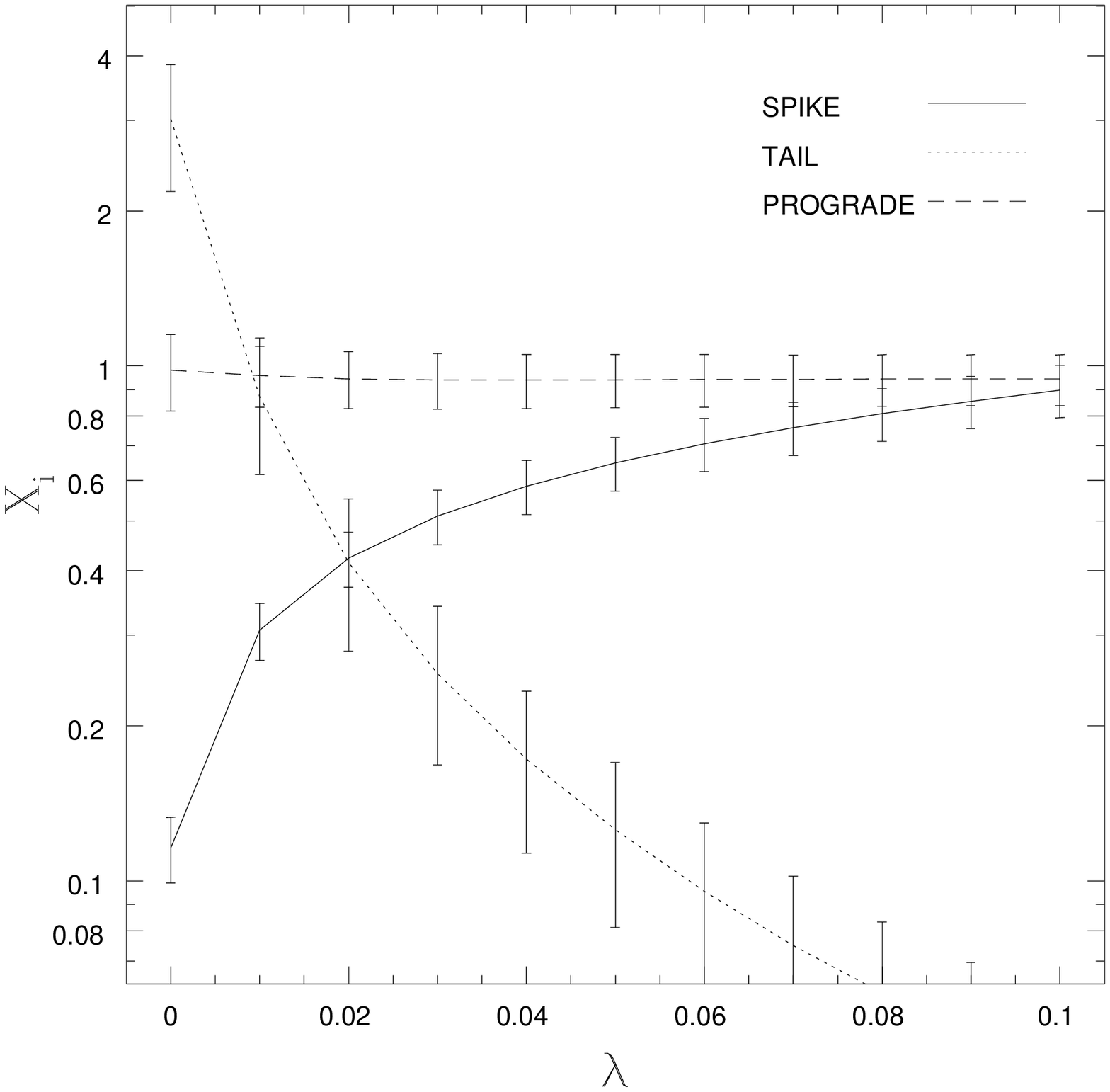,height=3in}}}}
\caption{The values of $X_i$ given a fixed fading probability
 $\lambda$ per apparition (model [b], Eq.~\ref{eq:fadeb}). For further
 details see the caption to Fig. \ref{fi:const_mv}.}
\label{fi:geo_mv}
\end{figure}

\begin{figure}[p]
\centerline{\vbox{\hbox{\psfig{figure=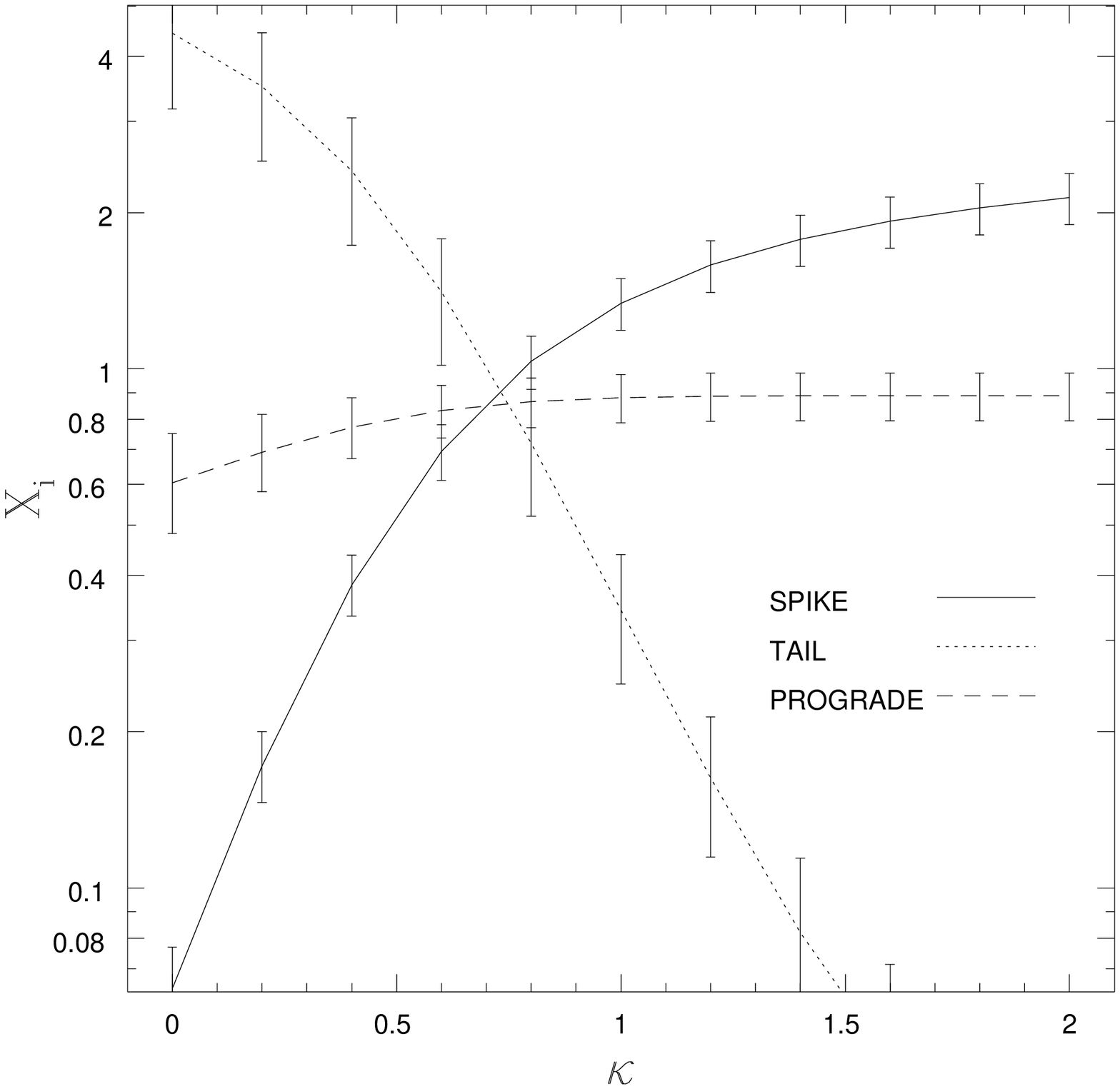,height=3in}
                        \psfig{figure=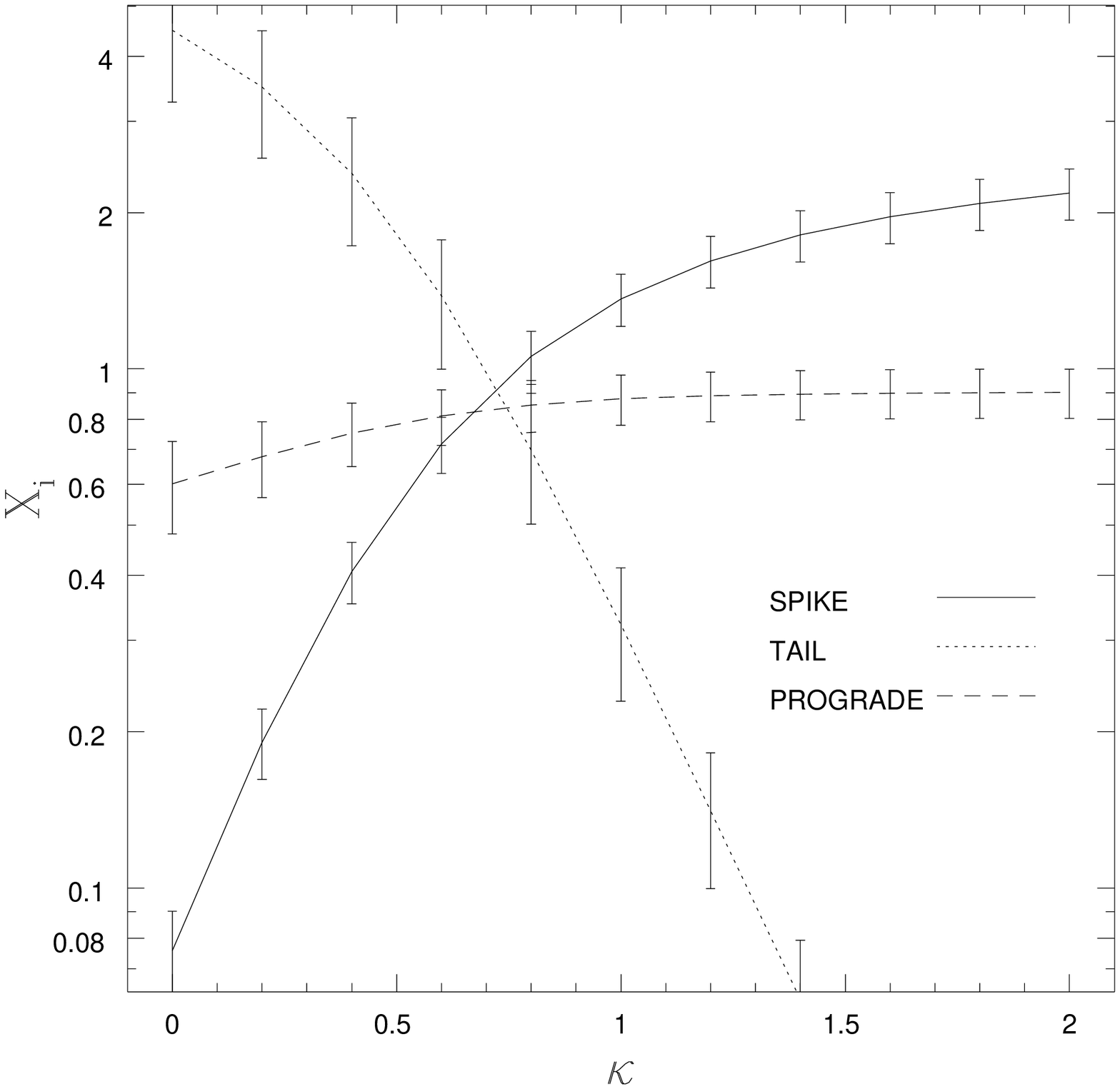,height=3in}}
                  \hbox{\psfig{figure=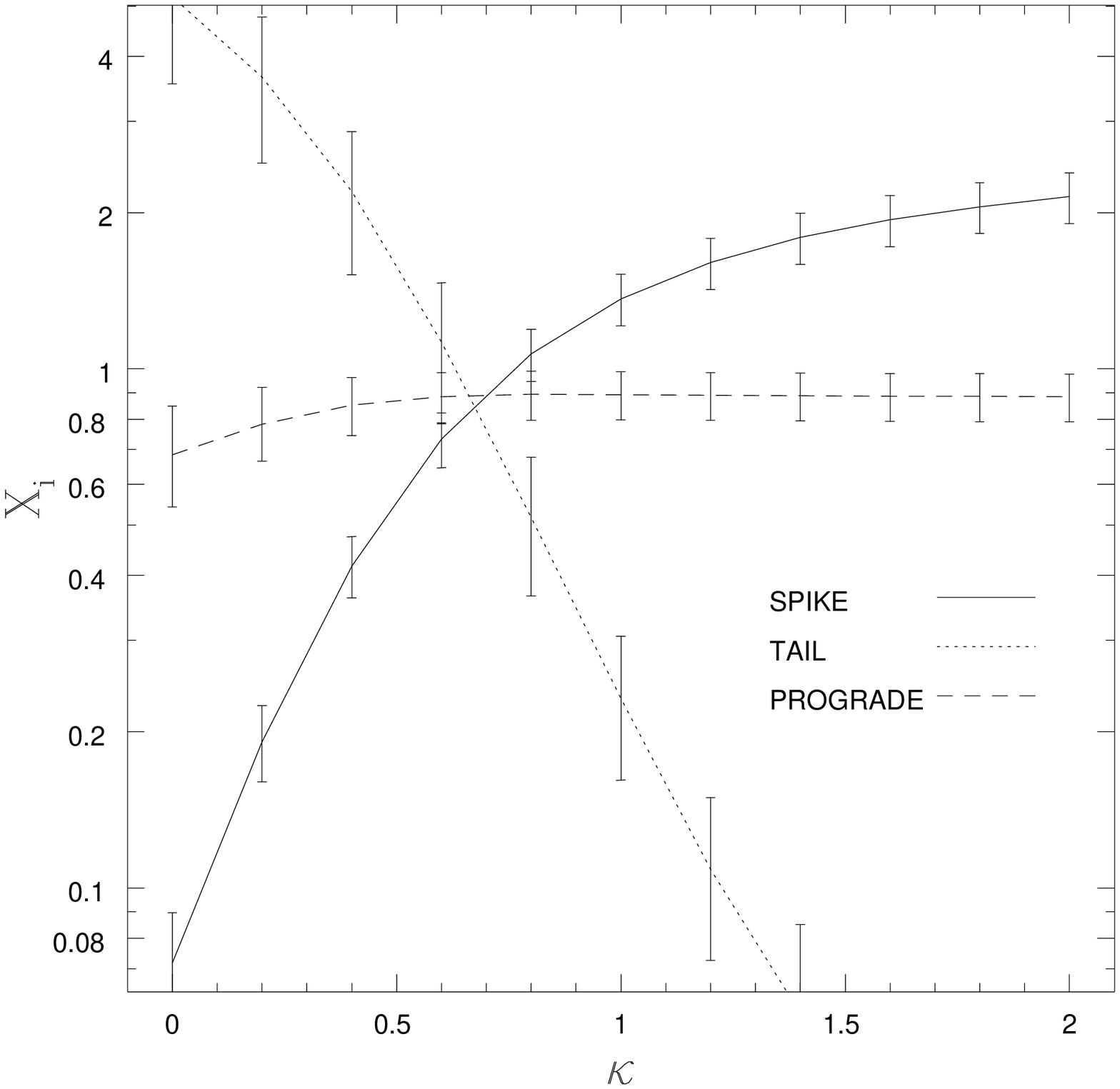,height=3in}
                        \psfig{figure=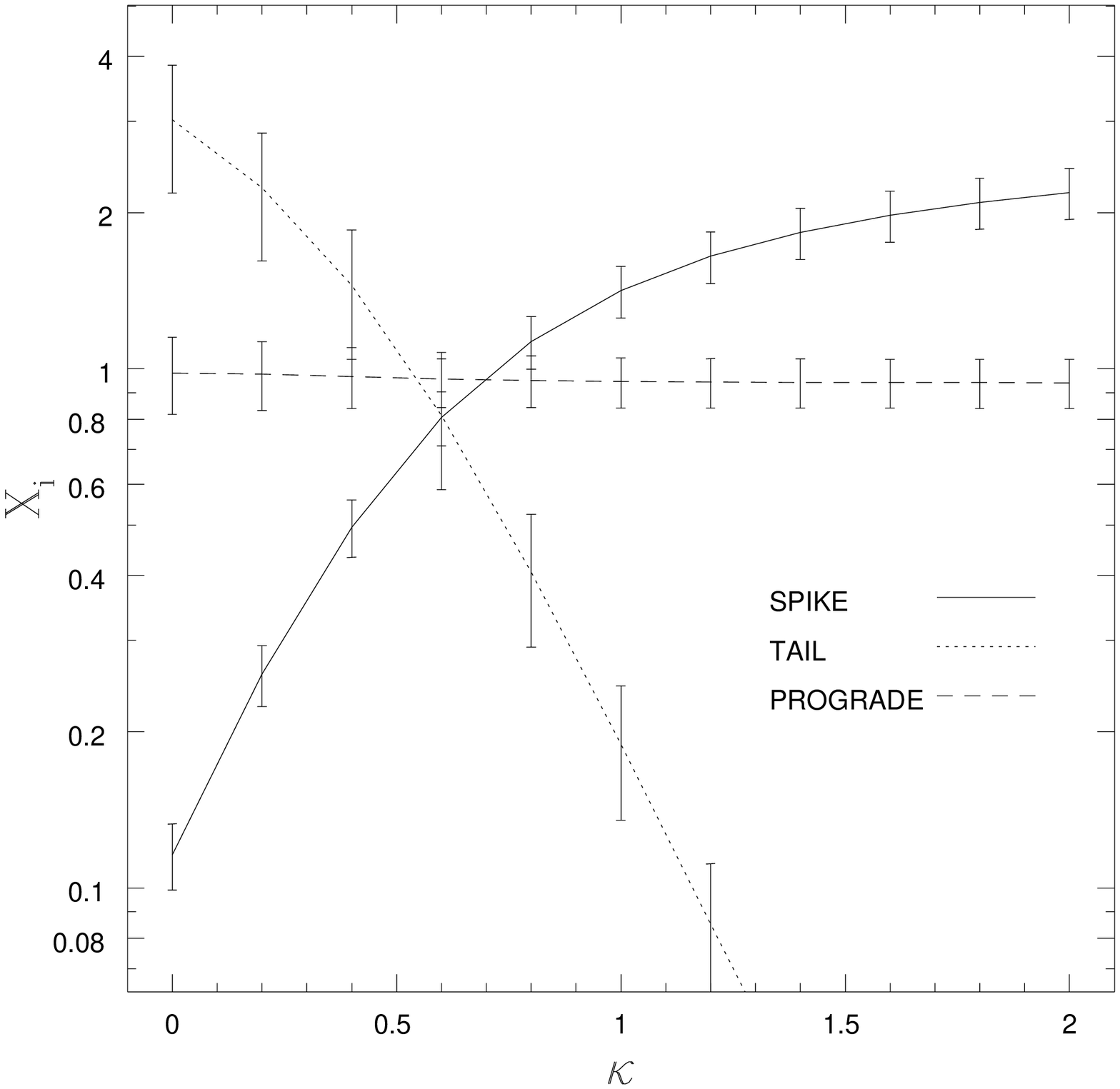,height=3in}}}}
\caption{The values of $X_i$ given a power-law fading function with
exponent $-\kappa$ (model [c], Eq.~\ref{eq:fadec}). For further
details see the caption to Fig. \ref{fi:const_mv}.}
\label{fi:pow_mv}
\end{figure}

The results from the fading laws (\ref{eq:fadea})--(\ref{eq:fadec}) are shown
in Figs.~\ref{fi:const_mv} to \ref{fi:pow_mv}.  The first of these figures
displays the $X$ parameters assuming LP comets have a constant lifetime in
apparitions (model [a]). The presence or absence of NG forces (bottom vs. top
panels), or the use of two different visibility criteria (left vs. right
panels) has very little effect on the results. The spike/total ratio matches
observations (i.e. $X_1=1$) at $m_v\simeq10$, but the tail/total ratio is far
too low at that point ($X_2\ll1$). The tail/total ratio is right at $m_v\simeq
100$, but $X_1$ is now too low.  The ratio $X_3$ is typically close to but
below unity. The model does not match the observations for any value of the
parameter $m_v$.

Figure~\ref{fi:geo_mv} displays the behaviour of the parameters $X_i$ given a
fixed fading probability $\lambda$ per apparition (model [b]). Once again, the
results are almost independent of NG forces and the visibility criterion, and
there is no value for the parameter $\lambda$ that matches the observations
($X_i=1$). 

\begin{figure}[p]
\centerline{\vbox{\hbox{\psfig{figure=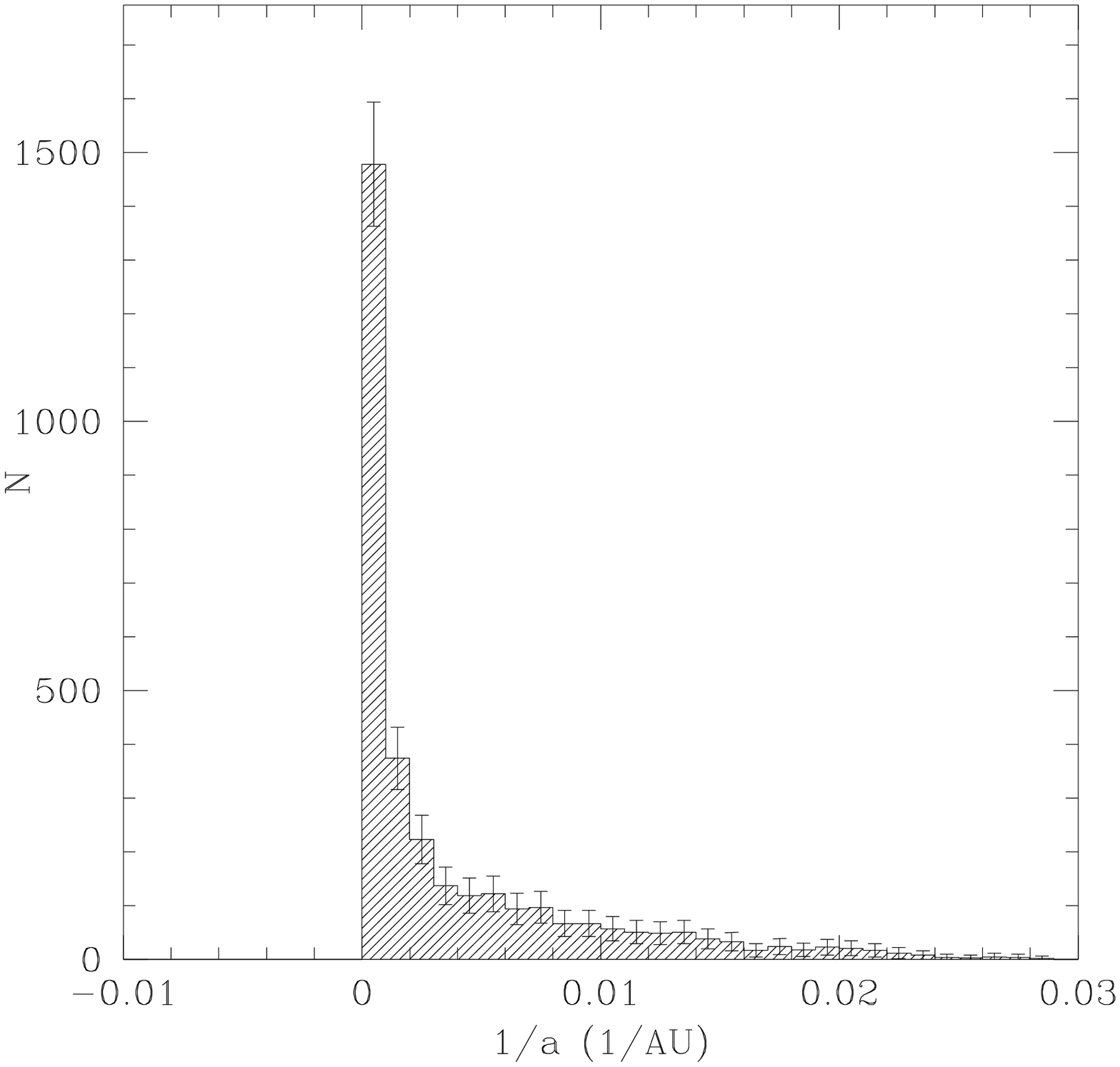,height=3in}
                        \psfig{figure=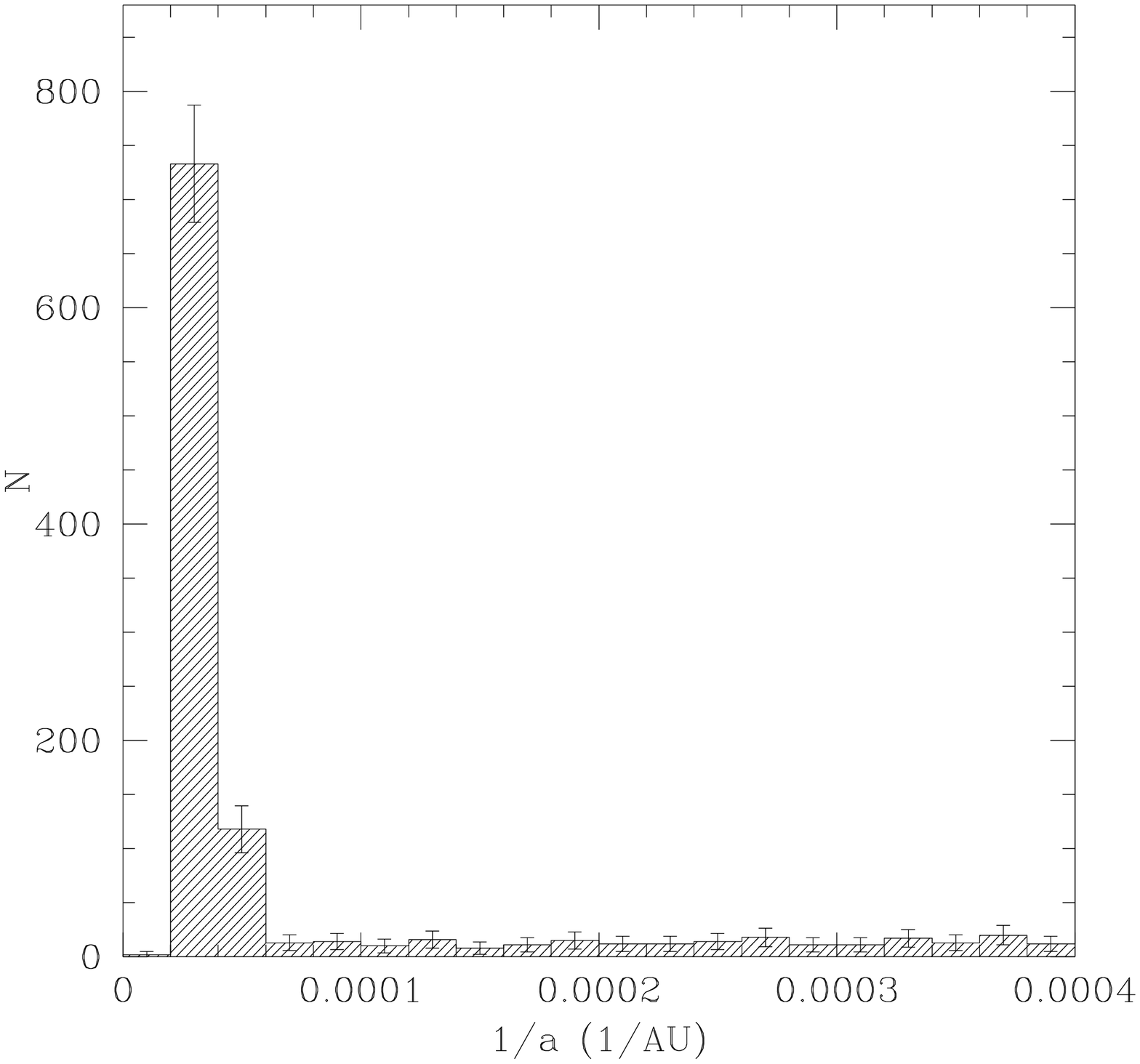,height=3in}}
                  \hbox{\psfig{figure=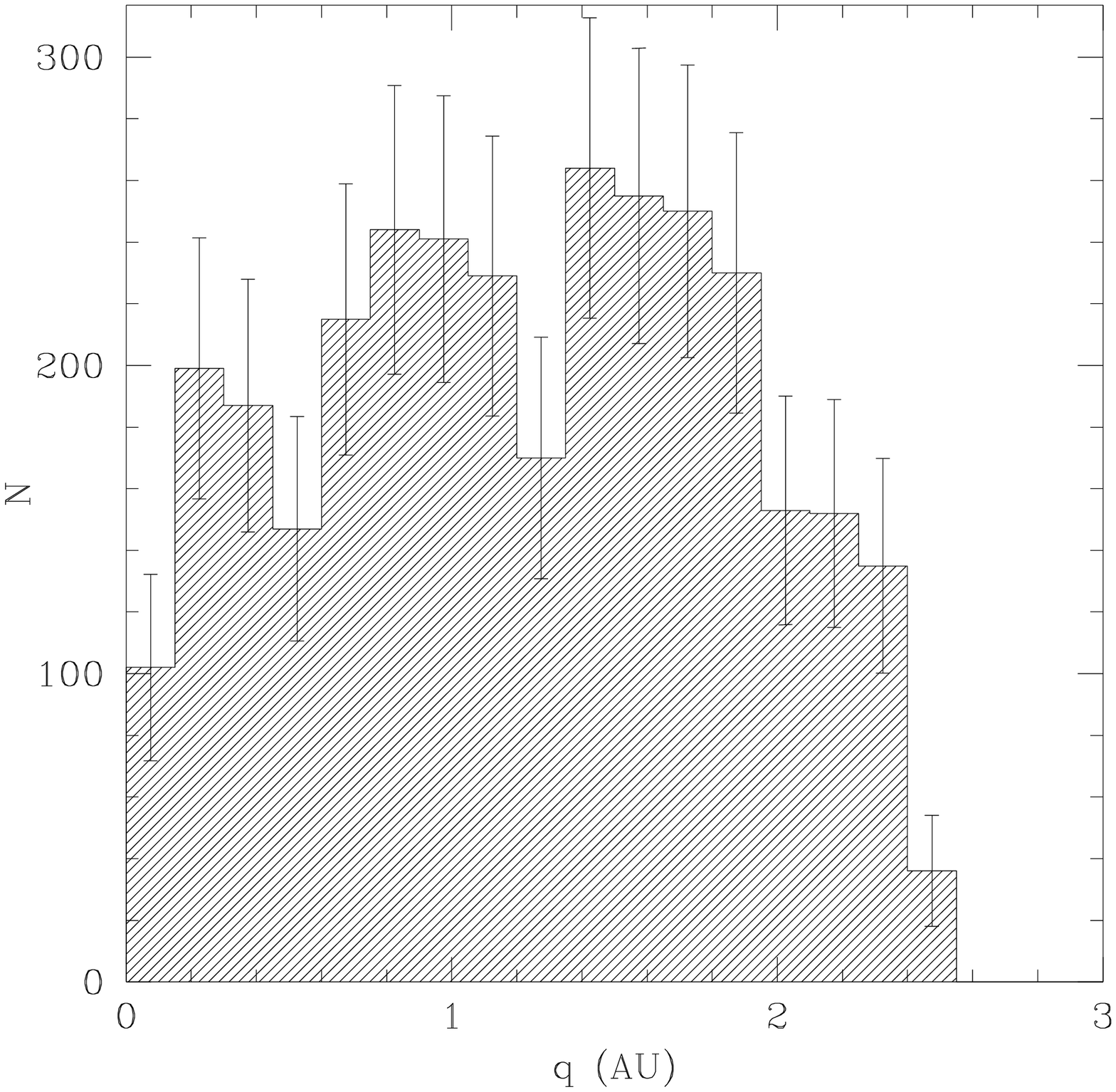,height=3in}
                        \psfig{figure=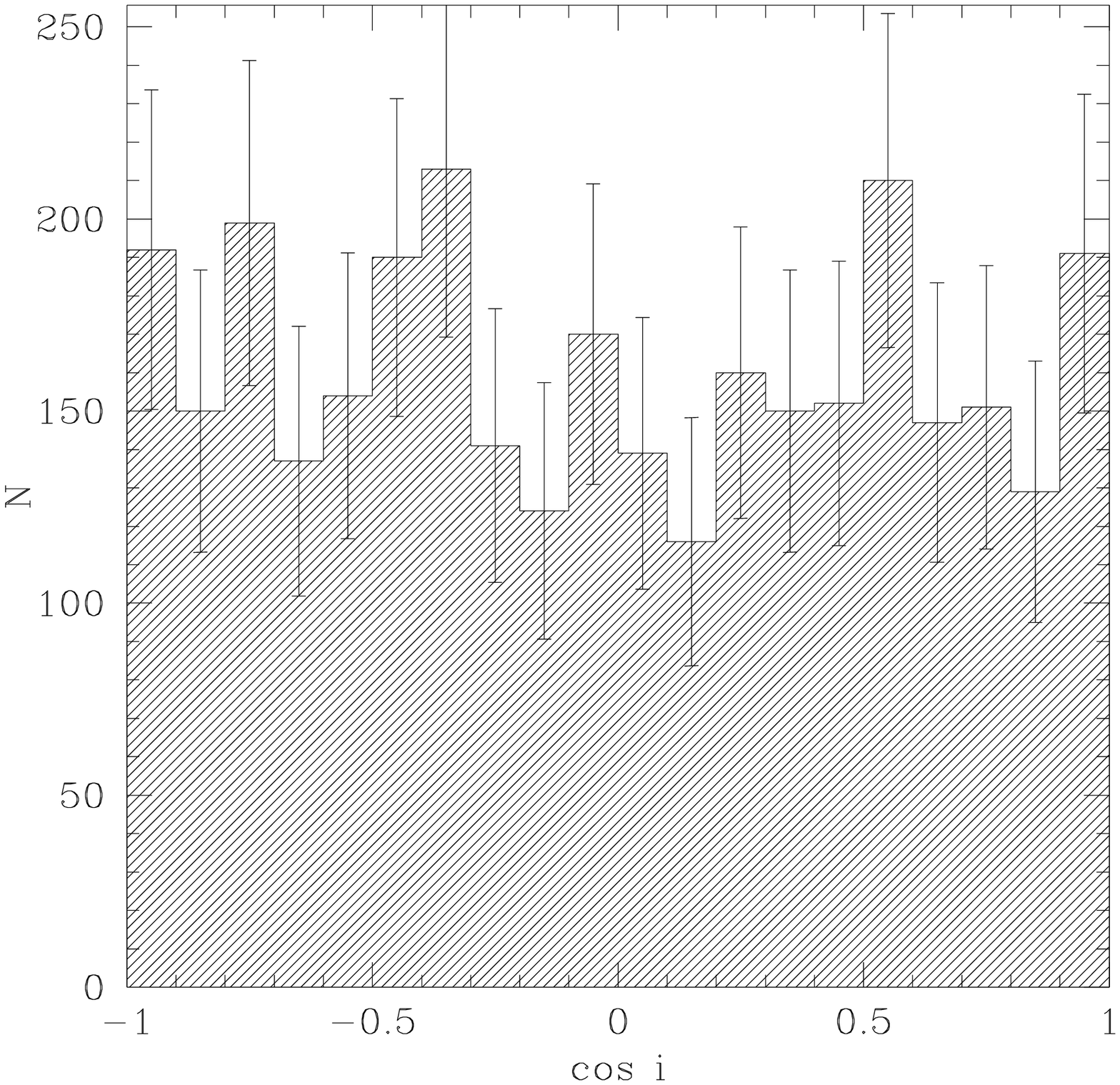,height=3in}}}}
\caption{The distribution of the inverse semimajor axis $1/a$,
perihelion distance $q$ and cosine of the ecliptic inclination $i$ for
a power-law fading function with exponent $\kappa = -0.6$
(Eq.~\ref{eq:fadec}). These simulations are based on the standard NG
model and the visibility probability in Eq.~\ref{eq:discoverprob2}.}
\label{fi:pow_mv4_elems}
\end{figure}

Figure~\ref{fi:pow_mv} shows the parameters $X_i$ for a power-law
fading function (model [c]). Although the match is not perfect,
an exponent $\kappa=0.6 \pm 0.1$ provides a
much better match than the previous two models:
$X_1 = 0.73 \pm 0.09$, $X_2 = 0.96 \pm 0.26$, and $X_3 = 0.95 \pm
0.12$ when the standard NG model and discovery
probability (Eq.~\ref{eq:discoverprob2}) are used. The distributions
of orbital elements are shown in Fig.~\ref{fi:pow_mv4_elems}, to be
compared with the observed distributions in \S~\ref{pa:observ}. For
$m\gg1$ this fading law is the same as an empirical law suggested by
\cite{whi62}, $\phi_m\equiv\Phi_m-\Phi_{m+1}\propto m^{-\kappa-1}$;
Whipple estimated $\kappa=0.7$.

\subsubsection{Other fading functions}

We have also examined several two-parameter fading functions:

\begin{description}
\item[d) Two populations] Suppose that the Oort cloud contains two populations
of comets, distinguished by their internal strength. The first and
more fragile set is disrupted after $m_v$ apparitions, while the  more
robust comets, comprising a fraction $f$ of the total, do not fade at all. Thus
\begin{equation}
\Phi_m=1,\quad m\le m_v,\qquad\qquad \Phi_m=f,\quad m>m_v.
\label{eq:modeldg}
\end{equation}
\item[e) Constant fading probability plus survivors] One population has a
fixed fading probability $\lambda$ per apparition, while the more robust
comets, comprising a fraction $f$ of the total, do not fade at all.
Thus 
\begin{equation}
\Phi_m=(1-f)(1-\lambda)^{m-1}+f.
\label{eq:weissfad}
\end{equation}
\item[f) Offset power law] The fading function is chosen to be
\begin{equation}
\Phi_m=[(m+ \beta)/(1+\beta)]^{-\kappa},
\end{equation}
\end{description}

\begin{figure}[p]
\centerline{\vbox{\hbox{\psfig{figure=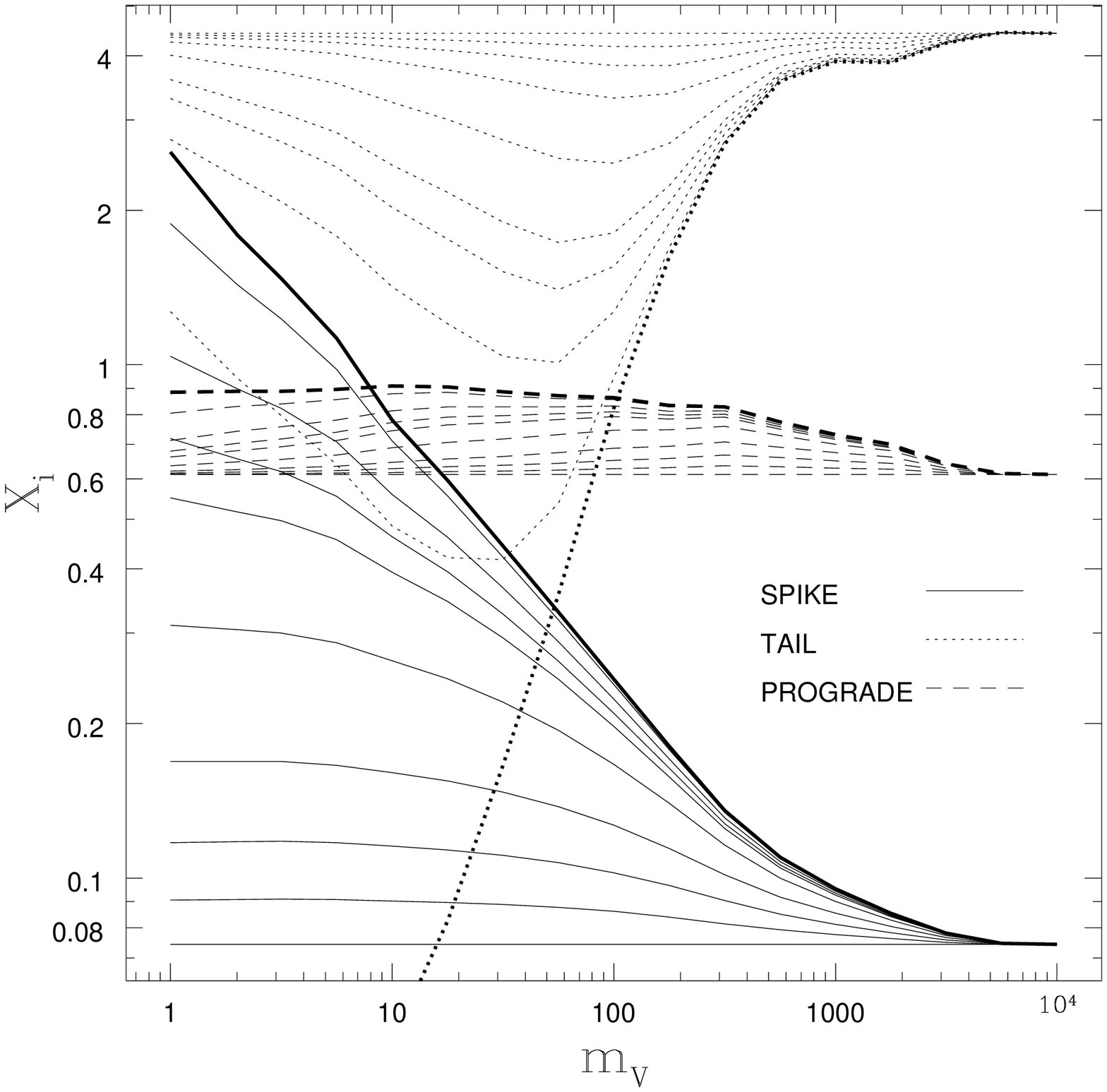,height=3in}
                        \psfig{figure=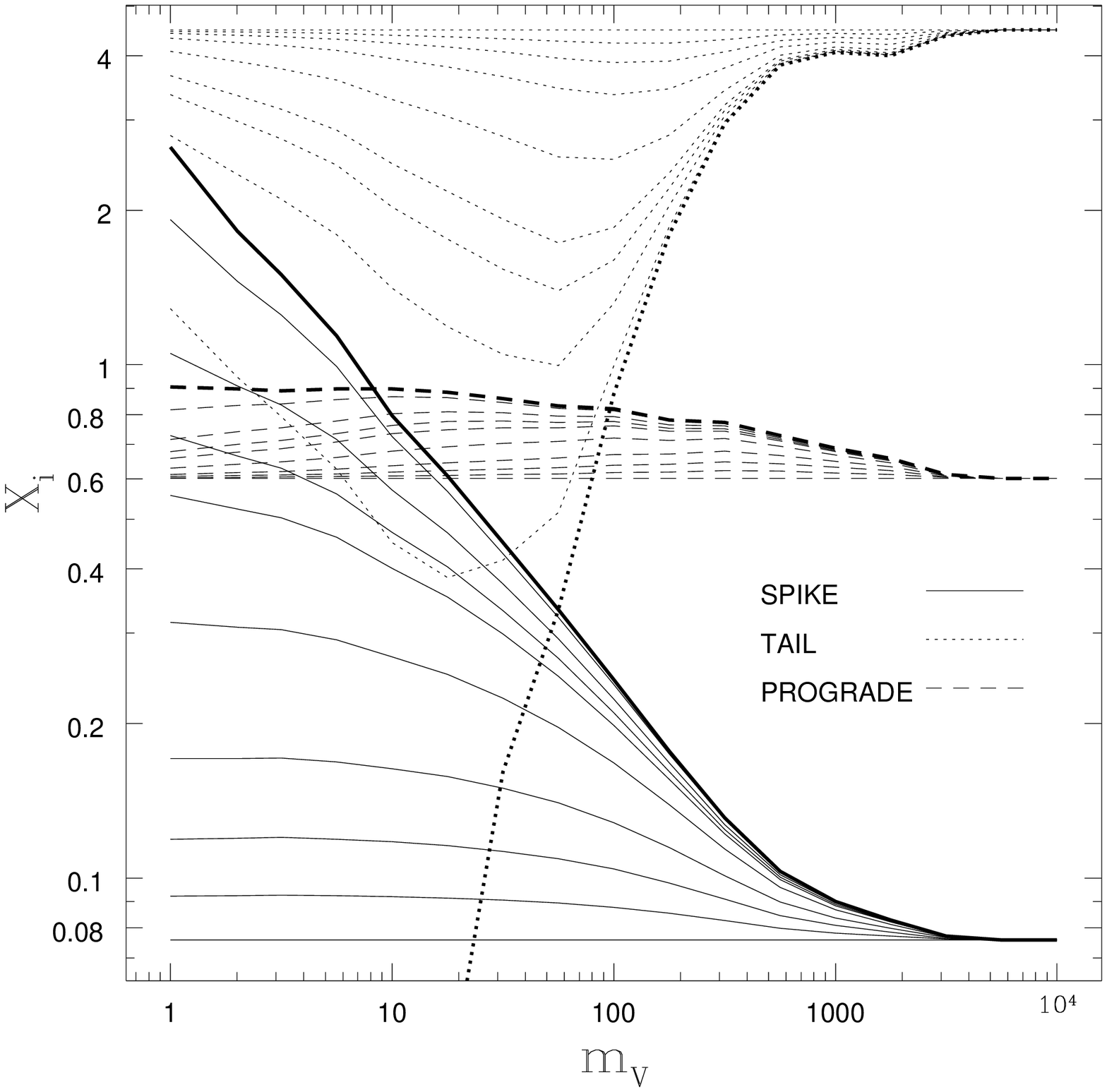,height=3in}}
                  \hbox{\psfig{figure=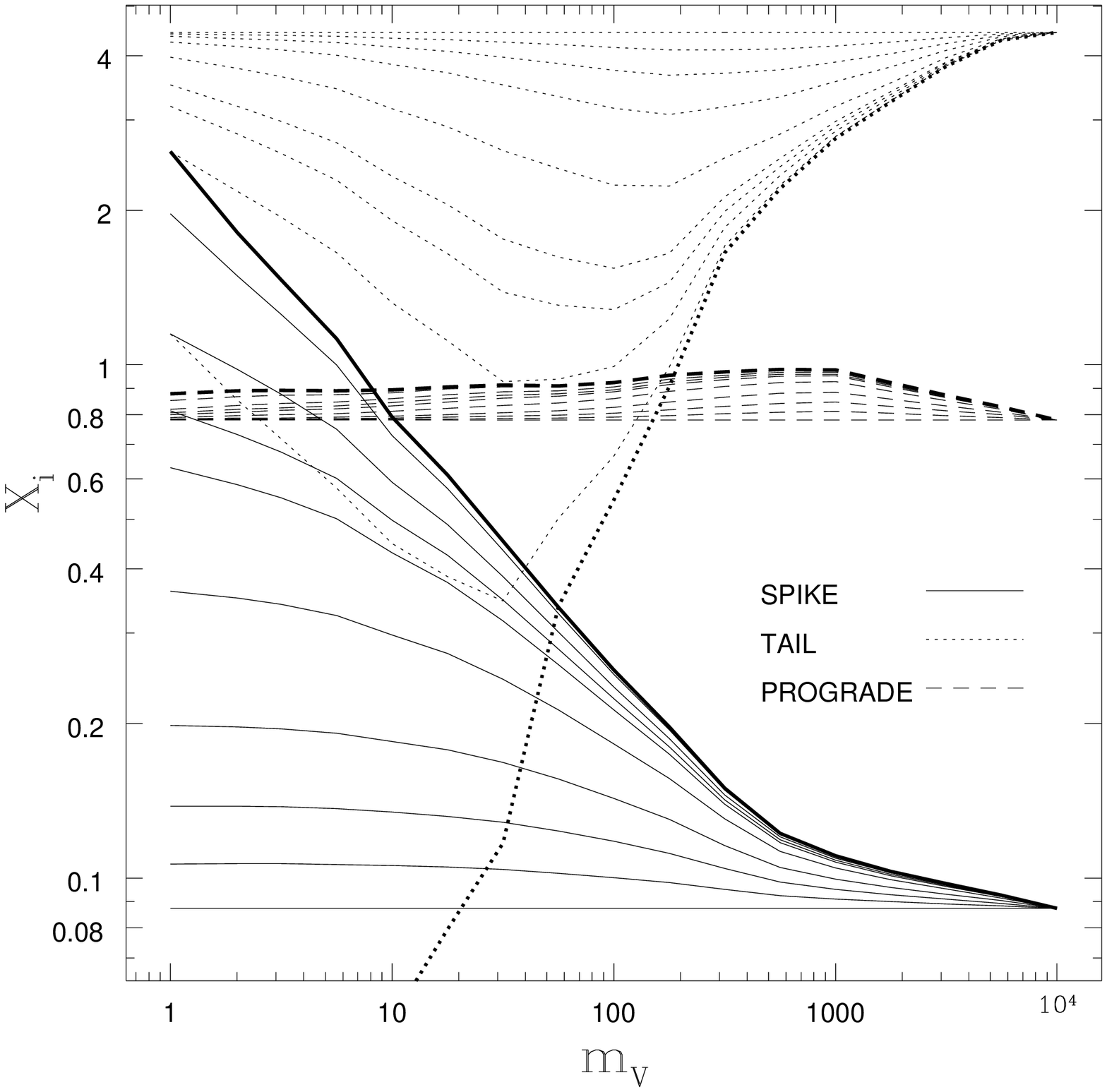,height=3in}
                        \psfig{figure=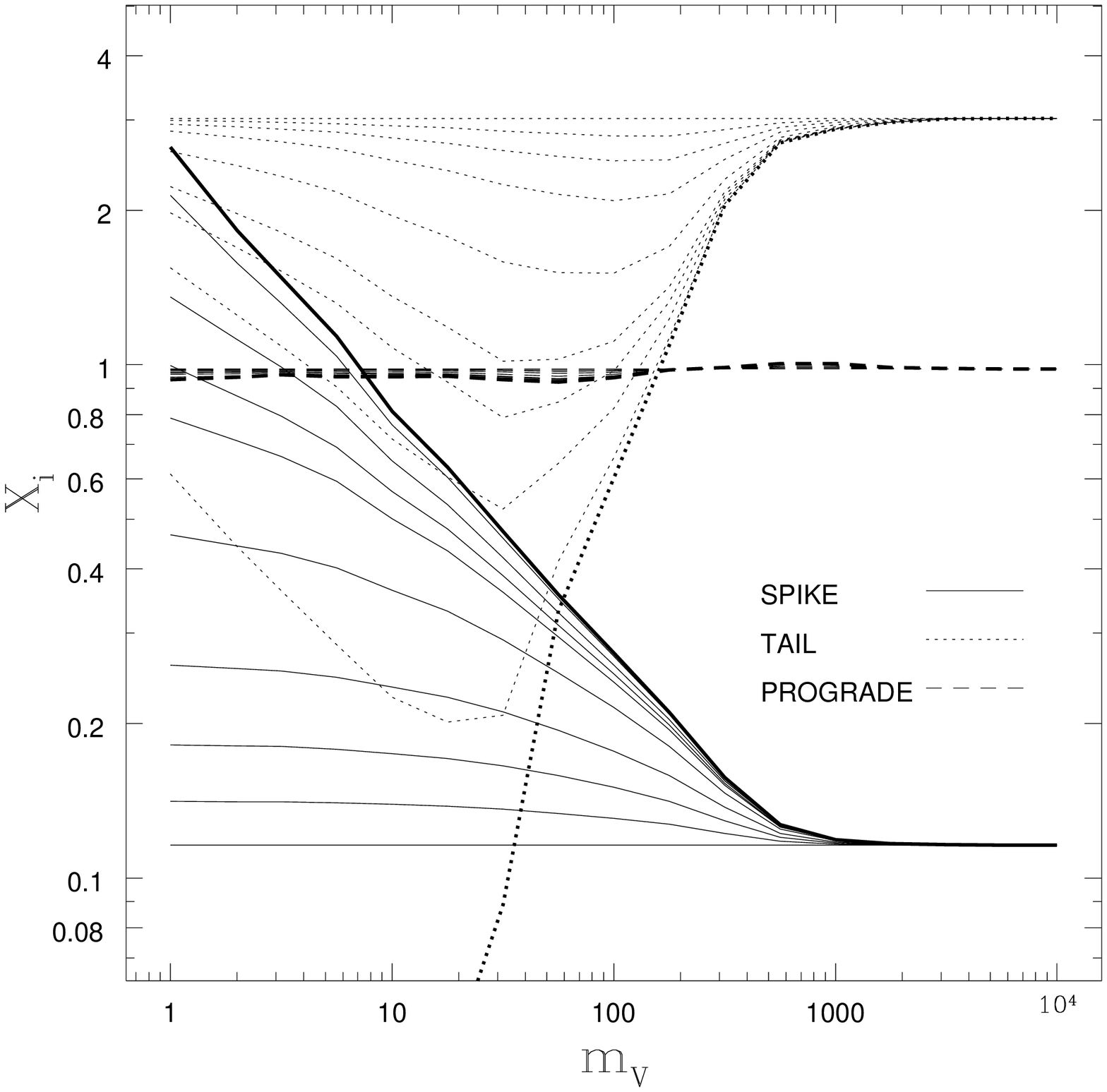,height=3in}}}}
\caption{The values of $X_i$ given a two-parameter fading function in which a
fraction $1-f$ survives for $m_v$ apparitions, while a fraction $f$ survives
forever (model [d], Eq.~\ref{eq:modeldg}).  The fractions $f$ for the
different curves are 0, 0.01, 0.04, 0.07, 0.1, 0.2, 0.4, 0.6, 0.8 and 1,
beginning with the heavy lines. For further details see the caption to Fig.
\ref{fi:const_mv}.}
\label{fi:p2_const_mv}
\end{figure}

The results of model (d) are shown in Fig.~\ref{fi:p2_const_mv}. In most
cases the fit is worse than in the one-parameter model (a), shown by the heavy
lines, because the prograde fraction described by $X_3$ is lower when some of
the comets do not fade.  The best match is for the standard NG model with
visibility probability (\ref{eq:discoverprob2}) (lower right panel). Here the
parameters $m_v=6$, $f=0.04$ yield $X_1=0.83 \pm 0.10$, $X_2=0.91 \pm 0.26$,
$X_3=0.96 \pm 0.11$,
slightly better than the match for model (a). This model is reminiscent of
Weissman's (1978) \nocite{wei78} favoured model, in which 85\% of LP comets
had significant fading probabilities while the reminder survived indefinitely.

Model (e) is a generalization of the one-parameter model (b) but ordinarily
does no better: the match to observations is usually best when the survivor
fraction $f$ is set to zero, and gets worse as $f$ increases. Model (f) also
does no better than its one-parameter counterpart, model (c).

Finally, we examine 
\begin{description}
\item[g) Other published fading functions:] In \S~\ref{sec:onedran}, we
described a
number of fading functions deduced in previous studies. 
\cite{oor50} took $\psi_1=0.8$, $\psi_m=0.014$ for $m>1$;
\cite{ken61} took $\psi_1 =0.8$, $\psi_m=0.04$ for $m>1$; \cite{whi62}
took $\phi_m\propto m^{-1.7}$; \cite{wei78} took $f=0.15$,
$\lambda=0.1$ (cf. Eq. \ref{eq:weissfad}); \cite{eve79} took
$\Phi_1=1$, $\Phi_m=0.2$ for $m>1$; \nocite{bai84} Bailey's (1984)
fading law is described by Eq. (\ref{eq:bailfad}); and \cite{emebai96}
assume $\Phi_m=0.3$ but add a probability $k^\ast=0.0005$ that the
comet is ``rejuvenated''.  In Table~\ref{ta:otherfade}, we have listed
the values of $X_i$ obtained for all these fading models (the results
in the Table are based on the model that includes the discovery
probability (\ref{eq:discoverprob2}) and standard NG forces; other
models give very similar results).  Many provide reasonable matches to
the data but none do as well as our best fits. 

\end{description}

\begin{table}
\centerline{
\begin{tabular}{|lccc|} \hline
Name & $X_1$ & $X_2$ & $X_3$ \\ \hline \hline
Oort             & $0.66 \pm 0.09$ & $1.21 \pm 0.44$ & $0.92 \pm 0.13$ \\
Kendall          & $0.99 \pm 0.12$ & $0.59 \pm 0.23$ & $0.92 \pm 0.11$ \\
Whipple          & $0.97 \pm 0.11$ & $0.58 \pm 0.16$ & $0.95 \pm 0.11$ \\
Weissman         & $0.50 \pm 0.07$ & $2.07 \pm 0.58$ & $0.97 \pm 0.14$ \\    
Everhart         & $0.47 \pm 0.07$ & $2.60 \pm 0.72$ & $0.97 \pm 0.15$ \\
Bailey           & $0.82 \pm 0.11$ & $1.68 \pm 0.63$ & $1.07 \pm 0.13$ \\
Emel'yanenko     & $0.69 \pm 0.08$ & $0.16 \pm 0.05$ & $0.94 \pm 0.10$ \\ \hline 
\end{tabular}}
\caption{The values of $X_i$ for the preferred fading models of
\protect\cite{oor50}, \protect\cite{ken61}, \protect\cite{whi62},
\protect\cite{wei78}, \protect\cite{eve79}, \protect\cite{bai84}, and
\protect\cite{emebai96}. The results are based on the model which
includes the discovery probability (\ref{eq:discoverprob2})
and the standard NG forces.}
\label{ta:otherfade}
\end{table}

\section{Summary} \label{pa:conclusions}

The LP comets provide our only probe of the properties of the Oort comet
cloud. The expected distribution of their orbital elements is only weakly
dependent on the properties of the Oort cloud and is straightforward---though
not easy---to predict if the distribution is in a steady state. Thus a central
problem in the study of comets is to compare the predicted and observed
distributions of the orbital elements of the LP comets.

We have simulated the dynamical evolution of LP comets from their
origin in the Oort cloud until the comets are lost or destroyed. We
have integrated the comet trajectories under the influence of the Sun,
the giant planets, and the Galactic tide. In some cases we have
included the effects of non-gravitational forces, a hypothetical
circumsolar disk or solar companion, and the disruption or fading of
the comet nucleus. We have not included the effects of passing stars
on the Oort cloud; these add a random component to the expected
distribution of LP comets which is more difficult to model but is not
expected to strongly affect the distribution except during rare comet
showers (cf. \S \ref{sec:passing}). Our conclusions from these
simulations include the following:

The Oort cloud presently contains roughly $5 \times 10^{11}(\Phi_{\rm
new}/3\hbox{ yr}^{-1})$ objects orbiting between 10~000 and
50~000~{\au} from the Sun (Eq.~\ref{eq:simlength}), assuming that the
cloud is in a steady state and that the number density in the cloud is
proportional to $r^{-3.5}$ \cite[]{dunquitre87}; here $\Phi_{\rm new}$
is the observed current of new comets with perihelion $<3\au$. This
estimate depends strongly on uncertain assumptions about the density
and extent of the inner Oort cloud; a more reliable parameter is that
the number of comets in the outer Oort cloud ($a>20~000\au$) is $4
\times 10^{11}(\Phi_{\rm new}/3\hbox{ yr}^{-1})$.

Over 90\% of the comets in the Oort spike ($1/a<10^{-4}\aui$) are
making their first apparition (\S\ref{pa:Vm}), and only 2\% of new
comets have energies outside the spike (\S\ref{pa:V1}). The Oort cloud
provides only a few percent of the observed short-period comets, and
even fewer if LP comets fade.  Thus another source, such as the Kuiper
belt, must provide the bulk of the short-period comets. On the other
hand, a significant fraction of the Halley-family comets may arise in
the Oort cloud; however, biases in and the small size of our both the
observed and simulated Halley-family comets render this estimate very
approximate.

LP comets collide with Jupiter and Saturn roughly once per $10^5\yr$ if
$\Phi_{\rm new}=3\hbox{ yr}^{-1}$ (\S~\ref{pa:plan_enc}).

This research does not explain the existence of comets on hyperbolic
original orbits (see Fig. \ref{fi:energy}). The excess velocities are
small, corresponding to roughly $-10^{-4}$~{\aui} in inverse semimajor
axis, but are larger than those produced by the Galactic tide ($\sim
-10^{-6}$~\aui ), by plausible non-gravitational forces ($\sim
-10^{-5}$~\aui) or by a circumsolar disk or solar companion small
enough to be compartible with the distribution of bound orbits.

Using simple models based on a one-dimensional random walk
(\S\ref{sec:onedran}), many investigators, starting with \cite{oor50}, have
concluded that the observed energy distribution of LP comets is incompatible
with the expected steady-state distribution, unless most new comets are
destroyed before their second or subsequent perihelion passage. 
We have shown that
this ``fading'' problem persists in a simulation that follows the comet orbits
in detail. 

Non-gravitational forces play a significant role in shaping the
distributions of the orbital elements of the LP comets, but are too small 
by at least two orders of magnitude to resolve the fading problem
(\S\ref{pa:nongrava}). Hypothetical additional components of the Solar System
such as a massive circumsolar disk or solar companion also do not resolve the
fading problem (\S\ref{pa:other_scen}). 

We can match the observed distribution of orbital elements to the expected
steady-state distribution with at least two fading functions: (a) a
one-parameter power-law (Eq.~\ref{eq:fadec}) with exponent $\kappa \simeq 0.6$
\cite[]{whi62}; (b) a two-population model (Eq.~\ref{eq:modeldg}) in which
approximately 95\% of comets survive for roughly six orbits and the remainder
do not fade (the latter model is also roughly consistent with the observed
splitting probabilities of dynamically new LP comets, approximately 0.1 per
orbit; see Weissman 1980). \nocite{wei80} The observation that the cratering
rate is roughly compatible with the rate expected from the current known
populations of comets and asteroids (\S~\ref{pa:fading}) suggests---within the
large uncertainties---that fading occurs through the fragmentation or
disruption of the comet nucleus rather than through the production of a single
``dead'' body.  We also note the lack of strong fading in new comets during
single perihelion passage, which might be expected if fading were due to loss
of volatiles from an intact nucleus. One possible model compatible with these
observations is that the nucleus of a new comet fragments during its first
apparition, but that the separation velocity of the fragments is small enough
that they remain within the coma until well past observability.

Although physically plausible, fading remains an {\it ad hoc} explanation for
the distribution of LP comet orbits which has not been independently
confirmed, and we should remain alert for other possible explanations. 

\bigskip

We thank Tom Bolton, Ray Carlberg, Martin Duncan,
Bob Garrison and Kim Innanen for helpful discussions and advice. This
research was performed at the University of Toronto and the Canadian
Institute for Theoretical Astrophysics, and has been supported in part
by the Natural Sciences and Engineering Research Council of Canada.

\bibliographystyle{natbib}
\bibliography{references}
\label{end}
\end{document}